\documentclass{article}
\usepackage[font=footnotesize]{caption}
\usepackage[textwidth=8em,textsize=small]{todonotes}
\usepackage{color}
\usepackage{amsthm,amsmath}
\usepackage{amssymb,bm}
\usepackage{enumerate,rotate}
\usepackage{amssymb,latexsym,mathrsfs}
\usepackage{algorithm}
\usepackage{tikz}
\usetikzlibrary{positioning}
\usepackage{graphicx}
\RequirePackage[OT1]{fontenc}
\usepackage[toc,page]{appendix} 
\usepackage[round]{natbib}
\usepackage{authblk}
\usepackage{cancel}
\usepackage{url}

\title{Modelling between- and within-season trajectories in elite athletic performance data\\
{\it Running title}: Seasonal effects in athletic performance data}
\author[2]{Jim E. Griffin}
\author[1]{Maria-Zafeiria Spyropoulou}
\author[1]{James Hopker}
\affil[1]{School of Sport and Exercise Sciences, University of Kent, U. K.}
\affil[2]{Department of Statistical Science, 
University College London, Gower Street, 
London WC1E 6BT,
U. K., email: j.griffin@ucl.ac.uk}
\date{}

\begin{document}
\maketitle

\begin{abstract}
Athletic performance follows a typical pattern of improvement and decline during a career. This pattern is also often observed within-seasons, as an athlete aims for their performance to peak at key events such as the Olympic Games or World Championships.  A Bayesian hierarchical model is developed to analyse the evolution of athletic sporting performance throughout an athlete's career and separate these effects whilst allowing for confounding factors such as environmental conditions. Our model works in continuous time and estimates both 
$g(t)$, the average performance level of the population at age $t$, and $f_i(t)$,
the difference of the $i$-th athlete from this average.
 We further decompose $f_i(t)$ into 
 a season-to-season trajectory and a within-season
 trajectory, which is modelled by
a restricted Bernstein polynomial. The model is fitted using 
 an adaptive Metropolis-within-Gibbs algorithm with a carefully chosen blocking scheme. The model allows us to understand seasonal patterns in athlete performance, how these differ between athletes, and provides individual fitted and trend performance trajectories. 
The properties of the model are illustrated using a simulation study and an application to 
 100 metres and 200 metres freestyle swimming for both female and male athletes.\\ \\ \textbf{Keywords: Longitudinal modelling; Restricted Bernstein polynomial; global-local prior; Bayesian inference; Splines.}
\end{abstract}

\section{Introduction}

The availability of large databases of elite athlete performance allows the modelling of performance levels over an athlete's career.  Results from these models can be used for
retrospective analysis (understanding how an athlete's performance level evolved over their career),
short-term predictions (such as the results of future events), or long-term predictions (such as talent spotting through prediction of the evolution of performance levels at the start of an athlete's career).
 Understanding the variation across athletes and time is critical to effectively performing these analyses. 
We will concentrate on retrospective analysis in centimetre, gram, and seconds (CGS) sports, where performance is measured using one of these units. 

\begin{figure}[!htbp]
\begin{center}
\begin{tabular}{cc}
\multicolumn{2}{c}{(a)}\\
\multicolumn{2}{c}{
\includegraphics[scale=0.24]{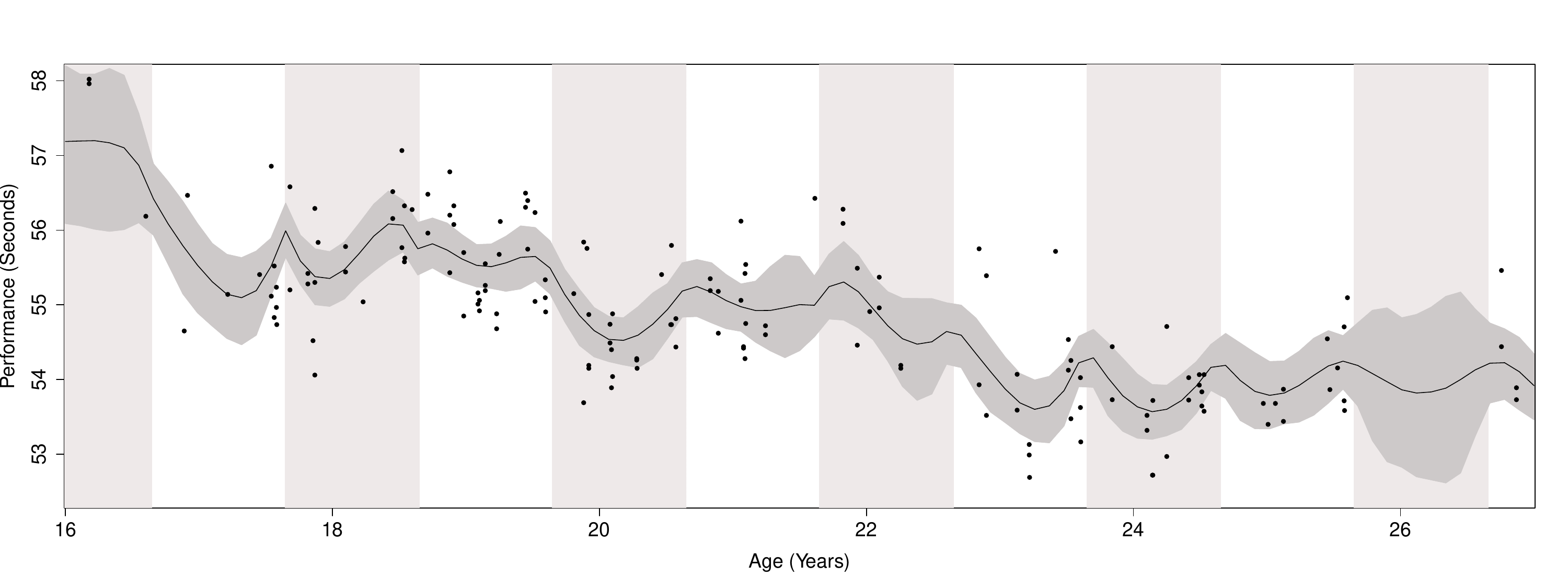}}\\
(b) & (c)\\
\includegraphics[scale=0.24]{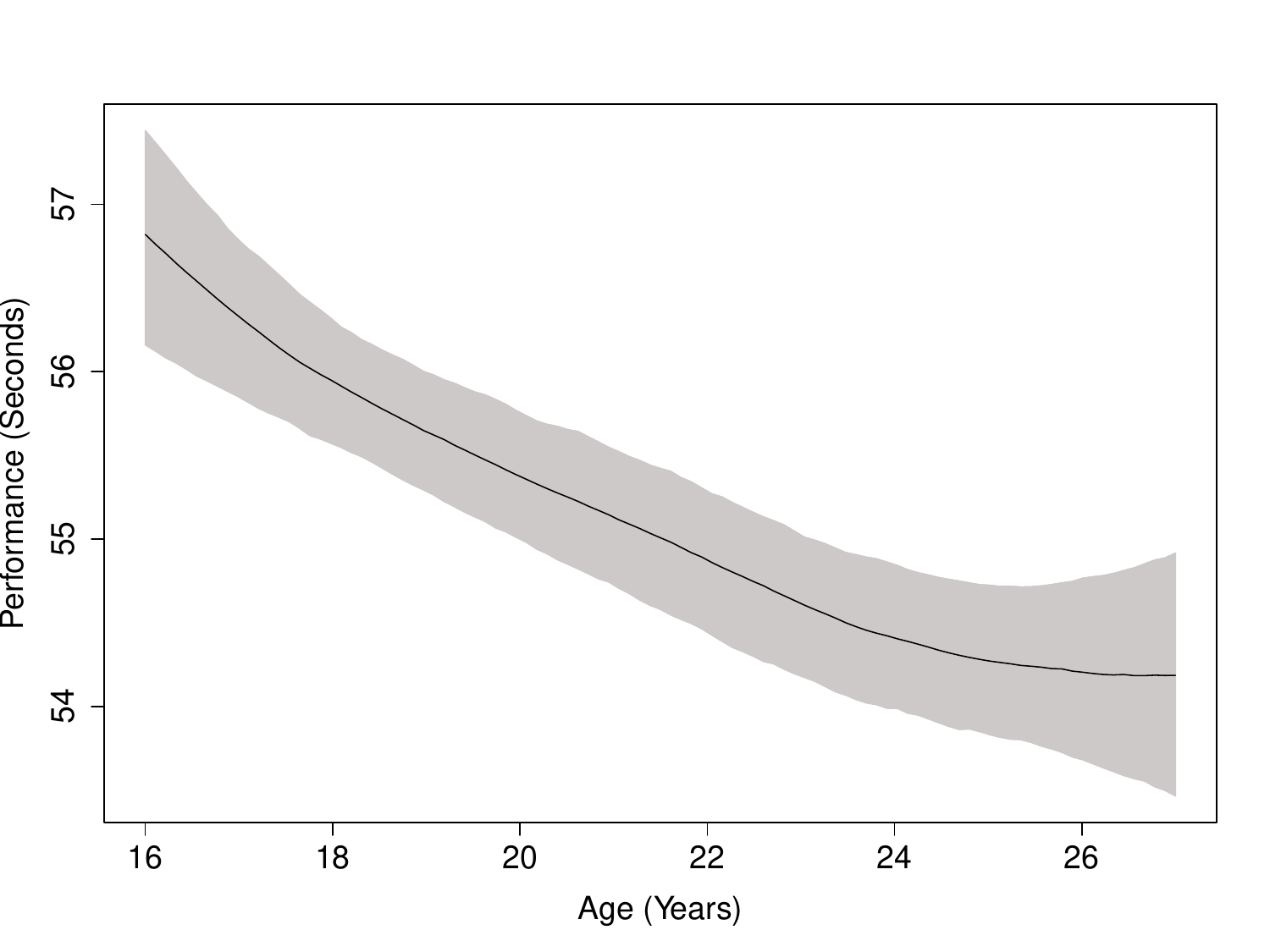} & \includegraphics[scale=0.24]{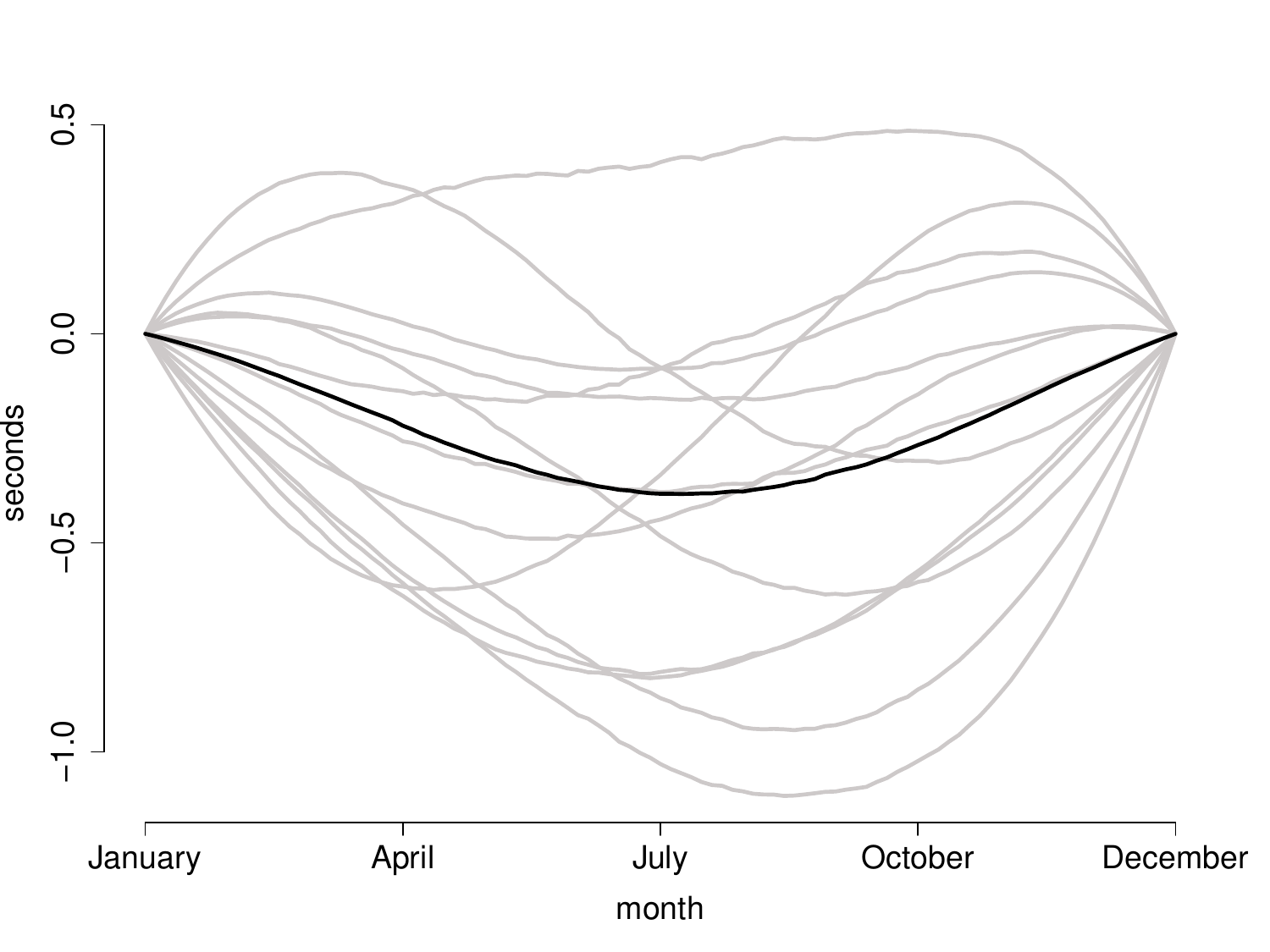}
\end{tabular}
\end{center}
\caption{Analysis of a 100 metres swimmer. Panel (a) shows the observed performances with individual performance trajectory (shown as posterior median (black line) and 95\% credible interval) with each calendar year indicated by alternating grey and white bands. Panel (b) shows the individual career trend trajectory (shown as posterior median (black line) and 95\% credible interval). Panel (c) shows posterior median within-season performance trajectories (light grey lines)
and the posterior median of the athlete's mean within-season performance trajectory (black line).}\label{data_plot}
\end{figure}
Figure~\ref{data_plot}(a) shows the performances of an elite  female 100 metres swimmer from ages 16 to 27 and a curve fitted through the data (which we refer to as the athlete's {\it individual performance trajectory}). 
Performance is  clearly improving over the athlete's career
but there are also  annual cyclical patterns. These follow from 
sports usually being organized into annual seasons with events of high importance (e.g. Olympics or World Championships) occurring at similar times in different years. Many athletes  will tailor their training to peak for these events.
Figures~\ref{data_plot}(b) and \ref{data_plot}(c) decompose the individual performance trajectory into two parts. 
 An {\it individual career trend trajectory} (Figure~\ref{data_plot}(b)) 
 which is a linear interpolation of the estimated performance level on the 1st January of each year
 and an effect for each year/season (Figure~\ref{data_plot}(c)), which we call
 {\it within-season performance trajectories}.  
Figure~\ref{data_plot}(b) shows the improvement in performance more clearly. Figure~\ref{data_plot}(c) shows that on average the swimmer peaks in August with a performance improvement of about 0.4 seconds compared to January.
However,  there are also substantial differences in the within-season performance trajectory from season-to-season ranging from 1 second improvement in some season and close to 0 seconds in others.
Being able to estimate the variability of 
within-season performance trajectories
 is important for understanding whether an athlete is able to peak at the same time and by the same amount in different years.

Mixed models with scalar athlete and season effects have been used to understand variation within and between seasons in athletic performance \citep{PatHop05, TreHopPyn04, BulGulMarRos09, pyne04}. However, this often leads to an overestimation of variability between consecutive performances because variability is calculated for all competitions within a season regardless of how the competitions were distributed within a given season \citep{MalHop14}. 
Longitudinal models were originally proposed for sporting performance data by 
\cite{Berry1999}, who build a linear mixed model with a common effect of age (usually called the ageing function or population performance trajectory), in addition to scalar athlete and season effects. 
The population performance trajectory reflects the usual effect of age on athletic performance which 
typically follows a ``u'' shape \citep{stival2023missing} with athlete's improving to a mid-career peak  followed by deterioration. The age of peak performance often occurs between the ages 23 to 28 but can differ between sports, gender and individuals \citep{griffin2022bayesian}. 

More recently, continuous-time longitudinal model have been used for the irregular observations in data sets of all  athlete performances. 
\cite{griffin2022bayesian} build a model which allows for time-varying athlete  effects and   confounders 
 such as meteorological factors ({\it e.g.} wind speed or temperature) or geographical factors ({\it e.g.} altitude). 
 The observational variation is modelled by a skew-$t$ distribution since 
athletes have a higher chance of underperforming rather than overperforming to the same degree.  
 This model was adequate for the disciplines considered in their paper (weightlifting and 100 metres sprint), but does not model  seasonal effects.  
In this paper, we develop this model to allow for changes in performance over a season using a restricted Bernstein polynomial \citep{WanGho12}. We also increase the flexibility of the model in two additional directions. Firstly, 
 we investigate the use of a distribution which allows for different tail heaviness in each tail. Secondly, we address the sparsity of observations for some athletes using global-local shrinkage priors in a normal hierarchical model. This encourages parts of the hierarchy to be strongly shrunk towards the mean unless the data supports differences.

  Similar to our approach is the model introduced by \cite{dolmeta2023bayesian}. The authors model performance with a non-linear function of time, a GARCH model for interseason changes and the effects of age, gender, and environmental covariates. 
  Our model uses a more structured, yet flexible, model of the variation within season, which allows us to understand both individual and population average intra-seasonal effects.

The paper is organized as follows: Section \ref{Data} discusses the available data for our application to 100 and 200 metres freestyle performances for female and male swimmers.
Section \ref{Model} explains how the model is formed. 
inference is discussed in Section \ref{infer}. 
Section \ref{Simulated} describes an application of our model to simulated data and a simulation study.
Section \ref{Results} discusses the application of  our model to 100 metres and 200 metres freestyle performances for both males and females. 
 Lastly, in Section \ref{Discussion}, a discussion is provided. Appendices present some derivations, the full Markov chain Monte Carlo (MCMC) for Bayesian inference, an exploratory data analysis, MCMC diagnostics for the real data applications, and 
  some additional results. R code to fit the model can be downloaded from 
\url{https://jimegriffin.github.io/website/}.

\section{Elite Swimming Data} \label{Data}

We use a large data base of performance data from 100 and 200 metres freestyle swimming for both females and males to illustrate our methods. We downloaded performance data from 19/3/2008 to 8/9/2023 from the World Aquatics website\footnote{https://www.worldaquatics.com} and 
 selected the 500 swimmers with the fastest personal bests for each combination of distance and gender. 
\begin{table}[h!]
\begin{center}
\begin{tabular}{|c|c|c|ccc|}\hline
Event & Number of & Number of  & \multicolumn{3}{c|}{Number of Performances} \\
& Performances & Athletes & \multicolumn{3}{c|}{per Athlete}\\\hline
& & & Median & Minimum & Maximum\\\hline
100 metres, female &23669 &500&37.5&5&267\\
100 metres, male& 23440&500&38&5&191\\
200 metres, female &21112&500&33&5&274\\
200 metres, male &19696&500&32&5&162\\\hline
\end{tabular}
\end{center}
\caption{Summaries of the number of performances, number of athletes and number of performances per athlete for the 100 metre and 200 metre freestyle data sets.}\label{summaries}
\end{table}
Summaries of the number of performances and athletes are provided in Table~\ref{summaries}. The data includes performances in both 25 and 50 metre  pools. Swimmers are able to achieve faster times in a 25 metre pool and so pool length is included  as a confounder. To make performances comparable when presenting data in the paper, we plot 25 metre pool performances adjusted to a 50 metre pool using the posterior mean of this pool effect. An exploratory data analysis of this data is presented in appendix C, which shows the effect of ages and seasonal effects within in the data.

\section{Model} \label{Model}

\subsection{Sampling Model}

We construct a continuous-time longitudinal model for athlete performance which allows for effects of age, the time of year and confounders building on \citet{griffin2022bayesian}.
Suppose that we have observations for $M$ athletes where the $i$-th athlete has $n_i$ performances $\boldsymbol{y}_{i} = (y_{i, 1}, \dots, y_{i,  n_i})$, which are observed at ages 
$\boldsymbol{a}_{i} = (a_{i, 1}, \dots, a_{i,  n_i})$ and calendar time $\boldsymbol{t}_{i} = (t_{i, 1}, \dots, t_{i,  n_i})$ (where 0 refers to the start of an athlete's first season)
with confounders $\boldsymbol{x}_{i} = (x_{i, 1}, \dots, x_{i, n_i})$. We assume that the $i$-th athlete competes in $S_i$ seasons and define
 $\boldsymbol{s}_i = (s_{i, 1}, \dots, s_{i, n_i})$ where $s_{i, k} \in \{1, \dots, S_i\}$ is the season of the $k$-th performance. We assume the basic model 
\begin{equation}
y_{i,k} = \mu_i(a_{i, k}, t_{i, k}) + x_{i,k}\,\zeta + \epsilon_{i, k}
\label{base_model_1}
\end{equation}
where $\mu_i(a, t)$ is the {\it individual performance trajectory} 
 of the $i$-th athlete at age $a$ and calendar time $t$ (which is defined to be 0 at the start of an athlete's first season),
$\zeta$ are the effects of the confounders,  and the $\epsilon_{i,k}$'s are i.i.d. observation errors. 
These errors will typically be  both skewed and heavy tailed since  athletes have a larger probabilities of extreme underperformance compared to overperformance. We first describe the distribution of the errors before discussing the form of $\mu_i(a, t)$.

 \cite{griffin2022bayesian} assume that the errors follow a skew-$t$ distribution but this assumes the same heaviness
 for both tails. 
  We consider a distribution which allows differences in the heaviness of the two tails 
  by
  generalising  the skew-$t$ distribution.
  We define $T\mathcal{N}_{R} \left(0, \mu,\sigma^2 \right)$
  to denote the normal distribution with mean $\mu$ and variance $\sigma^2$ truncated to the region $R$
  and 
   $\mathcal{IG}a(a, b)$ to denote an inverse gamma distribution with density 
$p(x) = \frac{b^a}{\Gamma(a)} x^{-a-1}\exp\{- b/x\}$. The  observation error is defined by
   $\epsilon_{i, k} = 
\epsilon^{\star}_{i, k} + \frac{\alpha}{\sqrt{1 + \alpha^2}}\,\kappa_{i,  k}$
where 
$\epsilon^{\star}_{i, k}\sim \mathcal{N}\left(0, \omega_{i,  k}\,\sigma^2_i \right)$, 
$\kappa_{i,  k}\sim T\mathcal{N}_{[0,\infty]} \left(0, \phi_{i,  k}\,\sigma^2_i \right)$,
$\omega_{i, k} \sim \mathcal{IG}\left(\frac{\nu_1}{2},\frac{\nu_1}{2}\right)$, and $\phi_{i, k} \sim \mathcal{IG}\left(\frac{\nu_2}{2},\frac{\nu_2}{2}\right)$ for $k = 1, \dots, n_i$, $i = 1,\dots, M$ and all elements are independent. This reverts to a skew-$t$ distribution if $\omega_{i, k} = \phi_{i, k}$ (and so $\nu_1 = \nu_2$).
If $\alpha > 0$,
the heaviness of the right-hand tail is controlled by the minimum of $\nu_1$ and $\nu_2$ and the heaviness of the left-hand tail by $\nu_1$ with smaller values representing heavier tails (with the effects on left- and right-hand tails reversed if $\alpha < 0$).

The model of the individual performance trajectory $\mu_i(a, t)$ is initially decomposed into population and individual parts
\begin{equation}
 \mu_i(a, t) 
= g(a) + \theta_i(a, t)
 \label{base_model_2}
\end{equation}
where $g(a)$ is the
 {\it population performance trajectory} and $\theta_i(a, t)$ is  the {\it individual excess performance}.
We follow 
\cite{griffin2022bayesian} by modelling $g(a)$ using a $d$-th order polynomial function $g(a) = \sum_{k=0}^d \delta_k \, (a - \bar{a})^k$ where $\delta_1, \dots, \delta_d$ are coefficients and $\bar{a}$ is the mean age of all observed performances (we find that $d=4$ is sufficiently flexible in our examples).

The individual excess performance is further decomposed into two parts
\begin{equation}
 \theta_i(a, t) = f_{i}(a)  + h_i(t) 
 \label{base_model_3}
\end{equation}
where $f_{i}(a)$ is called the {\it trend excess performance trajectory} for the $i$-th athlete, and
$h_{i}(t)$ is called the {\it seasonal performance trajectory}. 
We define
$
 g(a) + f_{i}(a)$ to be
the $i$-th athlete's {\it individual trend  performance trajectory},
 which adjusts the individual performance trajectory $\mu_i(a, t)$ for seasonal effects.

The trend excess performance trajectory is modelled by a piecewise linear function where the knots occur at the start/end of each season,
 \[
 f_{i}\left((s-1+r)\Delta\right) = \eta_{i, s}\,(1-z) + \eta_{i, s+1}\,z,  \qquad z\in (0, 1),\ s = 1,\dots, S_i.
 \]
where $\Delta$ is the length of each season and start on the same day of each year (although this could easily be changed in the model)
and  $\eta_{i, s}$ is the value at the start of the $s$-th season.

 The seasonal performance trajectory is modelled season-by-season. 
 Let the {\it within-season performance 
 trajectory} for the $i$-th athlete in the $s$-th season
 be
 $h^{\star}_{i, s}(z)$ for $0 < z < 1$. We define the seasonal performance trajectory to be
 \[
 h_i((s-1+z)\Delta) = h^{\star}_{i, s}(z),  \qquad z\in (0, 1),\ s = 1,\dots, S_i.
 \]
 We want a flexible form for $h^{\star}_{i, s}$ which is 
  constrained by $h^{\star}_{i, s}(0) = 0$ and $h^{\star}_{i, s}(1) = 0$. This allows us to identify the model 
in \eqref{base_model_3} since
\[
\theta_i(a, s\Delta) = f_i(a) + h_i(s\Delta) = 
f_i(a) + h^{\star}_{i, s}(1)
= f_i(a) + h^{\star}_{i, s + 1}(0) = 
f_i(a) = \eta_{i, s + 1}.
\]
Therefore, the individual excess performance $\theta_i(a, s\Delta)$ is equal to the trend excess performance $f_i(a)$ at the start and end of each season.

\subsection{Prior Distributions for $f_i(a)$ and $h^{\star}_{i, s}(r)$}

We wish to have flexible forms for both the individual trend excess performance
$f_i(a)$ and the within-season performance trajectories $h^{\star}_{i, s}(r)$.
 The values of $f_i$ at the start of each season are
$\eta_{i, 1}, \dots, \eta_{i, S_i+1}$.
These are modelled using a random walk with
 initial value $\frac{\eta_{i, 1}}{\sigma^2_\mu} \sim t_{\nu^{\mu}}$
  and increments 
 $\frac{\eta_{i,s+1} - \eta_{i, s}}{\sigma^2_\eta} \stackrel{i.i.d.}\sim t_{\nu^{\eta}}$ for $s = 1,\dots, S_i$. This specification allows for heavy tails in both the distribution of the initial level 
 $\eta_{i, 1}$  and possibly large changes from season to season. 

We want the functions $h^{\star}_{i, s}$ to have some specific features. 
Firstly, for identifiability, the functions $h^{\star}_{i, s}$ are constrained by
$h^{\star}_{i, s}(0) = 0$ and $h^{\star}_{i, s}(1) = 0$. Secondly, 
we want a hierarchical structure to allow sharing of information across different seasons for each athletes and between athletes. This allows the identification of those
athletes who are better able to peak and those who can achieve this consistently. Thirdly, 
we assume that the seasonal variation is largely caused by athletes peaking for particular events or parts of the year (for example, the summer season in athletics) and that this is shared across athletes ({\it i.e.} athletes are peaking for events in the same point of the year).

A Bayesian hierarchical model build using a Restricted Bernstein polynomials (RBPs) 
\citep{WanGho12} is a convenient way to achieve our three goals. An $n$-th order RBP $m(z)$ with $m_n(0) = m_n(1) = 0$ has the form
 \begin{equation}
 m_n(z) =  \sum_{v=1}^{n-1}\beta_{n, v} \, b_{n, v}(z), \qquad 0<z<1
\label{RBP1}
 \end{equation}
where 
 \[
 b_{n, v}(z) = \left(\begin{array}{c}
 n\\v\\ \end{array} \right)z^{v}(1-z)^{n-v},\quad v = 1,\dots, n-1.
 \]
To provide additional flexibility, we use the sum of RBPs of order 2 to $N$ giving the model 
 \begin{equation}
 h^{\star}_{i,s}(z) = \sum_{n=2}^N  \sum_{v=1}^{n-1}\beta^{(i,s)}_{n, v} \, b_{n, v}(z), \qquad 0<z<1.
\label{RBP2}
 \end{equation}
It is convenient to group the  coefficients of the RBP in $\boldsymbol{\beta^{(i,s)}} = \left( \beta^{(i,s)}_{2,1}, \beta^{(i,s)}_{3,1}, \beta^{(i,s)}_{3,2}, \dots, \beta^{(i,s)}_{N, 1}, \dots, \beta^{(i,s)}_{N, N-1} \right)$.

To allow sharing of information across seasons and athletes, we use independent  hierarchical priors for each coefficient in the RBP, which nest seasons within athletes. This leads to, for $i = 1,\dots, M$ and $s = 1,\dots, S_i$,
\[
\beta^{(i,s)}_{n, v} \stackrel{i.i.d.}{\sim} \mathcal{N}\left(\beta^{(i)}_{n, v},\lambda_i^2\, c_{n, v}^2\right), \quad \beta^{(i)}_{n, v} \stackrel{i.i.d.}{\sim} \mathcal{N} \left(\beta_{n, v}, \tau_i^2\,d_{n, v}^2\right), \quad v = 1, \dots, n-1,\ n = 2,\dots, N.
\]
The prior variances are a product of athlete-specific effects $\lambda_i^2$
 and $\tau_i^2$, and  coefficient-specific effects $c^2_{n, v}$ and $d^2_{n, v}$. The use of coefficient-specific variance parameters  allows for the variability of the functions $h^{\star}_i(z)$ and $h^{\star}(z)$ to change with $z$. The use of 
 individual-specific variance parameters $\lambda_i^2$ and $\tau^2_i$ allows for differences in the variability of the functions from athlete-to-athlete. The importance of allowing for differences is discussed in Section~\ref{infer}.
 In addition, since some athletes 
 only have a few performances, we use  
  global-local shrinkage priors \citep{bha19} which avoids overfitting and allows the effects to be shrunk towards the corresponding mean. 
Since the scale parameters $c_{n, v}^2$ and $\lambda_i^2$ enter multiplicatively (similarly, $d_{n, v}^2$ and $\tau_i^2$), we centre the $c^2_{n, v}$'s and the $d^2_{n, v}$'s around 1
to avoid problems of interpretation,  and  use gamma priors to encourage regularisation of the coefficients \citep[see][]{gribro10},
\[
c_{n, v}^{2} \sim \mathcal{G}a \left(5, 5\right), 
\quad d_{n, v}^2 \sim \mathcal{G}a \left(5, 5\right), \quad v = 1,\dots, n - 1,\ n = 2, \dots, N,
\]
\[
\quad\lambda_{i}^{2} \sim \mathcal{G}a \left(\lambda_0, \lambda_0/\lambda_1\right),
\quad \tau_i^{2} \sim \mathcal{G}a \left(\tau_0, \tau_0/\tau_1 \right), \quad i = 1,\dots, M
\]  
where $\mathcal{G}a(a, b)$ represents a gamma distribution with density 
$p(x) = \frac{b^a}{\Gamma(a)} x^{a-1}\exp\{- b\,x\}$ and 
so  $\lambda_1$ and $\tau_1$ are the prior means of $\lambda_i^2$ and $\tau_i^2$ respectively.

The hierarchical structure introduces parameters $\boldsymbol{\beta}^{(i)} =
 \left( \beta^{(i)}_{2,1}, \beta^{(i)}_{3,1}, \beta^{(i)}_{3,2}, \dots, \beta^{(i)}_{N, 1}, \dots, \beta^{(i)}_{N, N-1} \right)
$ and  $\boldsymbol{\beta} =  \left( \beta_{2,1}, \beta_{3,1}, \beta_{3,2}, \dots, \beta_{N, 1}, \dots, \beta_{N, N-1} \right)$  which can be interpreted as average coefficients over all seasons for the $i$-th athlete and average coefficients over all seasons and all athletes respectively.  
It is convenient to define functions formed by the coefficients 
$\boldsymbol{\beta}$, 
 \[
 h^{\star}(z) = \sum_{n=2}^N  \sum_{v=1}^{n-1}\beta_{n, v} \, b_{n, v}(z), \qquad 0 < z < 1,
 \] 
 is called the {\it average within-season performance trajectory at the population level}, and by the coefficients 
 $\boldsymbol{\beta}^{(i)}$,
\[
 h^{\star}_{i}(z) = \sum_{n=2}^N  \sum_{\nu=1}^{n-1}\beta^{(i)}_{n, v} \, b_{n, v}(z), \qquad 0 < z < 1.
 \]
is called the {\it average within-season performance trajectory for the $i$-th individual}.

Thirdly, the observation that athletes often peak at a particular time is included by constraining the within-season performance trajectory at the population level to have a single peak 
 (rather than at the individual or seasonal level). This allows for differences in peak time (or the number of peaks) for some athletes and some seasons, which allows athletes to have different goals in some seasons (for example, in Olympic and non-Olympic years).

The shape of the population-level
within-season performance trajectory is determined by the direction of improvement in the sport. In timed events such as swimming or running, the direction of improvement is negative since a better performances leads to a faster time whereas, in events such as weightlifting or throwing, the direction of improvement is positive since a better performance leads to a heavier lift or a longer throw. Therefore, we constrain the trajectory to be concave if the direction of improvement is positive and convex if the direction of improvement is negative.
 As discussed by \citet{WanGho12}, 
 the second derivative of the RBP in \eqref{RBP1} can be expressed as
\begin{align*}
\frac{m_n''(z)}{n(n-1)}
=&
\left(\beta_{n, 2} - 2\beta_{n, 1}\right) b_{n-2, 0}(z) +
\sum_{v=1}^{n-2}
\left(\beta_{n, v+2} - 2\beta_{n, v+1} + \beta_{n, v}\right) b_{n-2, v}(z)\\
&+ 
\left(\beta_{n, n-2} - 2\beta_{n, n-1} \right) b_{n-2, n-2}(z).
\end{align*}
Let ${\boldsymbol\beta}_n = (\beta_{n, 1}, \dots, \beta_{n, n-1})$ then $m_n(z)$ is convex if 
$\boldsymbol{D}_n {\boldsymbol\beta}_n  \geq \boldsymbol{0}_{n-1}$ and concave if $\boldsymbol{D}_n {\boldsymbol\beta}_n  \leq \boldsymbol{0}_{n-1}$ 
where 
\[
\boldsymbol{D}_n =
\left(
\begin{array}{cccccccc}
-2 & 1 & 0 & 0 & \cdots & 0 & 0 & 0\\ 
1 & -2 & 1 & 0 & \cdots & 0 & 0 & 0\\
0 & 1 & -2 & 1 & \cdots & 0 & 0 & 0\\
\vdots & \vdots & \vdots & \vdots & \ddots & \vdots & \vdots & \vdots\\
0 & 0 & 0 & 0 & \cdots & 1 & -2 & 1\\
0 & 0 & 0 & 0 & \cdots & 0 & 1 & -2
\end{array}
\right)_{(n-1)\times (n-1)}
\]
and ${\bf 0}_{n-1}$ represents an $(n-1)$-dimensional vector of 0's. Therefore, if the direction of improvement is positive, the prior is a  normal distribution 
with mean ${\bf 0}_{n-1}$ and covariance matrix $\mbox{diag}(\sigma^2_{\boldsymbol{\beta}, n})$ truncated to the region 
$ \{\boldsymbol\beta_n\mid \boldsymbol{D}_n \,\boldsymbol\beta_n \leq \boldsymbol{0}_{n-1}\}$ where 
$\mbox{diag}(\boldsymbol{a})$ represents a diagonal matrix with the elements of $\boldsymbol{a}$ on the diagonal, and
$\boldsymbol\sigma^2_{\beta, n}$ is an $(n-1)$-dimensional vector of variance parameters. If the direction of improvement is negative, the prior is truncated to the region 
$ \{\boldsymbol\beta_n\mid \boldsymbol{D}_n \,\boldsymbol\beta_n \geq \boldsymbol{0}_{n-1}\}$.
We apply the constraint for each value of $n$ in the sum of RBPs prior for $h^{\star}_{i, s}(z)$ to ensure that the sum is either convex or concave.

\subsection{Priors for other parameters}

The Bayesian model is completed by specifying priors for all other parameters. Following \cite{gribro10}, we 
centre the shape parameters $\lambda_0$ and $\tau_0$ on 1 and give vague priors to the means $\lambda_1$ and $\tau_1$ giving
$\lambda_0, \tau_0 \stackrel{i.i.d.}{\sim}
\mathcal{E}x(1)$,
where $\mathcal{E}x(\theta)$ represents an exponential distribution with mean $\frac{1}{\theta}$
and
$\lambda_1, \tau_1 \stackrel{i.i.d}{\sim} \mathcal{IG}(0.001, 0.001)$.
 The scale of the observation error for each athlete is given the population distribution $\sigma^{2}_i\sim \mathcal{IG}\left(\sigma^2_a, \sigma^2_a / \sigma^2_{m}\right)$,
and the skewness parameter is given the prior $\alpha \sim \mathcal{N} (0,3^2)$, which provides support for reasonable values of the skewness. 
The degrees of freedom parameters are given a prior mean of 20 and wide spread of possible values 
$\nu^{\mu} \sim \mathcal{G}a (2,0.1)$,
$\nu^{\eta} \sim \mathcal{G}a (2,0.1)$,
$\nu_1 \sim \mathcal{G}a (2,0.1)$, and 
$\nu_2 \sim \mathcal{G}a (2,0.1)$. All other parameters are given vague priors $p(\zeta, \delta)\propto 1$,
 $\sigma^2_{\eta} \sim \mathcal{IG}(0.001, 0.001)$,  
  $\sigma^2_{a} \sim \mathcal{IG} (0.001, 0.001)$, 
   $\sigma^2_{m} \sim \mathcal{IG} (0.001, 0.001)$, 
 and $\sigma^2_{\mu} \sim \mathcal{IG} (0.001, 0.001)$.

\section{Inference}\label{infer}

We use MCMC to fit the model and the full algorithm is described in Appendix~\ref{MCMC}. Blocking \citep{roberts1997updating, knorr2002block}
and interweaving \citep{YuMeng} are used to avoid slow mixing of some parameters.

The hierarchical model for the within-season performance trajectory allows us to estimate functional effects at the season, athlete and population level. Figure~\ref{data_plot}(c) plots  the within-season performance trajectory at the individual and seasonal levels. This shows substantial variation in the within-season performance trajectories for this individual. A univariate summary of the amount of variation is useful to compare athletes. Similarly, we define a univariate summary of the difference between an athlete's within-season performance trajectory and the population within-season performance trajectory.
To establish these summaries, we use the following results for the RBP in 
 \eqref{RBP2} (a proof is given in the appendix A). Suppose that $\epsilon(z) = \sum_{n=2}^N \sum_{v=1}^{n-1} a_{n, \nu}\, b_{n, v}(z)$ then 
\begin{enumerate}

\item[(a)] $\int_0^1 \epsilon(z)\,dz = \sum_{n=2}^N \frac{1}{n+1} \sum_{v=1}^{n-1} a_{n, v}$,

\item[(b)]
\[
\int_0^1 \epsilon(z)^2\,dz
=
\sum_{n_1=2}^N \sum_{v_1=1}^{n_1-1}
\sum_{n_2=2}^N \sum_{v_2=1}^{n_2-1}
 a_{n_1, v_1}  \,
 a_{n_2, v_2}  \,
B_{n_1, n_2, v_1, v_2} 
\]
where 
\[
B_{n_1, n_2, v_1, v_2} = \left(\begin{array}{c} 
n_1\\ v_1
\end{array}
\right)
\left(\begin{array}{c} 
n_2\\ v_2
\end{array}
\right)
\frac{(v_1 + v_2)! (n_1 + n_2 - v_1 - v_2)!}{(n_1 + n_2 + 1)!}.
\]
\end{enumerate}

The variability over different seasons for the $i$-th athlete can be measured by considering the differences between the within-season performance trajectory for the $s$-th season and the average within-season performance trajectory for the $i$-th athlete which is
\[
\psi_i(z) = h_{i, s}^{\star}(z) - h_i^{\star}(z) =
\sum_{n=2}^N \sum_{v = 1}^{n - 1}  \left(\beta^{(i, s)}_{n, v} - \beta^{(i)}_{n, v}\right) \, b_{n, v}(z).
\]
Taking the expectation of $\int_0^1 \psi_i^2(z)\, dz$ with respect to the prior distribution of $\beta^{(i, j)} - \beta^{(i)}$ leads to the expression 
\[
\Psi_i = \mbox{E}\left[\int_0^1 \psi_i^2(z) \,dz\right]
= \lambda_i^2 \sum_{n=2}^N \sum_{v = 1}^{n - 1}  
c_{n, v}^2\,\label{RBP2}
B_{n, n, v, v} 
\]
which we call the within-season variability for the $i$-th athlete. In a similar way, we can summarise the size of the difference in effect between an athlete's average within-season  performance trajectory and the average within-season  performance trajectory across all athletes using the measure
\[
\gamma_i(z) = h^{\star}_i(z) - h^{\star}(z)
=
\sum_{n=2}^N \sum_{v = 1}^{n - 1}  ( \beta^{(i)}_{n, v} - \beta_{n, v}) \, b_{n, v}(z).
\]
Taking the expectation of $\int_0^1 \gamma_i^2(z)\,dz$ with respect to the prior distribution of $\beta^{(i)} - \beta$ leads to the expression 
\[
\Gamma_i  = \mbox{E}\left[\int_0^1 \gamma^2(z) \,dz\right]
= \tau_i^2 \sum_{n=2}^N \sum_{v = 1}^{n - 1}  d_{k, v}^2\,
B_{n, n, v, v} 
\]
which we call the average effect size for the $i$-th athlete. We are often interested in the ordering of each of these measures across athletes to find athletes which have small or large variability in their season effects or athletes are far from the population average. This can be achieved by  just report $\lambda_i^2$ (in place of $\Psi_i$) and $\tau_i^2$ (in place of $\Gamma_i$).

\section{Simulated Data}\label{Simulated}

We use a simulated example and a simulation study to show how the model can capture individual career trend trajectories, within-season performance trajectories at the seasonal, individual and population levels, and differences in variation and the seasonal and individual levels. We generated data from 
 the full model in Section~\ref{Model} without  confounders. The $i$-th individual has observation for $S_i$ seasons were
\[
S_i \sim \left\{\begin{array}{ll}
\mathcal{P}o(4) & \mbox{with probability } p_1\\
\mathcal{P}o(8) & \mbox{with probability } 1 - p_1
\end{array}
\right..
\]
The parameter $p_1$ controls the proportion of athletes with more or less observed seasons. A larger value of $p_1$ leads to more athletes with a lower number of observed seasons.
In the $j$-th season, there were $n_{i, j} \sim \mathcal{U}(3, 11)$ performances. The age at the start of the first season was assumed to be distributed $\mathcal{U}(18, 22)$ and 
the population performance trajectory was 
$
g(t) = 40 + 0.1 (t - 26)^2$. The excess performances $\eta_{i, 1},\dots,\eta_{i, S_i + 1}$ were generated as $\eta_{i, 1}\sim\mathcal{N}(0, 4)$ and $\eta_{i, j + 1} - \eta_{i, j}\sim\mathcal{N}(0, 0.09)$ for $j = 1, \dots, S_i$. The population level within-season performance trajectory was $h^{\star}(t) = 4\left(t - \frac{1}{2}\right)^2 - 1$ for $0 <t < 1$, which has a minimum at $t = \frac{1}{2}$. The individual and seasonal  performance trajectories were parabolas with a maximum or minimum at different values for different individuals or seasons. The turning point for the individual trajectory was $p_i\stackrel{i.i.d.}{\sim} \mathcal{U}(0.5, 0.7)$ and the value at that point was
$
a_i \sim \mathcal{N}(0, \sigma^2_a).
$
The trajectory was 
\begin{equation}
h_i^{\star}(t)- h^{\star}(t) =
\left\{
\begin{array}{cc}
a_i \left(1 - \frac{(t - p_i)^2}{p_i^2}\right), & 0 < t\leq p_i\\
a_i  \left(1 - \frac{(t - p_i)^2}{1 - p_i^2}\right), & p_i < t < 1
\end{array}
\right..
\label{sim_a}
\end{equation}
The turning point for the individual trajectory was $r_{i, j}\stackrel{i.i.d.}{\sim} \mathcal{U}(0.5, 0.7)$ and the value at that point was
$
b_{i, j} \sim \mathcal{N}(0, \sigma^2_b).
$
The trajectory was 
\begin{equation}
h_{i,j}^{\star}(t) - h_i^{\star}(t) =
\left\{
\begin{array}{cc}
b_{i, j} \left(1 - \frac{(t - r_{i. j})^2}{r_{i, j}^2}\right), & 0 < t\leq r_{i, j}\\
b_{i, j} \left(1 - \frac{(t - r_{i, j})^2}{1 - r_{i, j}^2}\right), &  r_{i, j} <t < 1
\end{array}
\right..
\label{sim_b}
\end{equation}
The errors were generated with $\alpha = 3$, $\nu_1 = 30$ and $\nu_2 = 7$.

\begin{figure}[h!]
\begin{center}
\begin{tabular}{ccc}
Population performance  & Population within-season & Error density\\
trajectory & Performance trajectory\\
\includegraphics[scale=0.24]{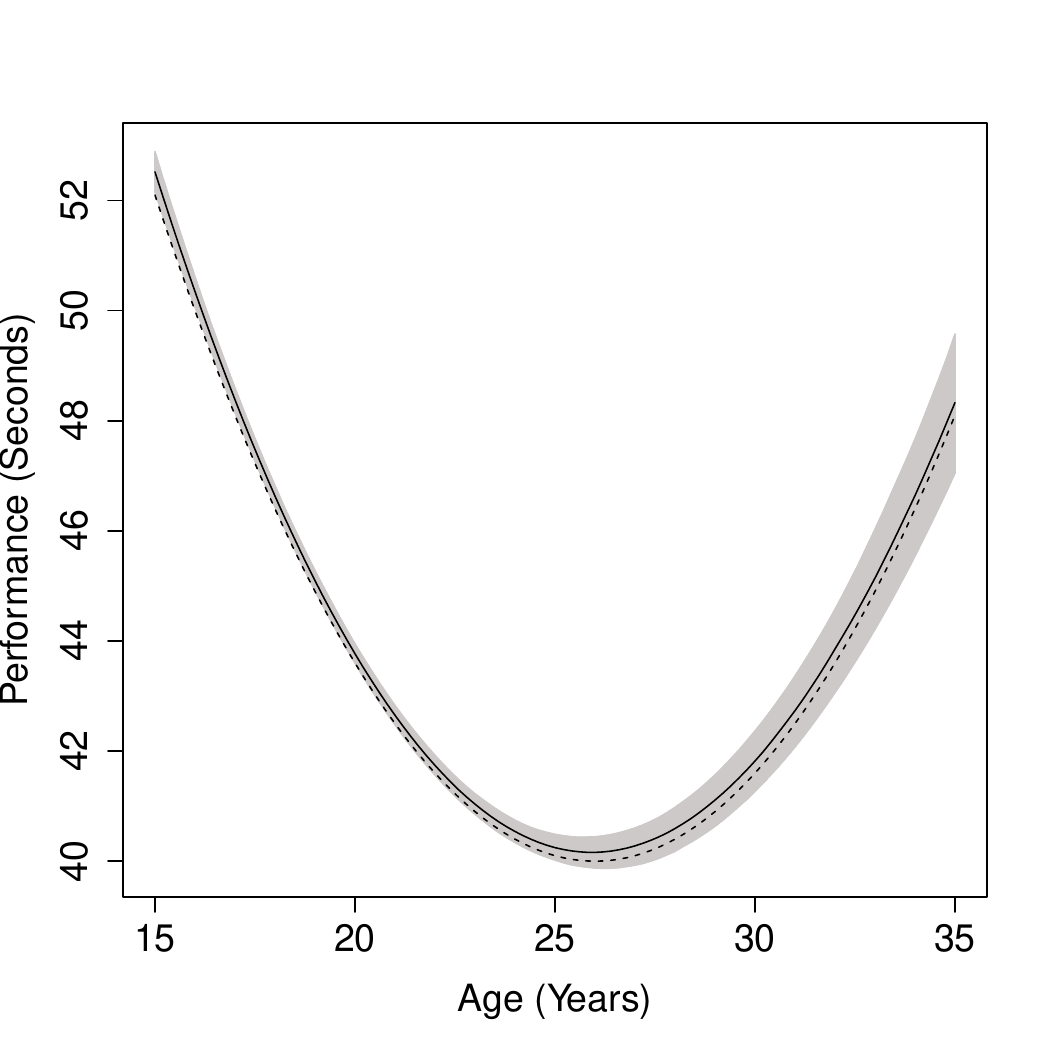} &
\includegraphics[scale=0.24]{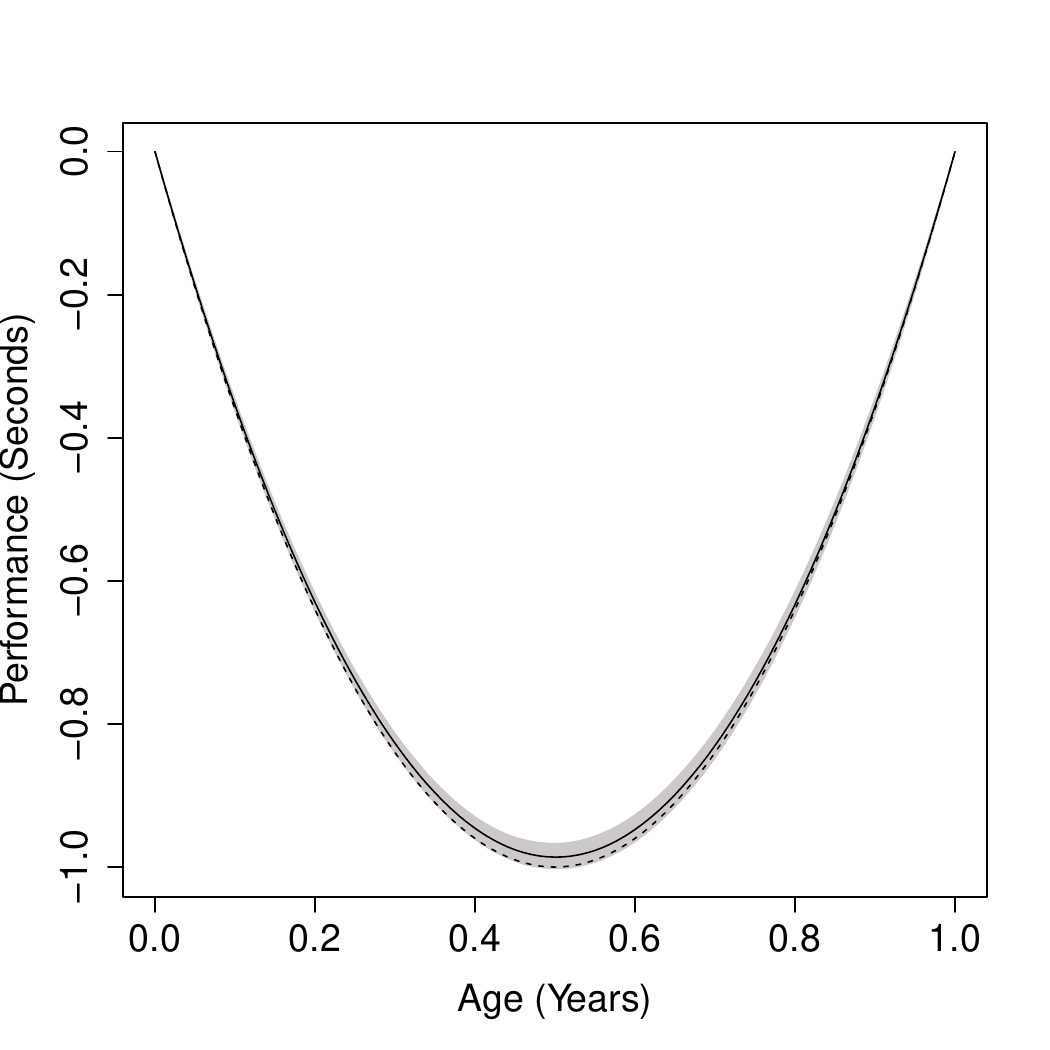} &
\includegraphics[scale=0.24]{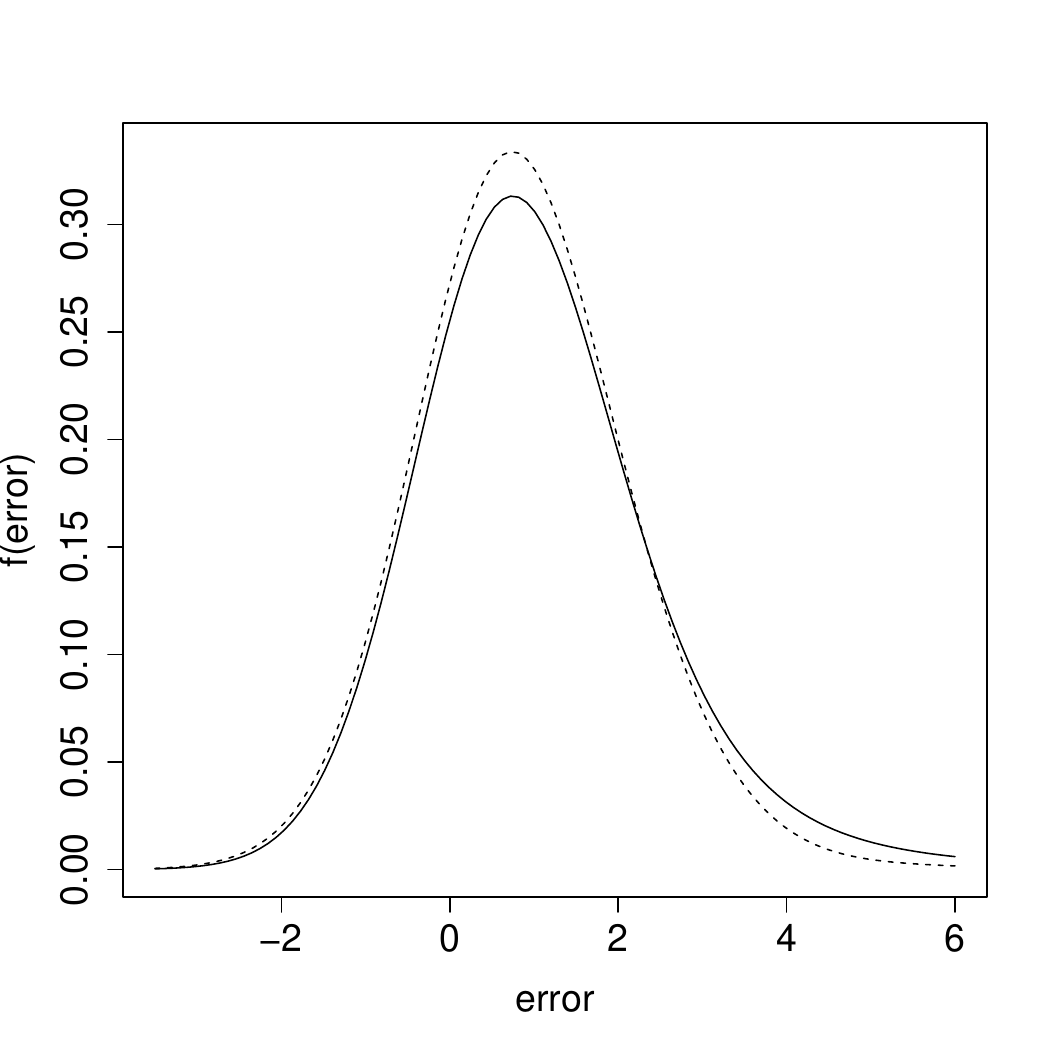} \\
$\eta$ & Individual within-season & Seasonal within-season\\
& Performance trajectory & Performance trajectory \\
\includegraphics[scale=0.24]{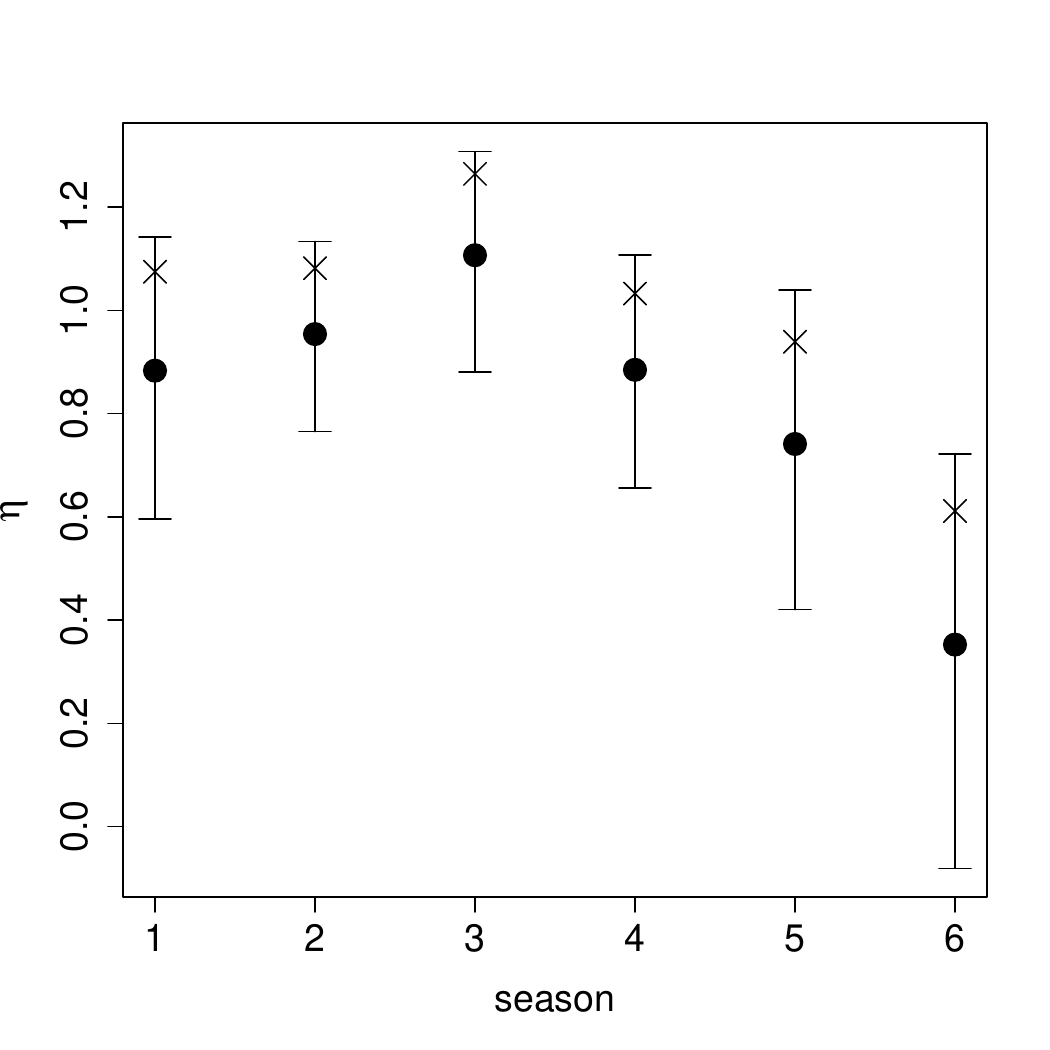} &
\includegraphics[scale=0.24]{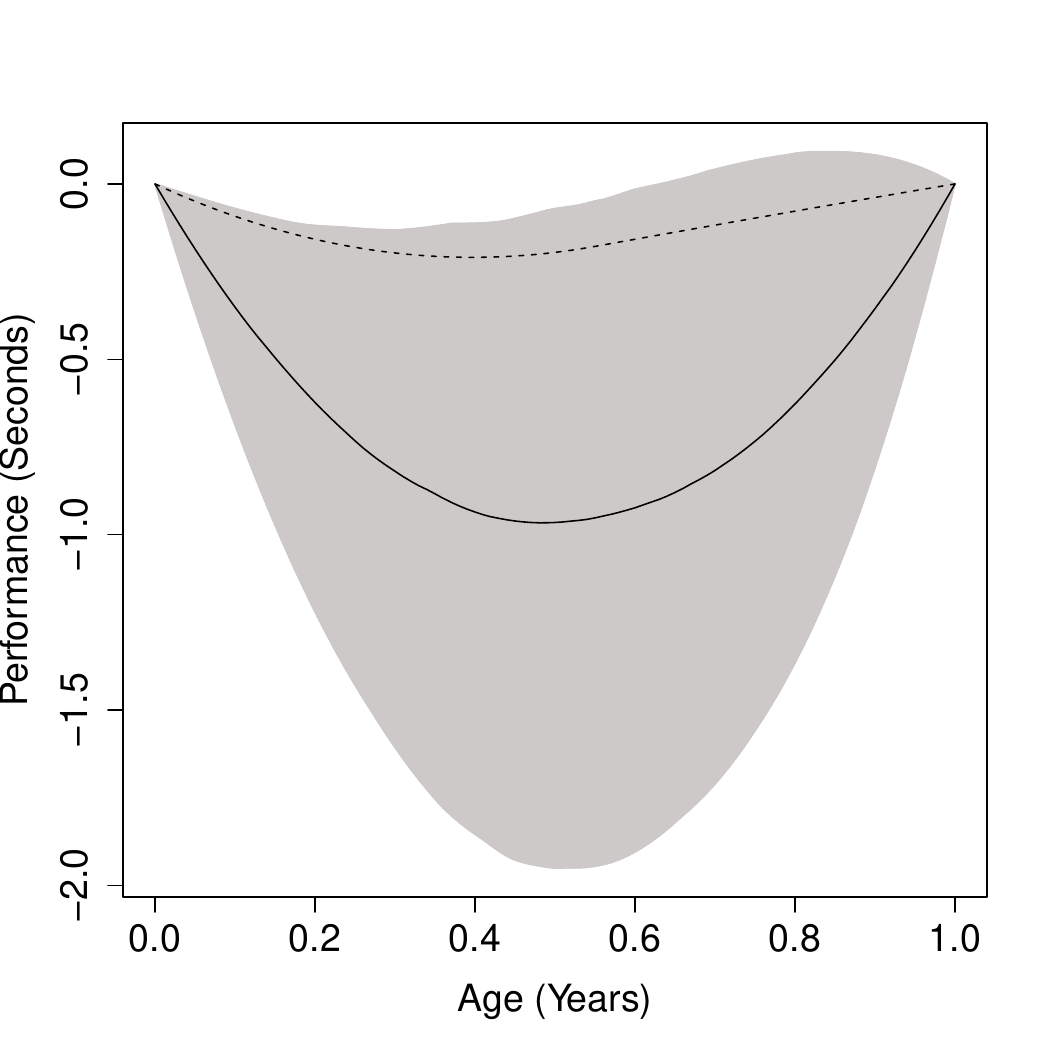} &
\includegraphics[scale=0.24]{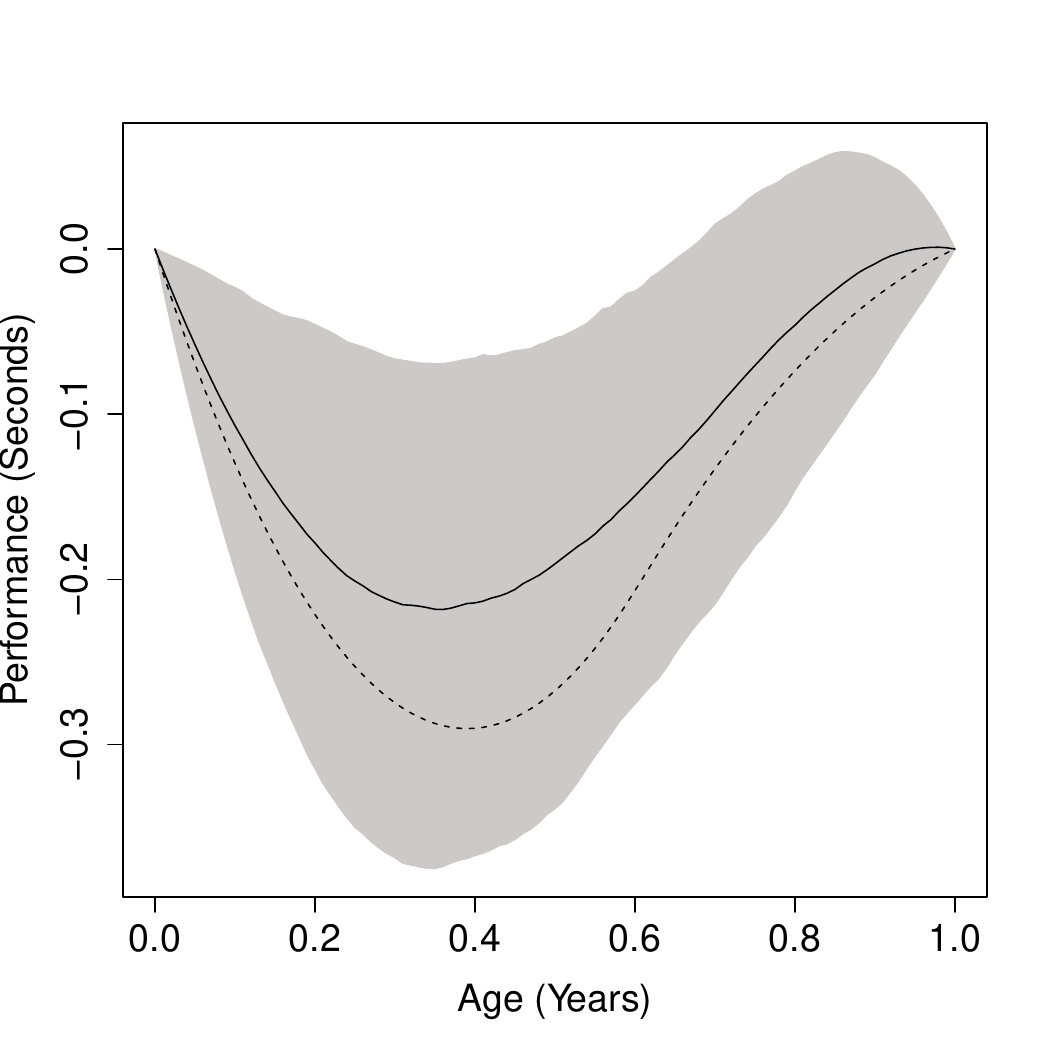}
\end{tabular}
\end{center}
\caption{Results for the simulated example. Results for Population performance trajectory, Population, Individual and Seasonal within-season performance trajectories are shown as true value (dotted line), posterior median (solid line) and 95\% credible interval. Results for the error density are shown as true value (dotted line) and posterior mean (solid line). The results for $\eta$ represent $\eta_{i, 1}, \dots, \eta_{i, 6}$ for one athlete and are shown as true value (block dot), posterior median (cross) with 95\%  credible interval.}
\label{f:sim_example}
\end{figure}
We simulated a data set of 500 individuals 
with $p_1 = 0.2$, 
$\sigma^2_a = 0.5$ and 
$\sigma^2_b = 0.25$. The results are shown in Figure~\ref{f:sim_example}. The population performance trajectory and the population within-season performance trajectory and error distribution are well-estimated. Estimates at the individual level are unsurprisingly less well-estimated.
The true values of the individual trend performance trajectory ($\eta$) and individual and seasonal within-season performance trajectories are all within the corresponding 95\% credible intervals.

To quantify the performance of the model, we conducted a simulation study.
We use 100 replications for 8 different combinations generated by varying four factors over 2 levels: 
$M = (200, 500)$, 
$p_1 = (0.2, 0.5)$,
$(\sigma^2_a, \sigma^2_b) = (0.1, 0.03), (0.5, 0.25)$. We use performance measures for seven parameters of interest: 
 the Population Performance Trajectory
$g(\cdot)$,  the Population Within-Season Performance Trajectory $h^{\star}$, Individual Within-Season Performance Trajectory $h^{\star}_i$, the Within-Season Performance Trajectory  $h^{\star}_{i, s}$, the  excess performance at the start of each season $\eta_1, \dots, \eta_M$, $\Psi_i$ and $\Gamma_i$. For the first four functional parameters, we define the root mean squared integrated error to be 
\[
\mbox{RMISE} = \int \left(\hat{f}(z) - f(z)\right)^2 \,dz
\]
where $f$ is the true value of the function and $\hat{f}$ is the posterior mean of $f$ and the limit of integration are 
 the possible range of $z$. 
The RMISE is averaged over all athletes for $h_i^{\star}(\cdot)$ and 
over all seasons and athletes for $h_{i, s}^{\star}(\cdot)$. For $\eta_i$, we calculate the average root mean squared error (AMRSE)
\[
\mbox{AMRSE} = \frac{1}{\sum_{i=1}^M S_i+1}
\sum_{i=1}^M \sum_{s=1}^{S_i+1} \left(\hat\eta_{i, s} - \eta_{i, s}\right)^2
\]
where $\hat\eta_{i, s}$ is the posterior mean of $\eta_{i, s}$. We would  like to use $\tau_i^2$ as a proxy for the squared difference between the individual and population within-season trajectory ($(h_i^{\star} - h^{\star})^2$).
 To understand the strength of this relationship, we calculate the Spearman's rank correlation coefficients between the posterior median of $\tau_1^2,\dots,\tau_M^2$ and 
 $\vert a_1\vert , \dots, \vert a_M\vert $ (as defined in \eqref{sim_a}) since these are not directly comparable. Similarly, to understand the relationship between $\lambda_i^2$ and the squared difference between the seasonal and individual within-season trajectory,
 we calculate the Spearman's correlation between $\bar{b}_i = \frac{1}{S_i} \sum_{s=1}^{S_i} \vert b_{i,s}\vert $ (where $b_{i, s}$ is given in \eqref{sim_b}) and $\lambda_i^2$.

\begin{figure}[h!]
\begin{center}
\begin{tabular}{cccc}
(a) & (b) & (c) & (d)
\\
\includegraphics[scale=0.2]{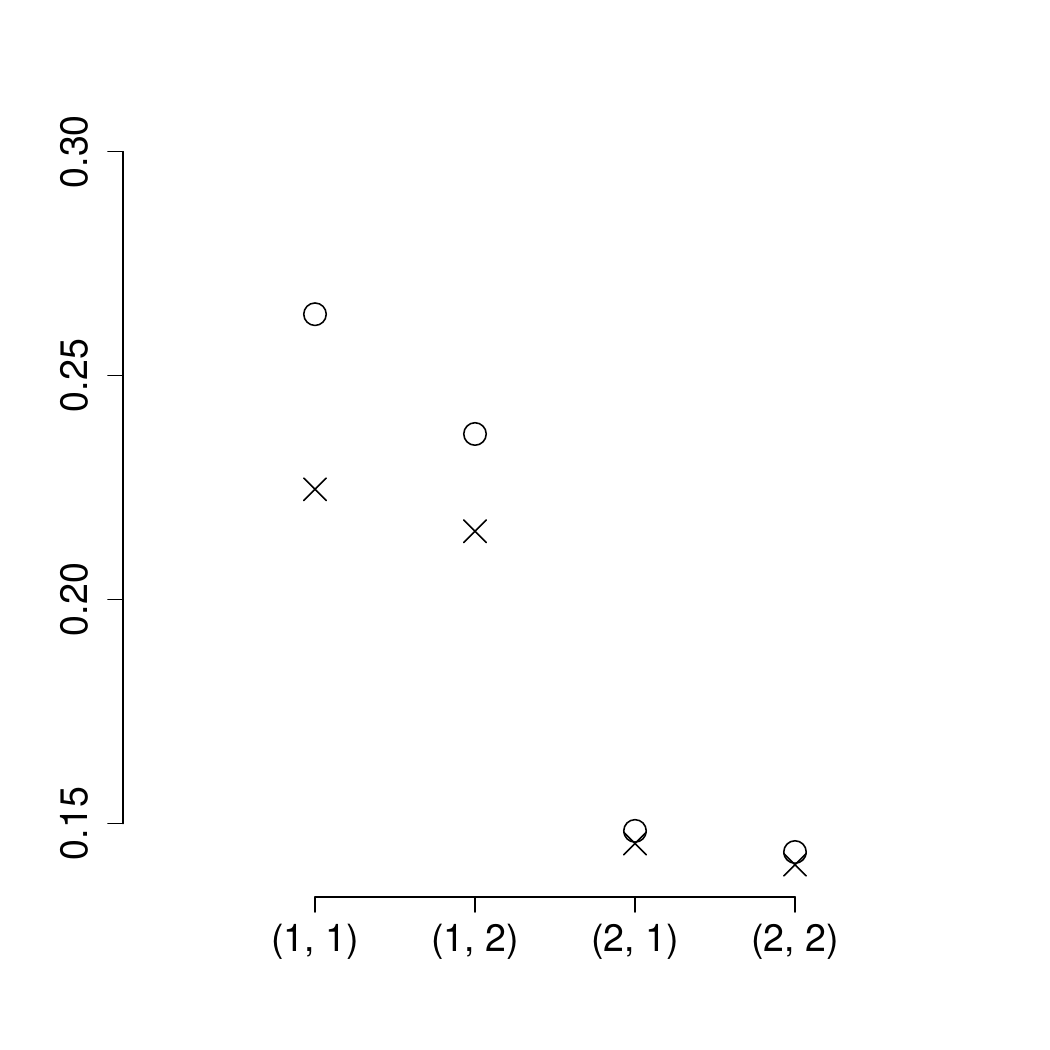} &
\includegraphics[scale=0.2]{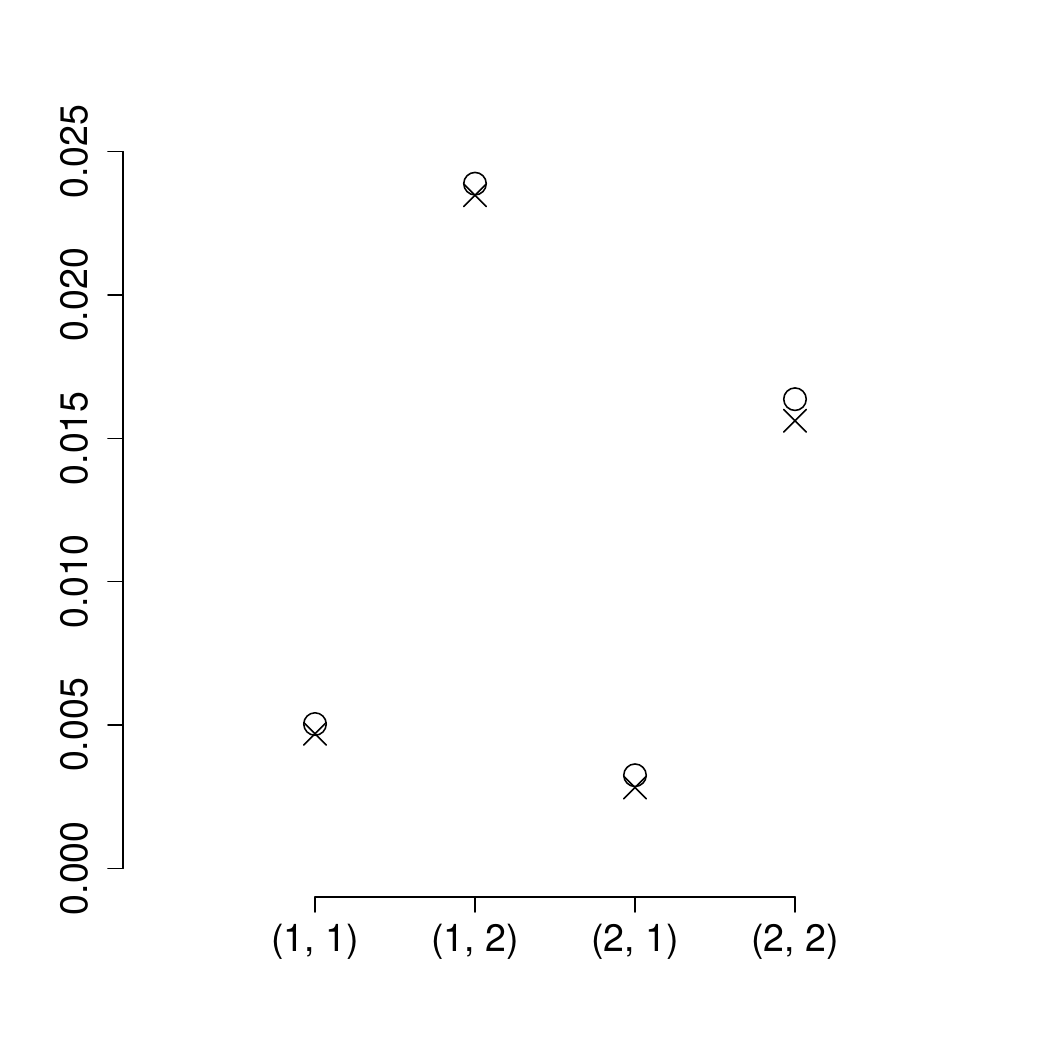} &
\includegraphics[scale=0.2]{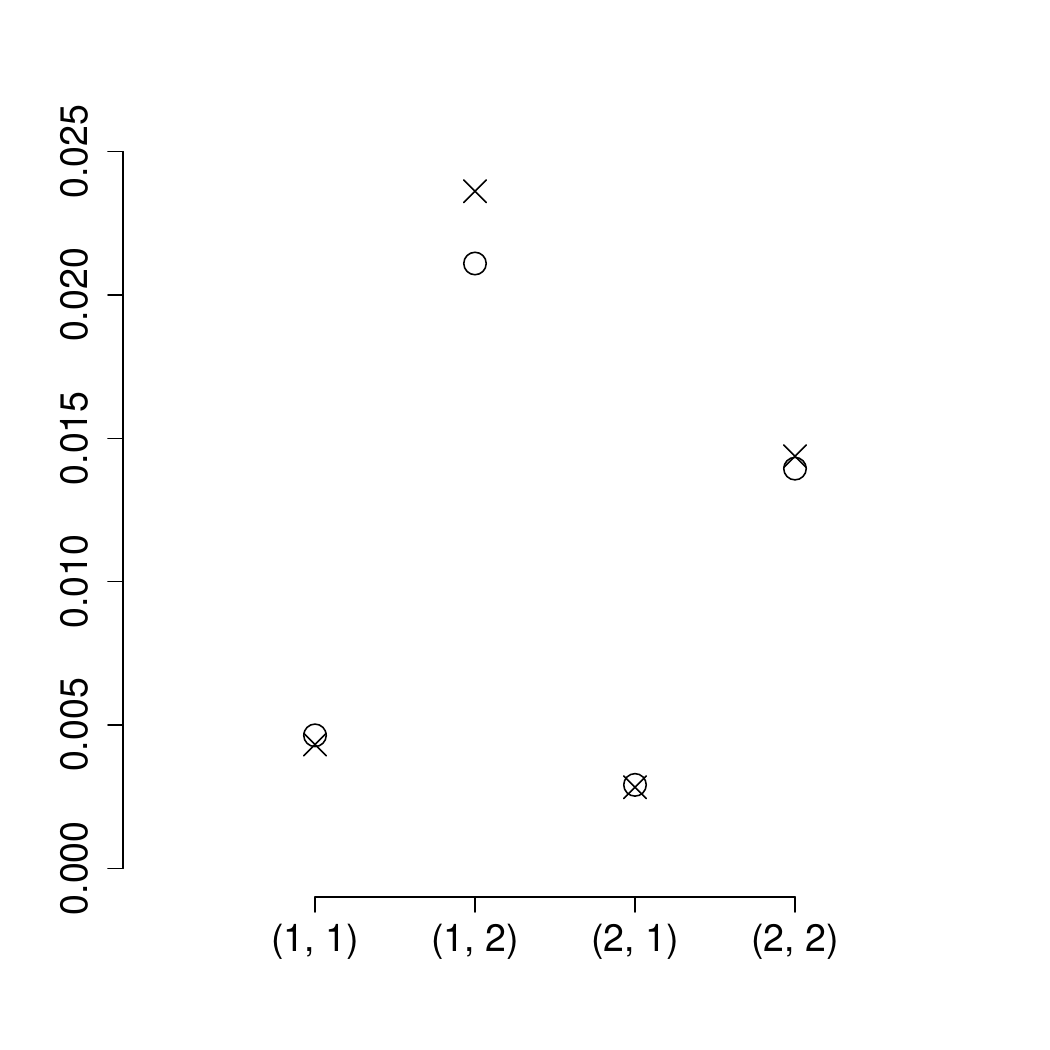} &
\includegraphics[scale=0.2]{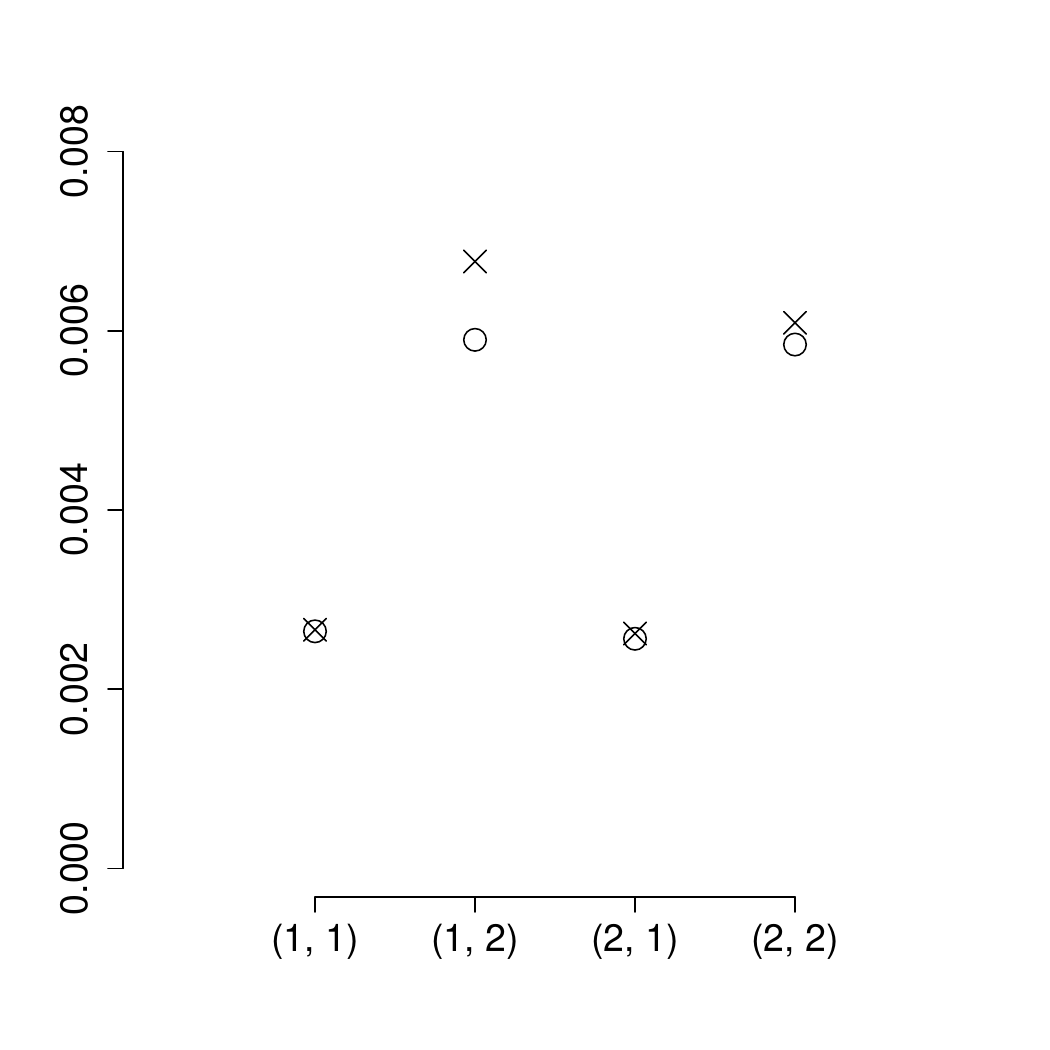} 
\end{tabular}
\end{center}
\caption{Simulation Study: The RMISE for 
(a) the Population Performance Trajectory $g(c\dot)$, (b) Population Within-Season Performance Trajectories $h^{\star}(\cdot)$, (c) Individual Within-Season Performance Trajectories $h^{\star}(\cdot)_i$, and (d) Seasonal  Within-Season Performance Trajectories $h^{\star}_{i, s}(t)$. For all plots, the x-axis labels are $(i, j)$ where $i$ is the level of $M$ and $j$ represents the level of $(\sigma^2_a, \sigma^2_b)$, cross represents $p_1 = 0.2$ and circle represents $p_1 = 0.5$}\label{f:simulated_2}
\end{figure}
The results for the functional parameters are shown in Figure~\ref{f:simulated_2}. A larger number of athlete, $M$, leads to lower RMISE's for each of these functions. The values of $\sigma^2_a$ and $\sigma^2_b$ have only a small effect on the estimation accuracy of the population performance trajectory. Conversely, $\sigma^2_a$ and $\sigma^2_b$ have a large effect on the estimation accuracy for all three within-season trajectories.

\begin{figure}[h!]
\begin{center}
\begin{tabular}{ccc}
(a) & (b) & (c) \\
\includegraphics[scale=0.2]{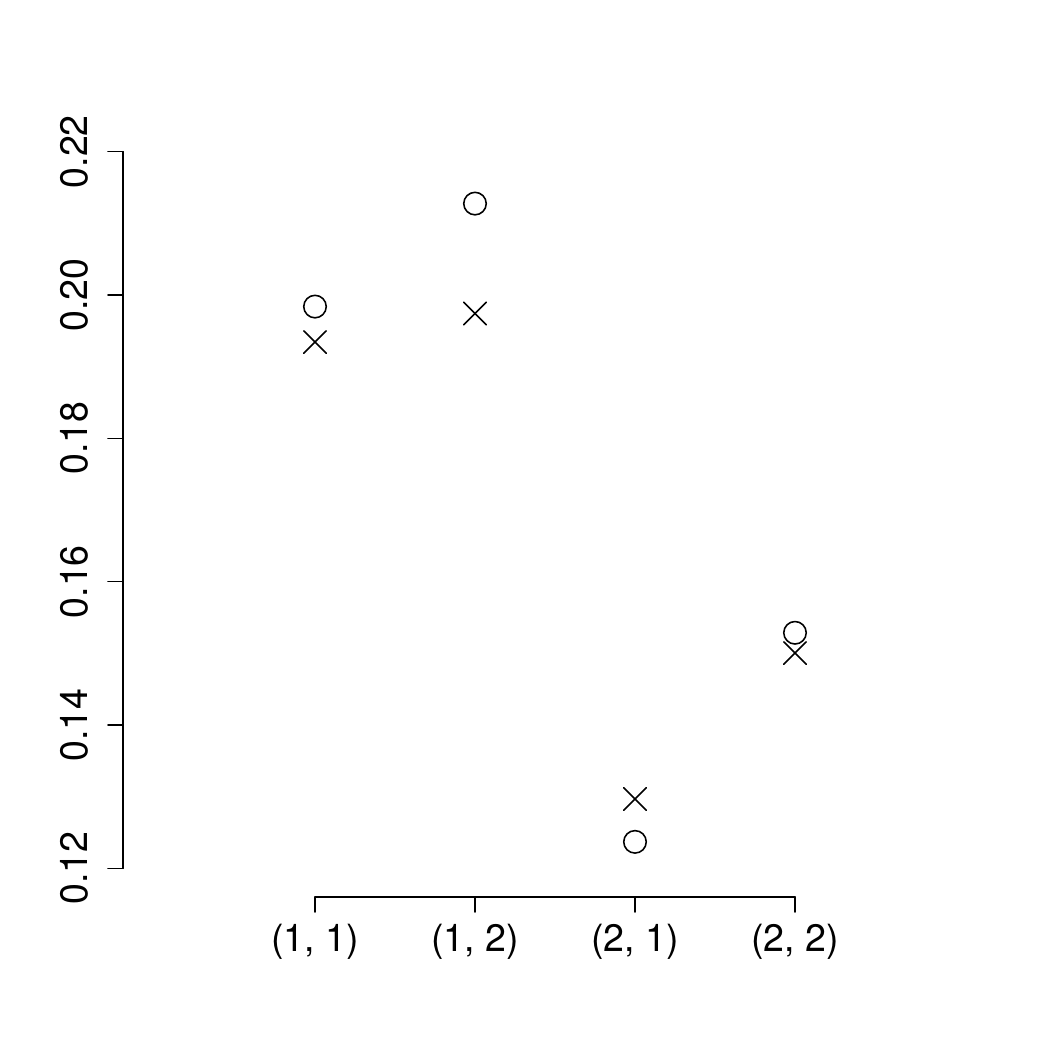} &
\includegraphics[scale=0.2]{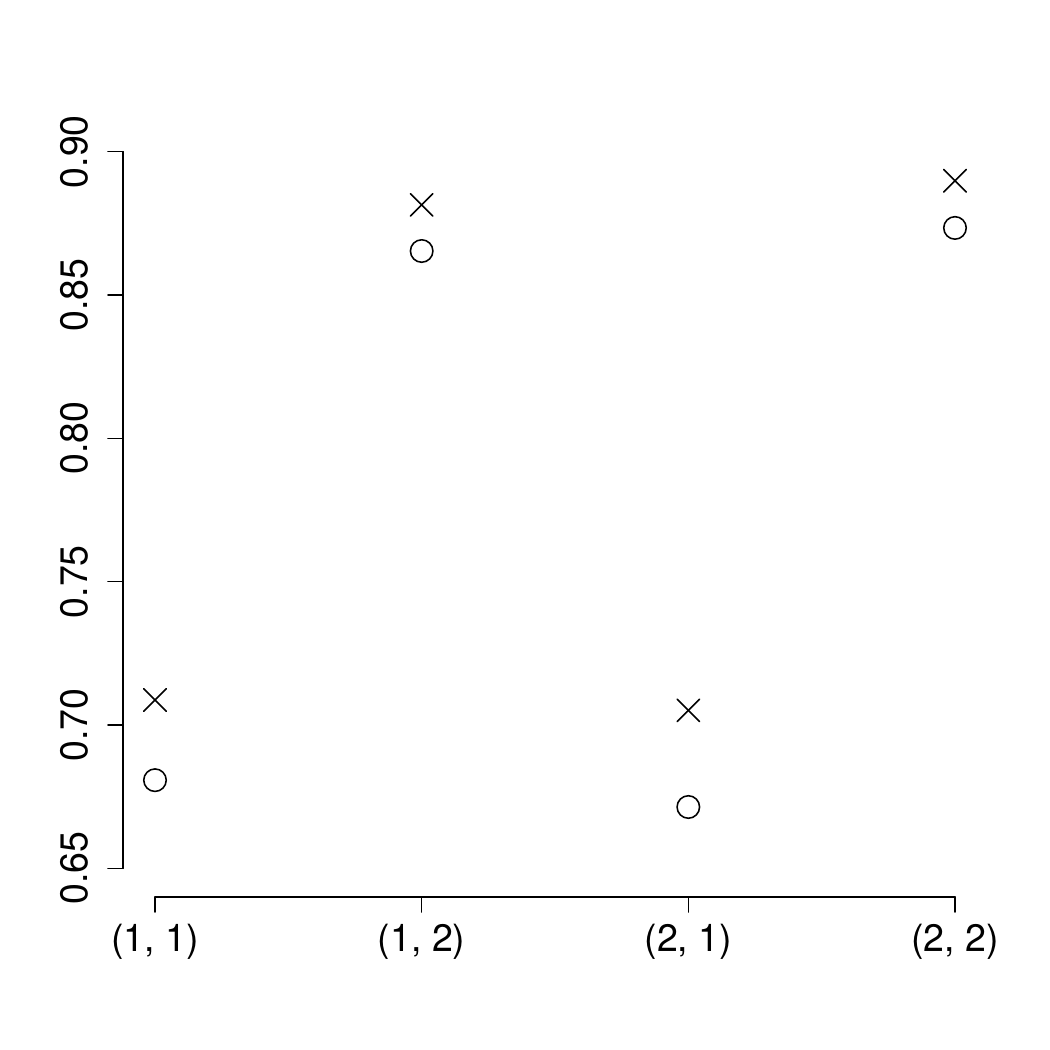} &
\includegraphics[scale=0.2]{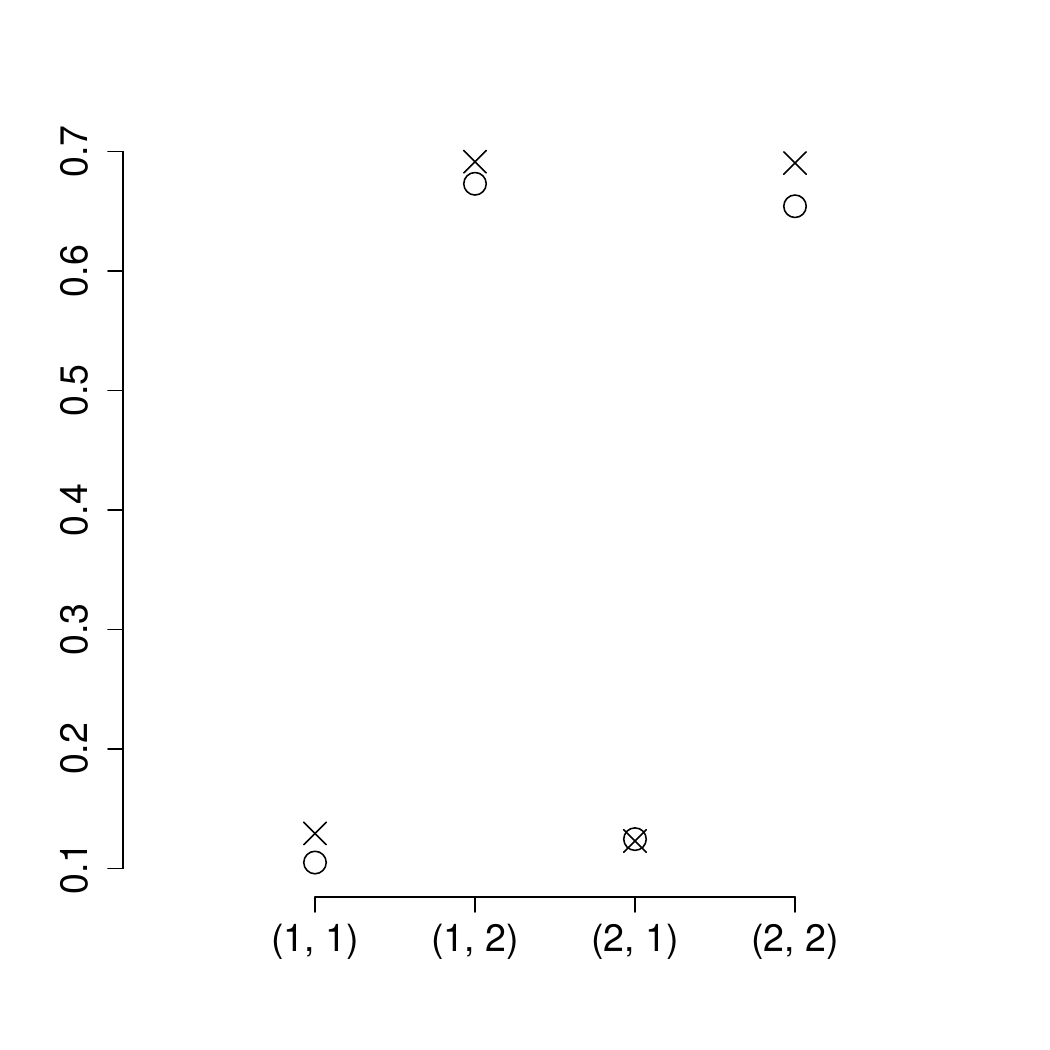} 
\end{tabular}
\end{center}
\caption{Simulation Study: (a) The ARMSE  of $\eta_{i, j}$,
(b) Spearman's rank correlation coefficient between  $a_i$ and $\tau_i^2$, and 
(c) Spearman's rank correlation coefficient between  $\bar{b}_i = \frac{1}{S_i}\sum_{j=1}^{S_i} \vert b_{i, j}\vert $ and $\lambda_i^2$.
 For all plots, the x-axis labels are $(i, j)$ where $i$ is the level of $M$ and $j$ represents the level of $(\sigma^2_a, \sigma^2_b)$, cross represents $p_1 = 0.2$ and circle represents $p_1 = 0.5$.}\label{f:simulated_3}
\end{figure}

The results for the other three parameters of interest are shown in Figure~\ref{f:simulated_3}. The estimation of $\eta$ is most strongly affected by the number of athletes $M$. 
The  two correlation measures increase with level of variation in the within-season trajectories but the other factors do not have a strong effect. We find that 
there is  a stronger correlation between $a_i$ and $\tau_i^2$ than $\bar{b}_i$ and $\lambda_i^2$. In fact, for the within-season correlation, the correlation is close to zero with the lower level of variation but is around 0.7 for the higher level of variation. Therefore, these are useful summaries of the variation in these trajectories if there is sufficient variation in the data.

\section{Application to elite swimming} \label{Results}

We applied our  model to performances for both female and male swimmers in the 100 and 200 metres freestyle. We fitted the model separately to each combination of gender and distance, and included pool length as a confounder.
  Some additional results  are provided in Appendix C. Two samplers were run for a total 50,000 iterations. The first 30,000 iterations were used as a burn-in and the subsequent 20,000 iterations were thinned every 20-th value to 1,000 posterior samples.  This took about ten hours to run with R using an Apple Mac with M4 chip. Some MCMC diagnostics for each data set are given in Appendix~\ref{MCMC_diagnostics}

\begin{figure}[h!]
\centering
\begin{tabular}{cccc}
\multicolumn{2}{c}{100 metres} &
\multicolumn{2}{c}{200 metres}\\
 Females &Males  & Females &Males \\
\includegraphics[scale=0.19]{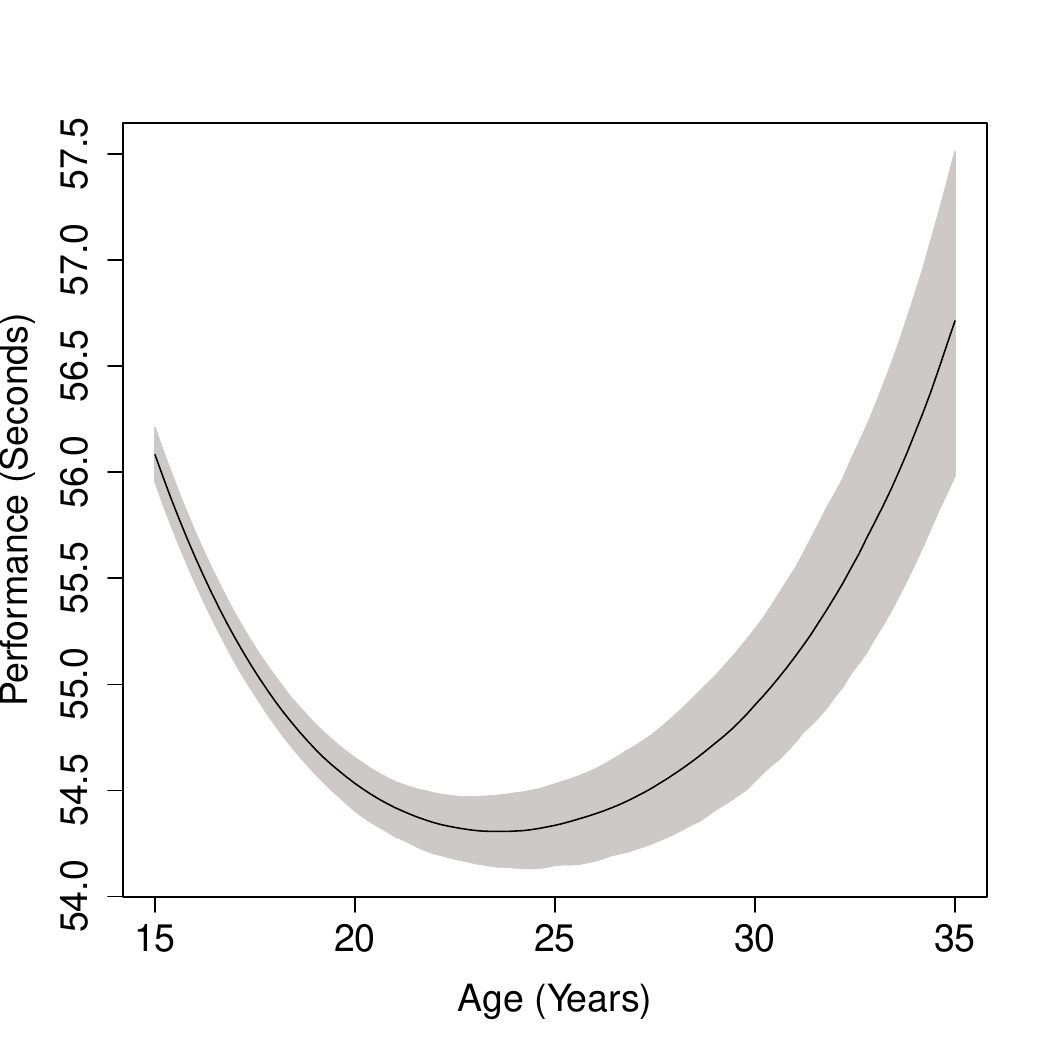}
&\includegraphics[scale=0.19]{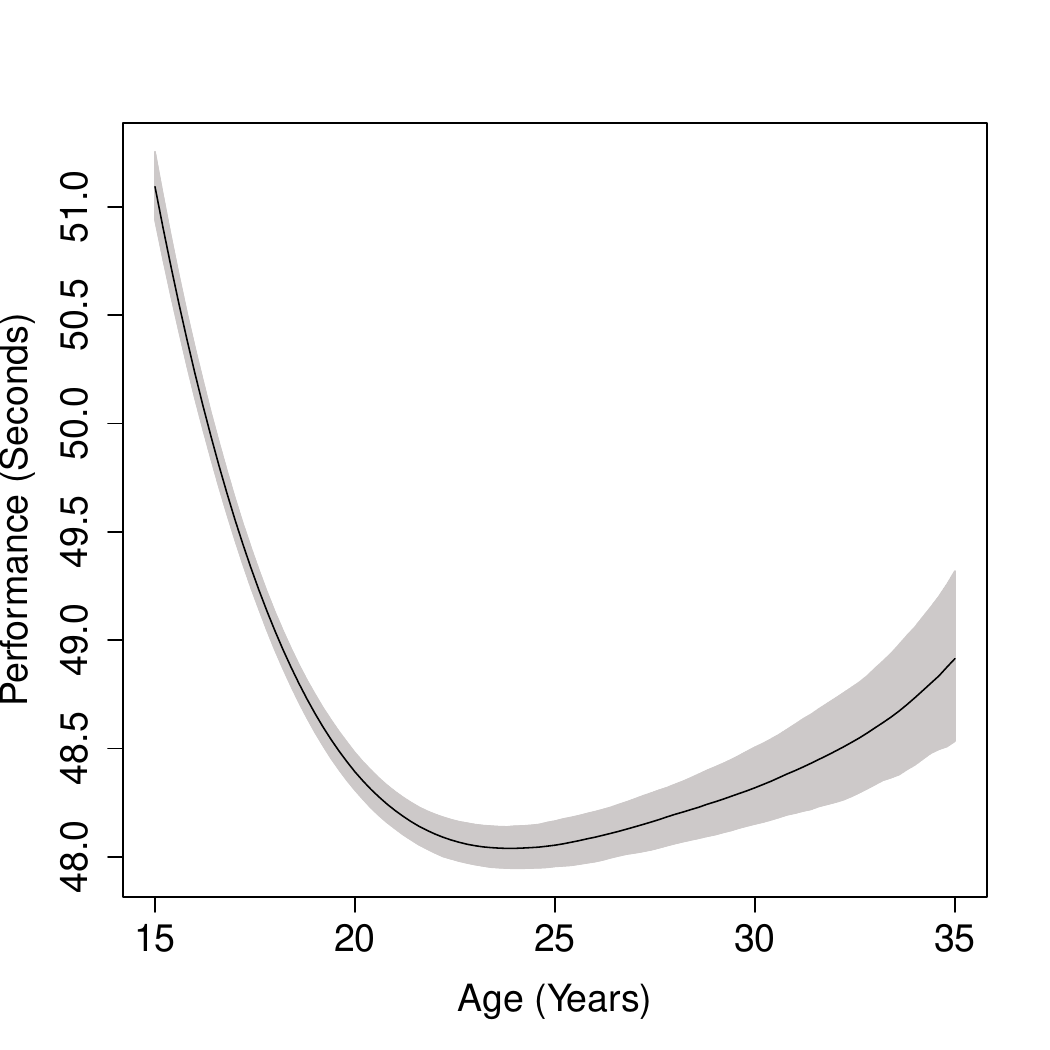}
&\includegraphics[scale=0.19]{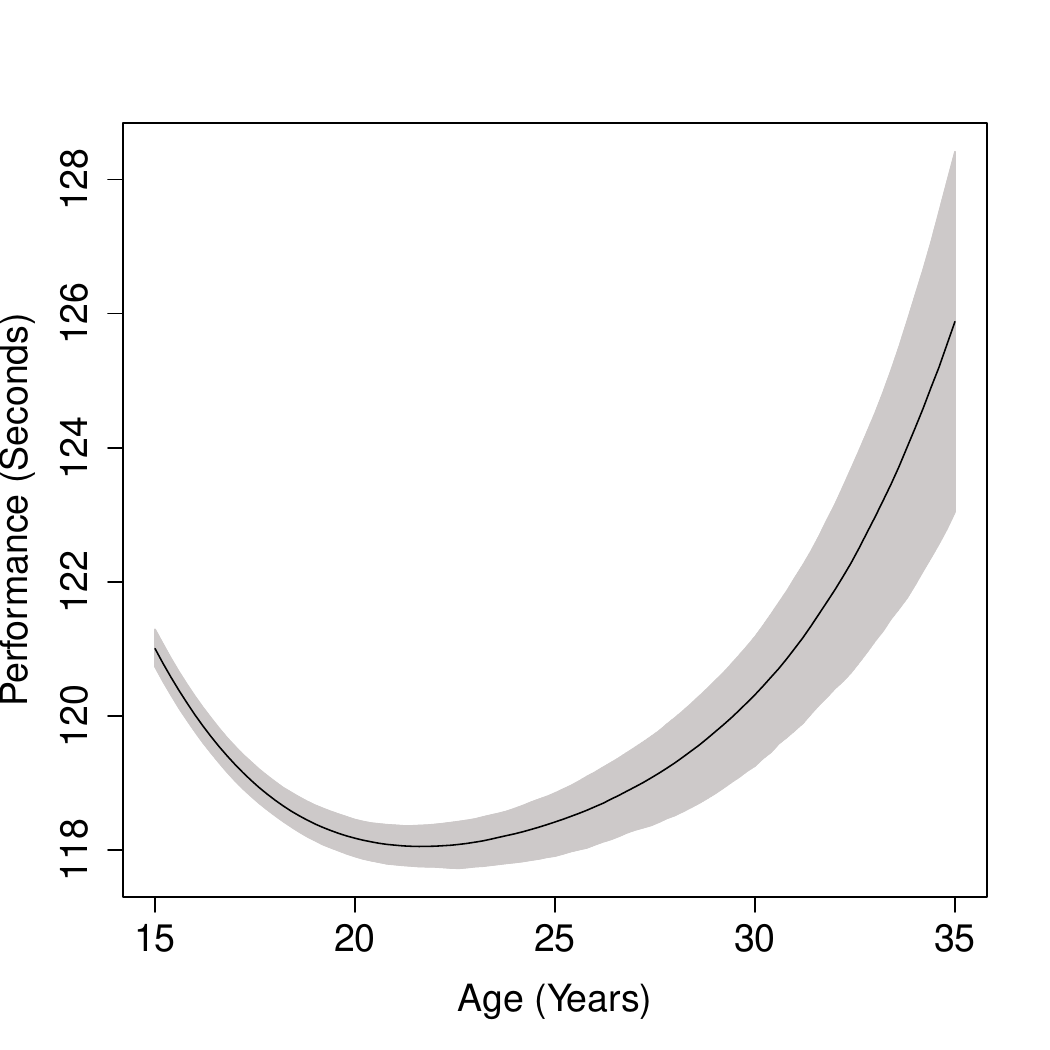}
&\includegraphics[scale=0.19]{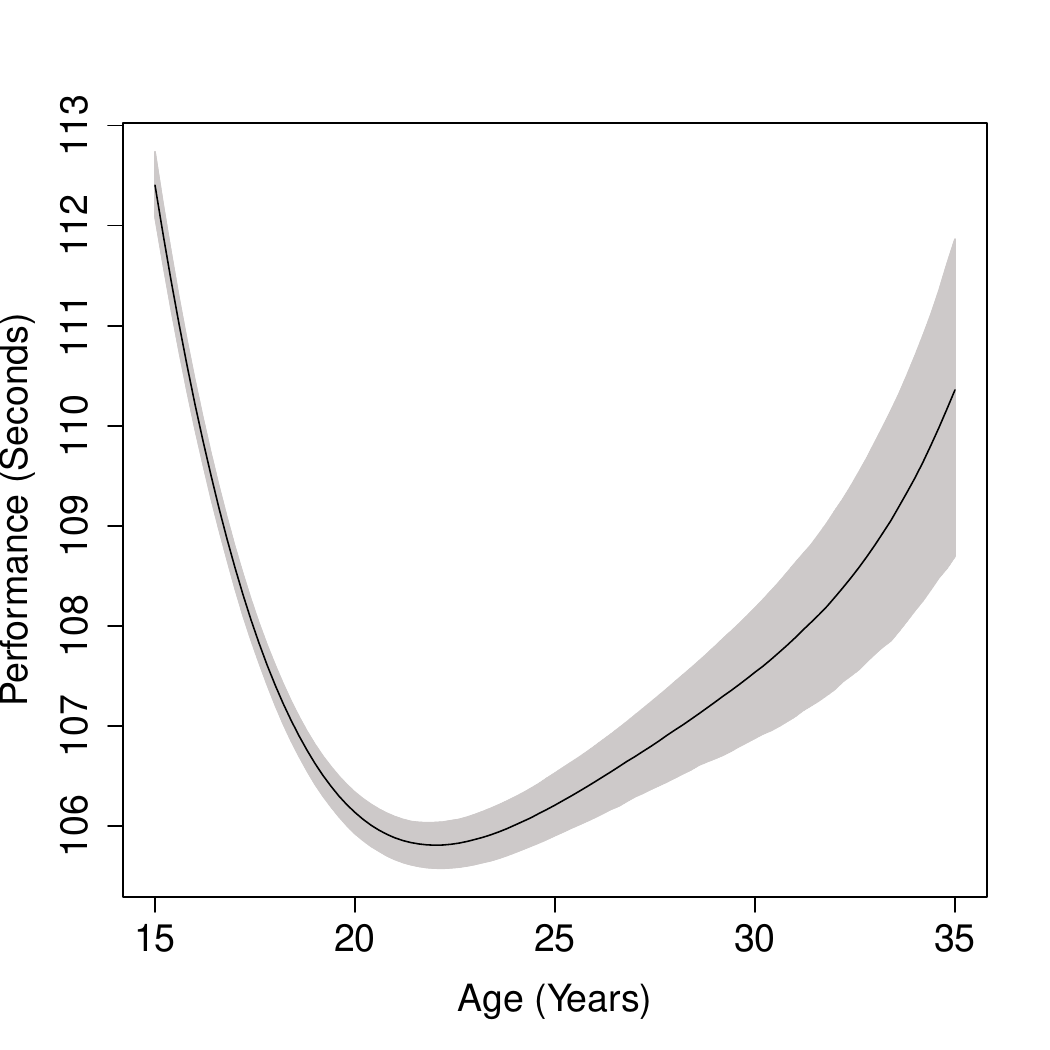}
\end{tabular}
\caption{\label{popul_fitted_curves} Estimated  population performance trajectory $g(\cdot)$ for both females and males in the 100 metre and 200 metre freestyle.
The trajectories are shown as posterior median (black line) and 95\% credible interval (grey shading).}
\end{figure}
 Figure \ref{popul_fitted_curves} shows the estimated
 population performance trajectories, which all have a reverse J-shape, as in \cite{griffin2022bayesian}, with performance rapidly improving between 15 and 20, peaking around 22 (in both men and women), and subsequently decreasing. There is a clear difference between men and women. Men show a much greater improvement in performance between 15 and 20 (for example, in the 100 metres, females improve by 1.5 seconds between 15 and 20 whereas males improve by 2.6 seconds. Men are also better able to maintain their performance in their late twenties (for example, female performance decreases by 2 seconds by 25 and 30 whereas men's performance decreases by 0.7 seconds).
\begin{figure}[h!]
\centering
\begin{tabular}{cccc}
\multicolumn{2}{c}{100 metres} &
\multicolumn{2}{c}{200 metres}\\
 Females &Males  & Females &Males \\
\includegraphics[scale=0.19]{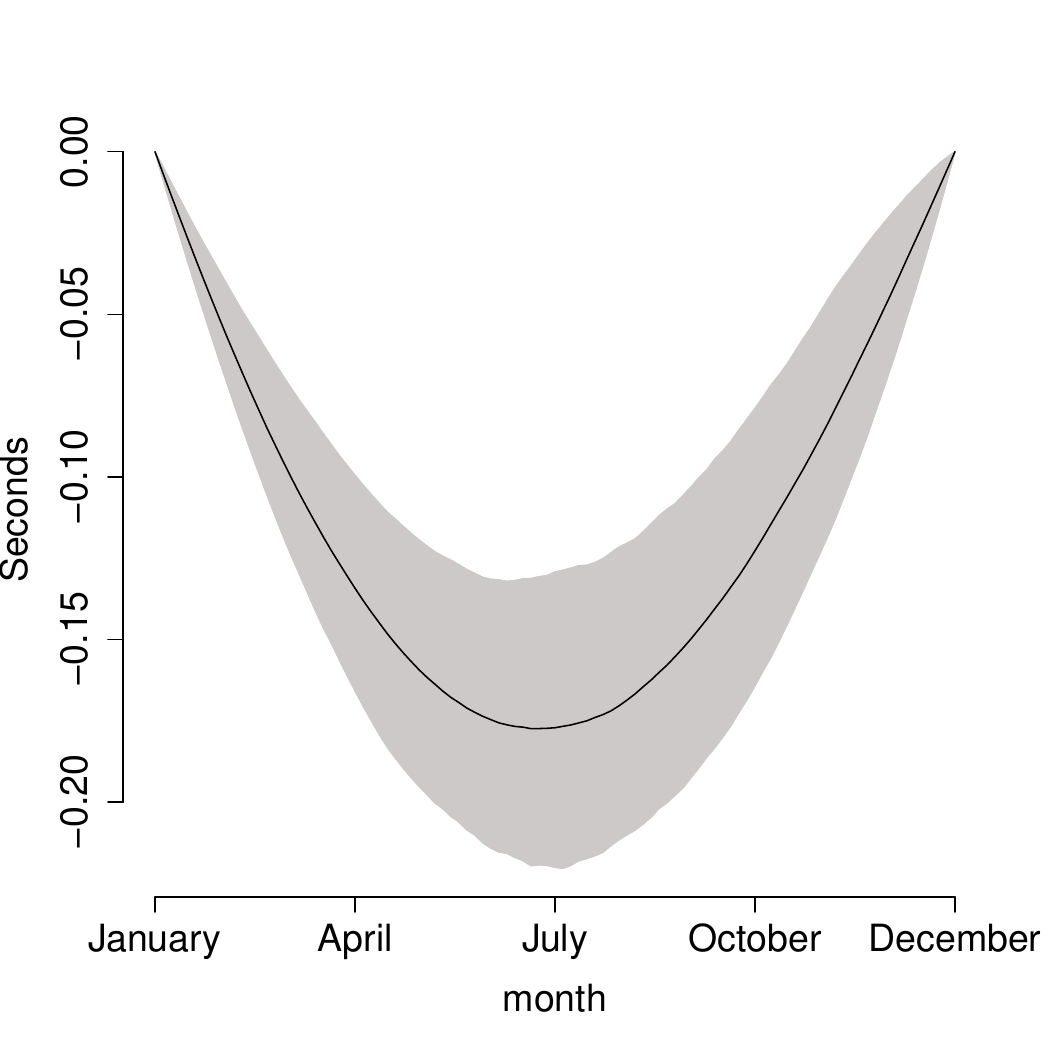}
&\includegraphics[scale=0.19]{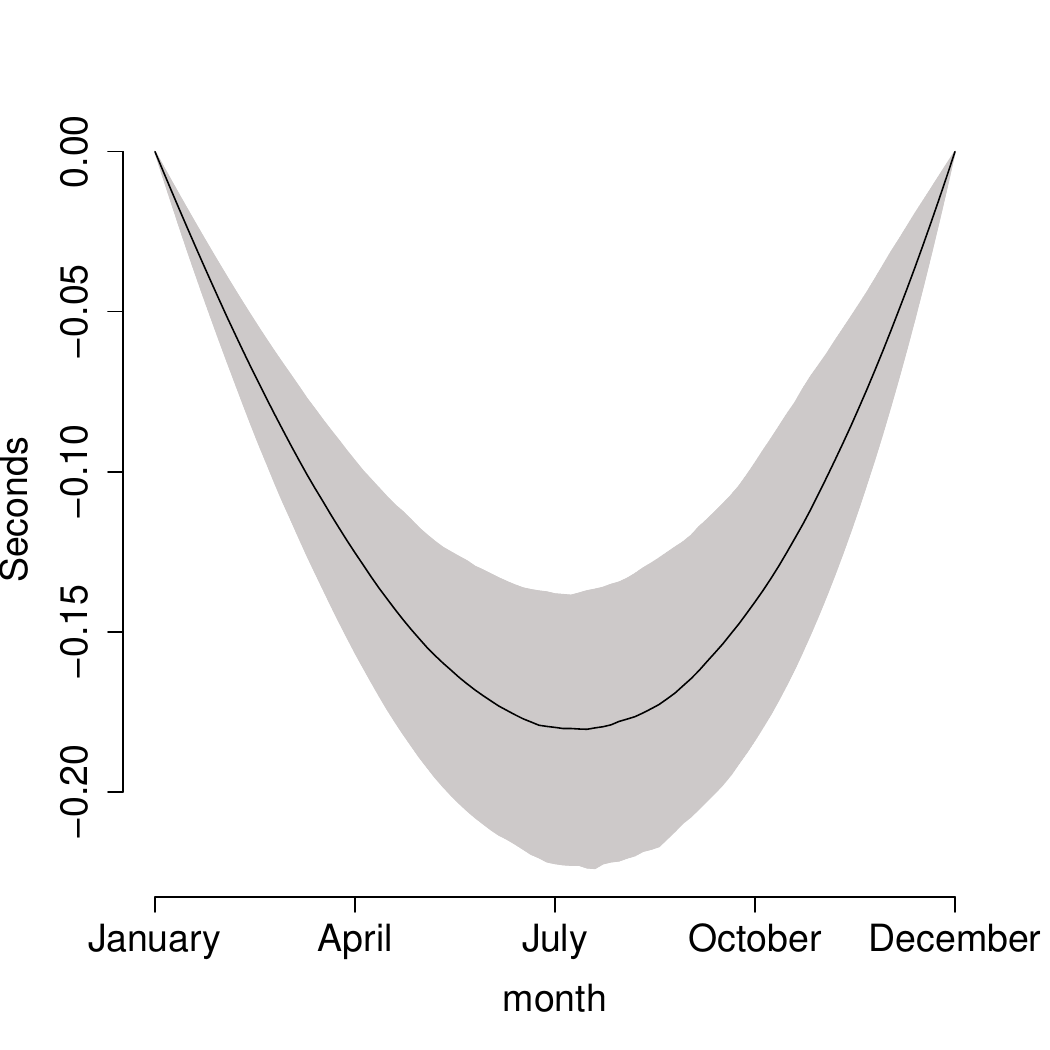} 
&\includegraphics[scale=0.19]{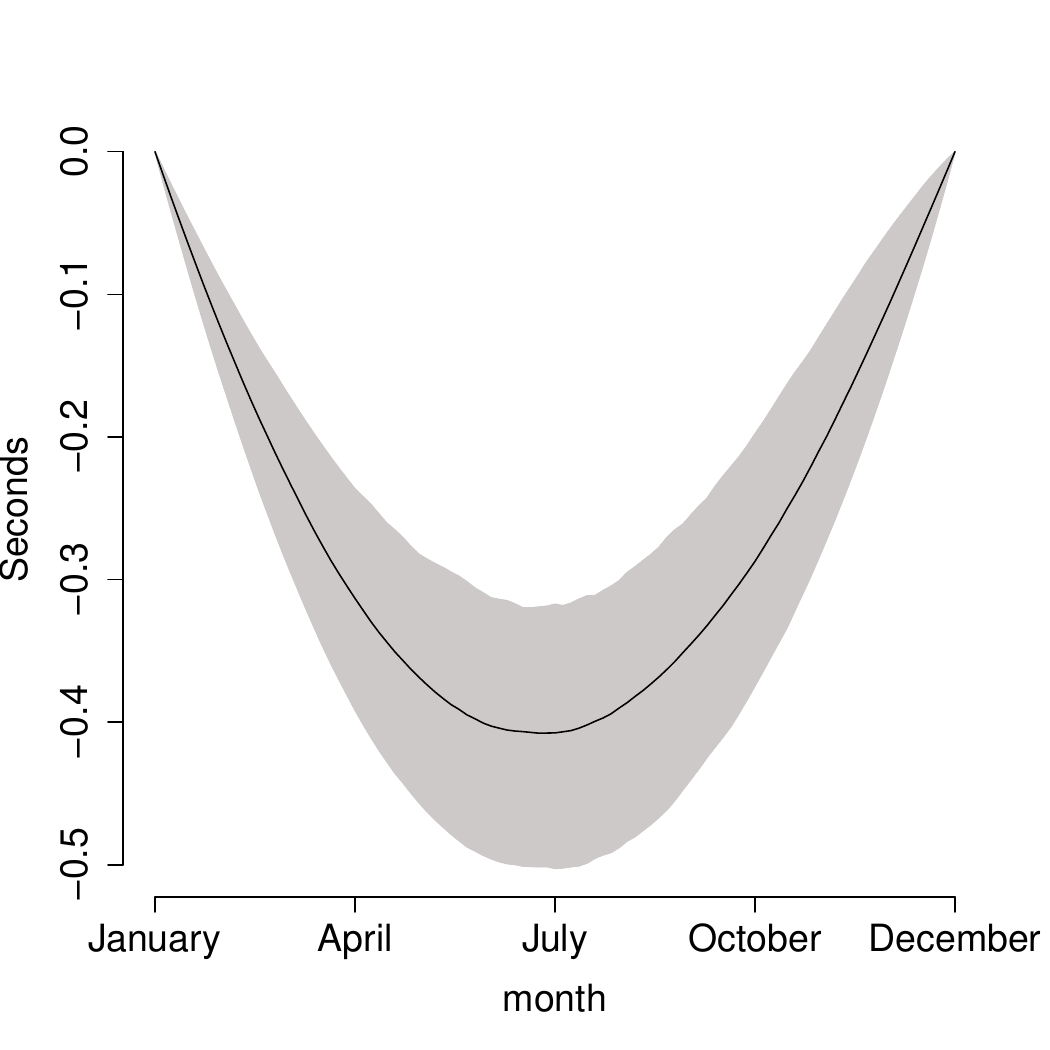}
&\includegraphics[scale=0.19]{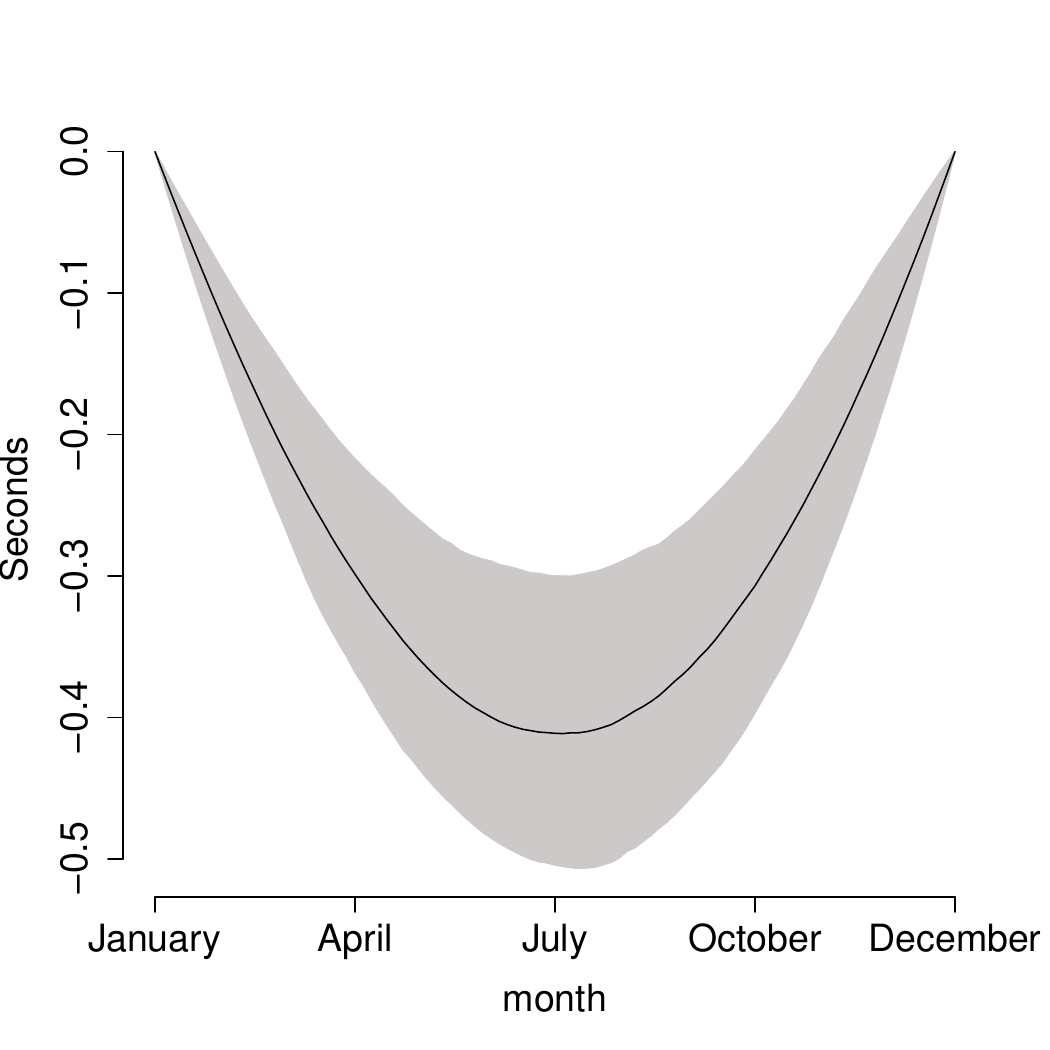}
\end{tabular}
\caption{\label{popul_intra_fitted_curves} Estimated  population  within-season performance trajectory $h^{\star}(\cdot)$ for both females and males in the 100 metre and 200 metre freestyle.
The trajectories are shown as posterior median (black line) and 95\% credible interval (grey shading).}
\end{figure}
The population level within-season trajectories are shown in Figure~\ref{popul_intra_fitted_curves}. The trajectories have very similar shapes across gender and distance with the within-season performance improvement peaking around July. The posterior median improvement is around 0.17 seconds in the 100 metres  and around 0.39 seconds in the 200 metres
for both males and females. 
The error distributions were found to be positively skewed with 
evidence of a large difference in the heaviness of the left-hand and right-hand tails. Results are presented in Appendix~\ref{Additional results}.

\begin{figure}[h!]
\begin{center}
\begin{tabular}{ccc}
\multicolumn{3}{c}{Female}\\
Swimmer 1 & Swimmer 2 & Swimmer 3\\
\includegraphics[scale=0.22]{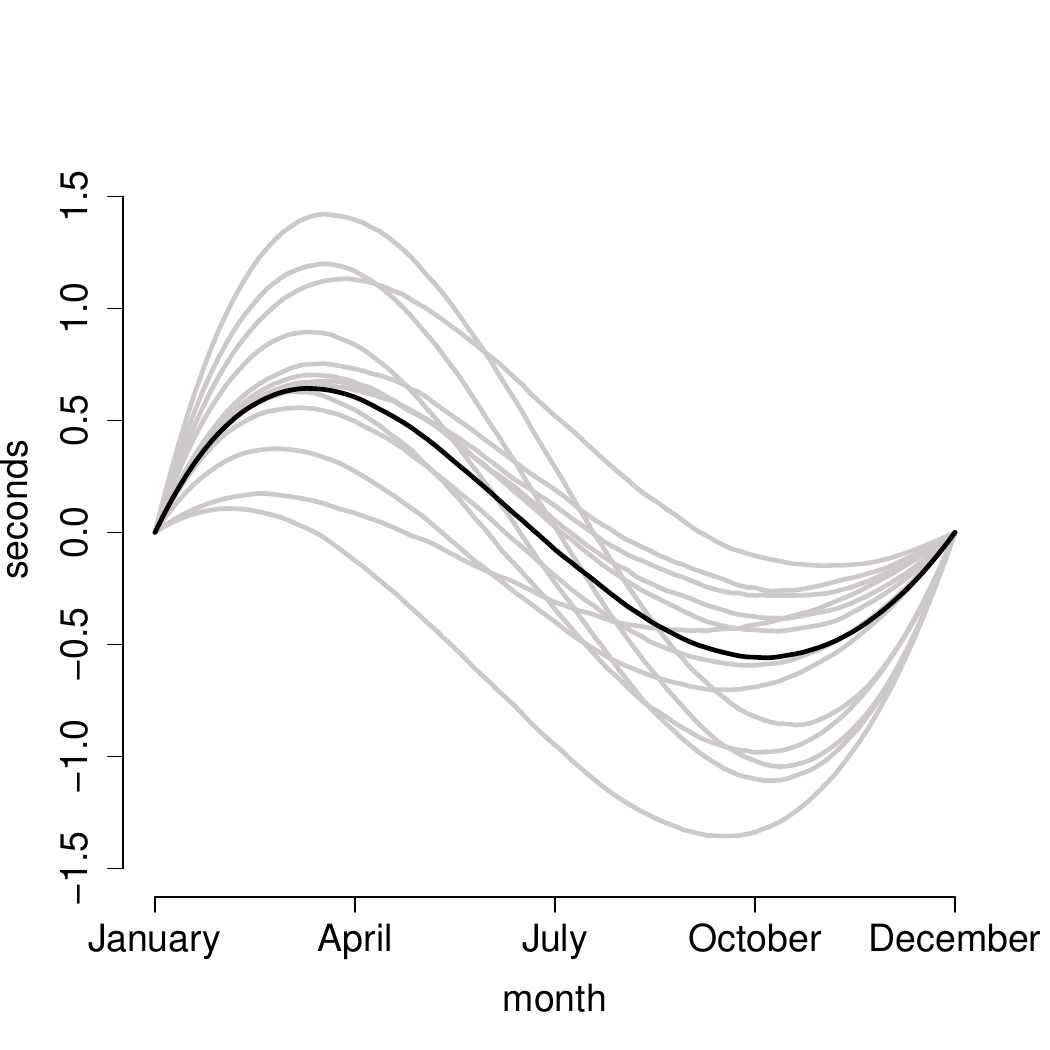}&
\includegraphics[scale=0.22]
{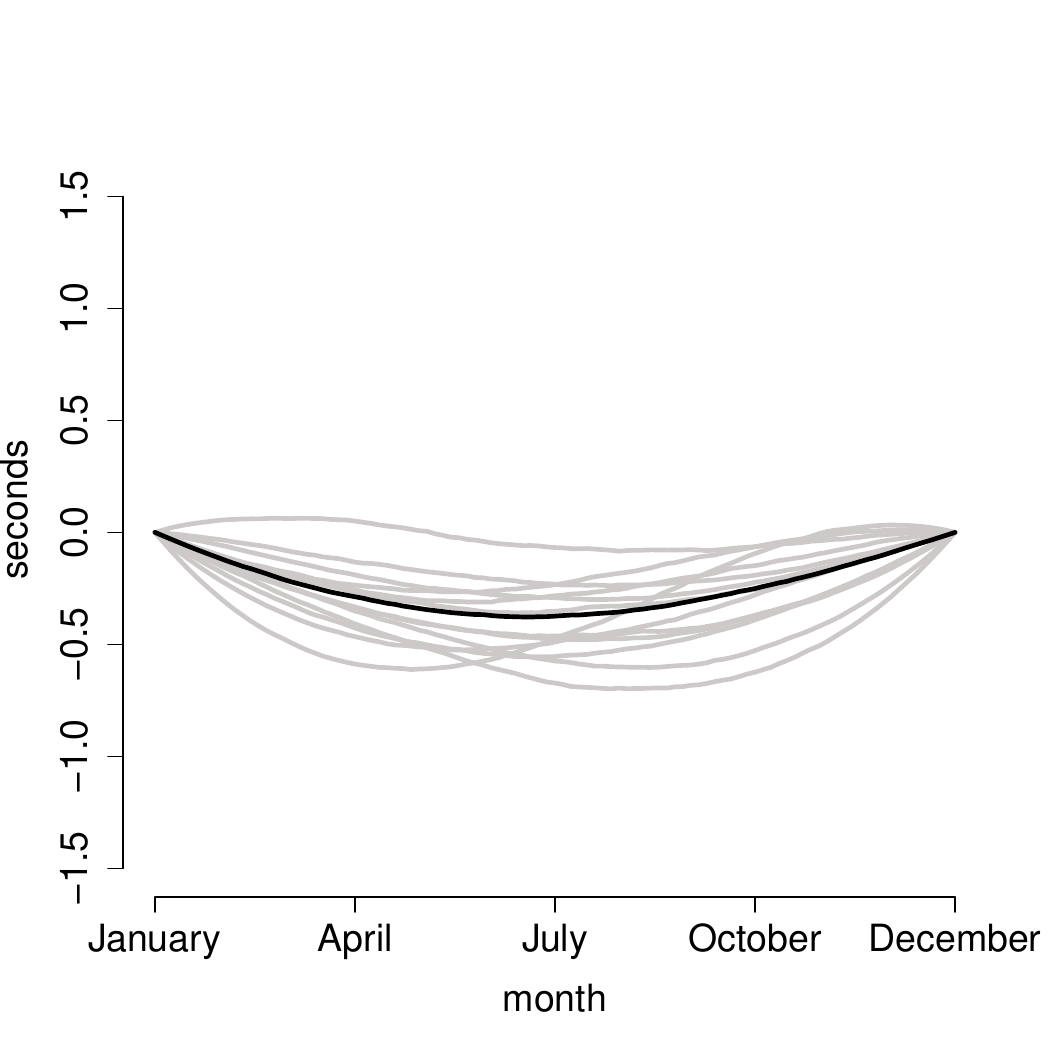 }&
\includegraphics[scale=0.22]{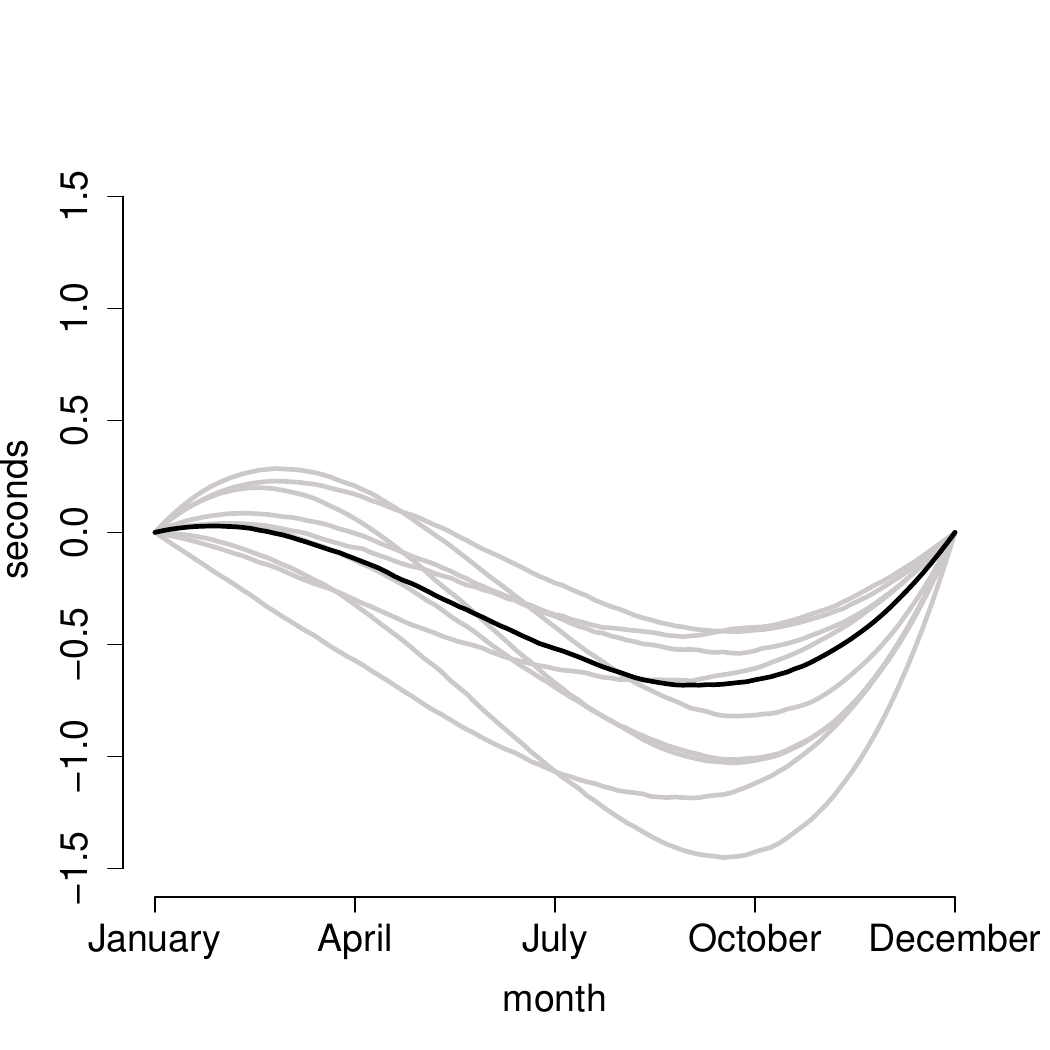}\\
\multicolumn{3}{c}{Male}\\
Swimmer 4 & Swimmer 5 & Swimmer 6\\
\includegraphics[scale=0.22]{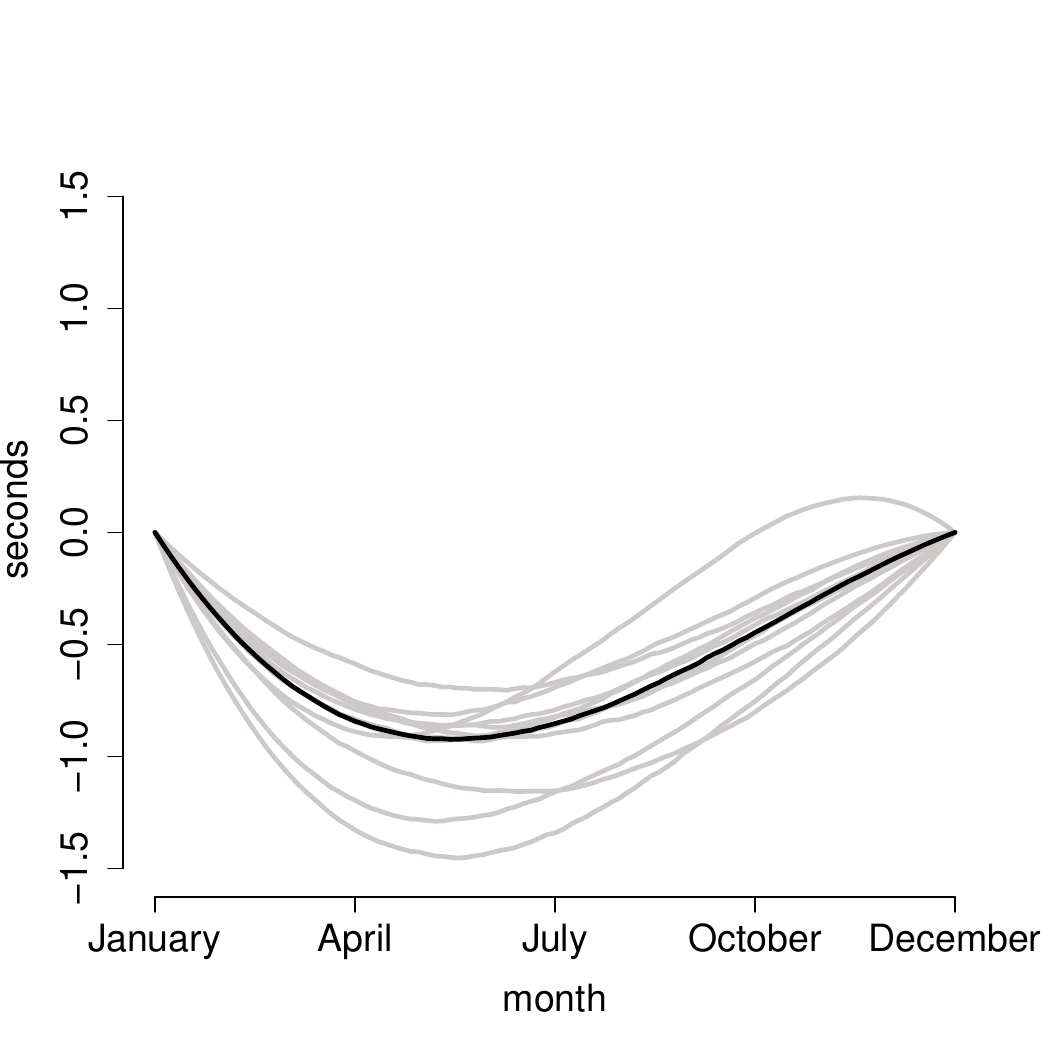}&
\includegraphics[scale=0.22]
{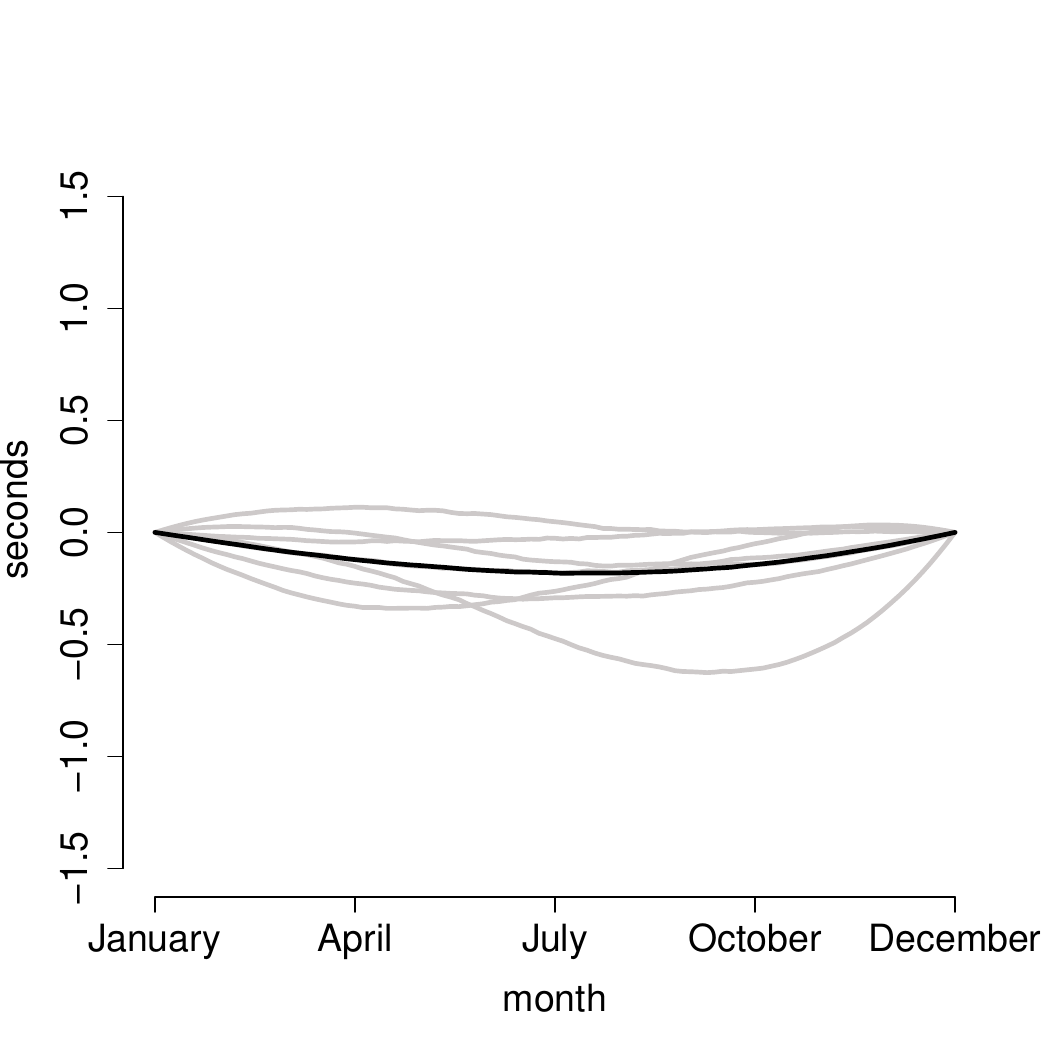}&
\includegraphics[scale=0.22]{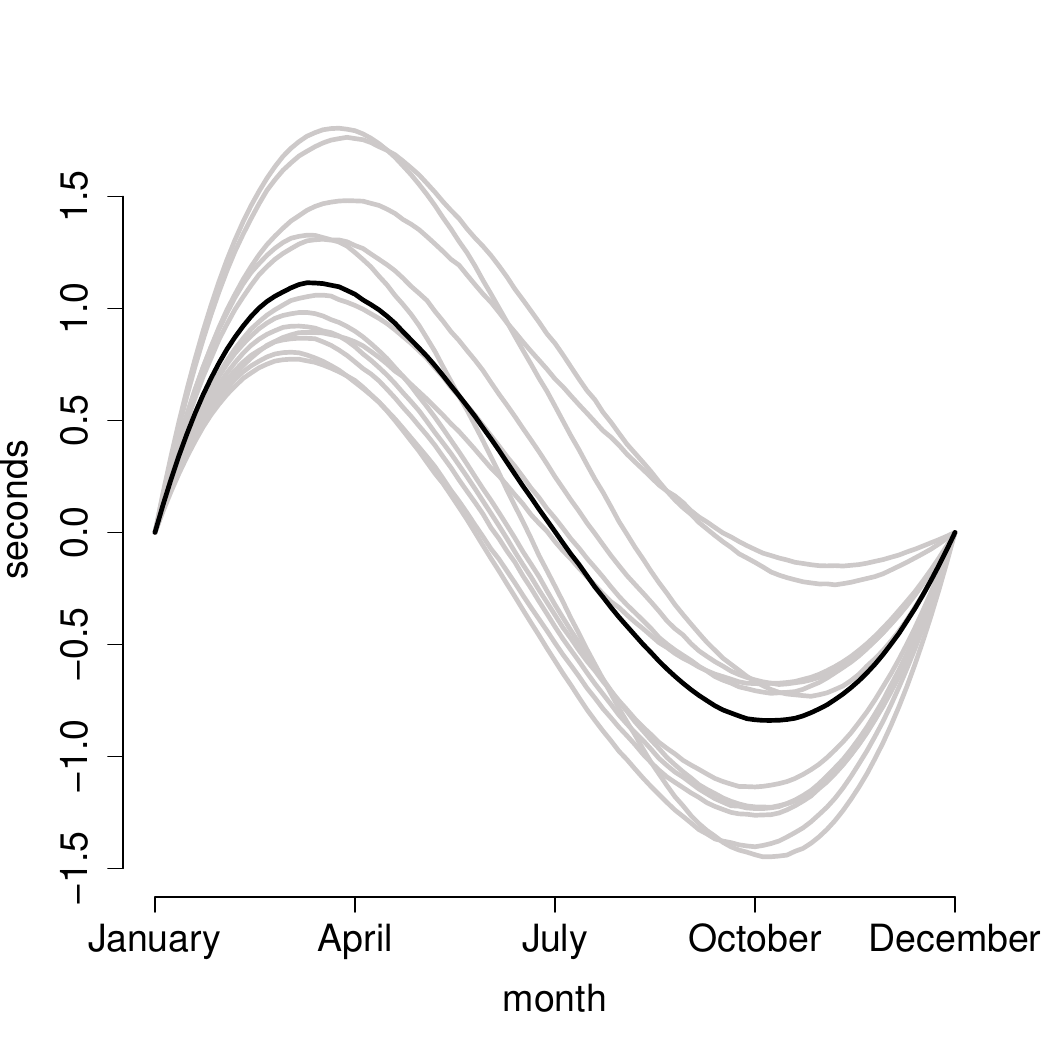}
\end{tabular}
\end{center}
\caption{\label{intra_seasonal} 
Estimated within-season trajectories.
 Posterior median athlete-level (black line) and  within-season performance trajectories for each career season (grey lines).}
\end{figure}




 Figure~\ref{intra_seasonal} shows within-season performance trajectories for six individuals in the 100 metres (three female and three male) with their average within-season performance trajectory.
 The athletes were chosen to show a range of behaviours where some individuals have a substantial difference between  their best and worst performance level within-season (Swimmer 1, 4 and 6) and others show much less variation over the season (Swimmers 2 and 5). 
There are also some clear differences in the shape of the curve. Swimmers 1, 3 and 6 peak in October, whereas Swimmers 2 and 5 peak in July and Swimmer 4 in May. This reflects differences in the aims of athletes. Some athletes will target events such as the Olympics whereas other athletes will target the winter rather than summer season. We can use the posterior median of $\tau_i^2$ to understand how these individual trajectories relate to the population trajectory. Swimmers 5 ($\tau_i^2 = 0.029$), 2 ($\tau_i^2 = 0.089$) and 3 ($\tau_i^2 = 0.175$) are closest to the population trajectory with a similar shape and level of performance improvement. There are larger differences for Swimmers 1 ($\tau_i^2 = 0.288$), who shows a different shape, and 4 ($\tau_i^2 = 0.274$), who shows a larger level of improvement. Swimmer 6 ($\tau_i^2 = 0.803$) shows differences in both the shape of trajectory and the level of performance improvement. 
There are also clear  differences in the consistency of the trajectories across seasons. Swimmers 2 ($\lambda_i^2 = 0.064$), 5 ($\lambda_i^2 = 0.077$) and 
4 ($\lambda_i^2 = 0.086$) show the lowest  level of variation in the seasonal trajectory. Swimmers 3 ($\lambda_i^2 = 0.1243$) and 6 show ($\lambda_i^2 = 0.154$)
show more variation and Swimmer 1 ($\lambda_i^2  = 1.083$) shows the most variation in trajectory.
 This may reflect several factors including changing priorities over an swimmer's career or injury problems.

\begin{figure}[!htbp]
\begin{center}
\begin{tabular}{cccc}
\multicolumn{2}{c}{100 metres} &
\multicolumn{2}{c}{200 metres}\\
 Females &Males  & Females &Males \\
\includegraphics[scale=0.19]{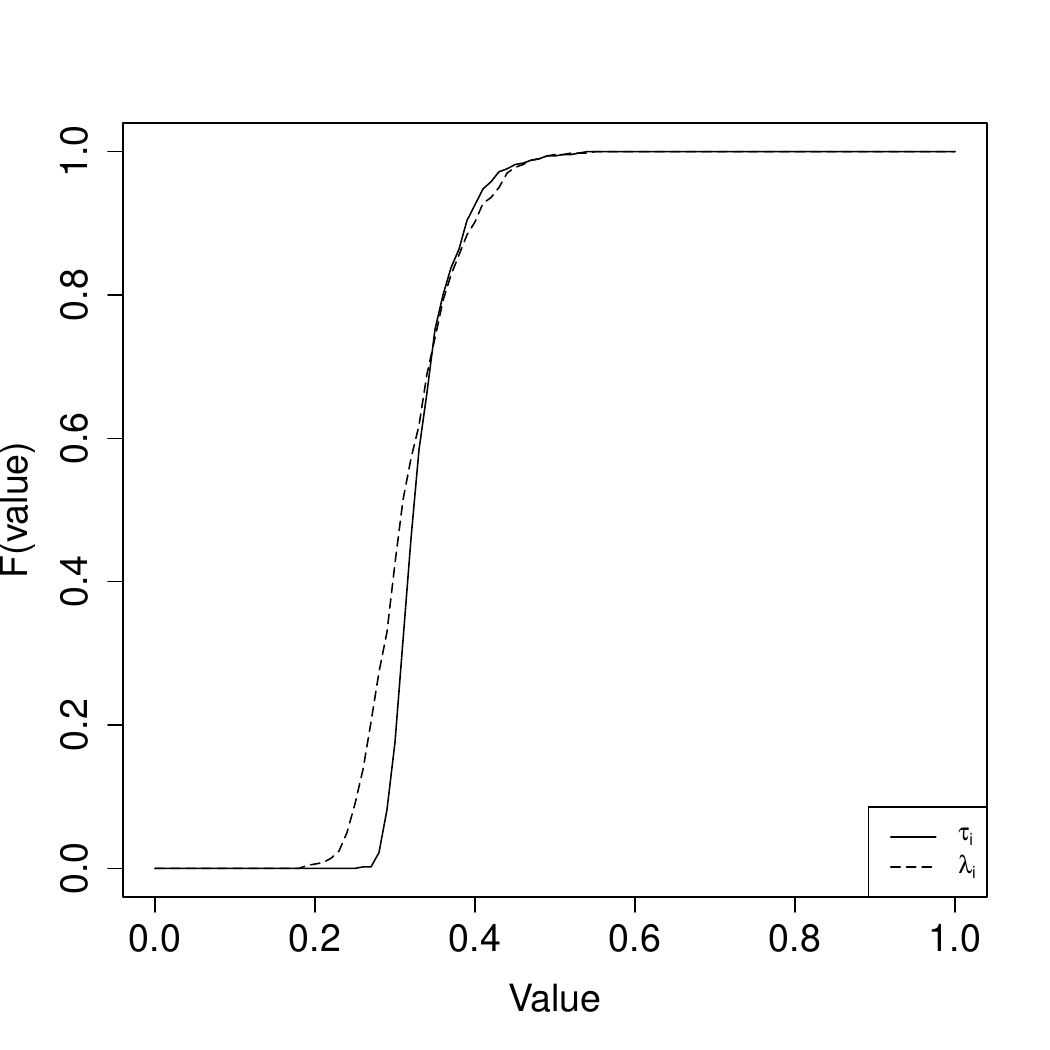}&
\includegraphics[scale=0.19]{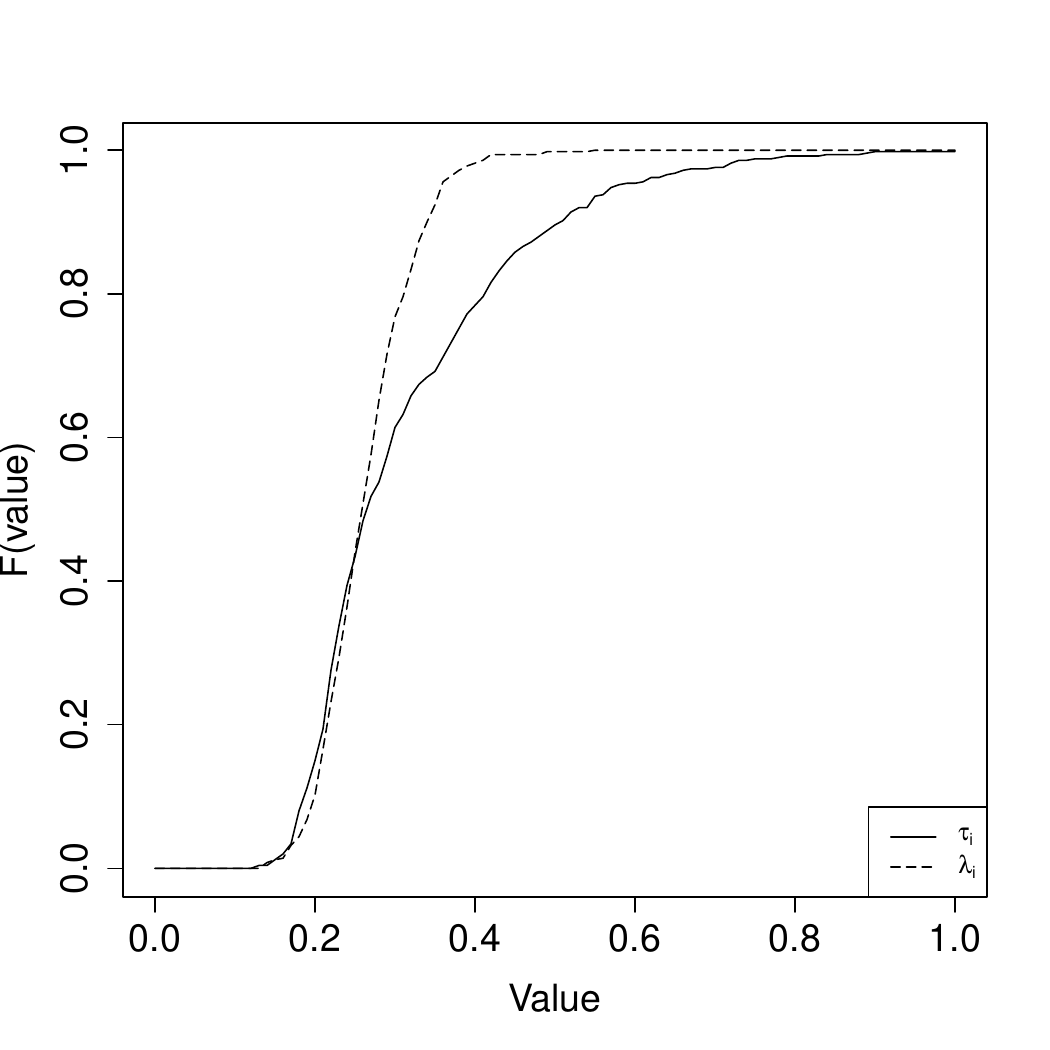} &
\includegraphics[scale=0.19]{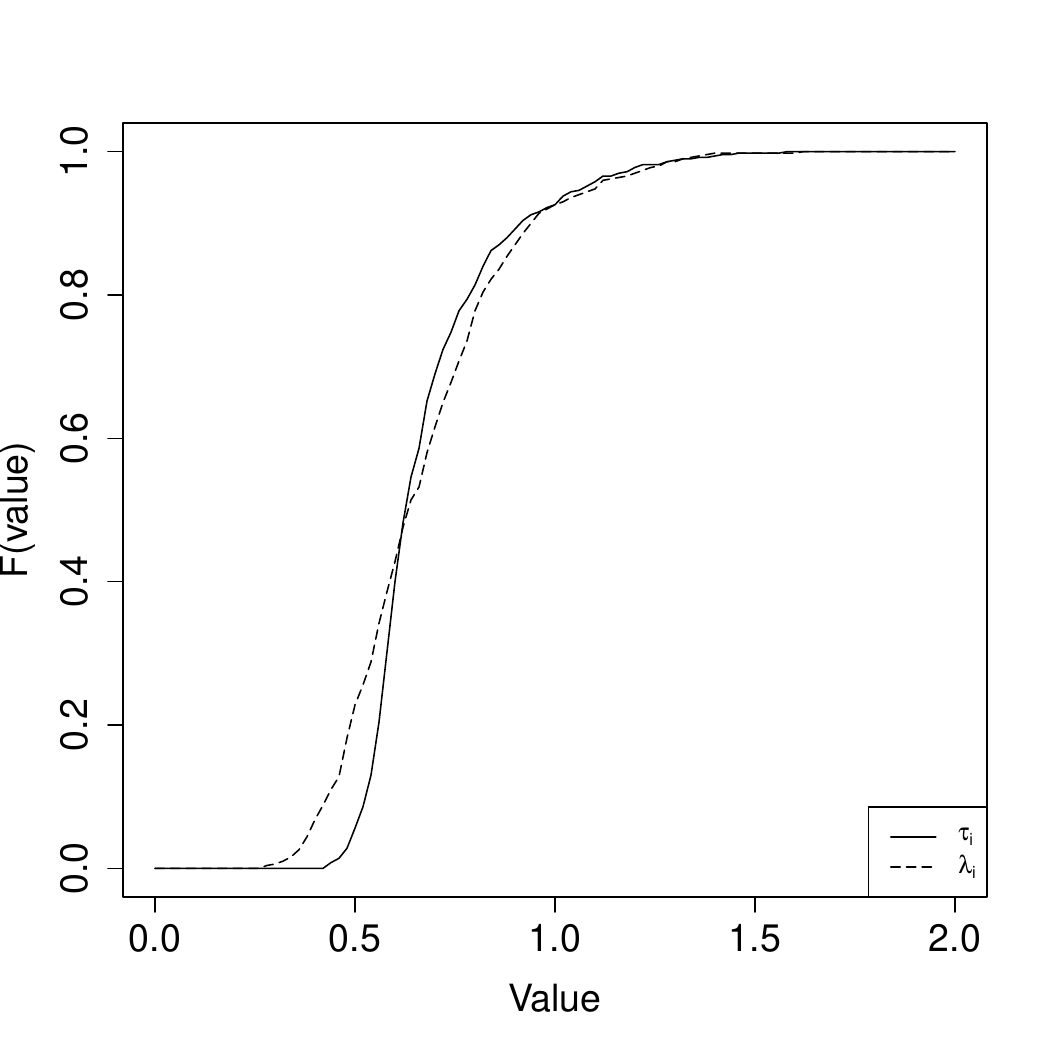}&
\includegraphics[scale=0.19]{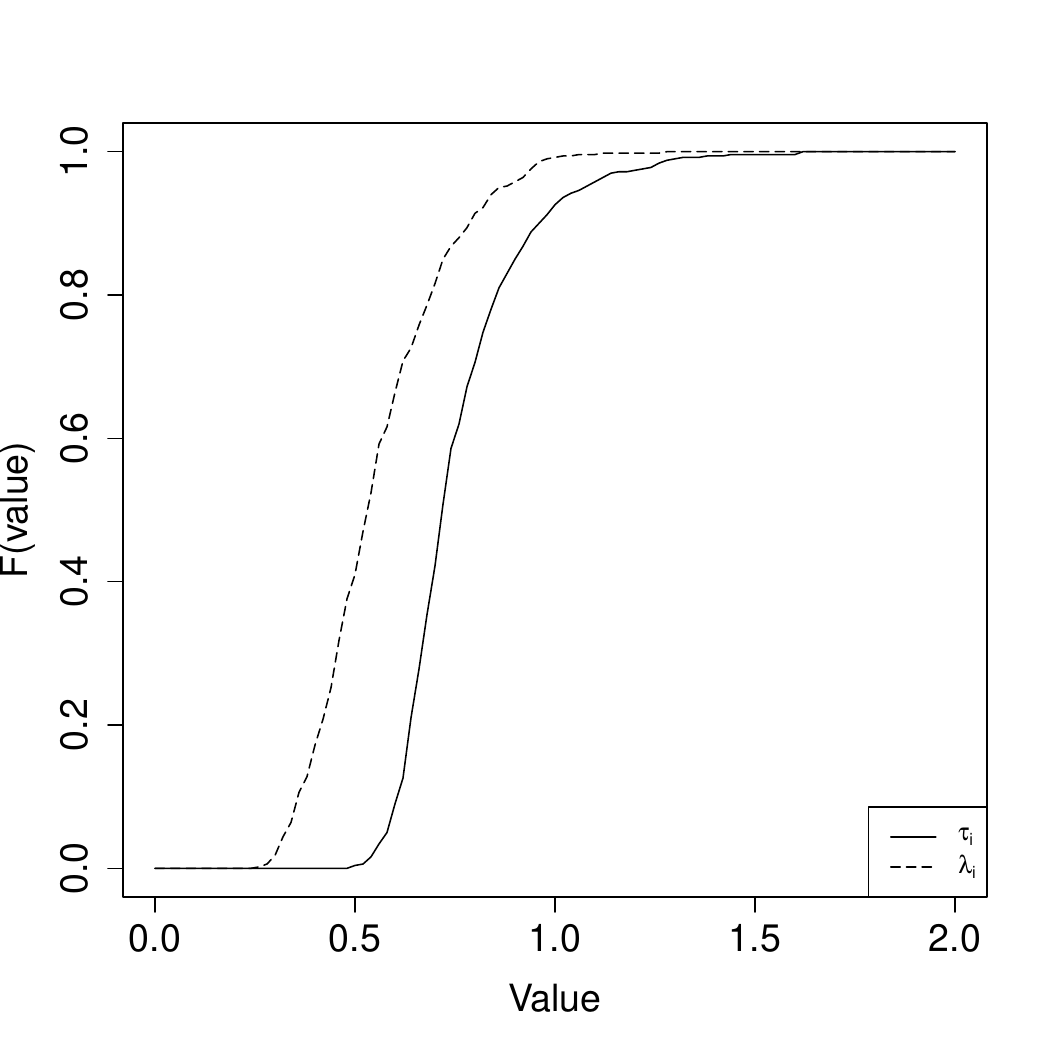}
\end{tabular}
\end{center}
\caption{\label{hist1} Empirical distribution function of the posterior median values of the within-season variability ($\Delta_i$) and the average effect size ($\Gamma_i$) for all swimmers (divided by gender) in the 100 and 200 metre freestyle.}
\end{figure}
Figure~\ref{hist1} shows the empirical distribution function of the 
posterior median values of the individual within-season variability ($\lambda_i$) and the individual between-season variability ($\tau_i$) for both females and males and the two distances. The distribution is similar for both females and males and so we concentrate on the results for female. Both distributions have a very heavy right-tail so that a few effects have much larger values of $\tau_i$ and $\lambda_i$ than others. The distribution of $\tau_i$ is shifted to the left of the distribution of $\lambda_i$ indicating that the within-season variability tends to be smaller than the average effect size. In other words, the season-on-season variability within a swimmer tends to be smaller than the variability between swimmers. This indicates that swimmers have some ability to control how their performance levels evolve over a season through the training process, and are able to replicate that improvement in different seasons.

\section{Discussion} \label{Discussion}
In this work we have developed a Bayesian longitudinal model 
which can account for the variation in performance change within a season across population-level, athlete-level and within-season (i.e. within athlete). An application to freestyle swimming data shows that there is substantial variation between swimmers and between seasons with some having a clear pattern of peaking for major events (e.g. Olympics Games and World Championships) which usually occur during the summer months (July - August). We use a $d$-th degree of polynomial  to model the population-level effect of age and an error distribution which allows for skewness and different heaviness of the left and right tail. We find that the population-level effect of age follows the expected reverse J shape in freestyle swimming with a difference between the improvement in performance of females and males in years 15 to 23. We find that the error distribution has a much lighter left tail than right tail.  The result suggests that swimmers are much less likely to have a performance that is substantially worse than expected, rather than one that is substantially better than expected. One explanation is that elite athletes generally perform close to their optimal level and so improvements are much harder to achieve than poor performances (which can be due to many factors including things such as poor race execution, illness and injury).

The model provides some interesting insights about athlete performance but there are some limitations. An individual's trend excess performance trajectory is assumed to follow a random walk, which is appropriate for  retrospective analysis. This approach may also be able to provide short-term prediction but the lack of structure will struggle to provide good long-term prediction performance. This would need additional structure to explain the evolution of career trajectories, which could include additional covariates. 
The model is currently restricted to a single discipline but athletes compete in multiple disciplines in some sports (across different distances in track running or distances and styles in swimming). It would be interesting to extend this model to  better understand differences in performance trajectories across disciplines for a single athlete.
The model also assumes a single season but the results show that athletes may follow different seasonal patterns. For example, Australian swimmers have different seasons to European swimmers. Our model is able to capture these differences but a more complicated model allowing for different seasonal patterns could provide better estimates.
The model also assumes independence across athletes conditional on the population performance trajectory and population within-season performance trajectory. However, there could be further sharing of information across athletes at other levels of the hierarchy, such as the individual average within-season performance trajectories.    \cite{colombi2025hierarchical} present a
an interesting Bayesian nonparametric approach to achieve this goal, which they use to stratify athletes by performance whilst allowing for difference between athlete and season.

\subsection*{Acknowledgements}

This research was supported by a Partnership for Clean Competition research grant awarded to JH (Grant: 514).

\bibliographystyle{chicago}
\bibliography{sample}

\appendix

\section{Proof of results for RBP}

We can show that
\begin{itemize}
\item $ \int_0^1 b_{n, \nu}(x) \,dx  = \frac{1}{n+1}$ 
\item 
\begin{align*}
\int_0^1 b_{n_1, \nu_1}(x) \,b_{n_2, \nu_2}(x) 
&= \left(\begin{array}{c} 
n_1\\ \nu_1
\end{array}
\right)
\left(\begin{array}{c} 
n_2\\ \nu_2
\end{array}
\right) \int x^{\nu_1 + \nu_2} (1 - x)^{n_1 + n_2 - \nu_1 - \nu_2}\,dx\\
&=
\left(\begin{array}{c} 
n_1\\ \nu_1
\end{array}
\right)
\left(\begin{array}{c} 
n_2\\ \nu_2
\end{array}
\right)
\frac{(\nu_1 + \nu_2)! (n_1 + n_2 - \nu_1 - \nu_2)!}{(n_1 + n_2 + 1)!}
\end{align*}

\end{itemize}

\section{MCMC sampler}\label{MCMC}

\subsection{Matrix form of the model}

We define $G = N(N-1)/2$ to be the total number of coefficient in the RBP. The following linear model representation  of our model is used to derive the Gibbs sampler
\begin{align}
\boldsymbol{y}_{i} &= \boldsymbol{A}_{i}\,\boldsymbol{\delta} + \boldsymbol{X}_{i}\,\boldsymbol{\zeta} +  \boldsymbol{Z}_{i}\,\boldsymbol{F}_i + \boldsymbol{C}_{i}\,\boldsymbol{\beta}_i
+  \frac{\alpha}{\sqrt{1+\alpha^2}} \boldsymbol{\kappa}_{i}
+\boldsymbol{\epsilon}^{\star}_{i} \label{lin_model}
\end{align}
where $\boldsymbol{y}_{i}$ is a $(n_{i} \times 1)$-dimensional column vector of all performances  of the $i$-th athlete, $\boldsymbol{A}_{i}$ is $(n_{i} \times (d+1))$-dimensional matrix whose $k$-th row is the polynomial terms  evaluated at the age, $a_{i, k}$,   $\boldsymbol{\delta}$ is a vector of length $d + 1$,  $\boldsymbol{X}_{i}$  is a $(n_{i} \times p)$-dimensional matrix of  observed confounders, the associated regression coefficients are the $p$-dimensional vector $\boldsymbol{\zeta}$, $\boldsymbol{Z}_{i}$ is a $(n_{i} \times (S_i + 1))$-dimensional matrix whose only non-zero entries in the $k$-th row are $1 - (t_{i, j} - s_{i, j})$ in the $s$-th column and 
$t_{i, j} - s_{i, j}$ in the $(s+1)$-th column and $\boldsymbol{F}_{i} = \left(\eta_{i, 1}, \eta_{i, 2}, \dots, \eta_{i, S_i+1}\right)$,
$\boldsymbol{C}_{i}$ is $(n_{i} \times S_i \, G)$-dimensional matrix, $\boldsymbol\beta_i$ is a $(S_i \, G\times 1)$-dimensional vector which stacks $\boldsymbol\beta^{(i, 1)}, \dots \boldsymbol\beta^{(i, S_i)}$. The only entries in the $k$-th row of $\boldsymbol{C}_i$ are in columns $(s_{i,k}-1)\,G + 1, \dots, s_{i,k}\, G$ and are the values of the Bernstein polynomials evaluated at $t_{i, k} - (s_{i,k} - 1)\Delta$.  This is a hierarchical linear mixed model where
$\boldsymbol{\delta}$ and  $\boldsymbol{\zeta}$  are fixed effects,  
and $\boldsymbol{F}_i$ and $\boldsymbol{\kappa}_i$ are random effects, and 
$\boldsymbol{\beta}_i$ is a random effect with hierarchical structure.
A representation of the model as a directed acyclic graph is given in Figure~\ref{f:DAG}
\begin{figure}[ht]
\centering
\begin{tikzpicture}
[roundnode/.style={circle, draw = black, fill = white},
squarednode/.style={rectangle, draw = black, fill = white, minimum size = 7mm},] 
\node[squarednode] at (0, 0)     (1) {$y_i$}; 
\node[squarednode] at (-2.8, -0.3)     (2) {$(\boldsymbol{A}_i, \boldsymbol{X}_i, \boldsymbol{Z}_i, \boldsymbol{C}_i, \boldsymbol{S}_i)$}; 
\node[squarednode] at (-6.9, 3) (32) {$\boldsymbol{D}$};
\node[roundnode] at (5, 1)     (3) {$\boldsymbol\delta$}; 
\node[roundnode] at (5, 0.15)     (4) {$\boldsymbol\zeta$}; 
\node[roundnode] at (5, -0.75)     (5) {$\alpha$}; 
\node[roundnode] at (-0.05, 3)     (6) {$\boldsymbol{\kappa}_i$}; 
\node[roundnode] at (1, 4.5)     (7) {$\boldsymbol\omega_i$}; 
\node[roundnode] at (-1.7, 1)     (8) {$\boldsymbol\beta^{(i, s)}$}; 
\node[roundnode] at (1.8, 3)     (9) {$\boldsymbol{F}_i$}; 
\node[roundnode] at (1.1, 6.5)     (10) {$\nu_1$}; 
\node[roundnode] at (0, 4.5)     (11) {$\boldsymbol\phi_i$}; 
\node[roundnode] at (0.1, 6.5)     (12) {$\nu_2$}; 
\node[roundnode] at (-3.5, 2)     (13) {$\boldsymbol{\beta^{(i)}}$}; 
\node[roundnode] at (-5.6, 3)     (14) {$\boldsymbol\beta$}; 
\node[roundnode] at (2.8, 4.7)     (15) {$\boldsymbol\omega_i^{\eta}$}; 
\node[roundnode] at (-1.8, 3.5)     (16) {$\sigma^2_i$}; 
\node[roundnode] at (-2.4, 6.5)     (17) {$\sigma^2_a$}; 
\node[roundnode] at (-1.3, 6.5)     (18) {$\sigma^2_m$}; 
\node[roundnode] at (-4, 3.5)     (19) {$\tau^2_i$}; 
\node[roundnode] at (-5.6, 5.5)     (20) {$\tau_0$}; 
\node[roundnode] at (-3.4, 4.5)     (21) {$\lambda^2_i$}; 
\node[roundnode] at (-5.6, 1)     (22) {$\boldsymbol{c}^2$}; 
\node[roundnode] at (-5.6, 2)     (23) {$\boldsymbol{d}^2$}; 
\node[roundnode] at (-5.6, 4.5)     (24) {$\tau_1$}; 
\node[roundnode] at (3.5, 3.45)     (25) {$\omega_i^{\mu}$}; 
\node[roundnode] at (5, 3.8)     (26) {$\nu^{\mu}$}; 
\node[roundnode] at (3.5, 6.5)     (27) {$\nu^{\eta}$}; 
\node[roundnode] at (5, 2.6)     (28) {$\sigma^2_{\mu}$}; 
\node[roundnode] at (2.4, 6.5)     (29) {$\sigma^2_{\eta}$}; 
\node[roundnode] at (-4.7, 6.5)     (30) {$\lambda_0$}; 
\node[roundnode] at (-3.7, 6.5)  (31) {$\lambda_1$}    ; 

\draw[->] (32) -- (14);
\draw[->] (16) -- (6);
\draw[->] (2) -- (1);
\draw[->] (3) -- (1);
\draw[->] (4) -- (1);
\draw[->] (5) -- (1);
\draw[->] (6) -- (1);
\draw[->] (7) -- (1);
\draw[->] (8) -- (1);
\draw[->] (9) -- (1);
\draw[->] (10) -- (7);
\draw[->] (11) -- (6);
\draw[->] (12) -- (11);
\draw[->] (13) -- (8);
\draw[->] (14) -- (13);
\draw[->] (15) -- (9);
\draw[->] (25) -- (9);
\draw[->] (16) -- (1);
\draw[->] (17) -- (16);
\draw[->] (18) -- (16);
\draw[->] (19) -- (13);
\draw[->] (20) -- (19);
\draw[->] (21) -- (8);
\draw[->] (22) -- (8);
\draw[->] (23) -- (13);
\draw[->] (24) -- (19);
\draw[->] (26) -- (25);
\draw[->] (27) -- (15);
\draw[->] (28) -- (9);
\draw[->] (29) -- (9);
\draw[->] (30) -- (21);
\draw[->] (31) -- (21);

\draw[black] (-4.8, -1.5) rectangle (4.25,5.7); 

\node at (-3.6, -1.2) (17) {$i=1,\dots, M$};
\end{tikzpicture}
\caption{Directed acyclic graph of the linear mixed model representation}
\label{f:DAG}
\end{figure}

It is convenient to define $\boldsymbol{W}_{i}$ to be a diagonal matrix whose diagonal elements are the $\omega_{i, k}$'s. For $s = 1, \dots, S_i$, we define the following sub-matrices. The rows of $\boldsymbol{C}^{\star}_{i, s}$ are the rows of $\boldsymbol{C}_{i}$ for which $s_{i, j} = s$. Similarly we form $\boldsymbol{A}^{\star}_{i, s}$ from 
$\boldsymbol{A}_i$, 
$\boldsymbol{X}^{\star}_{i, s}$ from 
$\boldsymbol{X}_i$,
$\boldsymbol{W}^{\star}_{i, s}$ from 
$\boldsymbol{W}_i$,
and 
$\boldsymbol{Z}^{\star}_{i, j}$ from 
$\boldsymbol{Z}_{i, j}$.
The elements of $\boldsymbol{y}^{\star}_{i,s}$ and 
$\boldsymbol{\kappa}^{\star}_{i,s}$
are the $y_{i, j}$'s and $\kappa_{i, j}$'s respectively
for which $s_{i, j} = s$.
Let $\boldsymbol{\Phi}_i$ and $\Psi$ be  $(S_i\times S_i)$-dimensional matrices where $\boldsymbol\Phi_i$ has non-zero elements $\Phi_{i, 1, 1} = 1$,
$\Phi_{i, j, j - 1} = -1$ and $\Phi_{i, j, j} = 1$
for $j = 2, \dots, S_i$, and 
$\boldsymbol\Psi_i$ is diagonal with non-zero elements 
$\omega_i^{\mu}, \omega_{i, 1}^{\eta}, \dots\omega_{i, S_i - 1}^{\eta}$.

The sampler uses a combination of joint updates and interweaving \citep{YuMeng} to achieve good performance. We provide the full conditional distribution if it has a known form,  otherwise, we provide the density of the full conditional. For these parameters, we use adaptive Metropolis-Hastings random walk updates \citep[see][for a review]{GriSte13}. For univariate parameters, we use an adaptive random walk tuned to an acceptance rate of 0.3
\citep{AtRo05}. For multivariate parameters, we initially use an adaptive random walk on each component of the parameter then we switch to the ASWAM algorithm \citep{AtFo10} which proposes from a random walk on whole parameter vector whose covariance matrix is a tuning parameter multiplied by the covariance matrix of the parameters estimated using the previous MCMC samples. We define the Generalized Inverse Gaussian distribution with parameters $\lambda$, $\chi$ and $\psi$, written $\mathcal{GIG}(\lambda, \chi, \psi)$ to have the density
\[
f(x) \propto x^{\lambda-1} \,\exp\left\{-\frac{1}{2}
\left(\frac{\chi}{x} + \psi\,x\right)\right\}.
\]
The steps of the Gibbs sampler are given below 
where $\cdot$ refers to all other parameters in a full conditional distribution.

\subsection*{Jointly update  $\boldsymbol{\beta^{(i)}}$, $\boldsymbol{F}_{i}$ and $\boldsymbol{\beta^{(i, 1)}}, \dots, 
\boldsymbol{\beta^{(i, S_i)}}$  } 

For $i = 1, \dots, M$, we jointly update
$\boldsymbol{\beta^{(i)}}, \boldsymbol{F}_{i}, \boldsymbol{\beta^{(i,1)}}, \dots, 
\boldsymbol{\beta^{(i,S_i)}}$ using the decomposition of the full conditional 
\[
p\left(\left.\boldsymbol{\beta^{(i)}}, \boldsymbol{F}_{i}, \boldsymbol{\beta^{(i,1)}}, \dots,
\boldsymbol{\beta^{(i,S_i)}}
\right\vert \cdot\right) 
= 
p\left(\left.
\boldsymbol{\beta^{(i)}},
\boldsymbol{F}_{i}\right\vert \cdot\right) 
\,
\prod_{j=1}^{S_i}
p\left(\left.\boldsymbol{\beta^{(i,j)}}\right\vert \boldsymbol{\beta^{(i)}}, \boldsymbol{F}_{i}, \cdot
\right).
\]
 Let
\begin{align*}
\boldsymbol{V}_{i,s} &= \sigma_i^2\,\boldsymbol{W}^{\star}_{i,s} + \lambda_{i}^2 \,{\boldsymbol{C}^{\star}_{i,s}}^T \,
\mbox{diag}(c_{1}^2, \dots, c_{Q}^2)\,
{\boldsymbol{C}^{\star}_{i,s}},\\
\boldsymbol{r}_{i,s} &= \boldsymbol{y}^{\star}_{i,s}- \boldsymbol{A}^{\star}_{i,s} \,\boldsymbol\delta - \boldsymbol{X}^{\star}_{i,s}\,\boldsymbol\zeta - \frac{a}{\sqrt{1+a^2}} \,\boldsymbol{\kappa}^{\star}_{i,s},\\
\boldsymbol{U}_{i,s} &= \left[
 \boldsymbol{C}^{\star}_{i,j}\quad \boldsymbol{0}_{n_{i,j}\times(j-1)} \quad \boldsymbol{Z}^{\star}_{i,j} \quad \boldsymbol{0}_{n_{i,j}\times(S_i-j-1)}\right],\\
 P_0 &= 
 \left(\begin{array}{cc}
\tau_i^{-2}\mbox{diag}\left(d^{\star\,-2}_1,\dots, d^{\star\,-2}_G\right) \beta_0\\
\boldsymbol{0}_{S_i \times 1}
\end{array}\right)
 ,\\
 Q_0 &= \left(\begin{array}{cc}
\tau_i^{-2}\mbox{diag}\left(d^{\star\,-2}_1,\dots, d^{\star\,-2}_G\right) 
&\boldsymbol{0}_{G \times S_i}
\\
\boldsymbol{0}_{S_i \times G} & 
\left(\boldsymbol{\Phi}^{-1}\right)^T \boldsymbol\Psi^{-1} \boldsymbol{\Phi}^{-1}
 \end{array}\right)
\end{align*}
The full conditionals are
\[
\left(\begin{array}{c}
 \boldsymbol\beta^{(i)}\\
 \boldsymbol{F}_{i}
 \end{array} \right) \sim \mathcal{N} \left( \boldsymbol{Q}_1^{-1} \boldsymbol{P}_i,  \boldsymbol{Q}_1^{-1} \right)
\]
where $
\boldsymbol{Q}_1 = \boldsymbol{Q}_0 + \sum_{s=1}^{S_i} \boldsymbol{U}_{i,s}^T \boldsymbol{V}_{i,s}^{-1} \boldsymbol{U}_{i,s}$ 
and 
$\boldsymbol{P}_1 = \boldsymbol{P}_0 + \sum_{s=1}^{S_i} \boldsymbol{U}_{i,s}^T \boldsymbol{V}_{i,s}^{-1} \boldsymbol{r}_{i,s}$,
and
\[
\boldsymbol{\beta^{(i,s)}}  \sim
\mathcal{N} \left( \boldsymbol{Q}_2^{-1}\left( {\lambda_i^2}\,\mbox{diag}\left(c_{1}^{-2}, \dots, c_C^{-2}\right)
\boldsymbol{\beta^{(i)}} + \left(\boldsymbol{C}^{\star}_{i,s}\right)^{T} \boldsymbol{W}_{i,s}^{-1} (\boldsymbol{r}^{\star}_{i,s}- \boldsymbol{Z}^{\star}_{i,s} \boldsymbol{F}_{i})\right), \boldsymbol{Q}_2^{-1}  \right), 
\qquad s = 1,\dots, S_i,
\]
and $
\boldsymbol{Q}_2 = \lambda_i^{-2}\,\mbox{diag}\left(c_{1}^{-2}, \dots, c_C^{-2}\right)
+ \boldsymbol{C}_{i,j}^T \boldsymbol{W}_{i,j}^{-1} \boldsymbol{C}_{i,j}$.

\subsection*{Update $\kappa_{i,j}$}

For $j = 1, \dots, n_i,\ i = 1,\dots, M$,
\[ 
\kappa_{i, j} \sim T\mathcal{N}_{[0,\infty)} \left( \frac{\frac{a}{\sqrt{1+a^2}}   \left(y_{i,j}- \boldsymbol{A}_{i,j}\,\boldsymbol\delta - \boldsymbol{X}_{i,j}\,\boldsymbol\zeta - \boldsymbol{Z}_{i,j}\,\boldsymbol{F}_{i}- \boldsymbol{C}_{i,j}\,\boldsymbol{\beta}_i\right) / \omega_{i,j} }{ 1 / \phi_{i,j} + a^2 /
((1+a^2)\omega_{i,j})},  \frac{\sigma_i^2}{ 1 / \phi_{i,j} + a^2 /  ((1+a^2)\omega_{i,j})}  \right).
\]

\subsection*{Jointly update  $\boldsymbol\delta$,   $\boldsymbol\beta$, and $\boldsymbol{F}_1, \dots, \boldsymbol{F}_M$} 

Let $\boldsymbol{h}_{i, s} = \boldsymbol\beta^{(i, j)} -\boldsymbol\beta$ 
then,
we can re-write the model in matrix form as
\[
\boldsymbol{y}^{\star}_{i, s} = \boldsymbol{A}^{\star}_{i, s}\,\boldsymbol{\delta} + \boldsymbol{X}^{\star}_{i, s}\,\boldsymbol{\zeta} +  \boldsymbol{Z}^{\star}_{i, s}\,\boldsymbol{F}_i 
+ \boldsymbol{C}^{\star}_{i, s}\,\boldsymbol{\beta}
+ \boldsymbol{C}^{\star}_{i, s}\,\boldsymbol{h}_{i, s}
+  \frac{\alpha}{\sqrt{1+\alpha^2}} \boldsymbol{\kappa}^{\star}_{i, s}
+\boldsymbol{\epsilon}^{\star}_{i, s} 
\]
where $\epsilon_{i, s}^{\star}$ is a vector whose elements are the $\epsilon^{\star}_{i, j}$ for which $s_{i, j} = s$. 
we jointly update
$ 
\boldsymbol{\delta},
\boldsymbol{\beta}, 
\boldsymbol{F}_{1}, \dots, \boldsymbol{F}_{M}$ using the decomposition of the full conditional 
\[
p\left(\left.
\boldsymbol{\delta},
\boldsymbol{\beta}, 
\boldsymbol{F}_{1}, \dots,
\boldsymbol{F}_{M}
\right\vert \cdot\right) 
= 
p\left(\left.
\boldsymbol{\delta},
\boldsymbol{\beta}
\right\vert \cdot\right) 
\,
\prod_{i=1}^{M}
p\left(\left.\boldsymbol{F}_i \right\vert 
\boldsymbol{\delta},
\boldsymbol{\beta}, \cdot
\right).
\]
The  full conditional distributions are 
\[ 
\left(\begin{array}{c}
\boldsymbol\delta\\
\boldsymbol\beta
\end{array} \right) \sim \mathcal{TN}_{\mathbb{R}^{d+1}\times \boldsymbol{D}^{\star}}\left(\left(\boldsymbol{P}^{\star}\right)^{-1}\boldsymbol{Q}^{\star},(\boldsymbol{P}^{\star})^{-1} \right) \]
where 
\begin{align*}
\boldsymbol{Q}_0^{\star} &=
\left(\begin{array}{cc}
\boldsymbol{0}_{(d+1)\times (d+1)} 
& \boldsymbol{0}_{(d+1)\times G}\\
\boldsymbol{0}_{G \times (d+1)} 
& \mbox{diag}\left(\boldsymbol\sigma^{-2}_\beta\right) 
\end{array}\right)
,\\
\boldsymbol{r}^{\star\star}_{i}& = \boldsymbol{y}^{\star}_{i} - 
\boldsymbol{X}^{\star}_{i}\, \zeta
-
\frac{a}{\sqrt{1+a^2}}\, \boldsymbol{z}^{\star}_{i}-
\boldsymbol{C}^{\star}_{i}\,\boldsymbol{h}_{i},\\
\boldsymbol{U}^{\star}_{i} &= [\boldsymbol{A}^{\star}_{i}\quad \boldsymbol{C}^{\star}_{i}],\\
\boldsymbol{R}_{i} &= 
\boldsymbol{Z}_i^T \,(\boldsymbol{W}_i)^{-1} \,\boldsymbol{U}_i^{\star},\\
\boldsymbol{\Psi}^{\star}_{i} &=  (\boldsymbol{\Phi}_i^{-1})^T\,\boldsymbol\Psi_i^{-1} \,\boldsymbol{\Phi}_i^{-1} + \boldsymbol{Z}_i^T\, (\boldsymbol{W}^{\star}_i)^{-1} \,\boldsymbol{Z}_i, \\
\boldsymbol{T}_{i} &= \boldsymbol{Z}_i^T\, (\boldsymbol{W}^{\star}_i)^{-1} \,\boldsymbol{r}_i^{\star\star},\\
\boldsymbol{P}^{\star} &=  \sum_{i=1}^M \left((\boldsymbol{U}_{i}^{\star})^T \,(\boldsymbol{W}_{i}^{\star})^{-1} \, \boldsymbol{U}_{i}^{\star} - \boldsymbol{R}_i^T \,\boldsymbol{\Psi}_i^{-1}\, \boldsymbol{R}_i \right), \\
\boldsymbol{Q}^{\star} &= \boldsymbol{Q}_0^{\star} + \sum_{i=1}^M \left((\boldsymbol{U}_{i}^{\star})^T \,(\boldsymbol{W}_{i}^{\star})^{-1} \, - \boldsymbol{R}_i^T\, \boldsymbol{\Psi}_i^{-1}\right).
\end{align*}
The use of the re-parameterisation in this interweaving step leads to deterministic updates 
\begin{align*}
\boldsymbol{\beta^{(i,j)}} &\leftarrow \boldsymbol{\beta^{(i,j)}} + \boldsymbol\beta - \boldsymbol\beta_{old},\\
\boldsymbol{\beta^{(i)}} &\leftarrow \boldsymbol{\beta^{(i)}} + \boldsymbol\beta - \boldsymbol\beta_{old}
\end{align*}
where $\boldsymbol\beta_{old}$ represents the value of $\boldsymbol\beta$ before updating in this step.

\subsection*{Update $\sigma_{i}^2$} 
\[ 
\sigma_{i}^{2} \sim \mathcal{IG} \left( \sigma^2_a +  n_{i}, b^{\star}\right) 
\]
where 
\[
b^{\star} = \frac{\sigma^2_a}{\sigma^2_m} +\frac{1}{2} \sum_{j=1}^{n_i}  \left [  \frac{\left(y_{i,j}- 
\boldsymbol{A}_{i,j}\boldsymbol\delta - \boldsymbol{X}_{i,j}\boldsymbol\zeta - \boldsymbol{Z}_{i,j} \boldsymbol{F}_{i} - \boldsymbol{C}_{i,j}\boldsymbol\beta_i - \frac{a}{\sqrt{1+a^2}} \kappa_{i,j}\right)^2}{\omega_{i,j}} + \frac{\kappa_{i,j}^2}{\phi_{i,j}}\right].
\]

\subsection*{Update $\sigma^2_a$}

The full conditional density is proportional to
\[ 
\left(\sigma^2_a \right)^{M \sigma^2_a-0.001-1} 
\Gamma(\sigma^2_a) ^{-M} \left( \prod_{i=1}^M \sigma_i^{-2} \right)^{\sigma^2_a} \exp \left \{ - \left(0.001+\frac{1}{\sigma^2_m} \sum_{i=1}^M \sigma_i^{-2} \right) \sigma^2_a \right\}.
\]

\subsection*{Update $\sigma^2_m$}  

\[
\sigma^2_m \sim \mathcal{IG} \left(0.001 + M\sigma^2_a, 0.001 + \sum_{i=1}^M \left(\frac{\sigma^2_a}{\sigma_i^2} \right)  \right)  \]

\subsection*{Update $\lambda_i^2$ and $c^2_{n, v}$} 

\[
\lambda_{i}^{2} \sim\mathcal{GIG}\left(\lambda_0 -
\frac{ G\,S_i}{2},  \sum_{s=1}^{S_i} 
 \sum_{n=2}^N \sum_{v=1}^{n-1}
\frac{\left(\beta^{(i,s)}_{n, v} - \beta_{n, v}^{(i)}\right)^2}
{c_{n, v}^2}, \frac{\lambda_0}{\lambda_1}\right)\qquad 
i = 1,\dots, M
\]
and
\[
c^2_{n, v} \sim\mathcal{IG}\left(5 + \frac{\sum_{i=1}^M S_i}{2}, 5 + \frac{1}{2} \sum_{i=1}^M \sum_{s=1}^{S_i}\frac{ \left(\beta^{(i,s)}_{n, v} - \beta^{(i)}_{n, v}\right)^2}{\lambda_i^2}\right),
 \qquad 
 v = 1, \dots, n - 1,\
n = 2,\dots, N.
\]

We use a working parameter $m_1^{\star}$ where we update $m_1^{\star}$ by
\[
m_1^{\star} \sim \mathcal{IG}(\alpha^{\star}, \beta^{\star})
\]
where 
\[
\alpha^{\star} = 0.001 + M\, \lambda_0 + 5\,G
\]
\[
\beta^{\star} = 0.001 + \lambda_1  \,m_1^{\star}\,
\sum_{i=1}^M \lambda_i^2  + 5\,m^{\star}_1\, \sum_{n=1}^{N}
\sum_{v=1}^{n-1} \frac{1}{c_{n, v}^2}.
\]
$\lambda_i\leftarrow \frac{\lambda_{i, old}}{m^{\star}_1}$ for $i = 1,\dots, M$
 and $c_{n, v} \leftarrow c_{n, v, old} \, m_1^{\star}$ for 
$v = 1, \dots, n - 1$ and $n = 2,\dots, N$.

\subsection*{Update $\tau_i^2$ and $d^2_{n, v}$} 

\[
\tau_i^2 \sim \mathcal{GIG} \left(\tau_0 - \frac{G}{2}, \sum_{n=2}^N \sum_{v=1}^{n-1}\frac{\left(\beta^{(i)}_{n, v}- \beta_{n, v}\right)^2}{d_{j, n}^{2}}, 2\,\frac{\tau_0}{\tau_1}\right),\qquad i = 1,\dots, M
\]
and
\begin{align*}
d^2_{n, v} \sim\mathcal{IG}\left(
5+\frac{M}{2}, 5 + \frac{1}{2}\sum_{i=1}^M\frac{(\beta^{(i)}_{n, v} - \beta_{n, v})^2}{\tau_i^2} \right), \qquad 
 v = 1, \dots, n - 1,\
n = 2,\dots, N.
\end{align*}
We use a working parameter $m_2^{\star}$ where we update $m_2^{\star}$ by
\[
m_2^{\star} \sim \mathcal{IG}(\alpha^{\star}, \beta^{\star})
\]
where 
\[
\alpha^{\star} = 0.001 + M\, \tau_0 + 5\,G
\]
\[
\beta^{\star} = 0.001 + \tau_1  \,m_2^{\star}\,
\sum_{i=1}^M \tau_i^2  + 5\,m_2^{\star}\, \sum_{n=1}^{N}
\sum_{v=1}^{n-1} \frac{1}{d_{n, v}^2}.
\]
$\tau_i\leftarrow \frac{\tau_{i, old}}{m_2^{\star}}$ for $i = 1,\dots, M$
 and $d_{n, v} \leftarrow d_{n, v, old} \, m_2^{\star}$ for 
$v = 1, \dots, n - 1$ and $n = 2,\dots, N$.

\subsection*{Update $\lambda_0$}

The full conditional density is proportional to
\[
\left(
\frac{\lambda_0}{\lambda_1}
\right)^{M \lambda_0}
(\Gamma(\lambda_0))^{-M}
\left(\prod_{i=1}^M \lambda_i^2\right)^{\lambda_0}
\exp\left\{- \lambda_0 \left[1 + 
   - \frac{1}{\lambda_1}  \sum_{i=1}^M \lambda_i^2\right]\right\}
\]

\subsection*{Update $\lambda_1$}

\[
\lambda_1 \sim \mathcal{IG}\left(0.001 + M\,\lambda_0,  0.001 + \lambda_0 \,\sum_{i=1}^M \lambda_i^2\right)
\]

\subsection*{Update $\tau_0$}

The full conditional density is proportional to
\[
\left(
\frac{\tau_0}{\tau_1}
\right)^{M \tau_0}
(\Gamma(\tau_0))^{-M}
\left(\prod_{i=1}^M \tau_i^2\right)^{\tau_0}
\exp\left\{- \tau_0 \left[1 + 
   - \frac{1}{\tau_1}  \sum_{i=1}^M \tau_i^2\right]\right\}
\]

\subsection*{Update $\tau_1$}

\[
\tau_1 \sim \mathcal{IG}\left(0.001 + M\,\tau_0,  0.001 + \tau_0 \,\sum_{i=1}^M \tau_i^2\right)
\]

\subsection*{Update $\boldsymbol\sigma^2_\beta$ } 

\[
\left(\sigma^2_\beta\right)_i \sim \mathcal{IG} \left(\frac{11}{2}, 5 + \frac{\beta_i^2}{2}  \right), \dots i = 1,\dots, G. 
\]

\subsection*{Update $\sigma^2_{\mu}$}
\[
\sigma^{2}_{\mu}\sim\mathcal{IG}\left(
0.001 + \frac{M}{2}, 0.001 + \frac{1}{2}\sum_{i=1}^M \frac{F_{i, 1}^2}{\omega_i^{\mu}}
\right)
\]

\subsection*{Update $\sigma^2_{\eta}$}
\[
\sigma^{2}_{\eta}\sim\mathcal{IG}\left(0.001 +
\frac{1}{2}\sum_{i=1}^M (S_i - 1), 0.001 +
\frac{1}{2}\sum_{i=1}^M \sum_{s=1}^{S_i-1}\frac{ (F_{i,s+1} - F_{i, s})^2
}{\omega^{\eta}_{i,s}}\right)
\] 

\subsection*{Jointly update $\nu_1$ and $\omega_{i, j}$}

We first update $\nu_1$ using an adaptive Metropolis-Hastings random walk and then update $\omega_{i, j}$ for all possible values of $i$ and $j$.
The full conditional density of $\nu_1$ is proportional to
\begin{align*}
&\left[\left(\frac{\nu_1}{2}\right)^{\nu_1/2}\frac{\Gamma\left(\frac{\nu_1+1}{2}\right)}{\Gamma\left(\nu_1/2\right)}\right]^{\sum_{i=1}^M \sum_{s=1}^{S_i} n_{i,s}}
\nu_1\exp \left\{ -0.1\,\nu_1 \right\} \\  &\prod_{i=1}^M
\prod_{j=1}^{n_{i}}
\left(\frac{ \left(y_{i,j}- \boldsymbol{A}_{i,j}\boldsymbol\delta - \boldsymbol{X}_{i,j}\boldsymbol\zeta - \boldsymbol{Z}_{i,j}\boldsymbol{F}_{i}- \boldsymbol{C}_{i,j}\boldsymbol\beta_i-\frac{a}{\sqrt{1+a^2}} \kappa_{i,j}\right)^2 }{\sigma_i^2}+ \nu_1\right)^{-(\nu_1+1)/2}.
\end{align*}
For $i = 1,\dots, M$ and $j = 1,\dots, n_i$,
sample
\[ 
\omega_{i,j} \sim \mathcal{IG} \left( 
\frac{1}{2}(\nu_1+1), \frac{1}{2}\left(\frac{ \left(y_{i,j} - \boldsymbol{A}_{i,j}\boldsymbol\delta - \boldsymbol{X}_{i,j}\boldsymbol\zeta - \boldsymbol{Z}_{i,j}\boldsymbol{F}_{i}- \boldsymbol{C}_{i,j}\boldsymbol\beta_i -\frac{a}{\sqrt{1+a^2}} \kappa_{i,j}\right)^2 }{\sigma_i^2}+ \nu_1\right)\right).
\]

\subsection*{Update $\nu_2$ and $\phi_{i, j}$}

We first update $\nu_2$ using an adaptive Metropolis-Hastings random walk and then update $\phi_{i,s,k}$ for all possible values of $i$ and $j$.
The full conditional density of $\nu_2$ is proportional to
\[
\left[\left(\frac{\nu_2}{2}\right)^{\nu_2/2}\frac{\Gamma\left((\nu_2+1)/2\right)}{\Gamma(\nu_2/2)}\right]^{\sum_{i=1}^M \sum_{s=1}^{S_i} n_{i, s}}
 \nu_2 \exp \left\{-0.1\,\nu_2 \right\} 
 \prod_{i=1}^M \prod_{j=1}^{n_i}
 \left(\frac{1}{2}\left(\nu_2 + \frac{\kappa_{i,j}^2}{\sigma^2_i}
  \right)
 \right)^{-(\nu_2+1)/2}.
\]
For  $i = 1,\dots, M$, and $j = 1,\dots, n_i$, sample
\[
\phi_{i,j} \sim \mathcal{IG} \left(\frac{1}{2}(\nu_2+1), \frac{1}{2} \left(\frac{\kappa_{i,j}^2}{\sigma_i^2}+\nu_2\right) \right).
\]

\subsection*{Update $\alpha$} 

The full conditional density is proportional to
\[ 
\exp \left \{-\frac{1}{2} \left[\frac{\left( y_{i,j}- \boldsymbol{A}_{i,j}\,\boldsymbol\delta - \boldsymbol{X}_{i,j}\,\boldsymbol\zeta - \boldsymbol{Z}_{i,j}\,\boldsymbol{F}_{i}- \boldsymbol{C}_{i,j}\,\boldsymbol{\beta}_i - \frac{\alpha}{\sqrt{1+\alpha^2}} \kappa_{i,j} \right)^2  }{\sigma_i^2 \,\omega_{i,j}} + 
\frac{\alpha^2}{3^2} \right] \right\}.
\]

\subsection*{Update $\nu^{\mu}$ and $\omega_1^{\mu}, \dots, \omega_M^{\mu}$}

We first update $\nu^{\mu}$ using an adaptive Metropolis-Hastings random walk and then update $\omega_{i}^{\mu}$ for  $i = 1,\dots, M$.
The full conditional density of $\nu^{\mu}$ is proportional to
\[
\left(\nu^{\mu}\right)^{M\nu^{\mu}/2 + 1}
  \exp \left\{-0.1\,\nu^{\mu} \right\} \left[\frac{\Gamma\left((\nu^{\mu}+1) / 2\right)}{\Gamma(\nu^{\mu}/2)}
 \right]^ {M}
 \prod_{i=1}^M
 \left(\nu^{\mu} + \frac{F_{i,1}  ^2}{\sigma^2_\mu}
 \right)^{-(\nu^{\mu}+1)/2}.
 \]
For $i=1,\dots, M$, sample
\[
\omega_i^{\mu}\sim\mathcal{IG}
\left(\frac{1}{2}(\nu^{\mu} + 1), \frac{1}{2}\left(\nu^{\mu} + \frac{F_{i, 1}^2}{\sigma^2_{\mu}}\right)\right).
\]

\subsection*{Update $\nu^{\eta}$ and $\omega^{\eta}_{1, 1}, \dots, \omega^{\eta}_{1, S_i-1}, \dots, \omega^{\eta}_{M, 1}, \dots, \omega^{\eta}_{M, S_M-1}$}

We first update $\nu^{\eta}$ using an adaptive Metropolis-Hastings random walk and then update $\omega_{i}^{\eta}$ for all possible values of $i$ and $s$.
The full conditional density of $\nu^{\mu}$ is proportional to
\begin{align*}
\nu_\eta^{\nu_\eta\,\sum_{i=1}^M (S_i-1) / 2 + 1}
 \exp \left\{-0.1\,\nu_\eta \right\} \left[\frac{\Gamma\left((\nu_\eta+1) / 2\right)}{\Gamma(\nu_\eta/2)}
 \right]^ {\sum_{i=1}^M (S_i - 1)}
\prod_{i=1}^M \prod_{s=1}^{S_i-1}
 \left(\nu_\eta + \frac{(F_{i,s + 1}-F_{i, s})^2}{\sigma^2_\eta}
 \right)^{-(\nu_\eta+1)/2}.
\end{align*}
For $i = 1,\dots, M$ and $s = 1,\dots, S_i - 1$, sample
\[
\omega^{\eta}_{i, s} \sim \mathcal{IG}\left(\frac{\nu^{\eta}+1}{2}, \frac{1}{2} \left( \nu^{\eta} +  \frac{(F_{i, s + 1} - F_{i, s})^2}{\sigma^2_\eta} \right)\right).
\]

\section{Exploratory Data Analysis}

\begin{figure}[h!]
\begin{center}
\begin{tabular}{ccc}
& (a) & (b) \\
\raisebox{0.9in}{100m Female} &
\includegraphics[scale = 0.25]{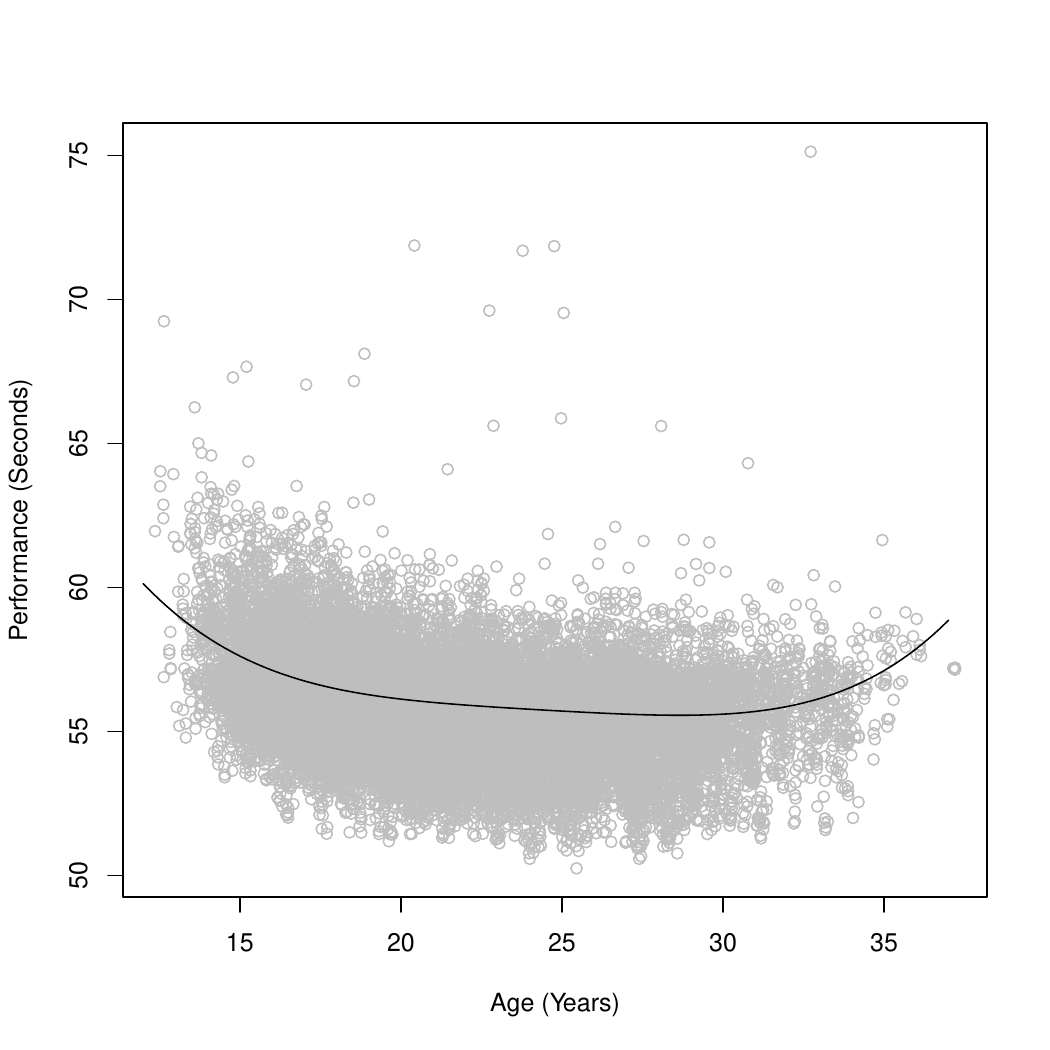}
& 
\includegraphics[scale = 0.25]{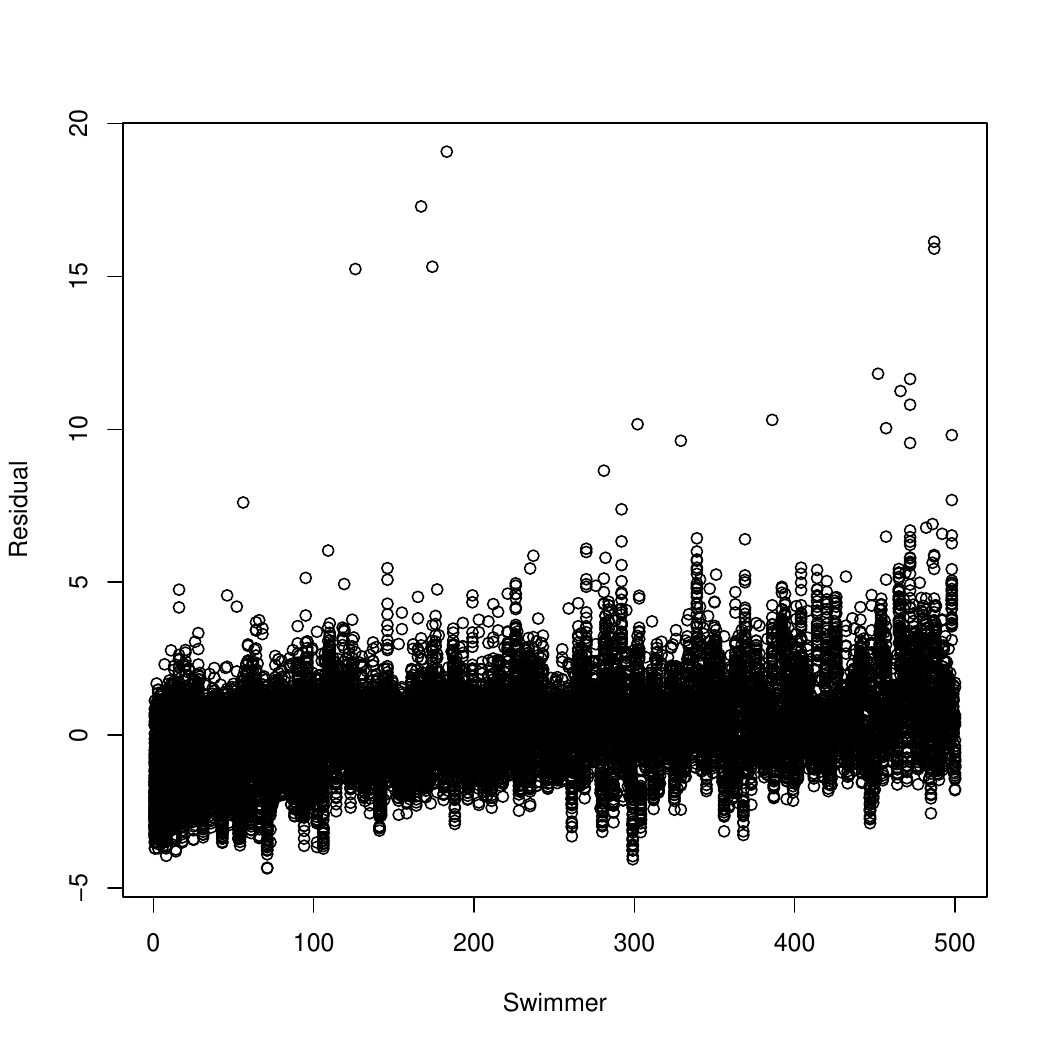}
\\
\raisebox{0.9in}{100m Male} &
\includegraphics[scale = 0.25]{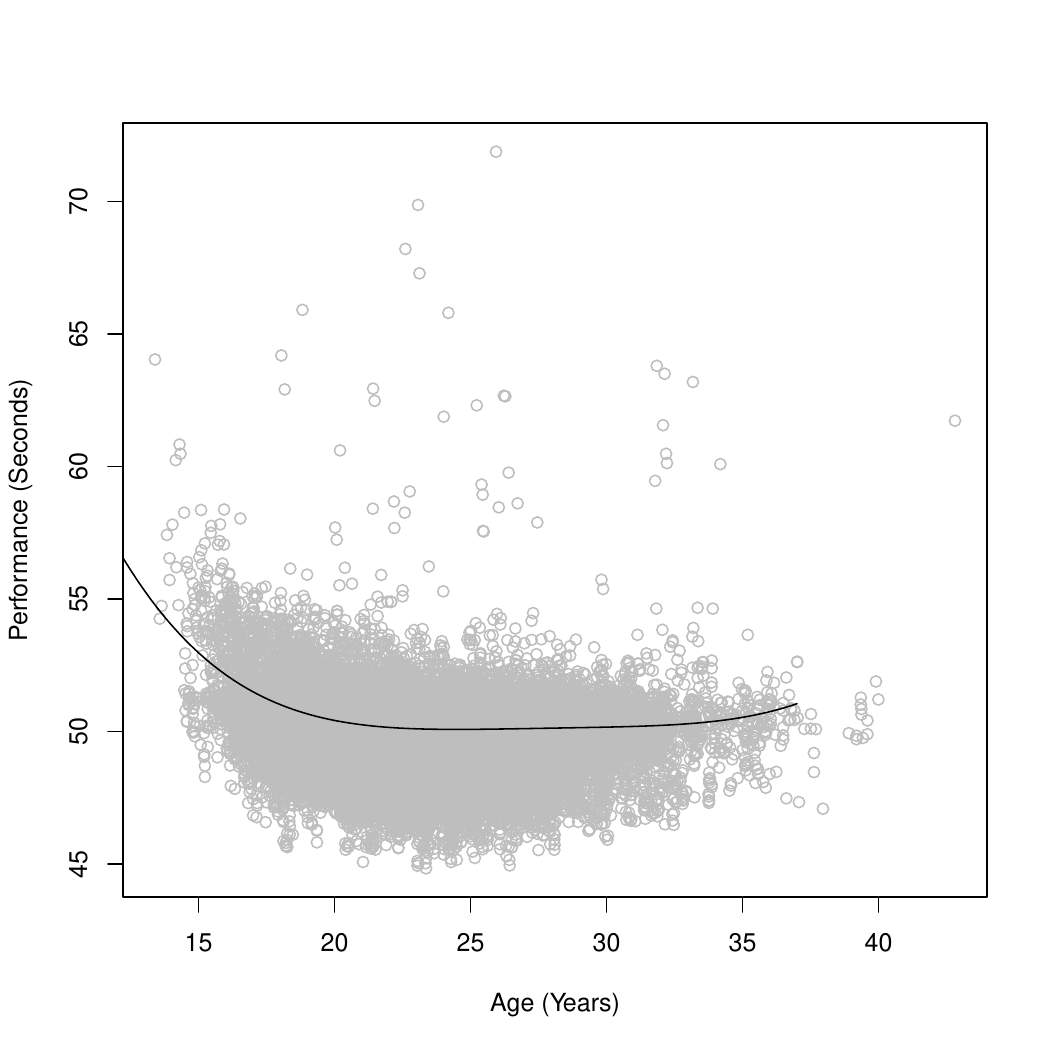}
& 
\includegraphics[scale = 0.25]{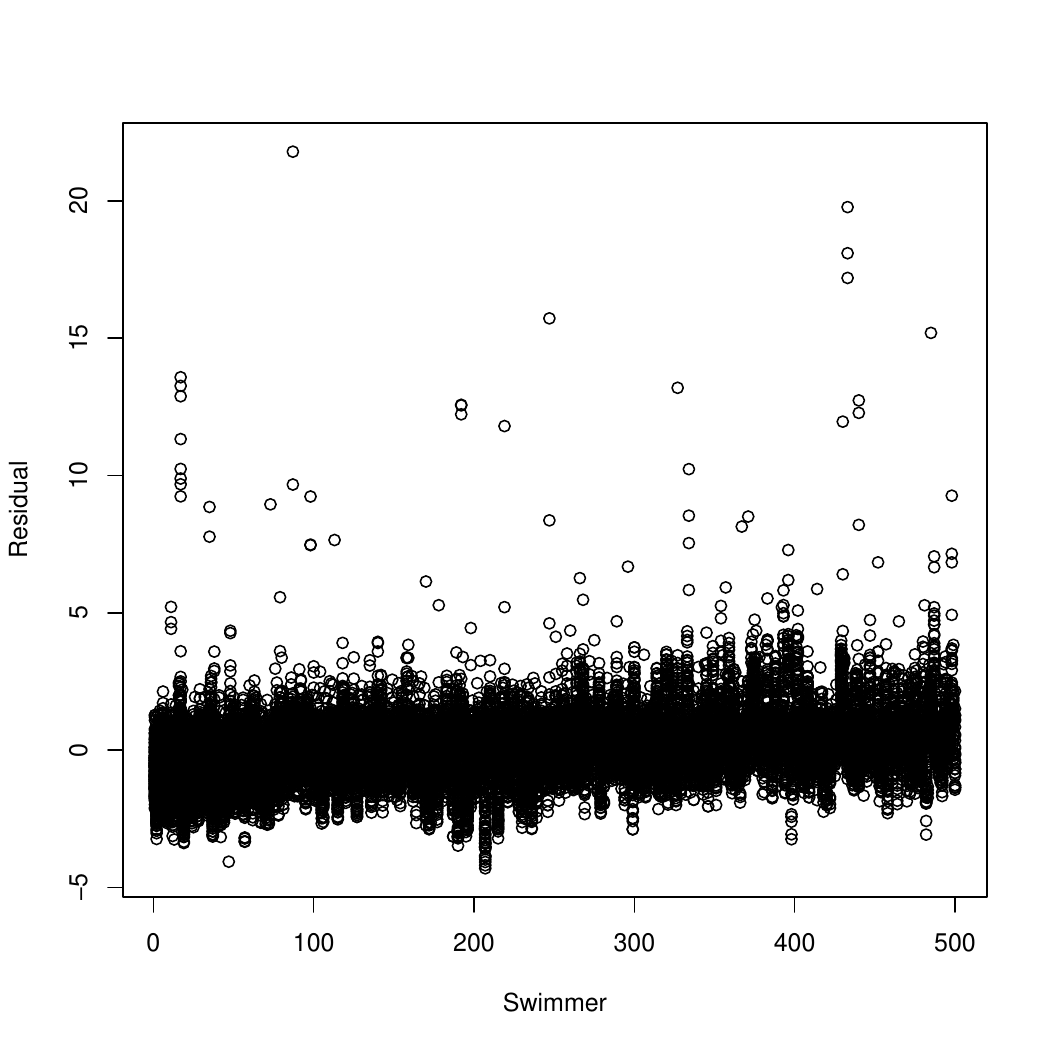}\\
\raisebox{0.9in}{200m Female} &
\includegraphics[scale = 0.25]
{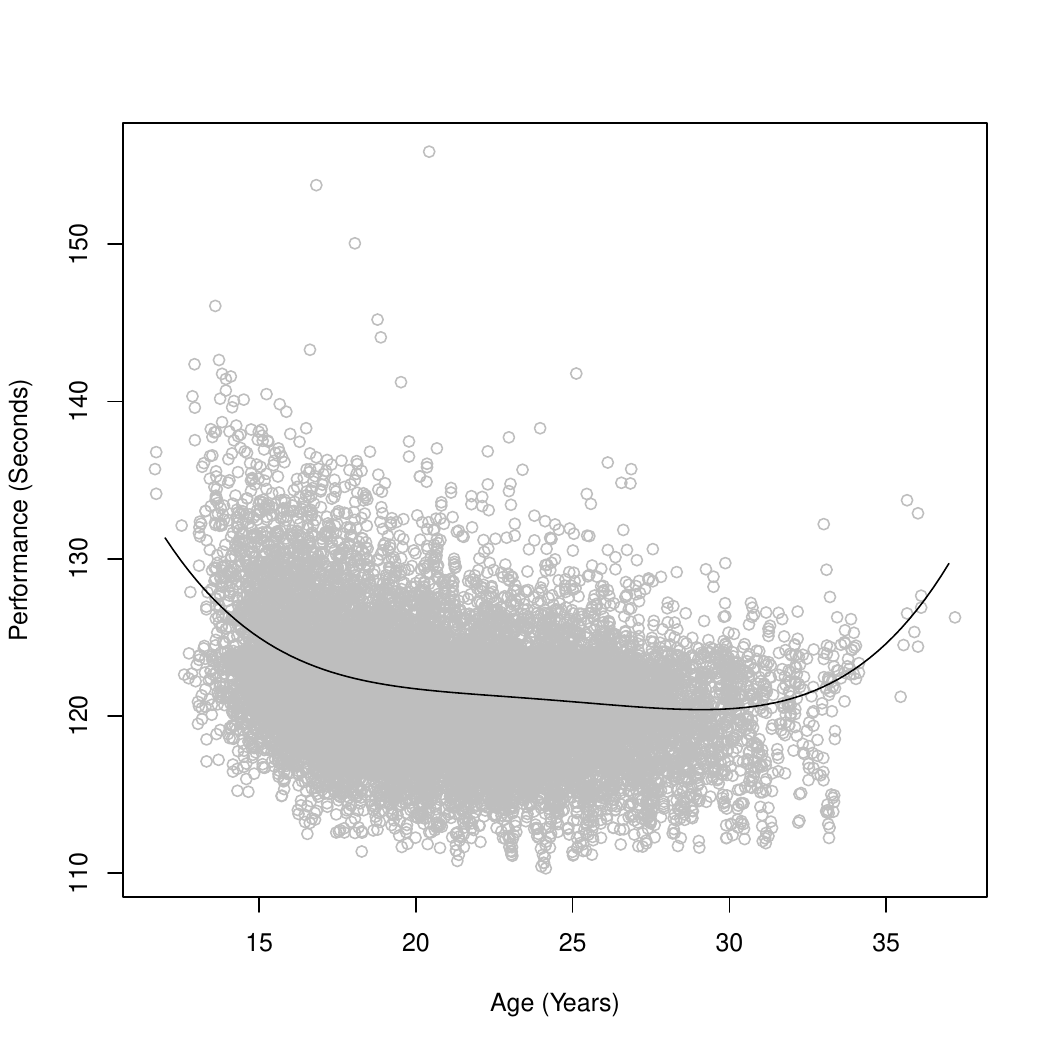}
& 
\includegraphics[scale = 0.25]{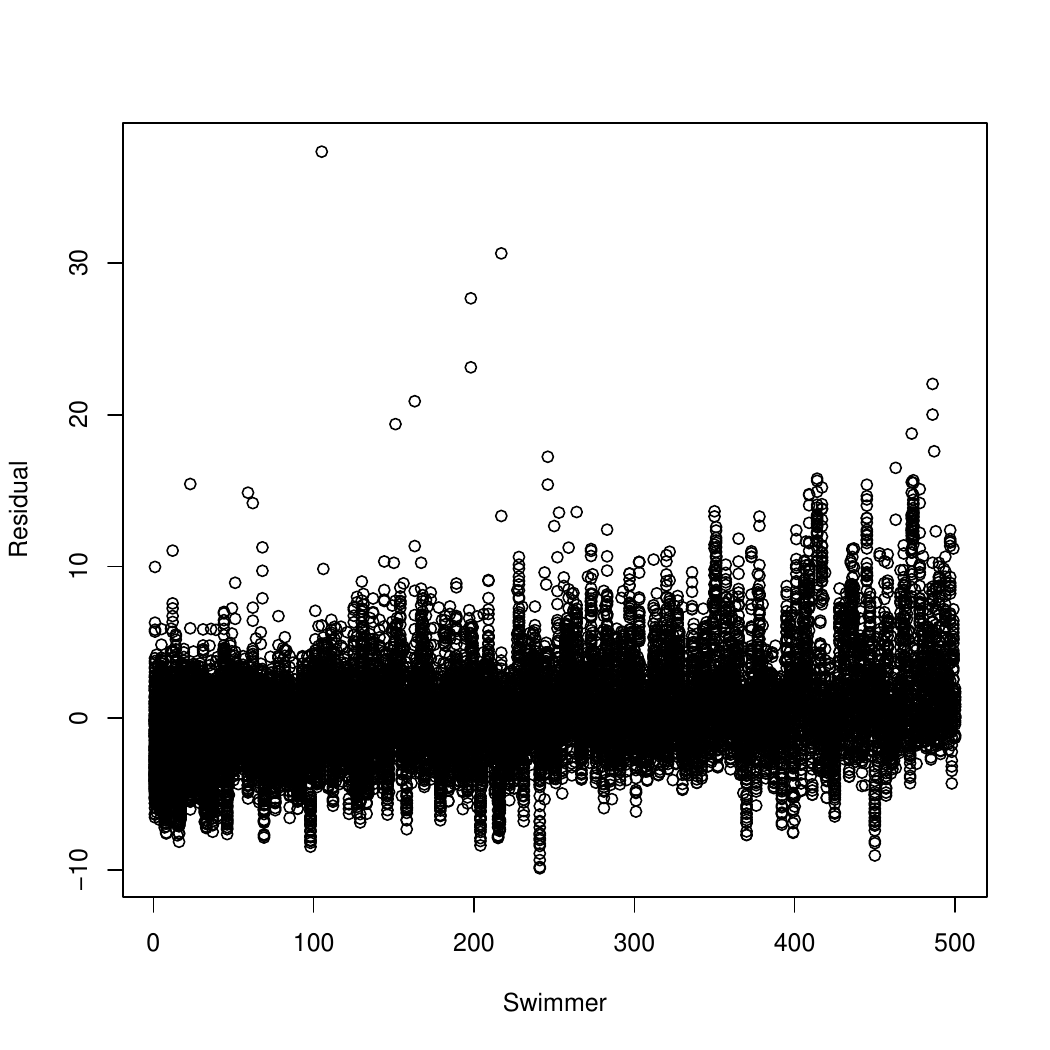}
\\
\raisebox{0.9in}{200m Male} &
\includegraphics[scale = 0.25]{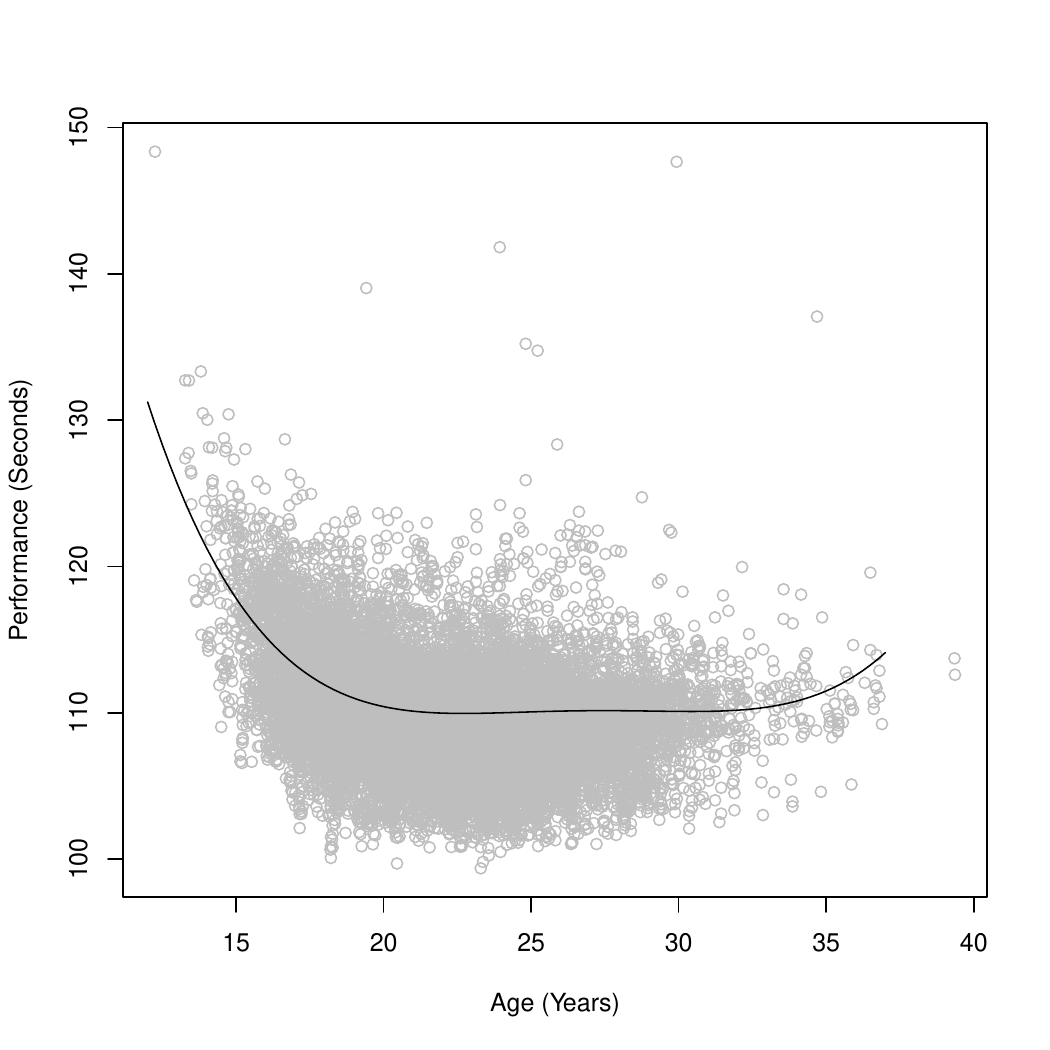}
& 
\includegraphics[scale = 0.25]{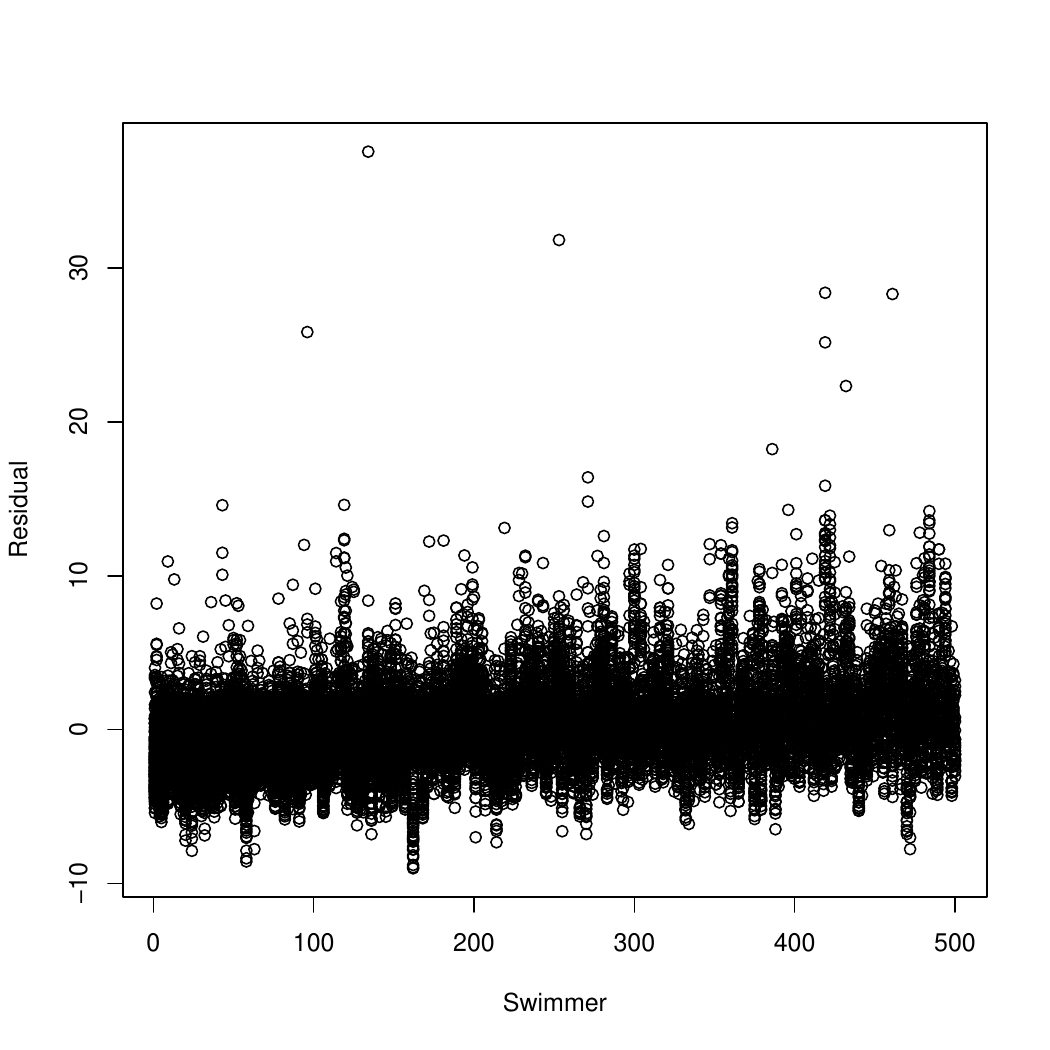}
\\
\end{tabular}
\end{center}
\caption{EDA-1: (a) all data and fitted regression lines, and (b) all residuals against swimmer number.}\label{EDA_Model_1}
\end{figure}
As an exploratory data analysis, we fit several regression models to illustrate the effect of age, within-season evolution and the shape of the error distribution. The first model fitted (EDA-1)
is
\[
\mbox{Performance} = \alpha + \beta_1 \,\mbox{Age} + \beta_2 \,\mbox{Age}^2 + \beta_3\, \mbox{Age}^3 + \beta_4 \mbox{Age}^4 + \mbox{Pool}
\]
where Age represents centred age and Pool is a binary variable with (25 metre coded as 0 and  50 metre coded as 1). Some results are shown in 
Figure~\ref{EDA_Model_1}. For each combination of gender and distance, there is a clear improvement in performance in the late teens and a reduction in performance after age 30. The residuals are skewed and heavy-tailed and show a clear effect of the individual athlete on performance.

\begin{figure}[h!]
\begin{center}
\begin{tabular}{ccc}
& (a) & (b) \\
\raisebox{0.9in}{100m Female} &
\includegraphics[scale = 0.25]{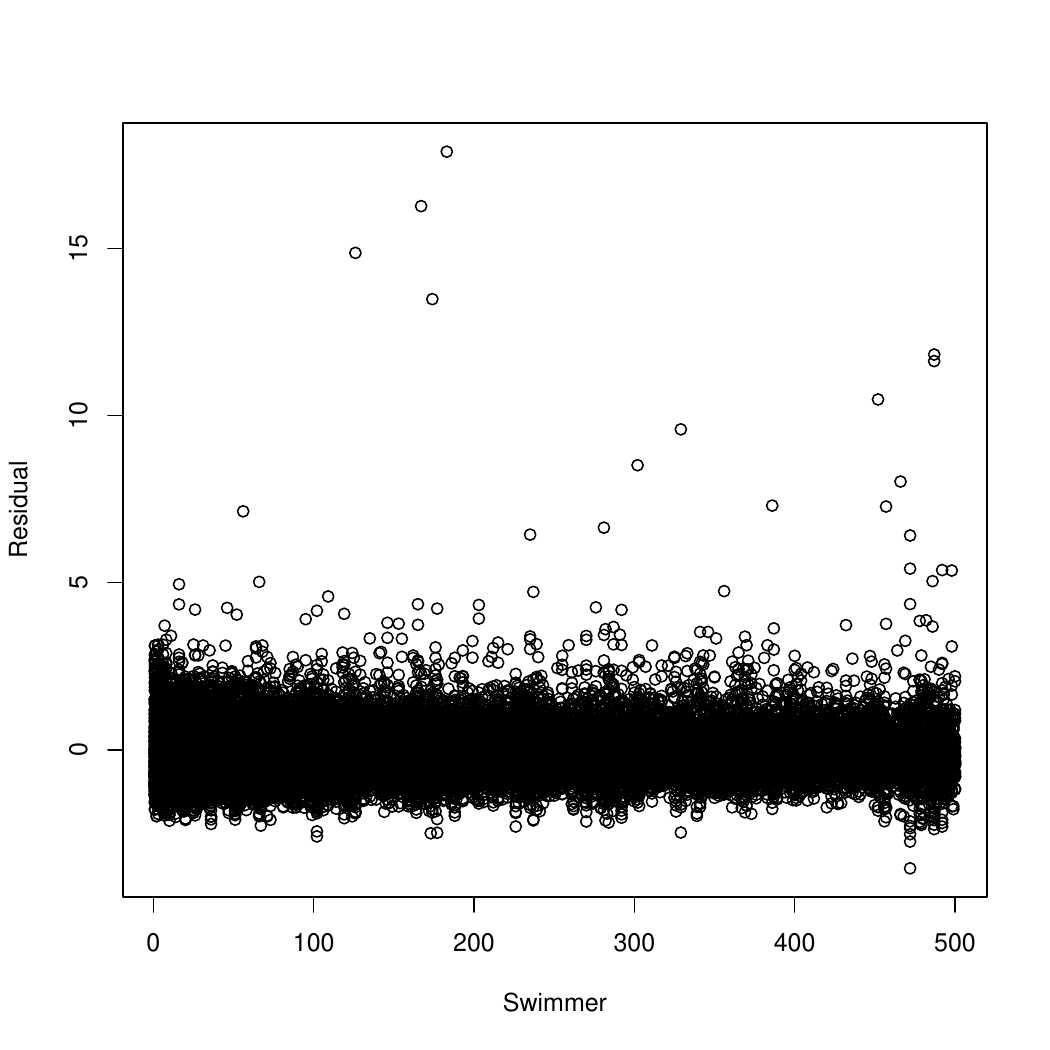}
&
\includegraphics[scale = 0.25]{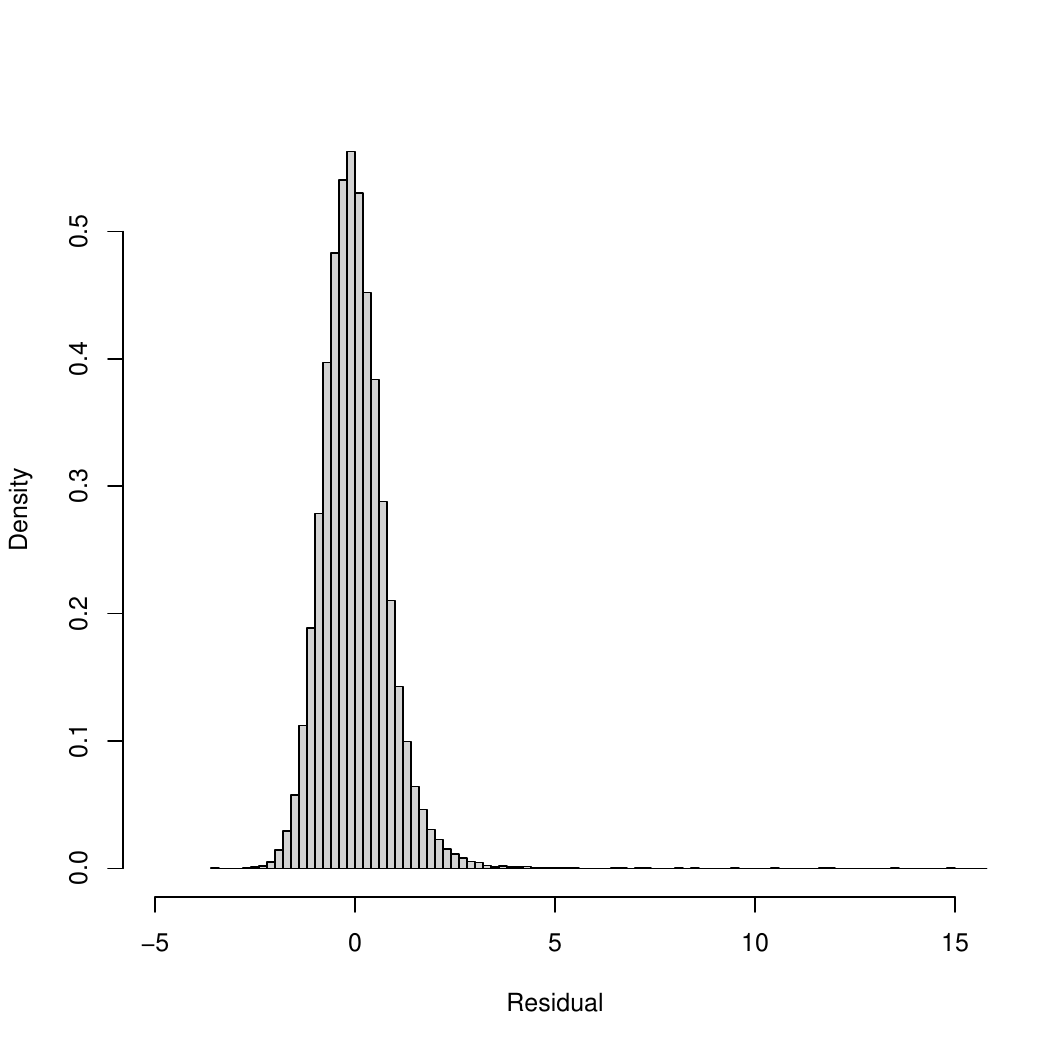}

\\
\raisebox{0.9in}{100m Male} &
\includegraphics[scale = 0.25]{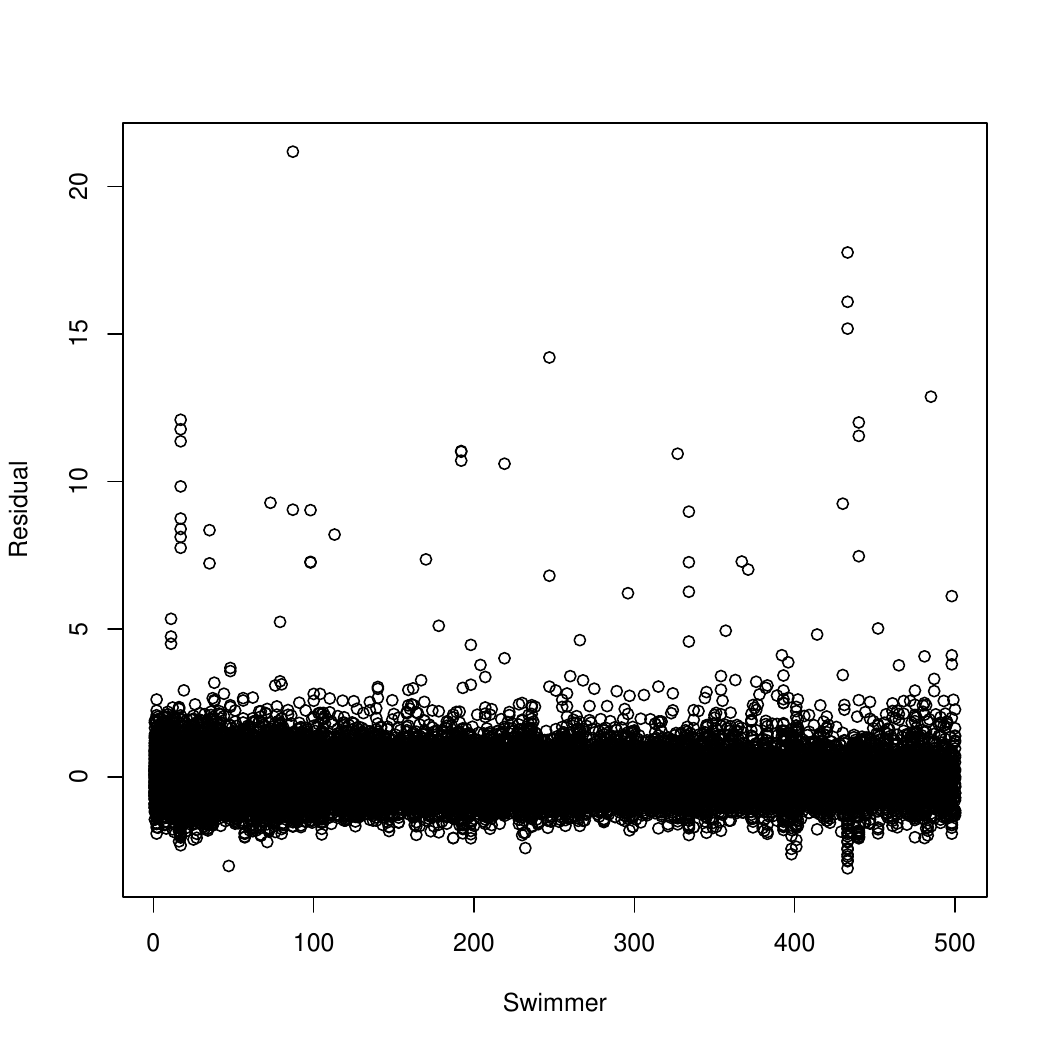}
& 
\includegraphics[scale = 0.25]{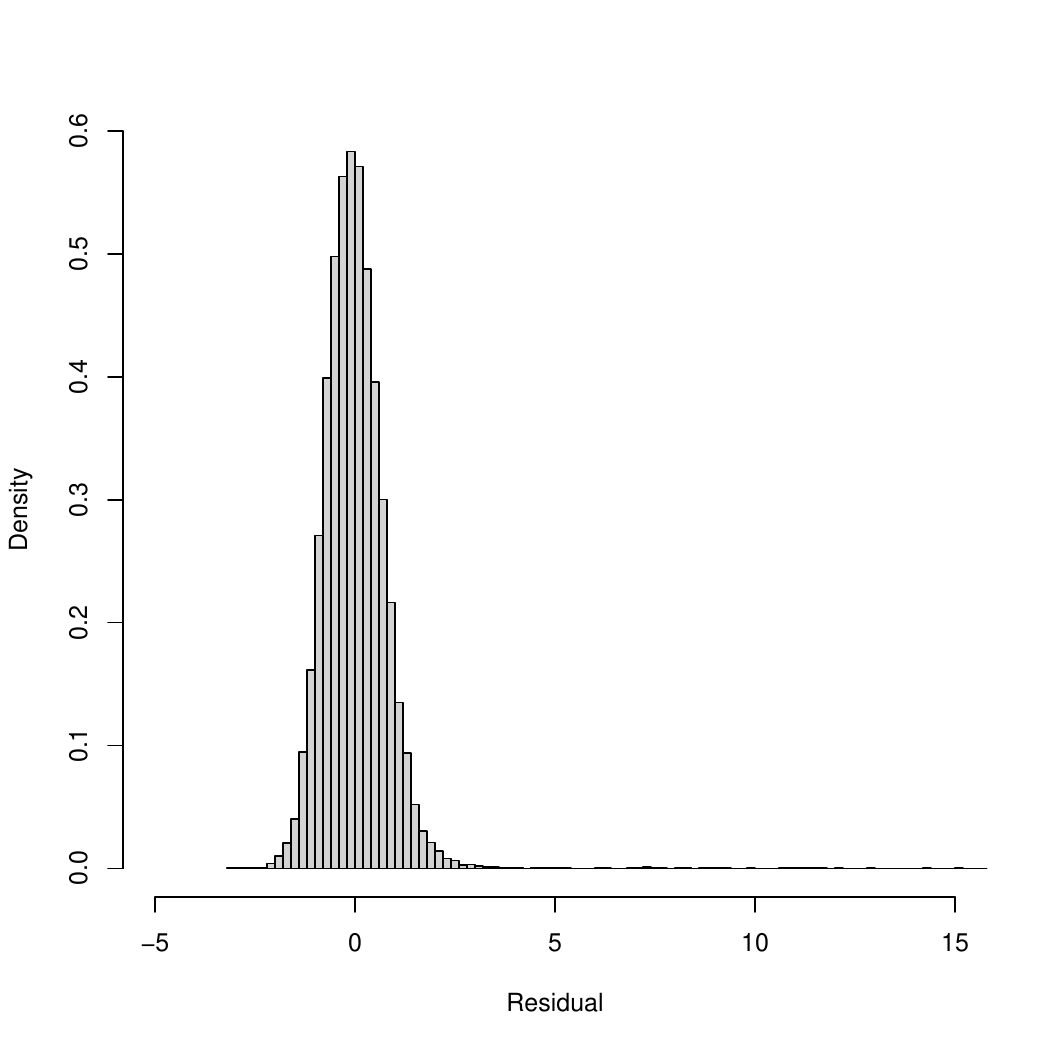}\\
\raisebox{0.9in}{200m Female} &
\includegraphics[scale = 0.25]
{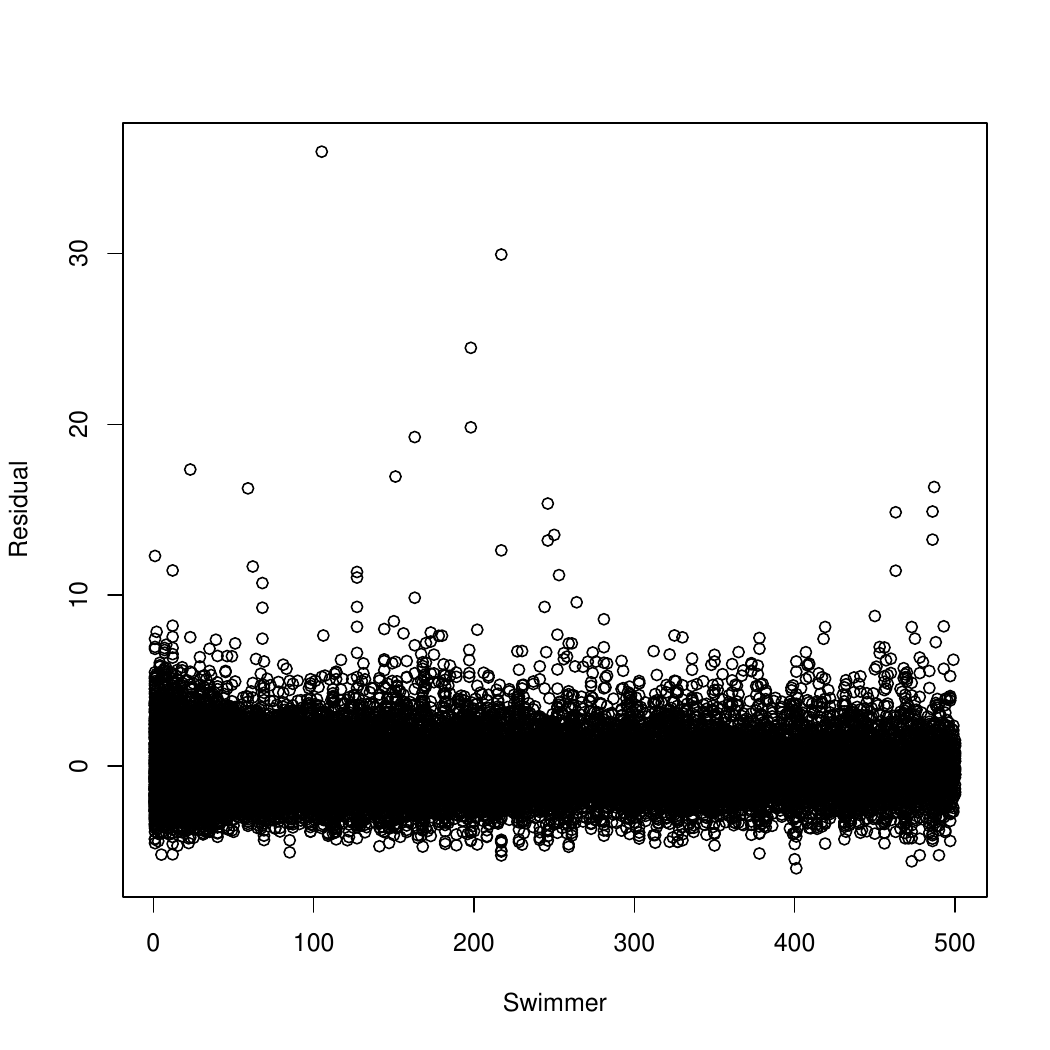}
& 
\includegraphics[scale = 0.25]{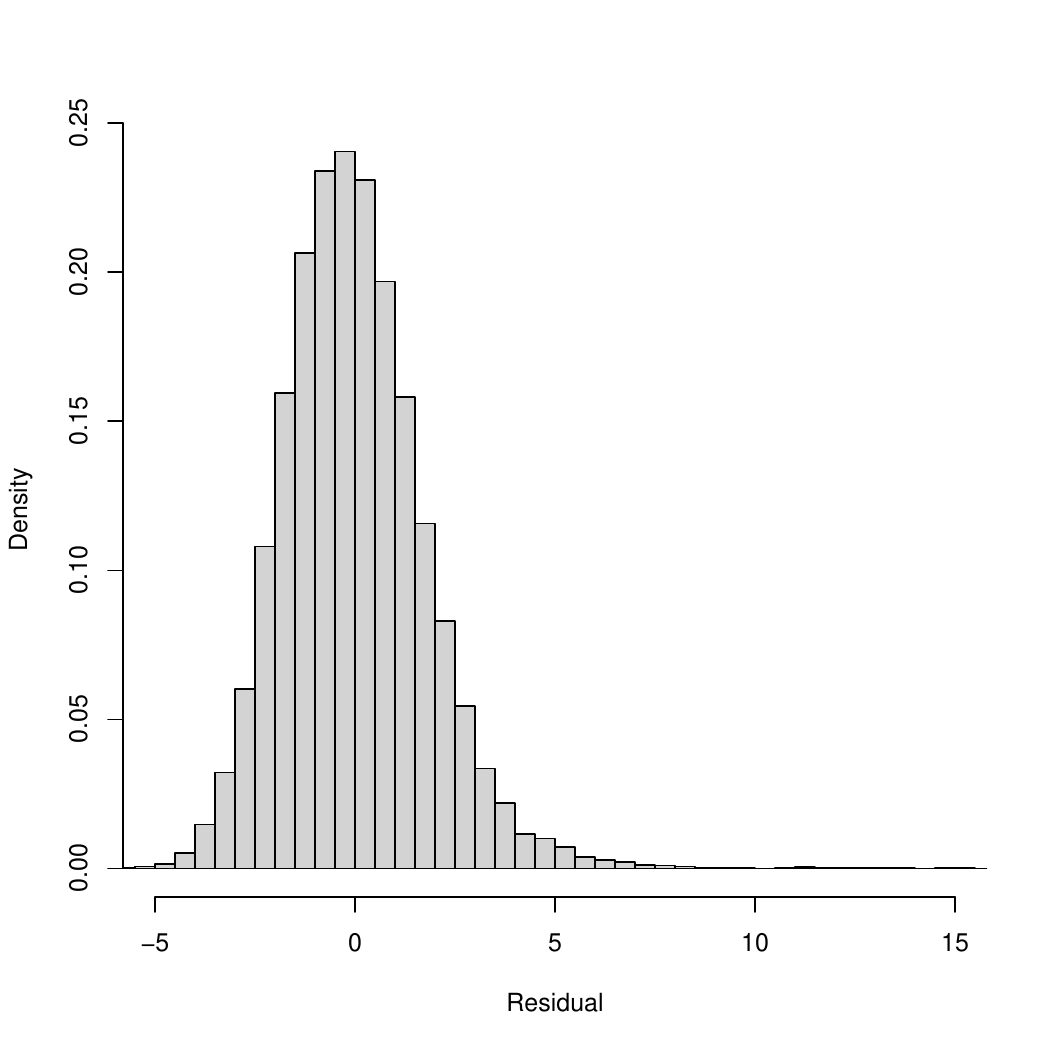}
\\
\raisebox{0.9in}{200m Male} &
\includegraphics[scale = 0.25]{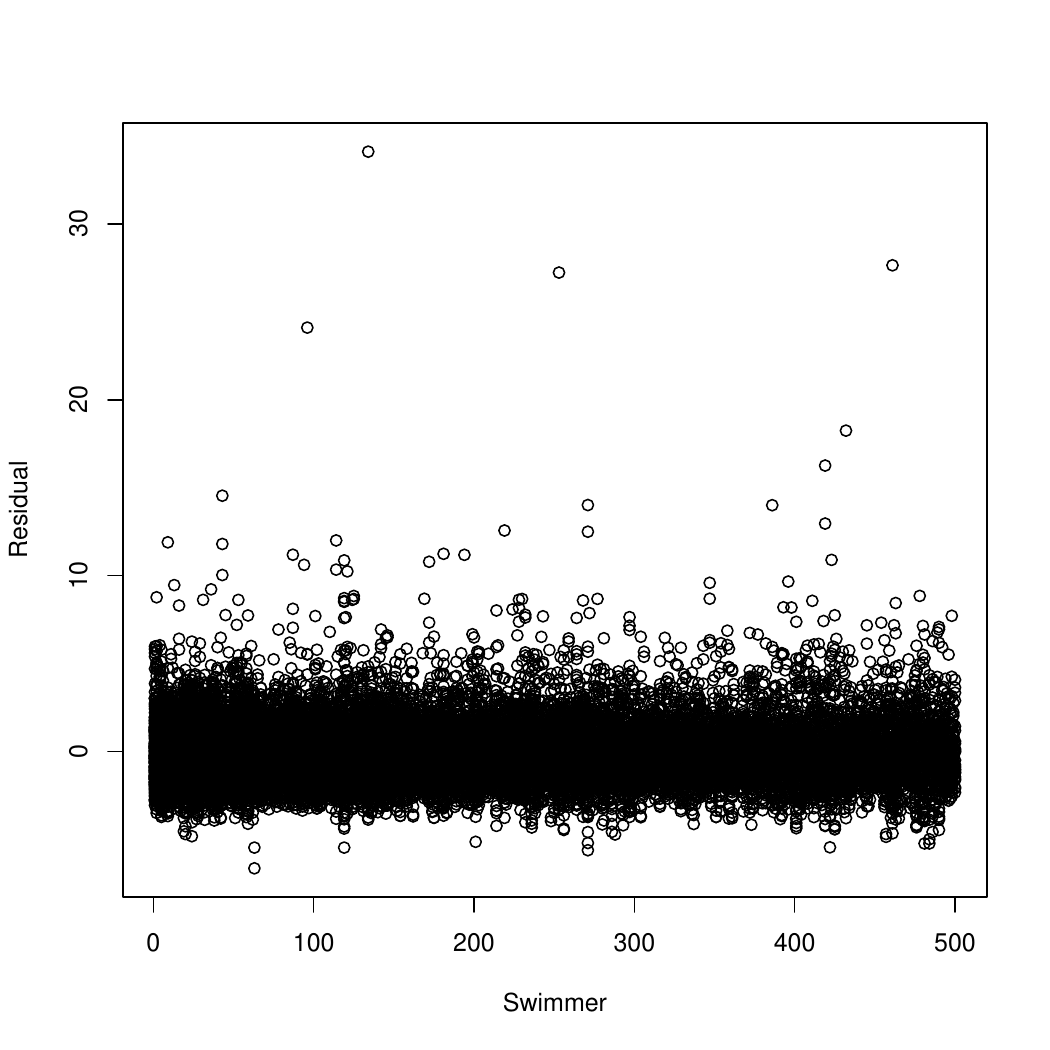}
& 
\includegraphics[scale = 0.25]{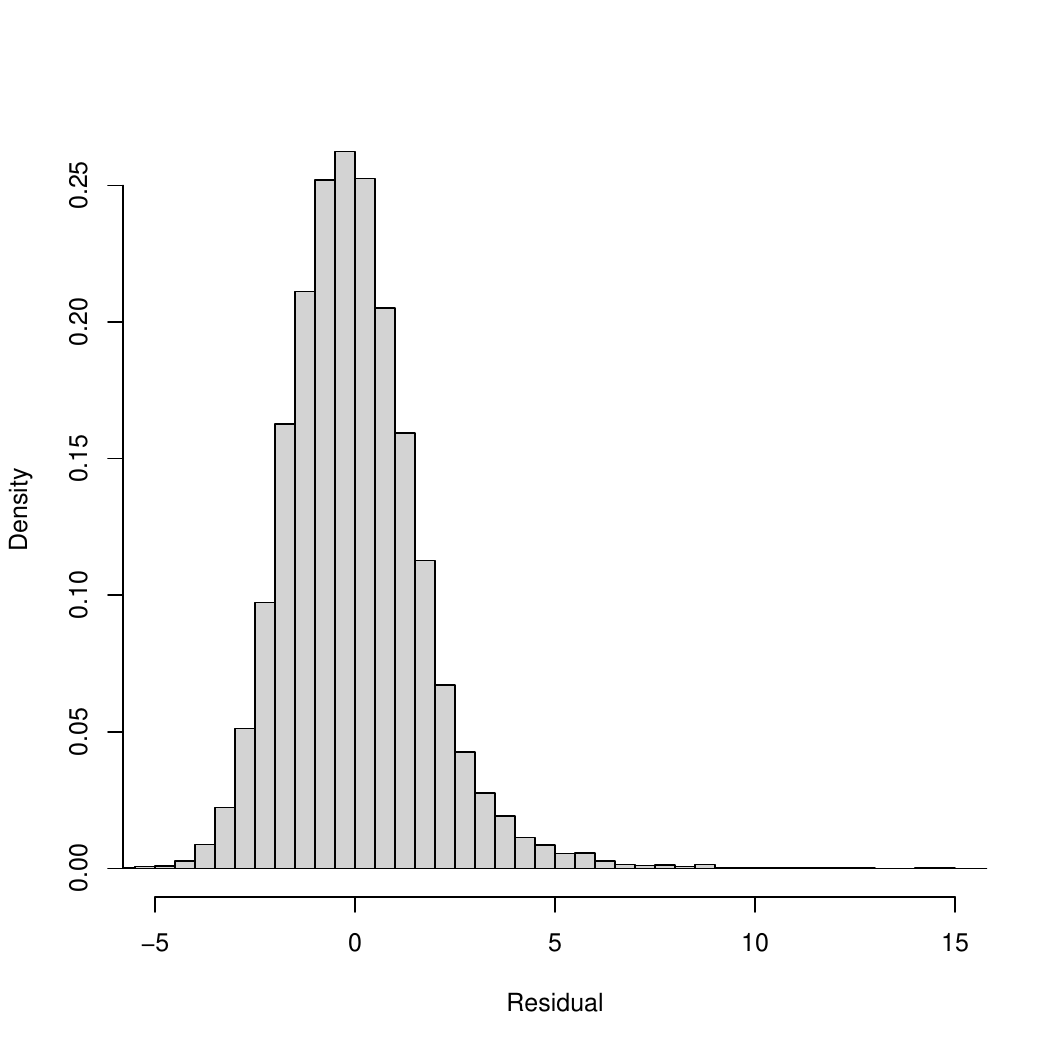}
\\
\end{tabular}
\end{center}
\caption{EDA-2: (a) 
 all residuals against swimmer number and (b) histogram of all residuals.}\label{EDA_Model_2}
\end{figure}
To address the effect of individuals, we extend the previous regression model by including each swimmer's ID as a factor
(EDA-2). Some plots of the residuals from this model are shown in Figure~\ref{EDA_Model_2}. These show that the skewness and heavy tails remain in the residuals after adjusting for the effects of swimmers.

\begin{figure}[h!]
\begin{center}
\begin{tabular}{cc}
Swimmer 1 & Swimmer 2\\
\includegraphics[scale=0.4]{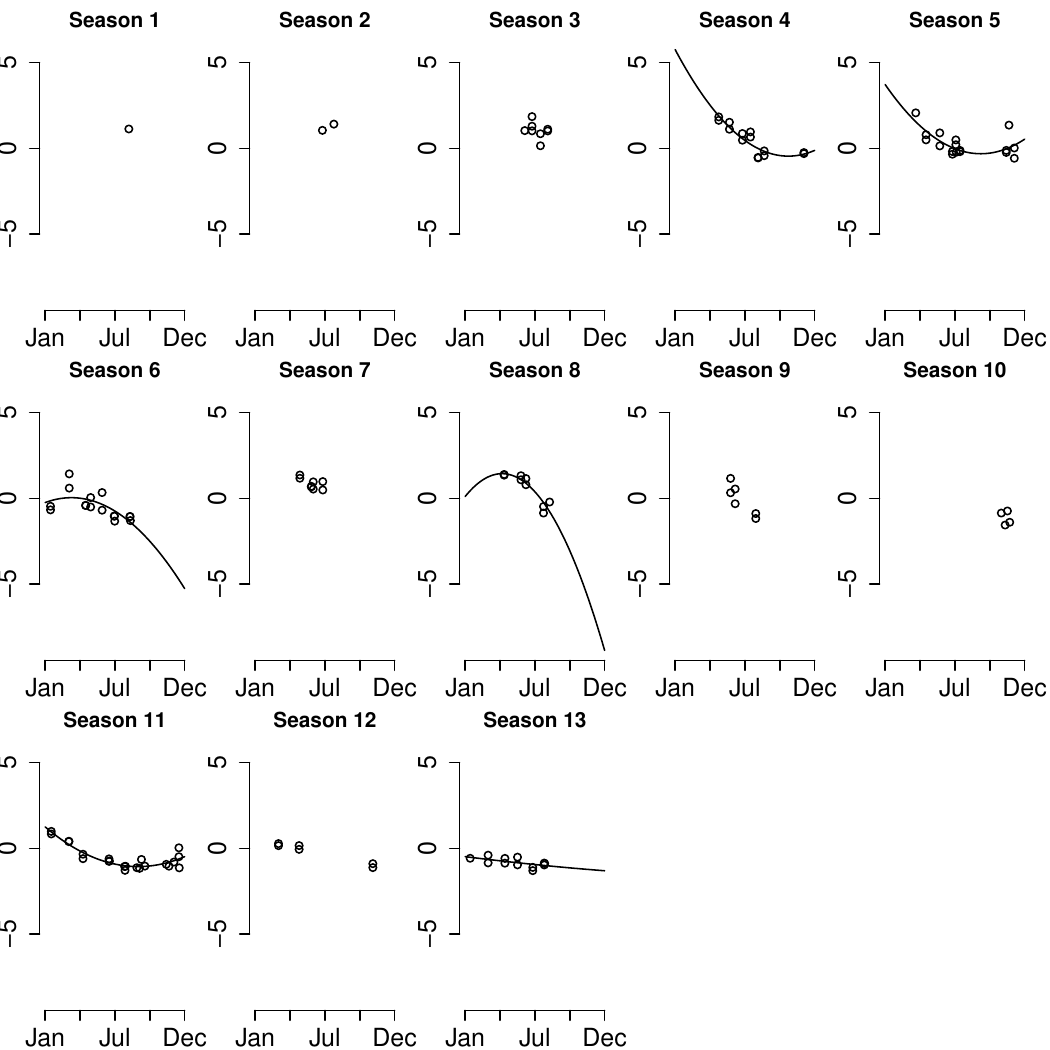} &
\includegraphics[scale=0.4]
{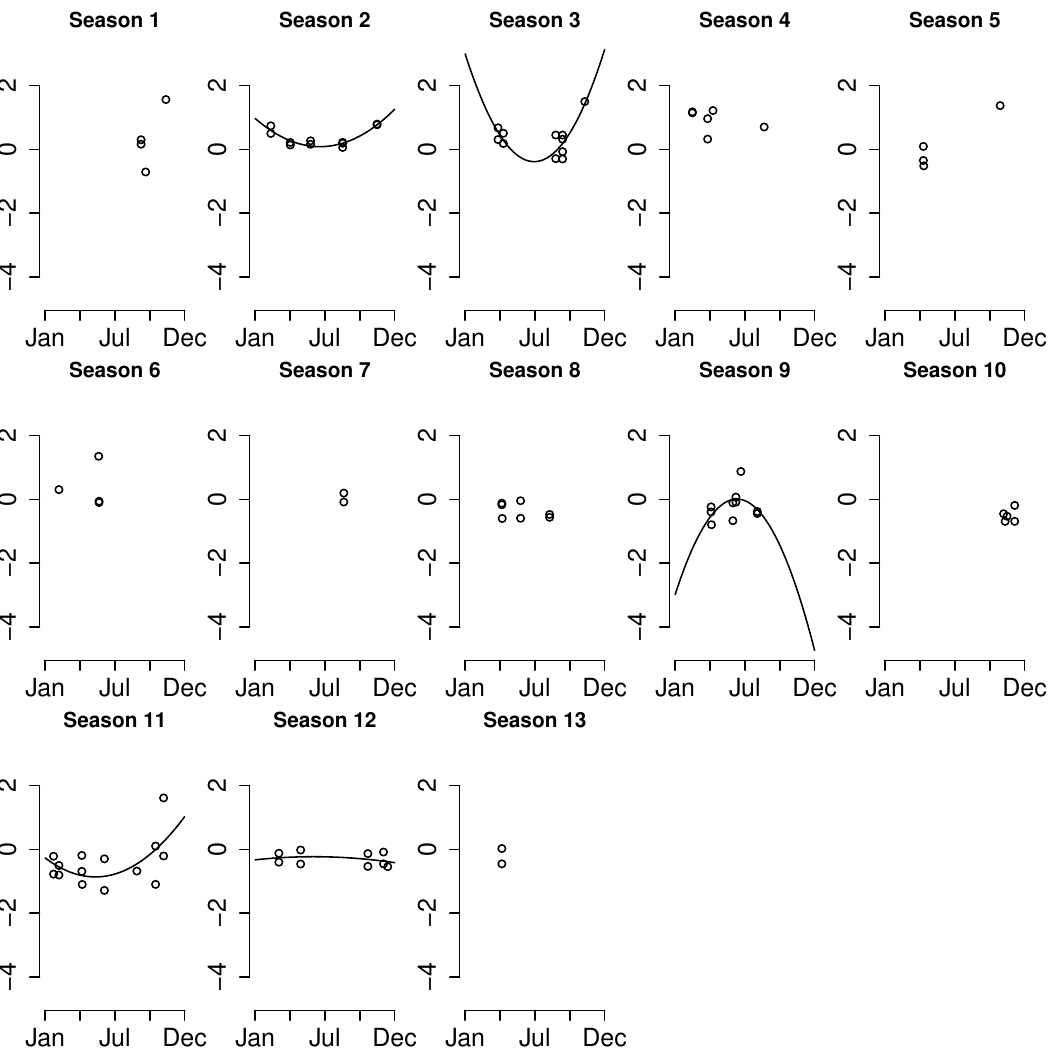 }\\
\\
\multicolumn{2}{c}{Swimmer 3}\\
\multicolumn{2}{c}{\includegraphics[scale=0.4, trim = 0mm 60mm 0mm 0mm, clip]{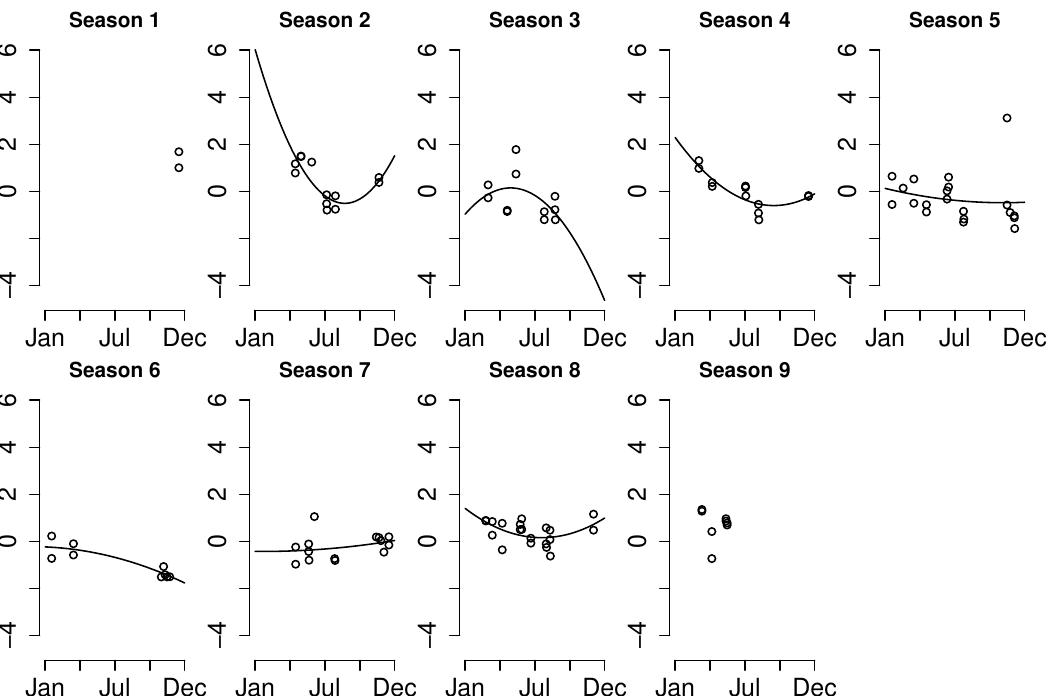}}
\end{tabular}
\end{center}
\caption{EDA-2, female 100 metre freestyle swimmers: Plots of residuals from regression 2 for each season with a fitted quadratic regression line if there are performance on more than 4 unique dates.}\label{EDA_Model_3}
\end{figure}
\begin{figure}[h!]
\begin{center}
\begin{tabular}{cc}
Swimmer 4 & Swimmer 5\\
\includegraphics[scale=0.4, trim = 0mm 60mm 0mm 0mm, clip]{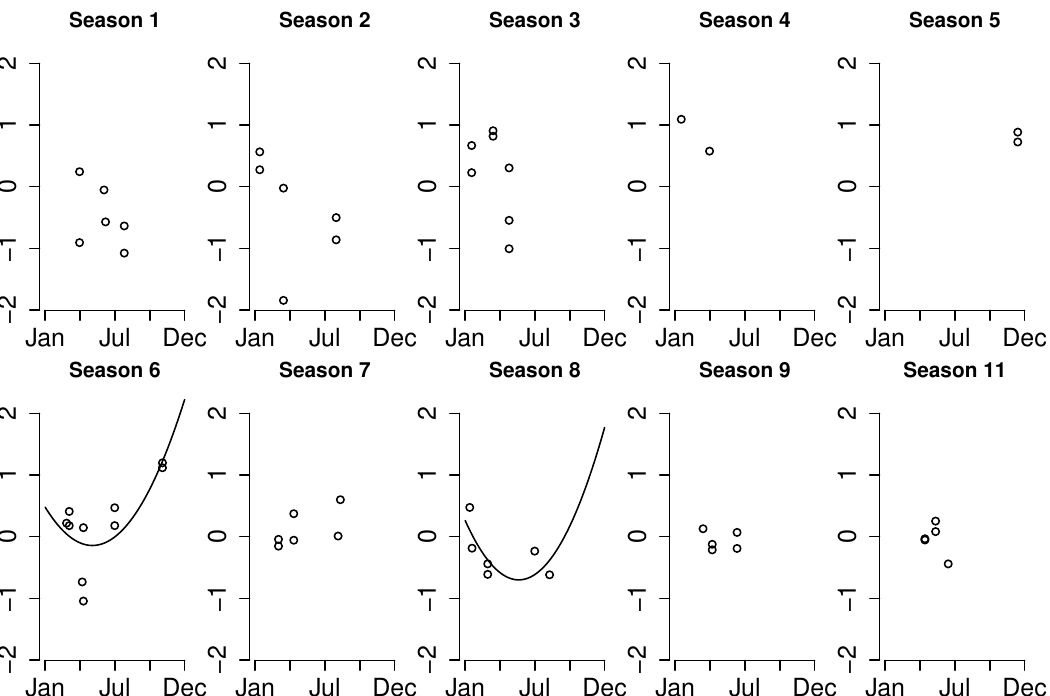} &
\includegraphics[scale=0.4, trim = 0mm 60mm 0mm 0mm, clip]
{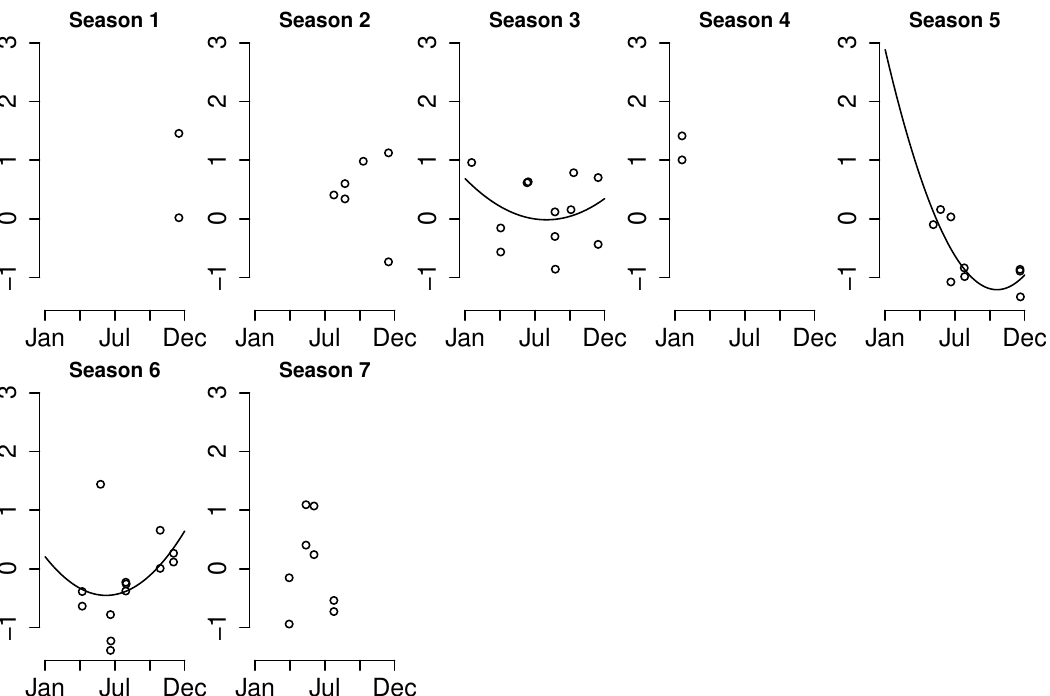 }\\
\\
\multicolumn{2}{c}{Swimmer 6}\\
\multicolumn{2}{c}{\includegraphics[scale=0.4]{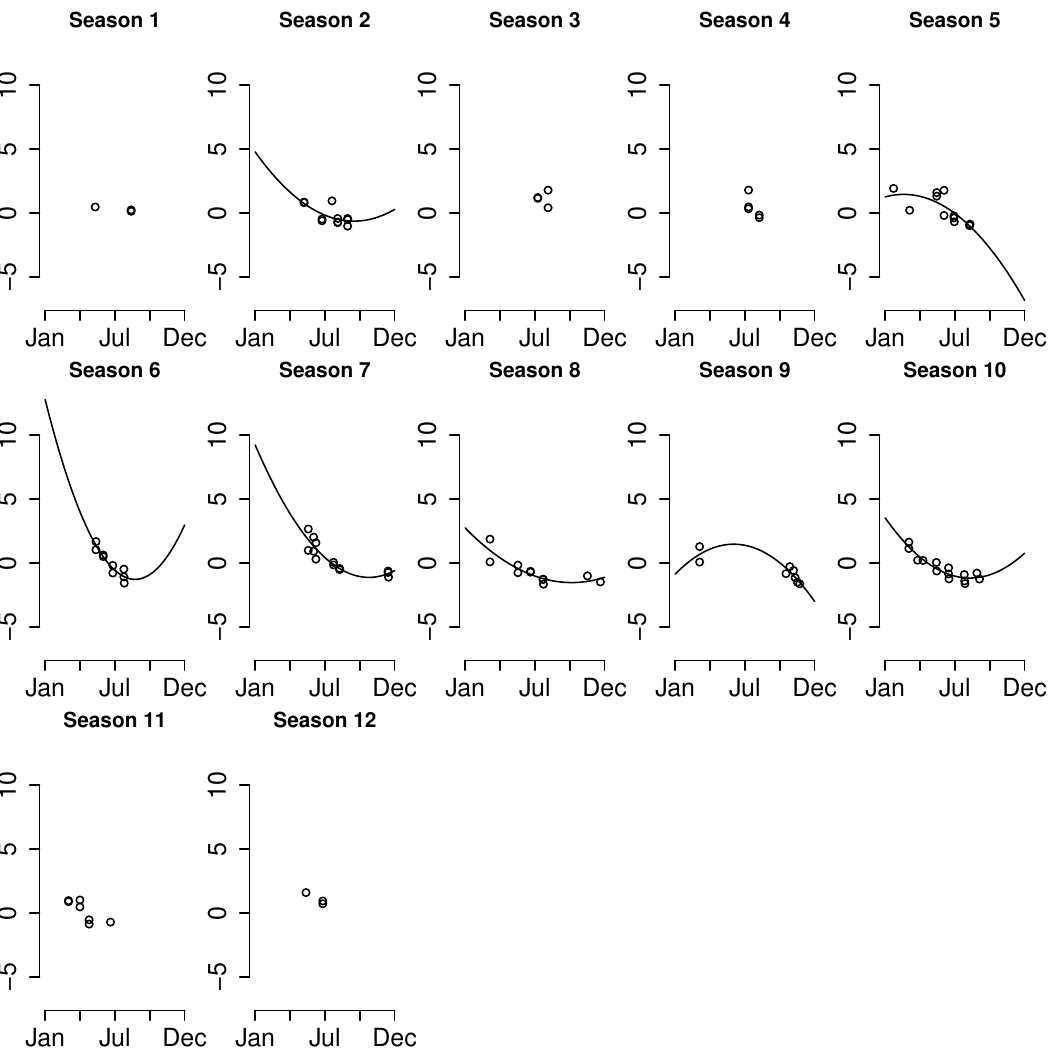}}
\end{tabular}
\end{center}
\caption{EDA-2, male 100 metre freestyle swimmers: Plots of residuals from regression 2 for each season with a fitted quadratic regression line if there are performance on more than 4 unique dates.}\label{EDA_Model_4}
\end{figure}
\begin{figure}[h!]
\begin{center}
\begin{tabular}{cc}
Swimmer 4 & Swimmer 5\\
\includegraphics[scale=0.4]{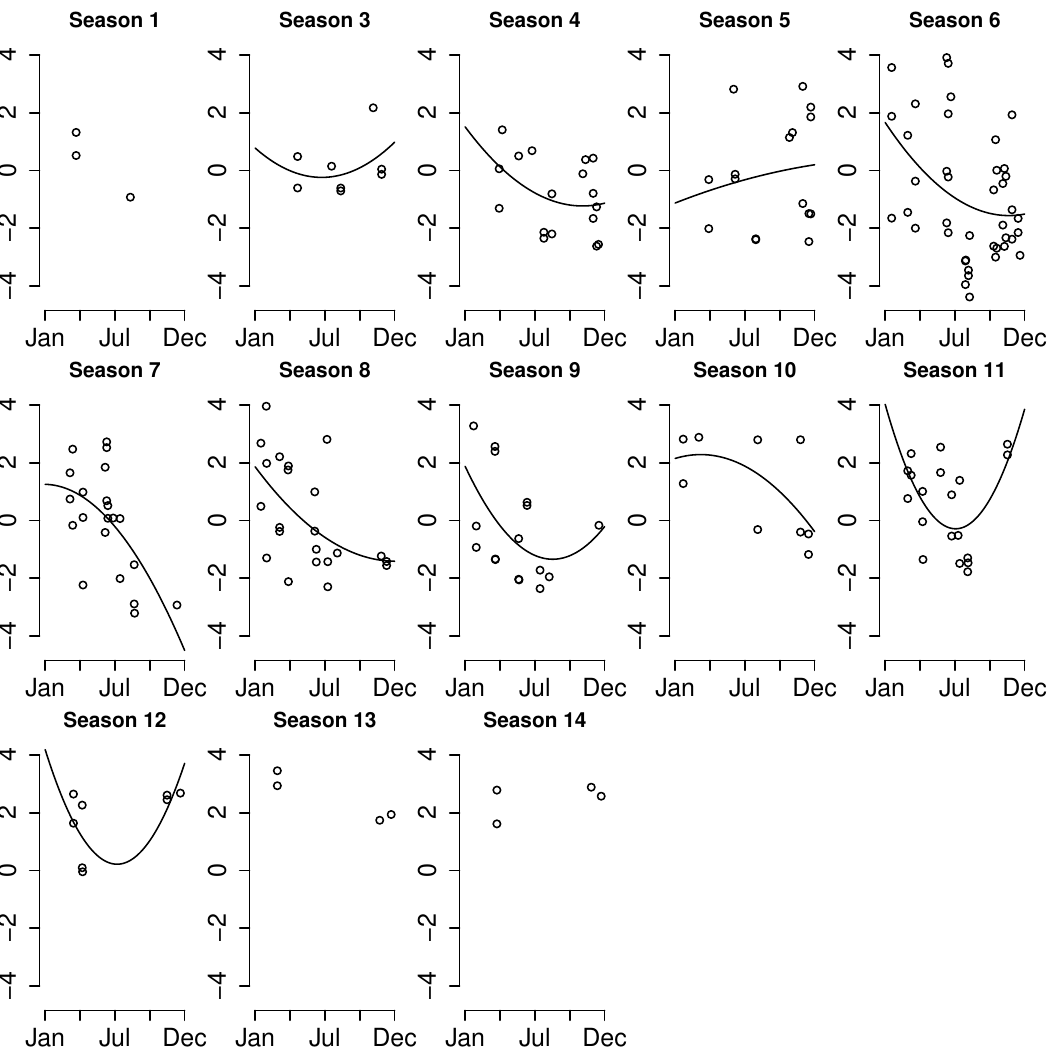} &
\includegraphics[scale=0.4]
{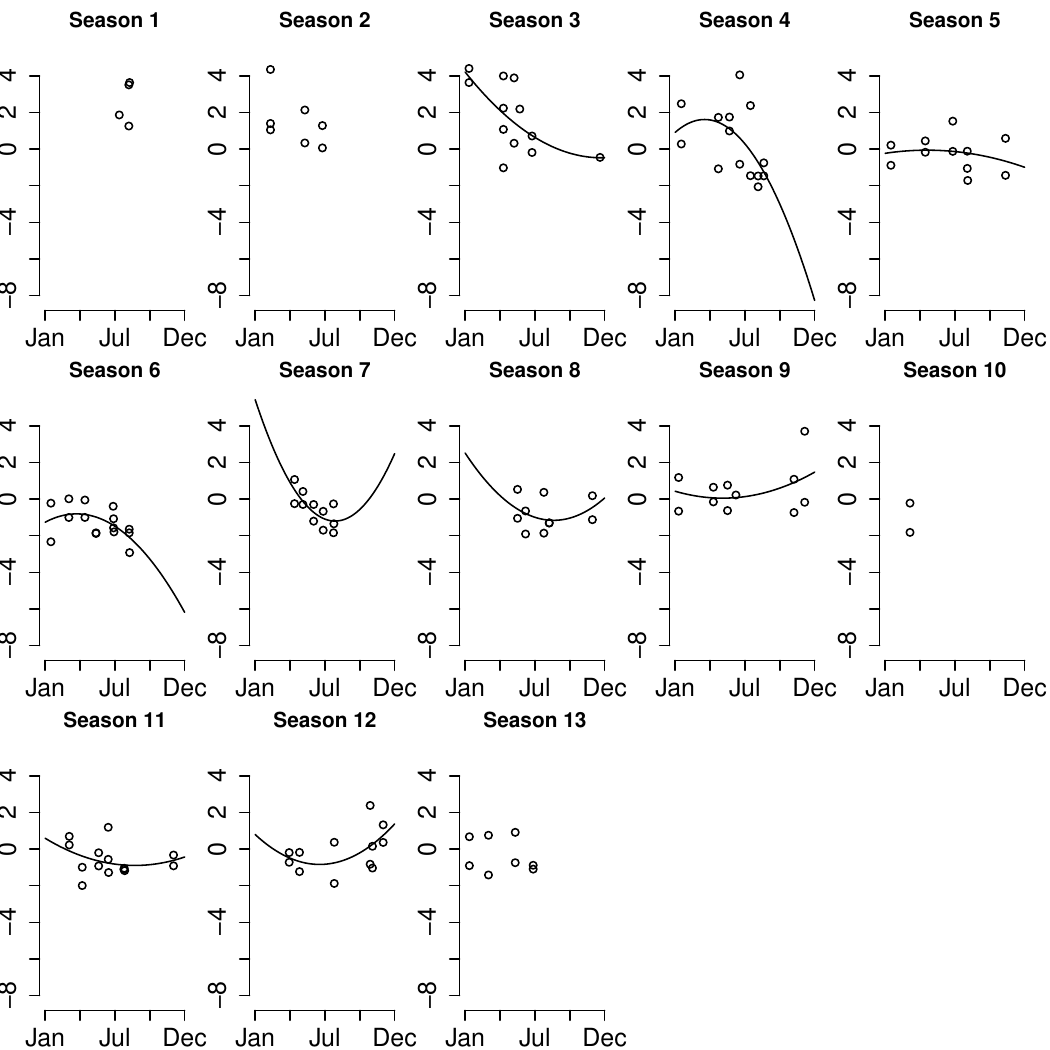 }\\
\\
\multicolumn{2}{c}{Swimmer 6}\\
\multicolumn{2}{c}{\includegraphics[scale=0.4, trim = 0mm 60mm 0mm 0mm, clip]{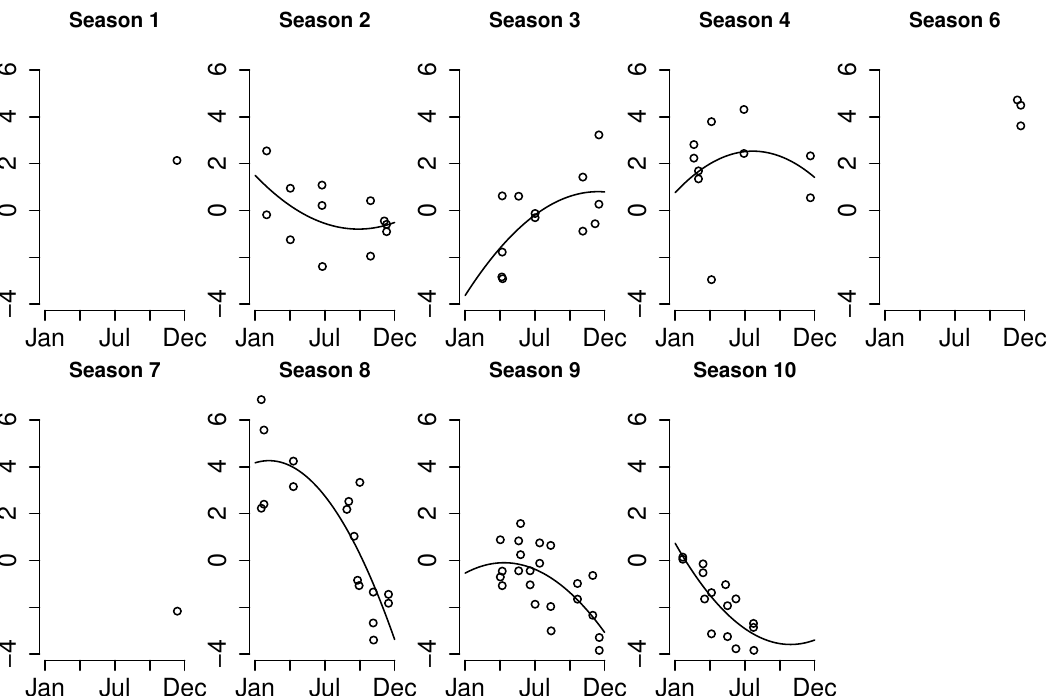}}
\end{tabular}
\end{center}
\caption{EDA-2, female 200 metre freestyle swimmers: Plots of residuals from regression 2 for each season with a fitted quadratic regression line if there are performance on more than 4 unique dates.}\label{EDA_Model_5}
\end{figure}
\begin{figure}[h!]
\begin{center}
\begin{tabular}{cc}
Swimmer 4 & Swimmer 5\\
\includegraphics[scale=0.4]{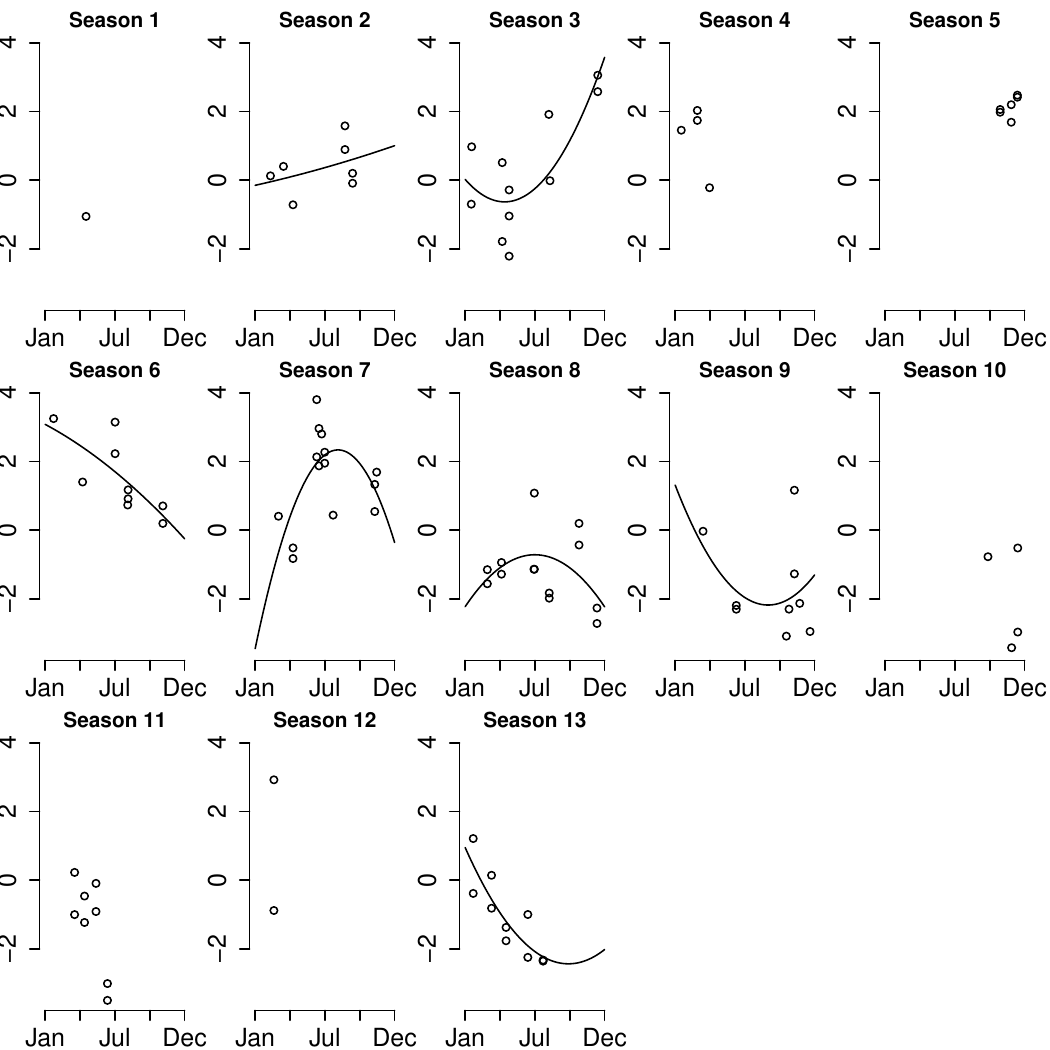} &
\includegraphics[scale=0.4]
{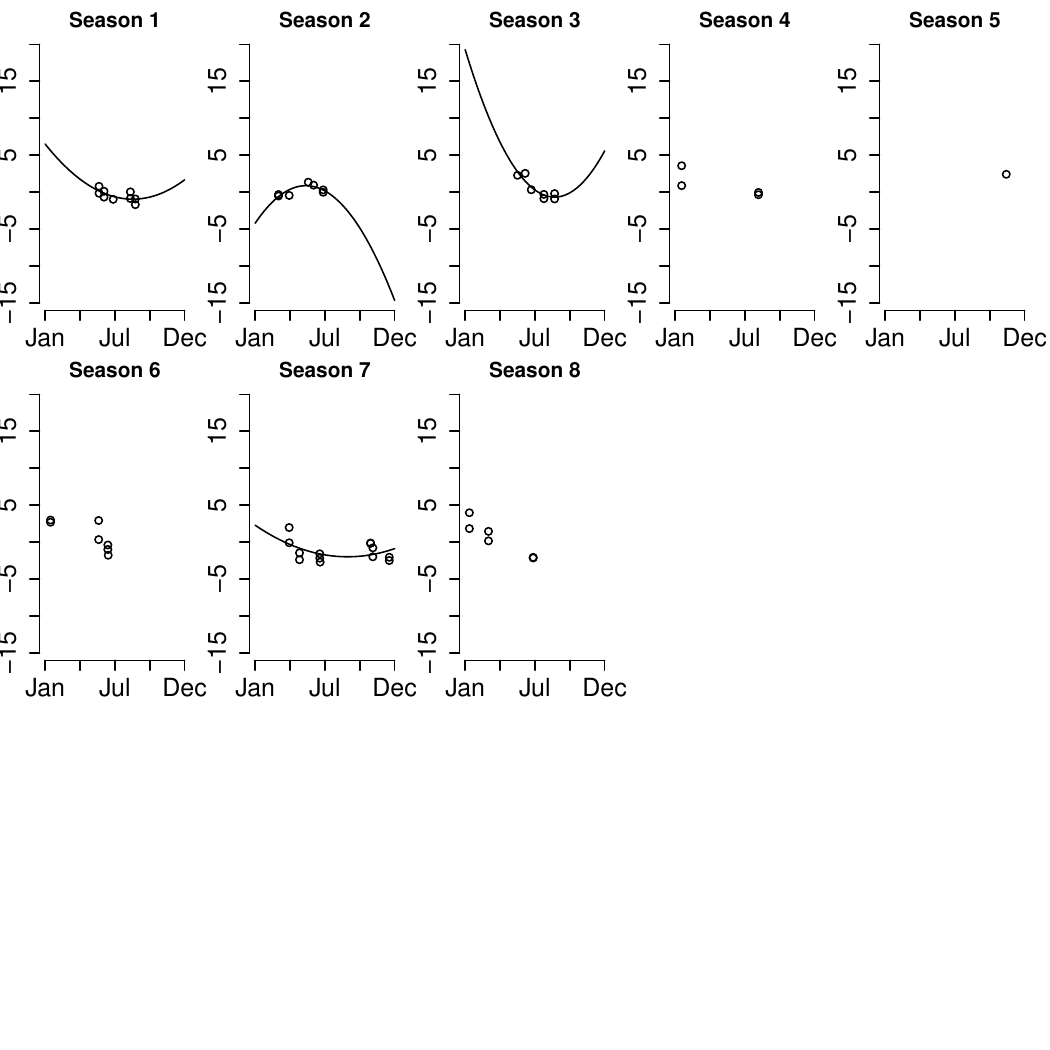 }\\
\\
\multicolumn{2}{c}{Swimmer 6}\\
\multicolumn{2}{c}{\includegraphics[scale=0.4, trim = 0mm 60mm 0mm 0mm, clip]{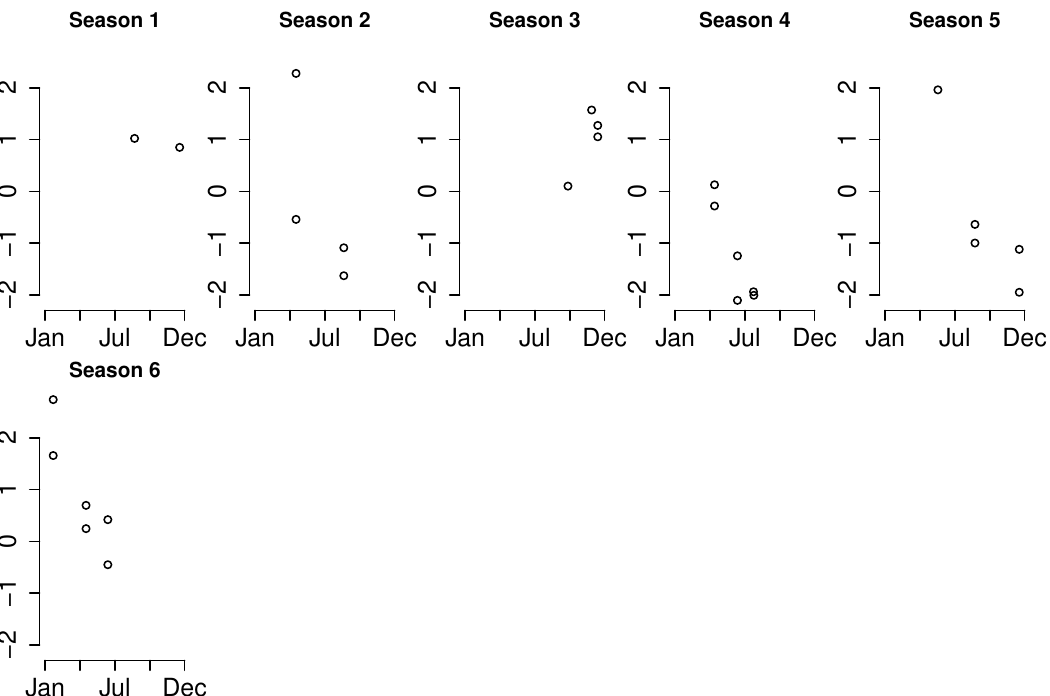}}
\end{tabular}
\end{center}
\caption{EDA-2, male 200 metre freestyle swimmers: Plots of residuals from regression 2 for each season with a fitted quadratic regression line if there are performance on more than 4 unique dates.}\label{EDA_Model_6}
\end{figure}

Figures \ref{EDA_Model_3}, \ref{EDA_Model_4}, \ref{EDA_Model_5} and 
\ref{EDA_Model_6} plots the residuals from model EDA-2 plotted against the day of the year for all seasons in which each athlete competed. In the swimmers shown are those used in the ``Additional results for Individual Athletes'' section. In seasons with more than 4 unique performance days, a quadratic regression line is fitted. These regression lines often show a ``u''-shape with performance improving to a peak performance in the middle of the year with differences in the strength of effect over seasons and athletes. There is also evidence of instability in the estimation when performance days are close and/or there are few performances. These motivate the use of a hierarchical model to allow for differences in the level of performance over  seasons and borrow strength.

\section{MCMC Diagnostics}\label{MCMC_diagnostics}

The convergence and mixing of the chain are assessed using the \verb+CODA+ package in R \citep{CODA}. Convergence is assessed using the Potential Scale Reduction Factor (PSRF) calculated using all recorded values and mixing is assessed using the Effective Sample Size (ESS). The PSRF should be close to 1 and a large ESS is preferred. Table~\ref{t:MCMC_diag} presents the results for the 100 metres and 200 metres swimming for both men and women. If a parameter is not univariate, the maximum PSRF and minimum ESS are presented as these represents the least favourable values.
In all cases, the PSRF is close to zero and the ESS is large for many parameters.

\begin{table}[h!]
\begin{center}
\begin{tabular}{|c|lr|lr|lr|lr|}\hline
 & \multicolumn{2}{c|}{100m women} & \multicolumn{2}{c|}{100m men}  & \multicolumn{2}{c|}{200m women} & \multicolumn{2}{c|}{200m men}\\\hline 
  Parameter & PSRF & ESS & PSRF  & ESS & PSRF  & ESS & PSRF  & ESS\\\hline
$\sigma^2_a$ & 1 & 1761 & 1 & 1940 & 1 & 2000 & 1 & 2000\\ 
$\sigma^2_m$ & 1 & 1805 & 1.01 & 318 & 1 & 2026 & 1.02 & 1912 \\
$\alpha$ & 1 & 862 & 1.07 & 127 & 1 & 1622 & 1 & 1266\\
$\zeta$ & 1 & 1898 & 1.01 & 1460 & 1 & 1749 & 1 & 2000\\
$\beta_{n, v}$ & 1.01 & 1887 & 1.01 & 1094 & 1.01 & 1861 & 1 & 1874\\
$\beta_{n, v}^{(i)}$ & 1.04 & 1430 & 1.05 & 1310 & 1.03 & 1194 & 1.03 & 1281\\
$\beta_{n, v}^{(i, s)}$ & 1.03 & 663 & 1.03 & 1183 & 1.03 & 504 & 1.03 & 1117\\
$\lambda_i^2$ & 1.05 & 890 & 1.04 & 754 & 1.06 & 854 & 1.06 & 1247\\
$c^2_{n, v}$ & 1.03 & 164 & 1.03 & 333 & 1.03 & 231 & 1.06 & 317\\
$\tau_i^2$ & 1.07 & 638 & 1.15 & 1002 & 1.08 & 751 & 1.07 & 929\\
$d^2_{n, v}$ & 1.01 & 763 & 1.02 & 673 & 1.02 & 604 & 1.01 & 926\\
$\eta_{i, s}$ & 1.03 & 1261 & 1.02 & 1144 & 1.03 & 1175 & 1.03 & 1400\\
$\sigma^2_i$ & 1.02 & 1602 & 1.02 & 1460 & 1.02 & 1549 & 1.04 & 1521\\
$\sigma^2_{\mu}$ & 1 & 2000  & 1.01 & 2000 & 1 & 2000 & 1 & 2000\\
$\sigma^2_{\eta}$ & 1 & 796 & 1 & 530 & 1.02 & 160 & 1 & 1461 \\
$\nu_1$ & 1 & 2501 & 1.04 & 2000 & 1 & 2000 & 1.01 & 2000 \\
$\nu_2$ & 1 & 1036 & 1.01 & 1480 & 1 & 1405 & 1 & 1583 \\
$\nu^\mu$ & 1.01 & 1941 & 1.04 & 2000 & 1 & 2000 & 1.01 & 1851 \\
$\nu^\eta$ & 1 & 814 & 1 & 602 & 1.02 & 155 & 1 & 809 \\
$\lambda_0$ & 1.05 & 170  & 1.01 & 171 & 1.02 & 151 & 1 & 167 \\
$\lambda_1$ & 1.05 & 261 & 1 & 273 & 1.01 & 259 & 1.01 & 231 \\
$\tau_0$ & 1.04 & 486 & 1 & 427 & 1.03 & 378 & 1.02 &  273 \\
$\tau_1$ & 1.02 & 546 & 1.02 & 593 & 1.02 & 444 & 1 & 432 \\\hline
\end{tabular}
\end{center}
\caption{The value of the Potential Scale Reduction Factor (PSRF) and the Effective Sample Size (ESS) for all parameters in the 100 metres and 200 metres for both men and women. If parameter are not univariate, the maximum GR and minimum ESS are reported.}\label{t:MCMC_diag}
\end{table}

\section{Additional results}\label{Additional results}

\begin{figure}[!htbp]
\begin{center}
\begin{tabular}{cccc}
\multicolumn{2}{c}{100 metres} &
\multicolumn{2}{c}{200 metres}\\
 Females &Males  & Females &Males \\
\includegraphics[scale=0.19]{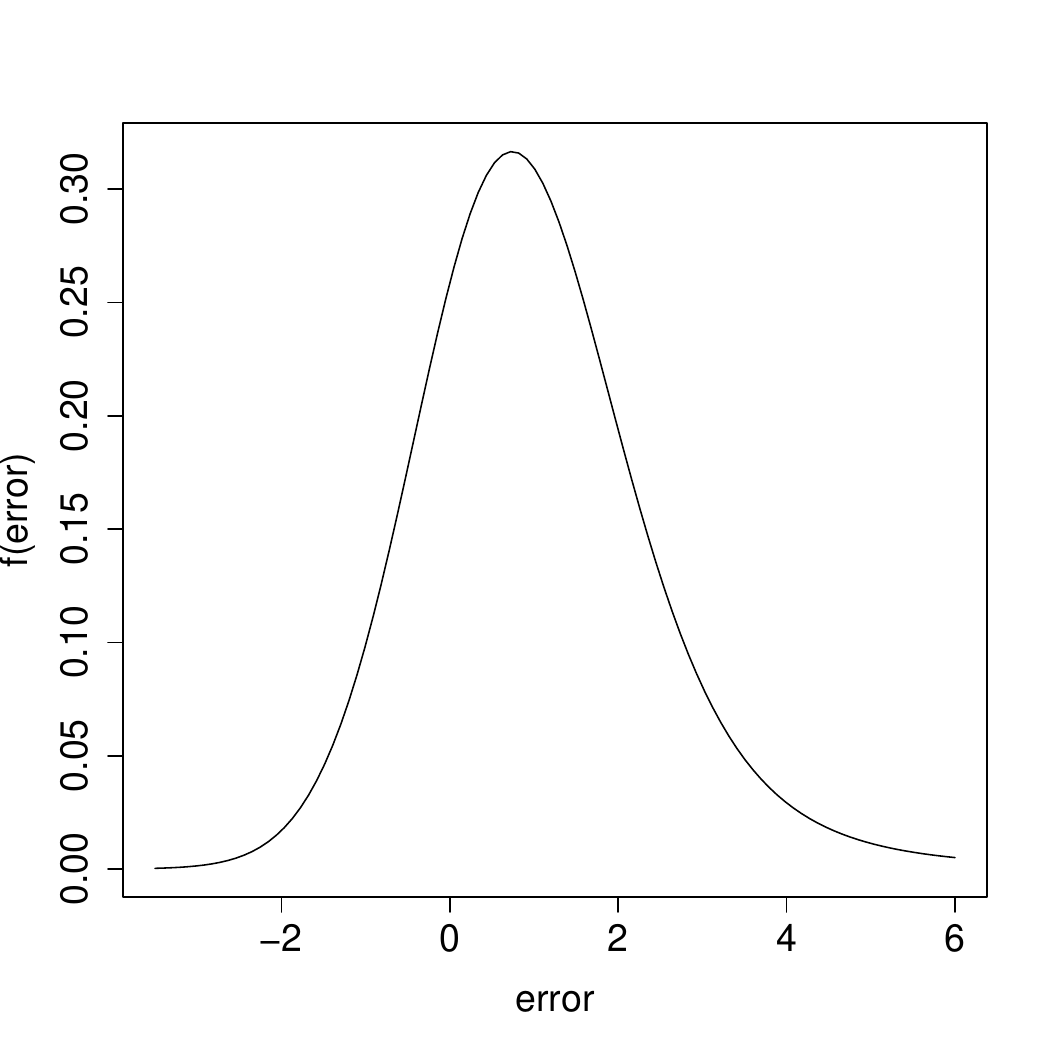}&
\includegraphics[scale=0.19]{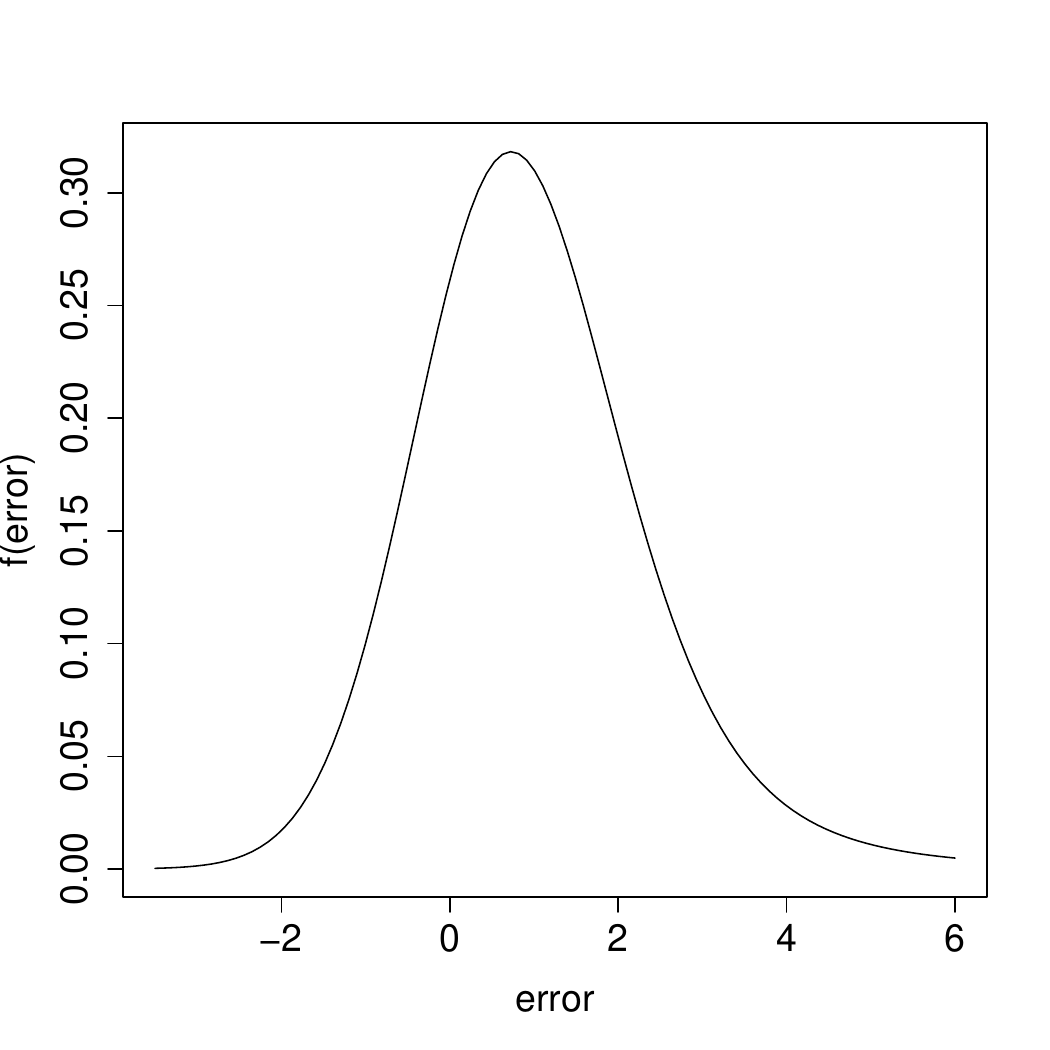}&
\includegraphics[scale=0.19]{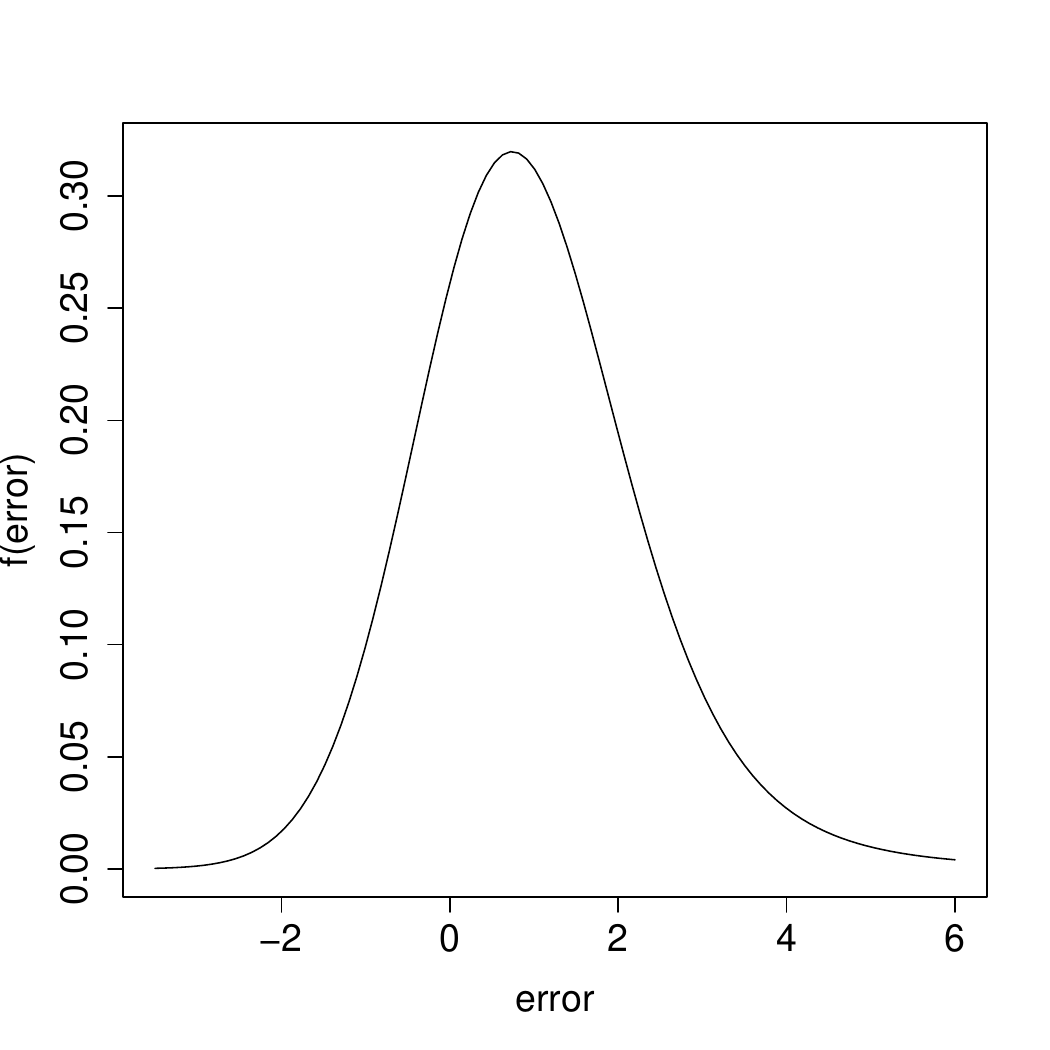}&
\includegraphics[scale=0.19]{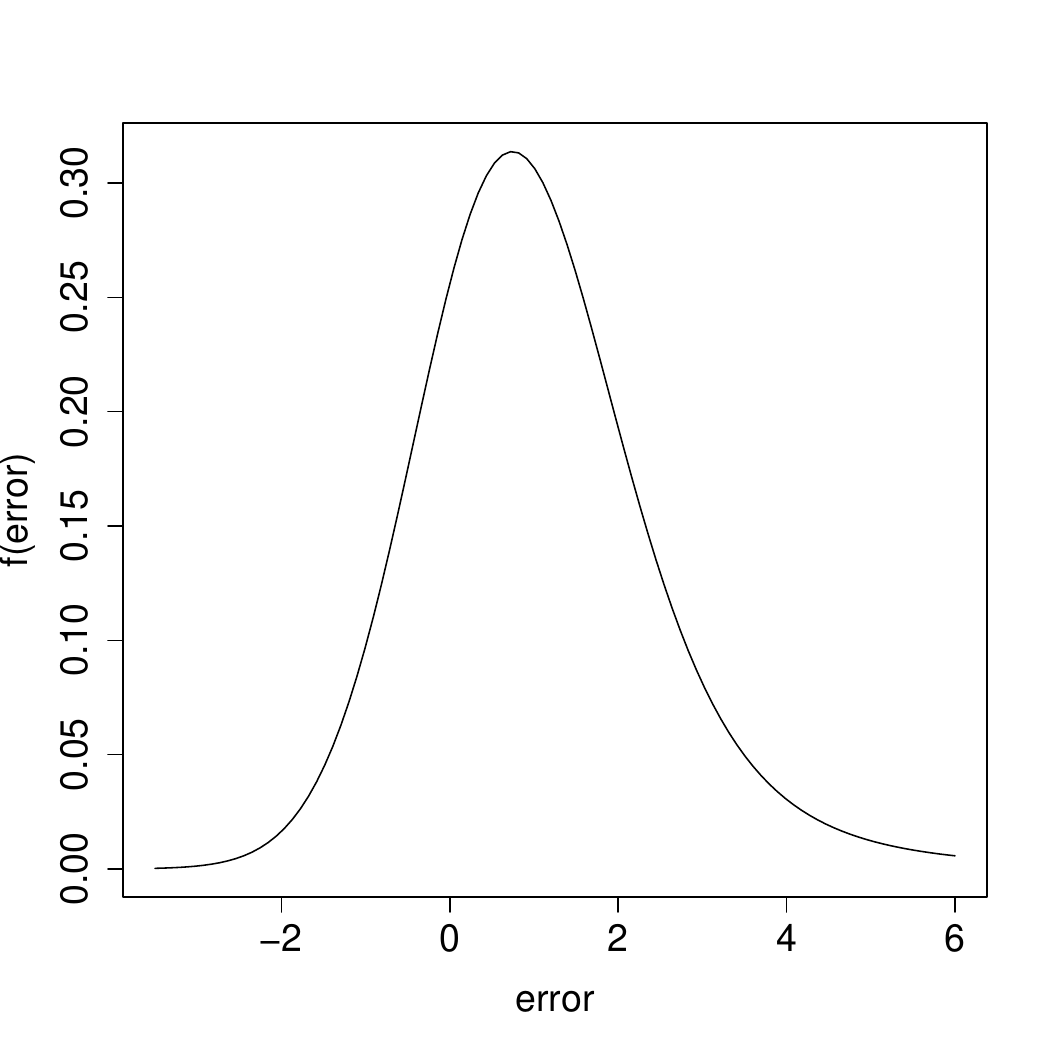}
\\
\includegraphics[scale=0.19]{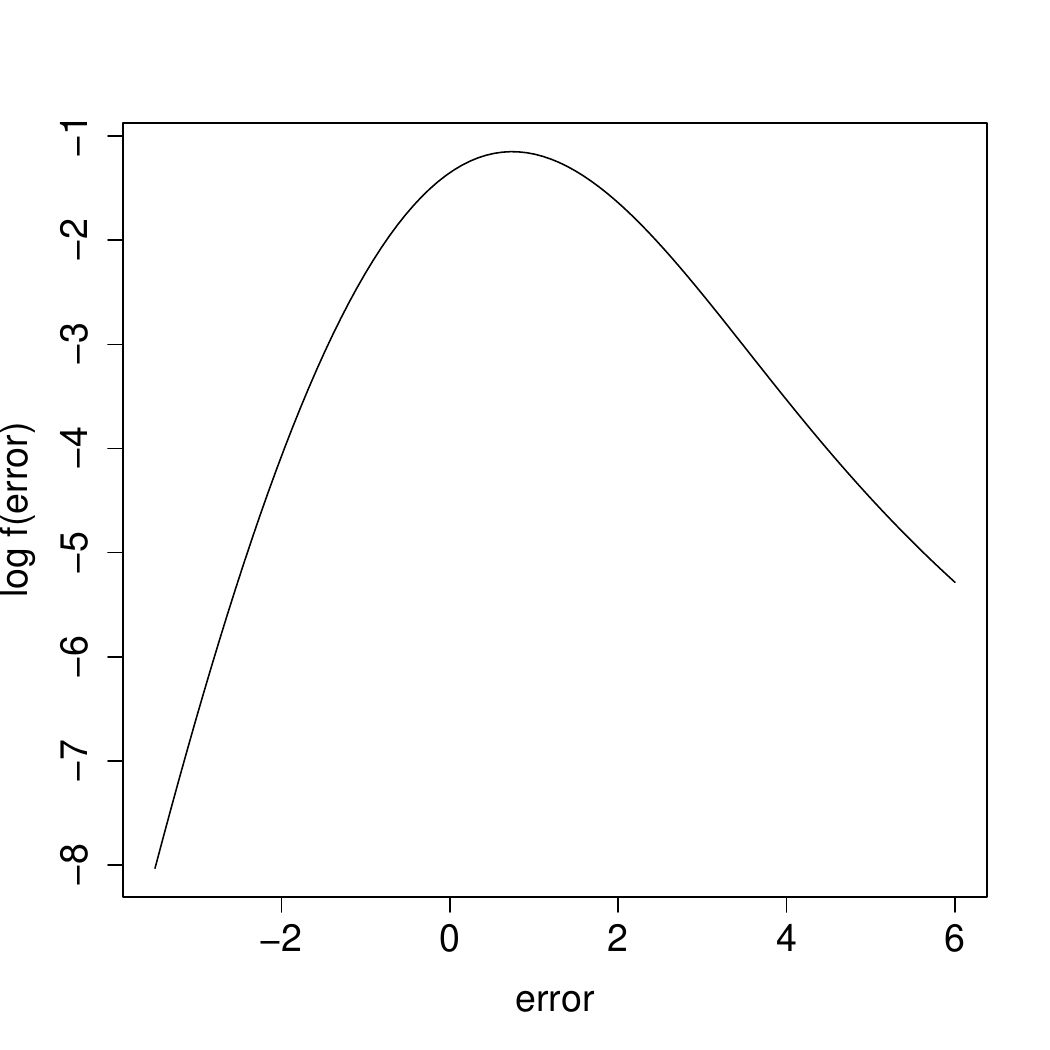}&
\includegraphics[scale=0.19]{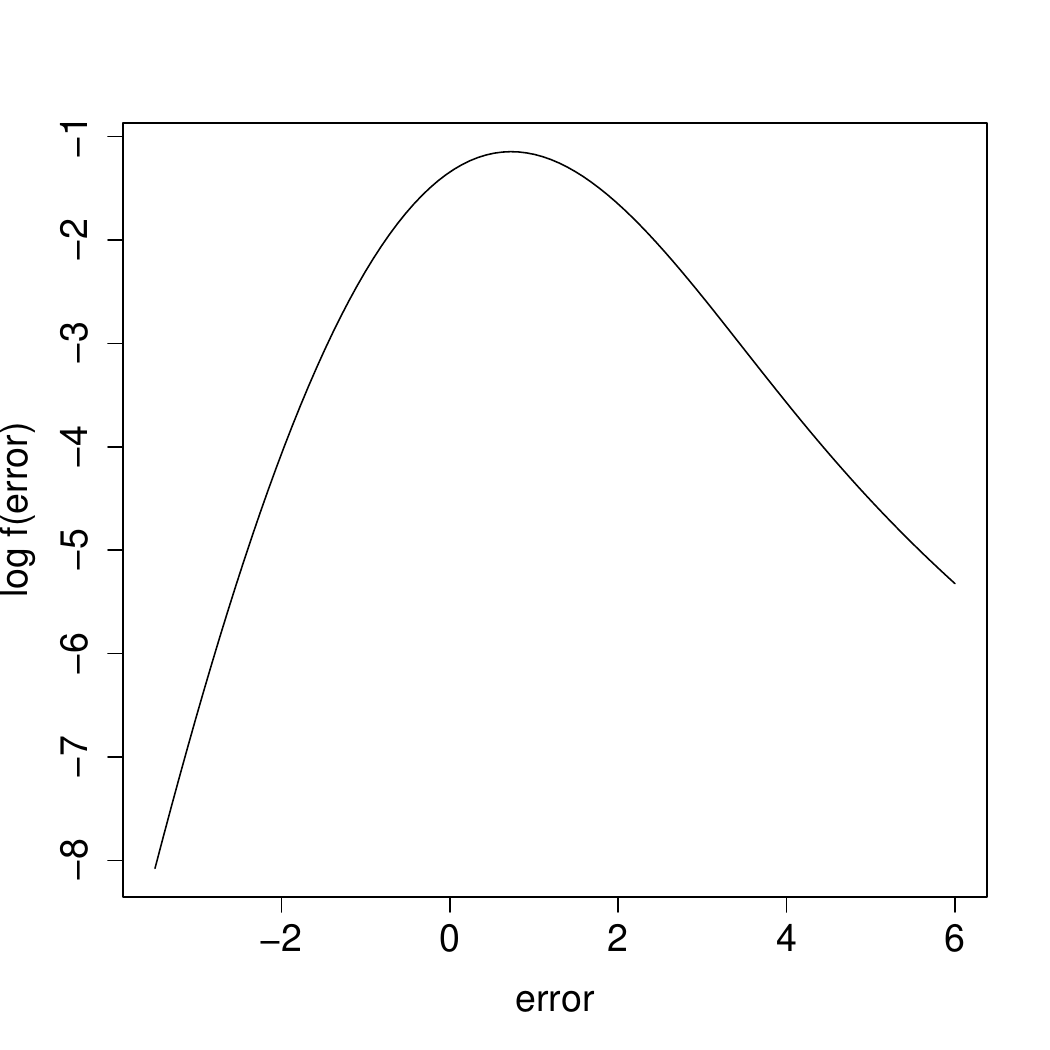}
&
\includegraphics[scale=0.19]{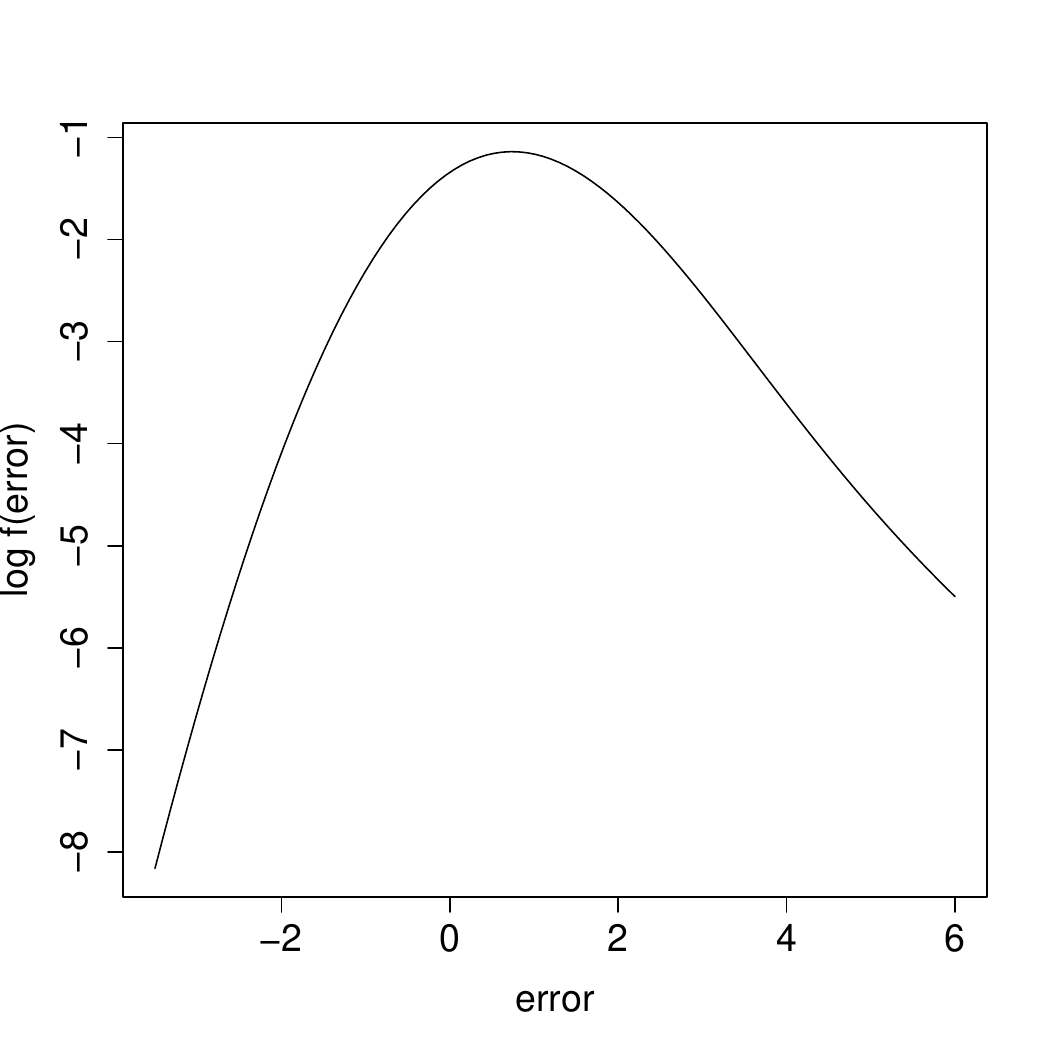}&
\includegraphics[scale=0.19]{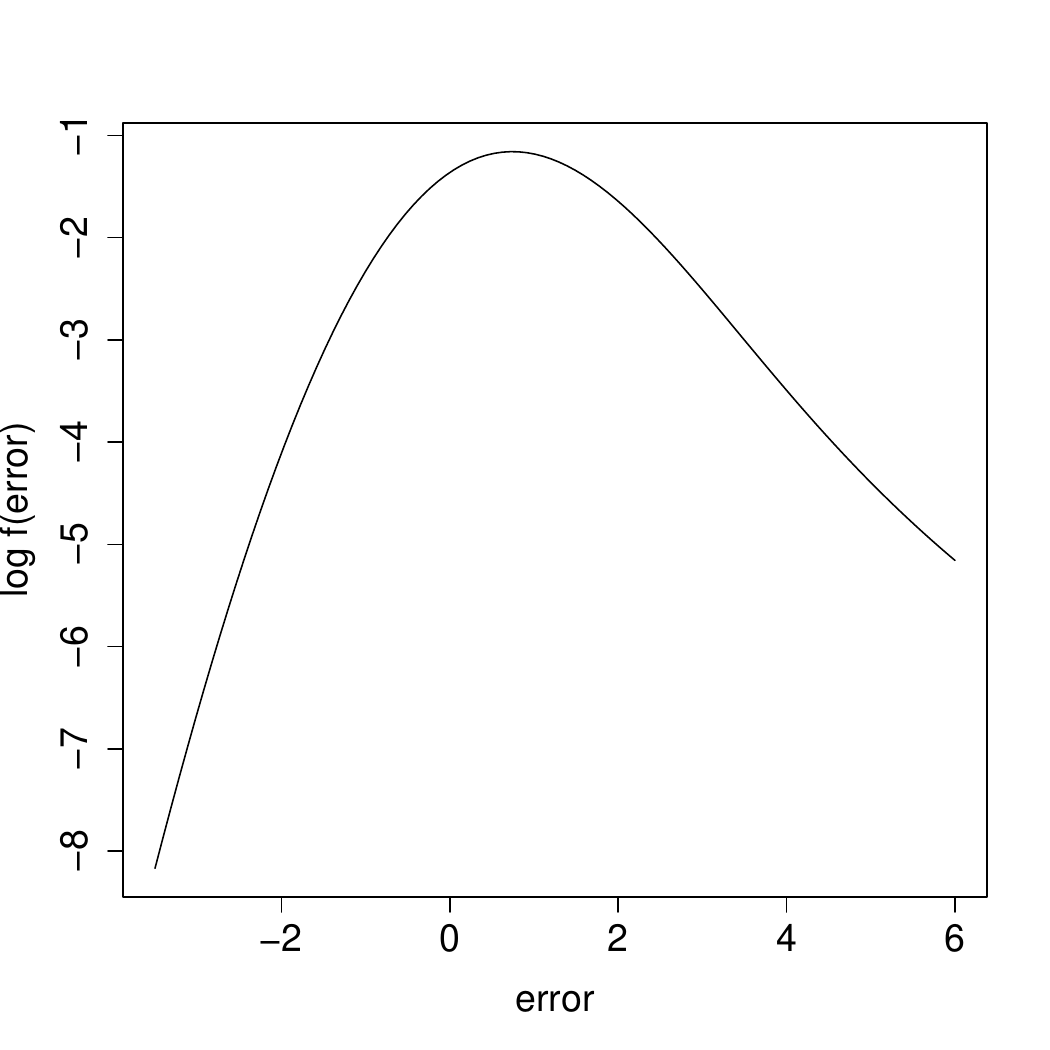}
\end{tabular}
\end{center}
\caption{\label{error_dist1} Probability density function of the errors  
for both females and males in the 100 metre and 200 metre freestyle. Posterior mean  (top row) and logarithm of the posterior mean (bottom row).}
\end{figure}
Figure~\ref{error_dist1} shows the posterior mean estimated error densities for females and males at both distances. The logarithm of the posterior mean more clearly shows the differences in the weight of the two tails. The right-hand tail is clearly much heavier than the left-hand tail which is close to normal (a quadratic decay).

\begin{figure}[!htbp]
\begin{center}
\begin{tabular}{c c c}
Swimmer 1 & Swimmer 2 & Swimmer 3\\
\includegraphics[scale=0.24]{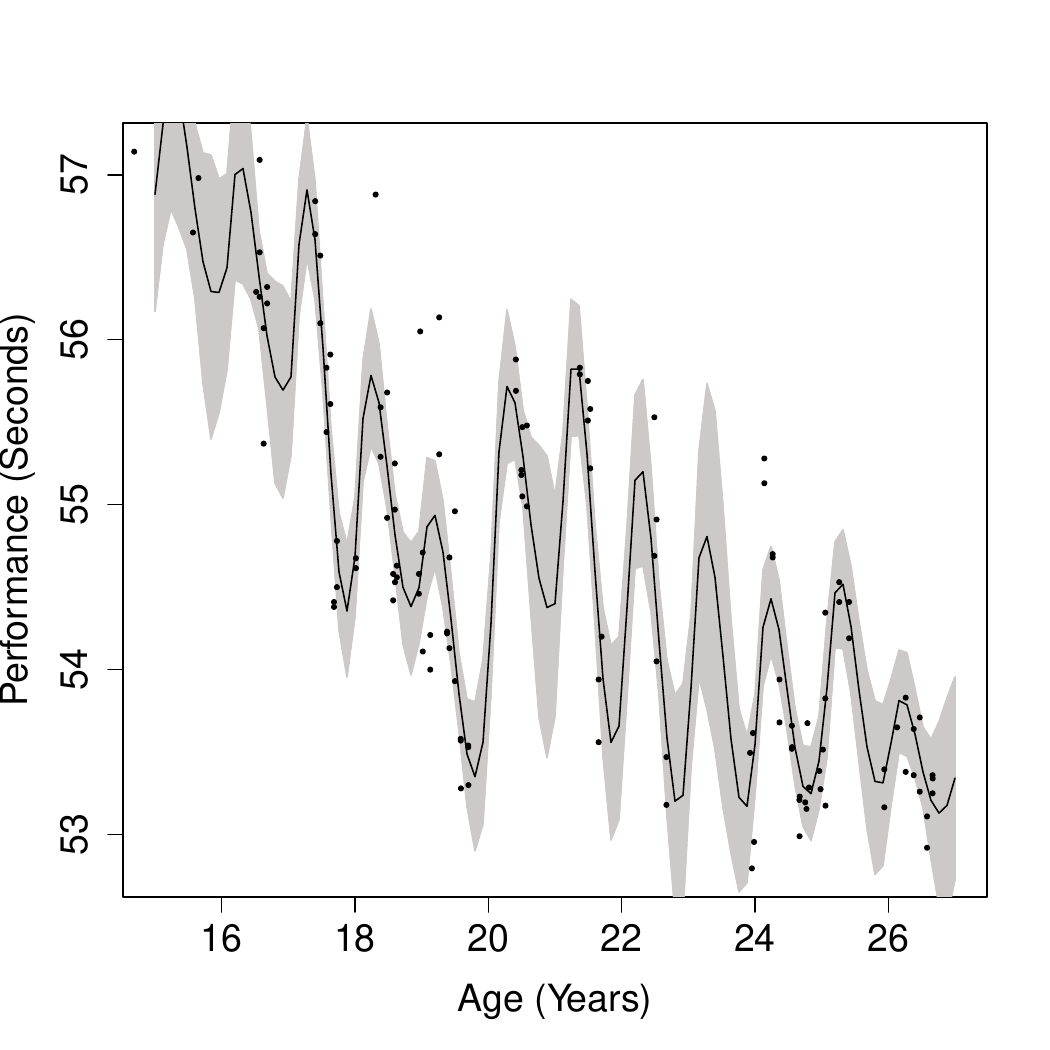}&
\includegraphics[scale=0.24]{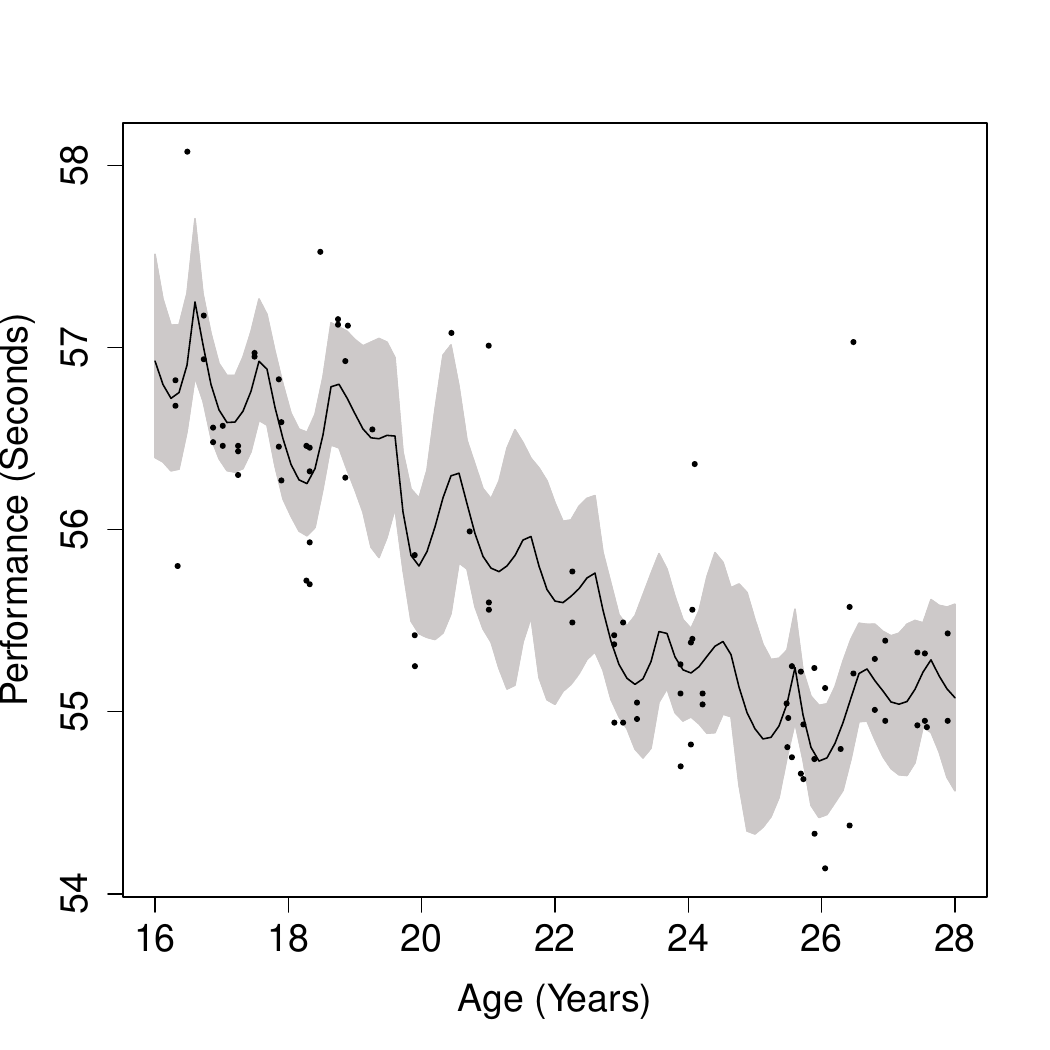}
&\includegraphics[scale=0.24]{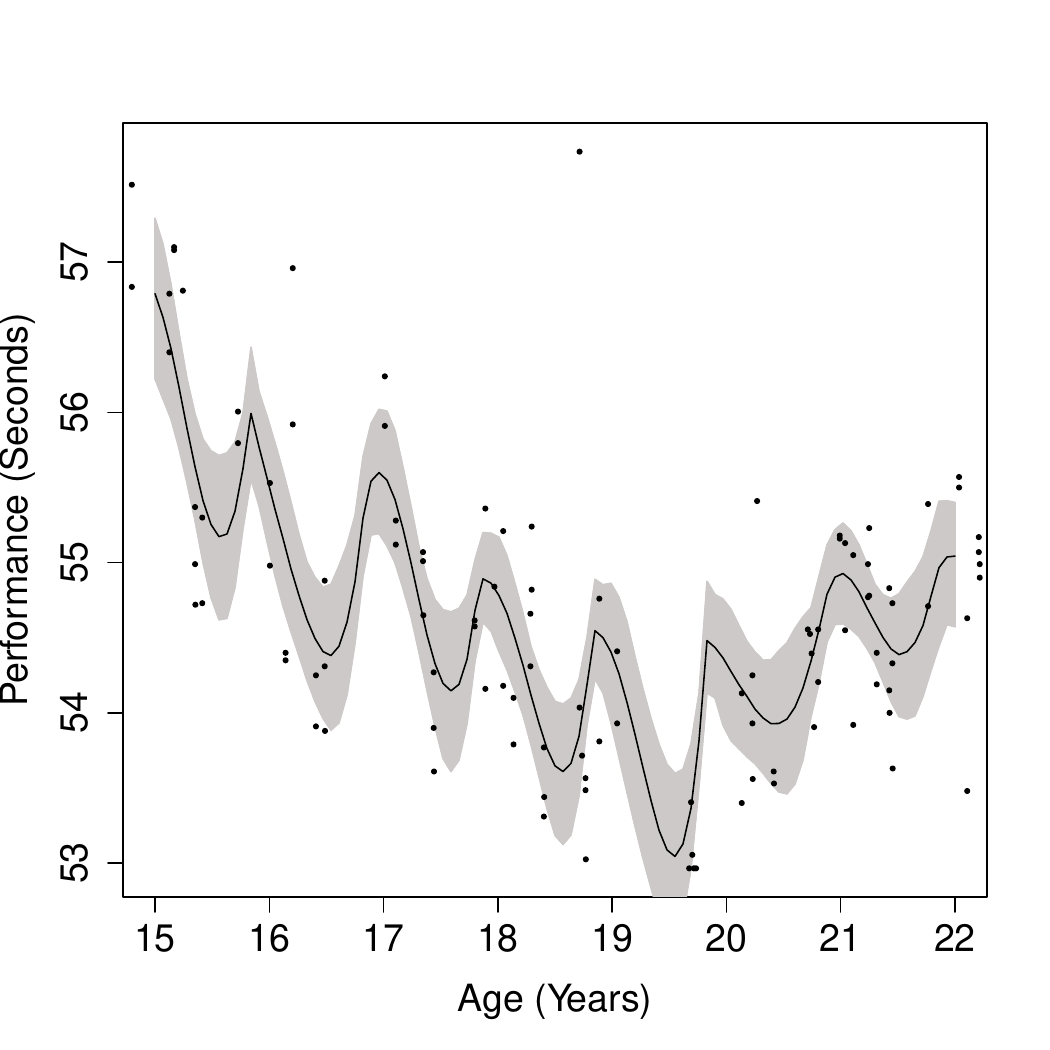}\\
\includegraphics[scale=0.24]{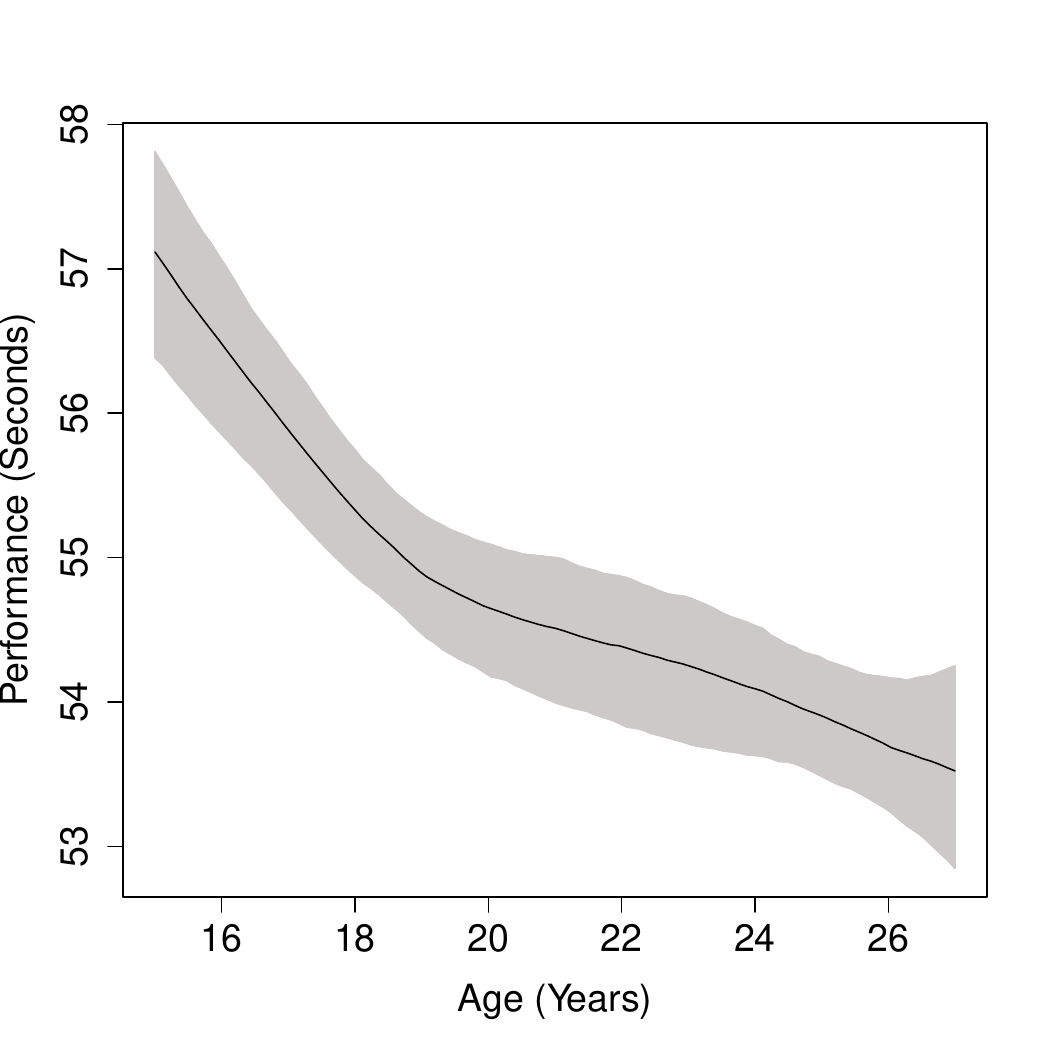} &
\includegraphics[scale=0.24]{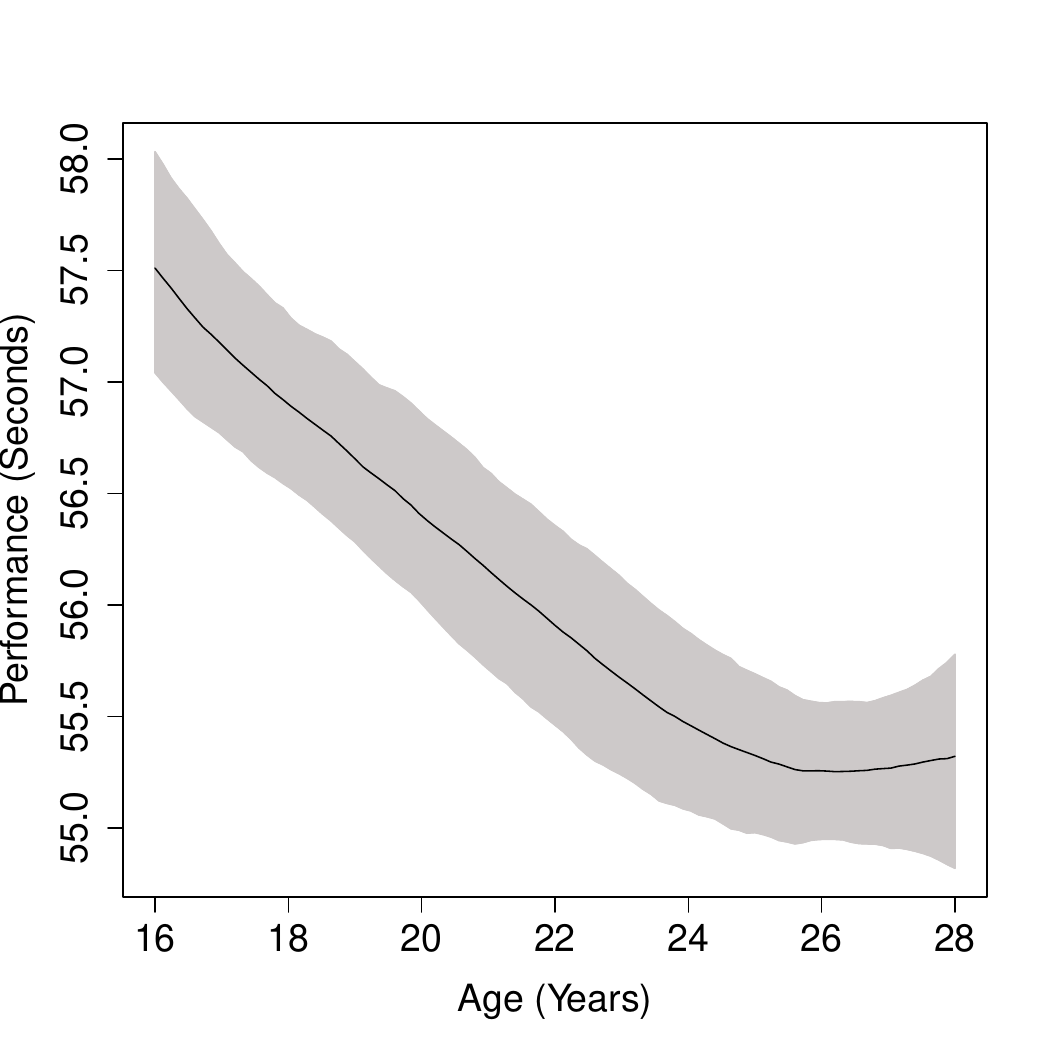} &
\includegraphics[scale=0.24]{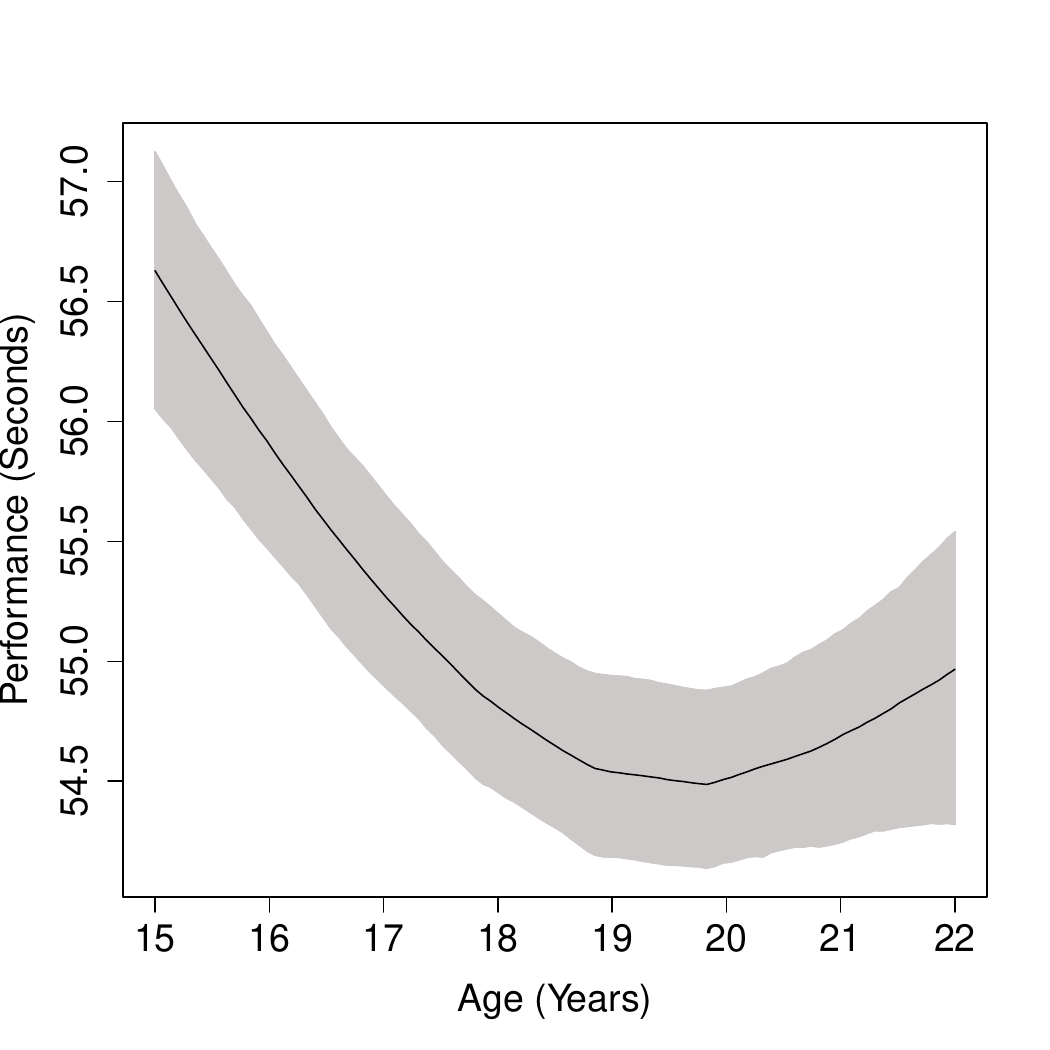}\\
\includegraphics[scale=0.24]{100m_Female_ID_168446_within2.pdf}&
\includegraphics[scale=0.24]
{100m_Female_ID_66604_within2.pdf }&
\includegraphics[scale=0.24]{100m_Female_ID_5021_within2.pdf}
\end{tabular}
\end{center}
\caption{\label{female1} 
 Female 100 metre freestyle swimmers:
Individual performance trajectories (top row) and estimated individual trend performance trajectories (middle row). The trajectories are shown as posterior median (black line) and 95\% credible interval (grey shading)
with the observed performances (dots).
 Bottom row: posterior median athlete-level (black line) and  within-season performance trajectories for each career season (grey lines).
}
\end{figure}

\begin{figure}[!htbp]
\begin{center}
\begin{tabular}{c c c}
Swimmer 4 & Swimmer 5 & Swimmer 6\\
\includegraphics[scale=0.24]{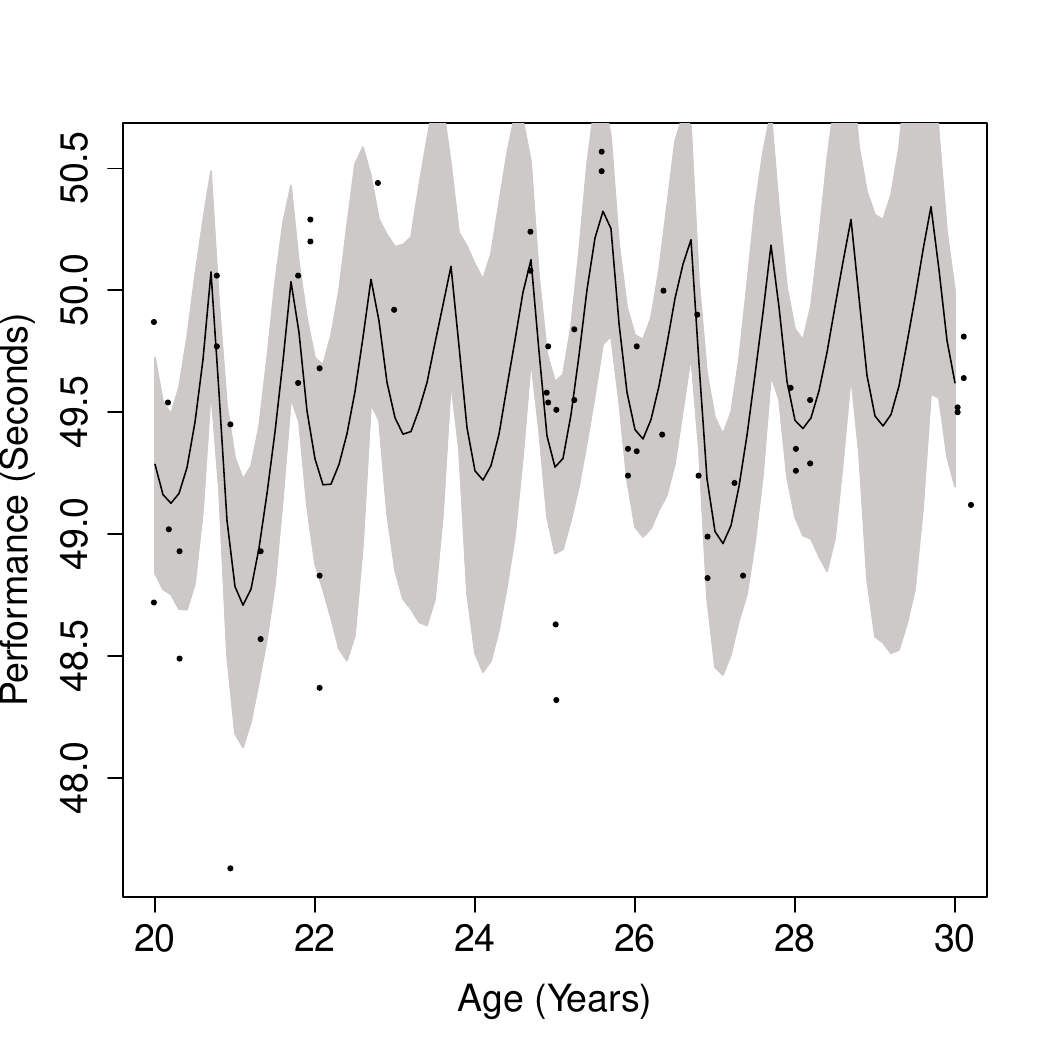}&
\includegraphics[scale=0.24]{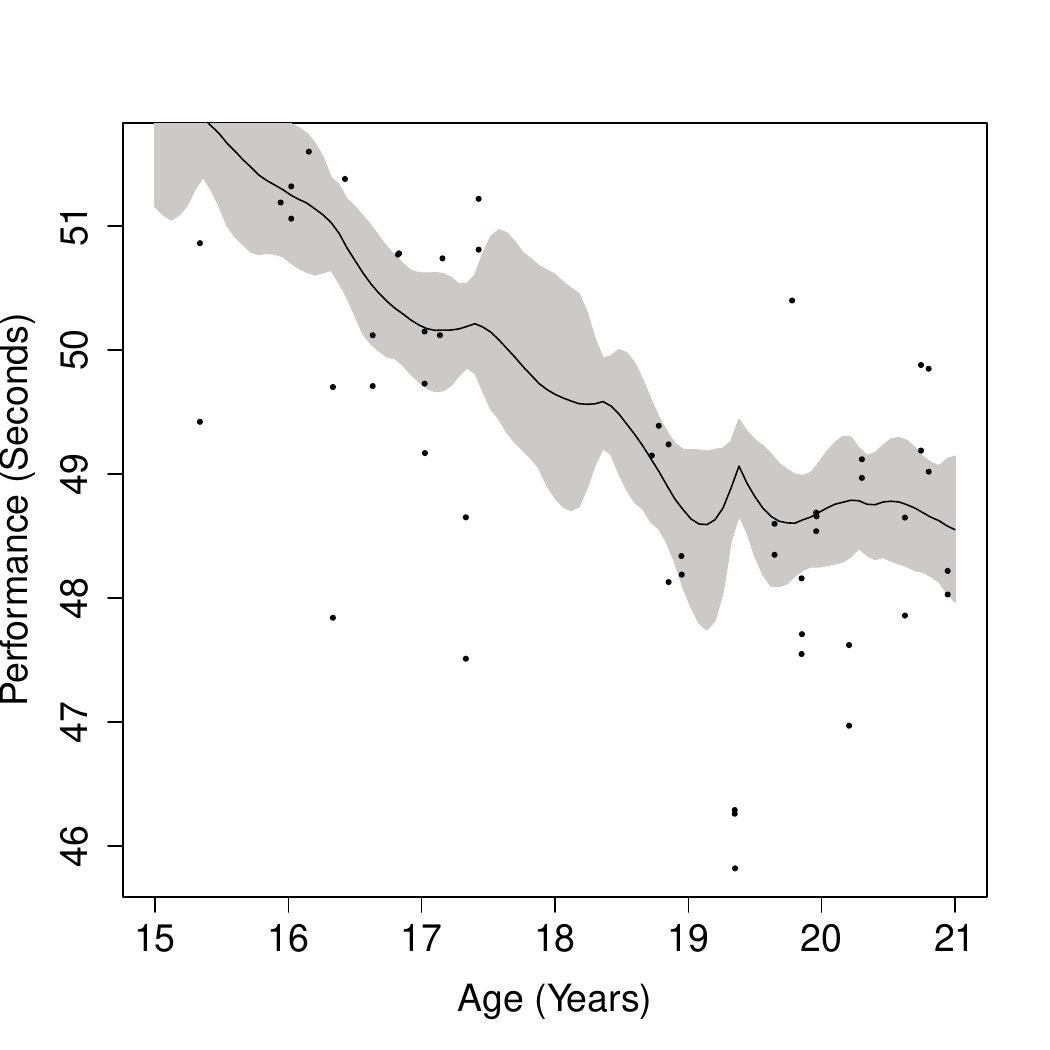}&
\includegraphics[scale=0.24]{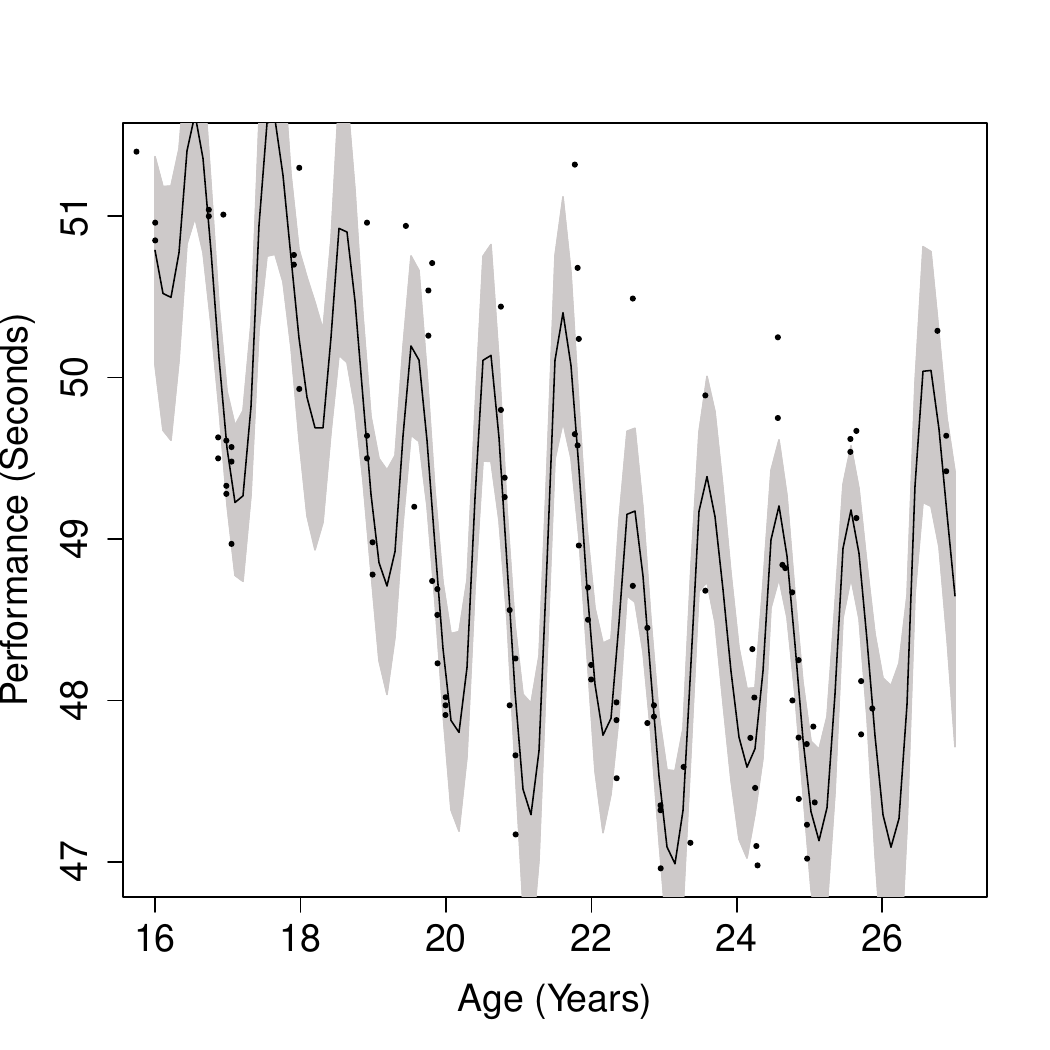}\\
\includegraphics[scale=0.24]{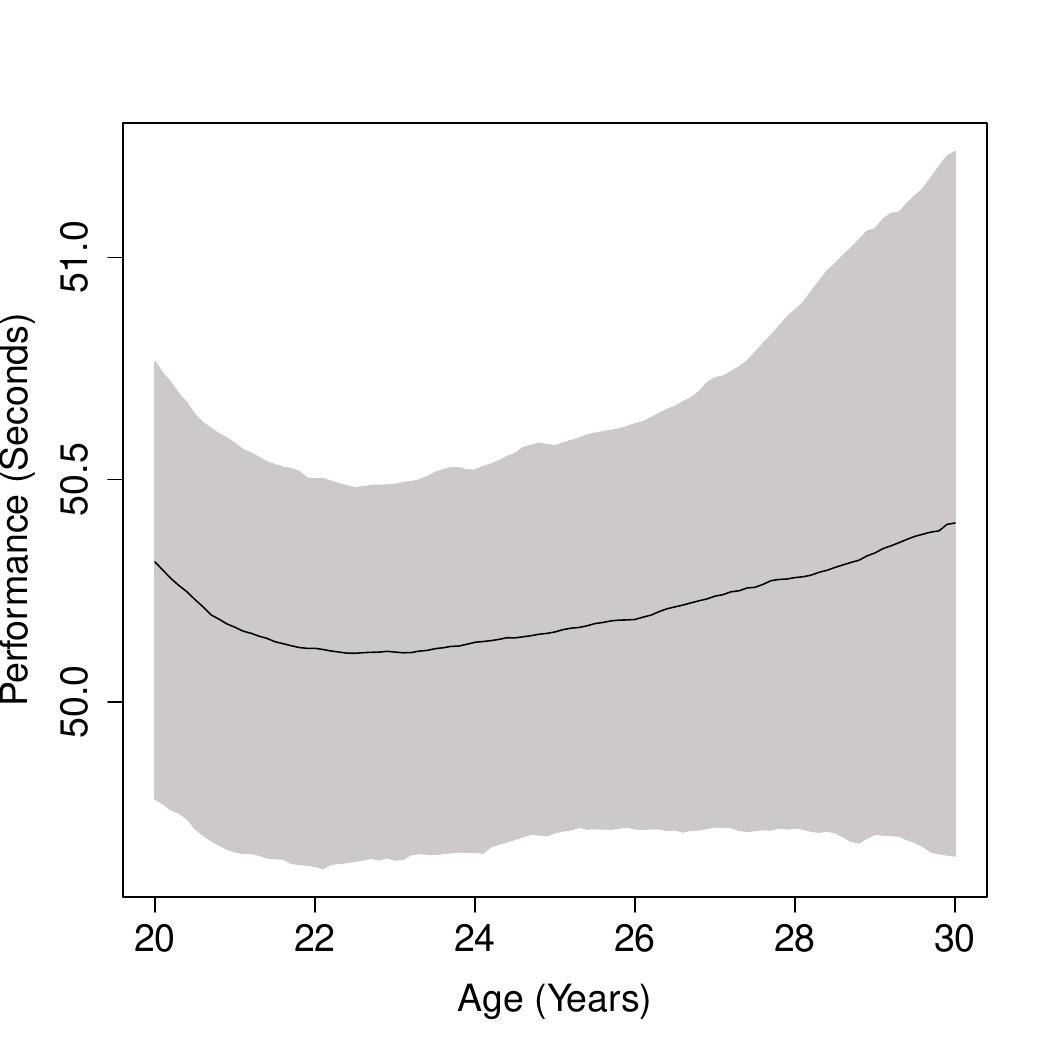} &
\includegraphics[scale=0.24]{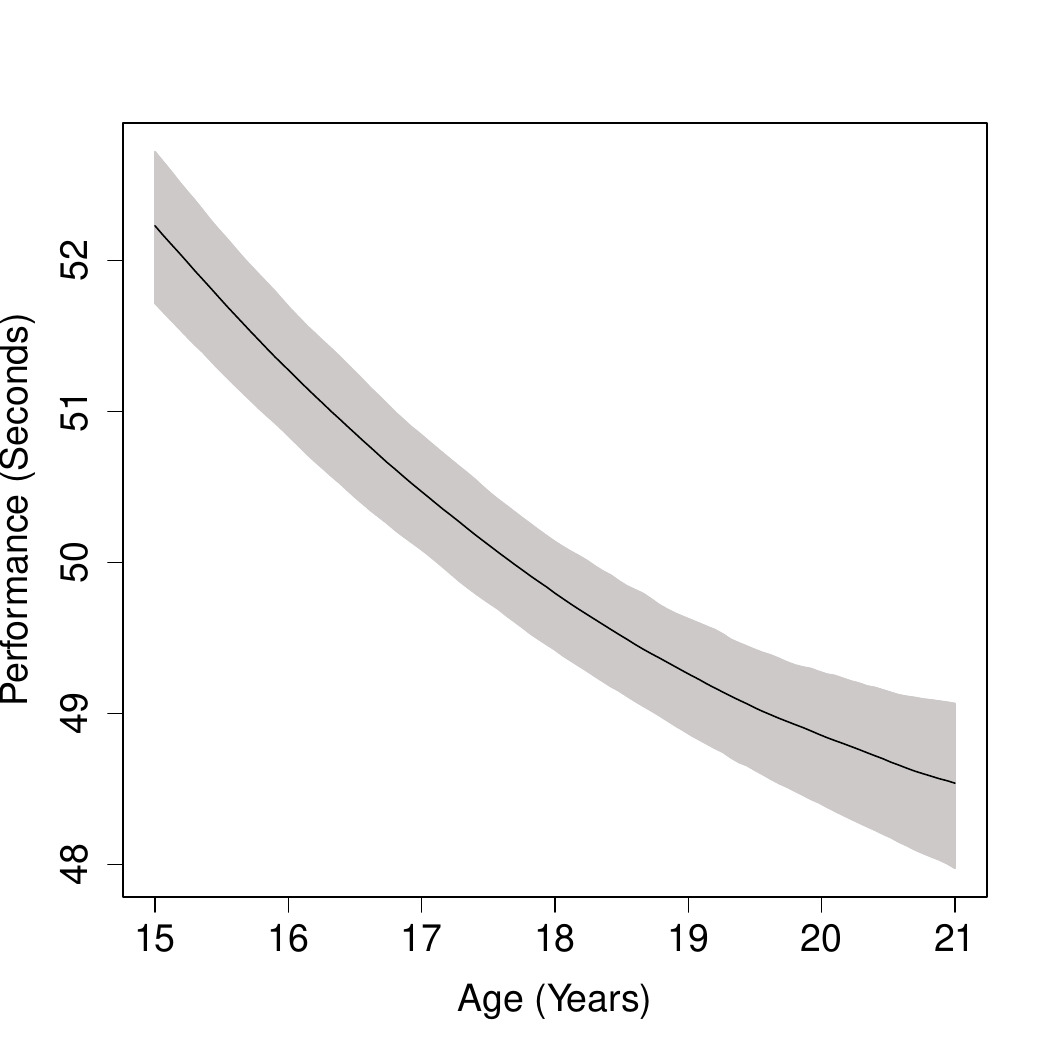} &
\includegraphics[scale=0.24]{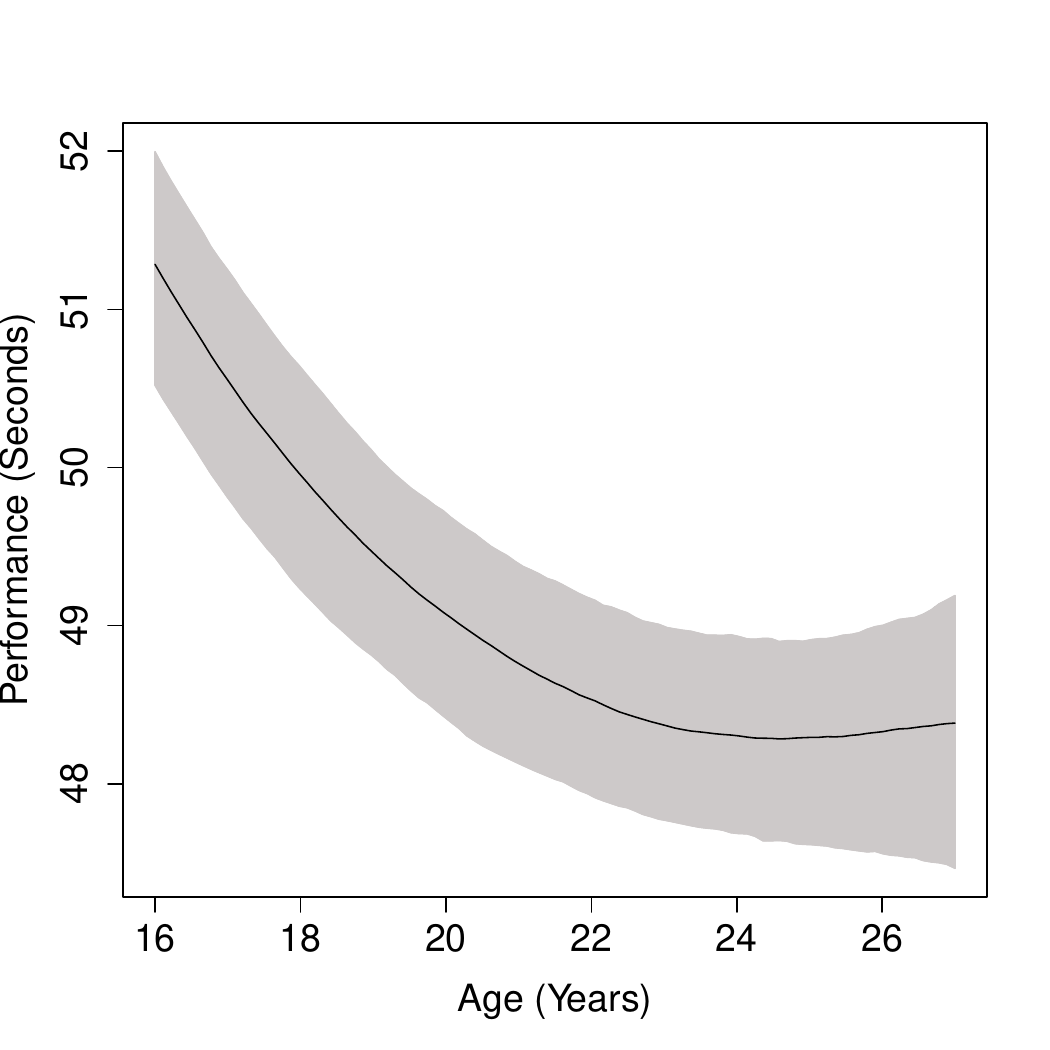}\\
\includegraphics[scale=0.24]{100m_Male_ID_131527_within2.pdf}&
\includegraphics[scale=0.24]
{100m_Male_ID_89567_within2.pdf}&
\includegraphics[scale=0.24]{100m_Male_ID_38649_within2.pdf}
\end{tabular}
\end{center}
\caption{\label{male1} 
Male 100 metre freestyle swimmers:
Individual performance trajectories (top row) and estimated individual trend performance trajectories (middle row). The trajectories are shown as posterior median (black line) and 95\% credible interval (grey shading)
with the observed performances (dots).  
Bottom row: posterior median athlete-level (black line) and  within-season performance trajectories for each career season (grey lines).}
\end{figure}

\begin{figure}[!htbp]
\begin{center}
\begin{tabular}{c c c}
Swimmer 4 & Swimmer 5 & Swimmer 6\\
\includegraphics[scale=0.24]{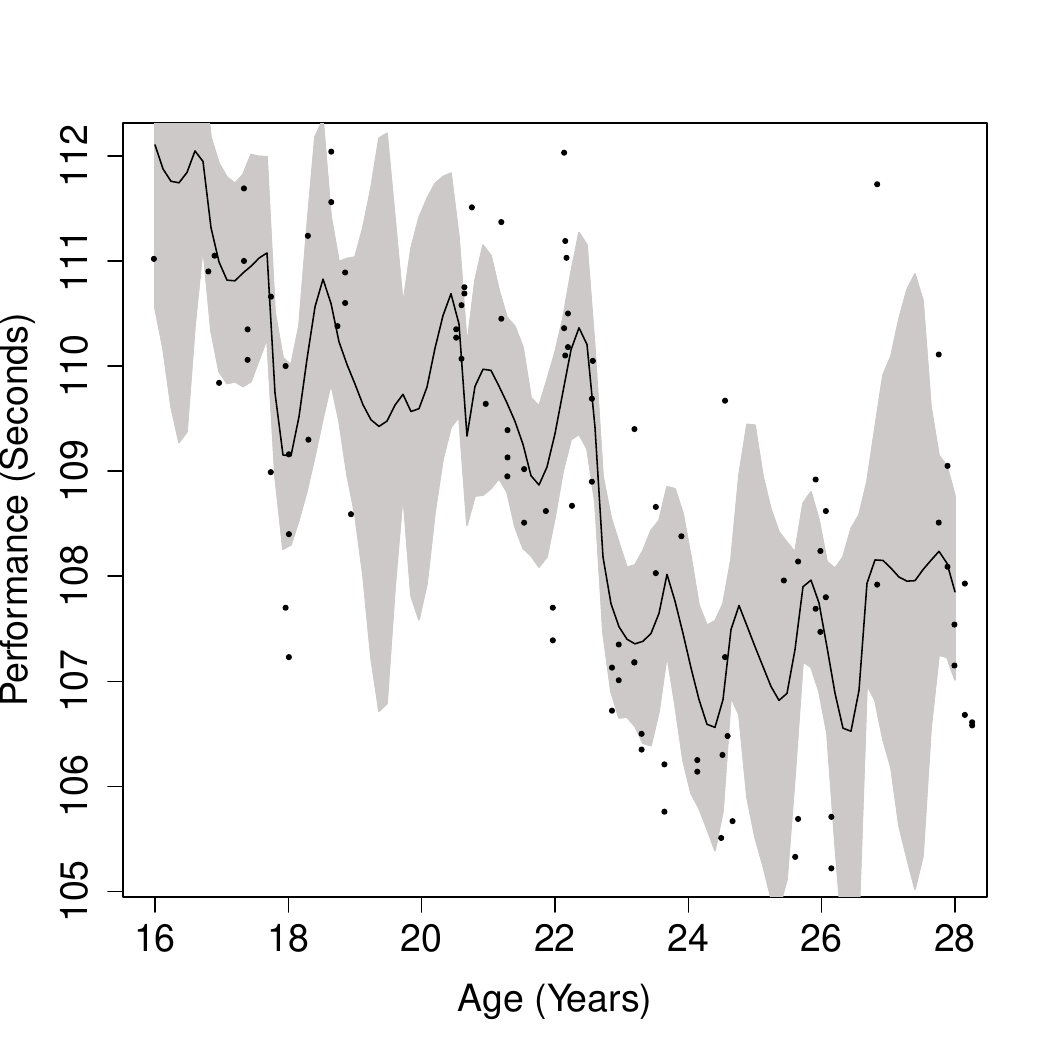}&
\includegraphics[scale=0.24]{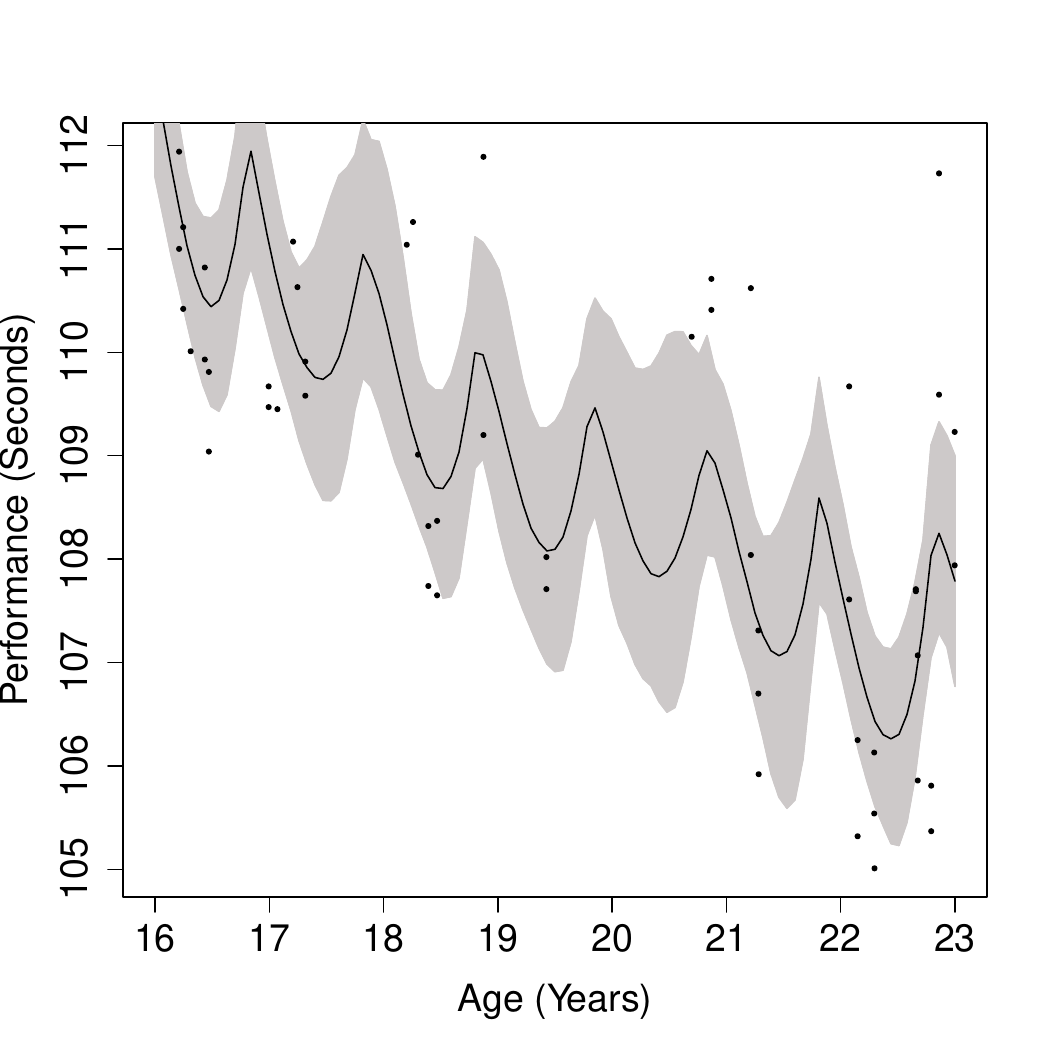}&
\includegraphics[scale=0.24]{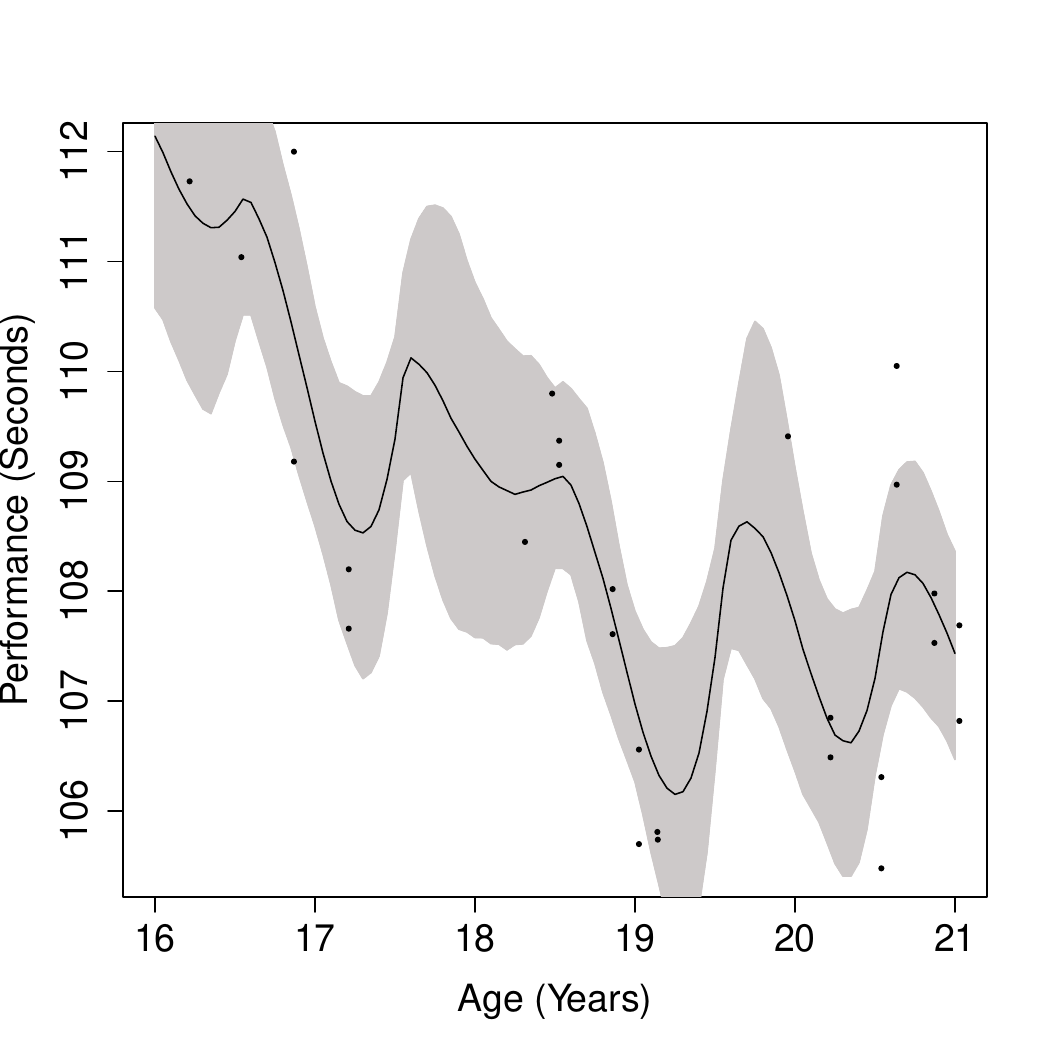}\\
\includegraphics[scale=0.24]{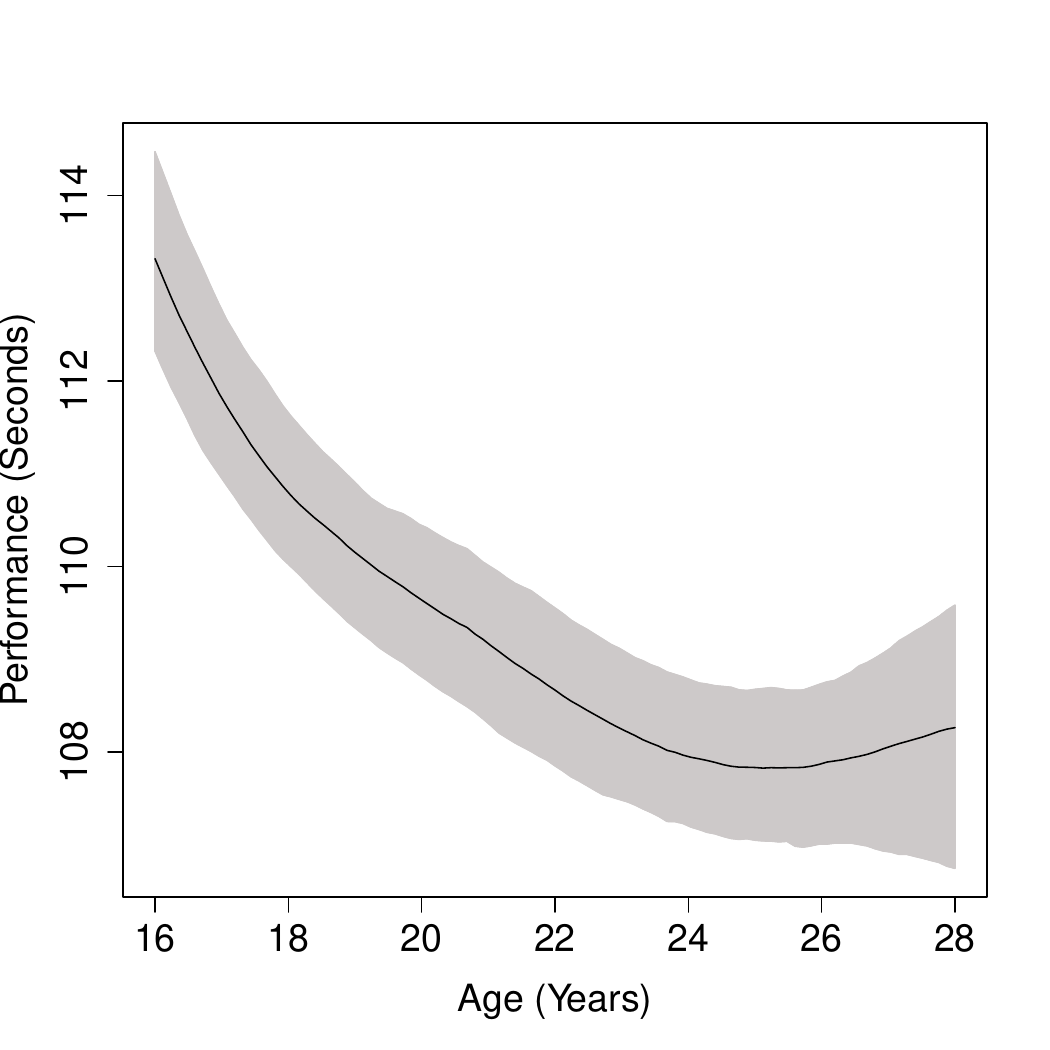} &
\includegraphics[scale=0.24]{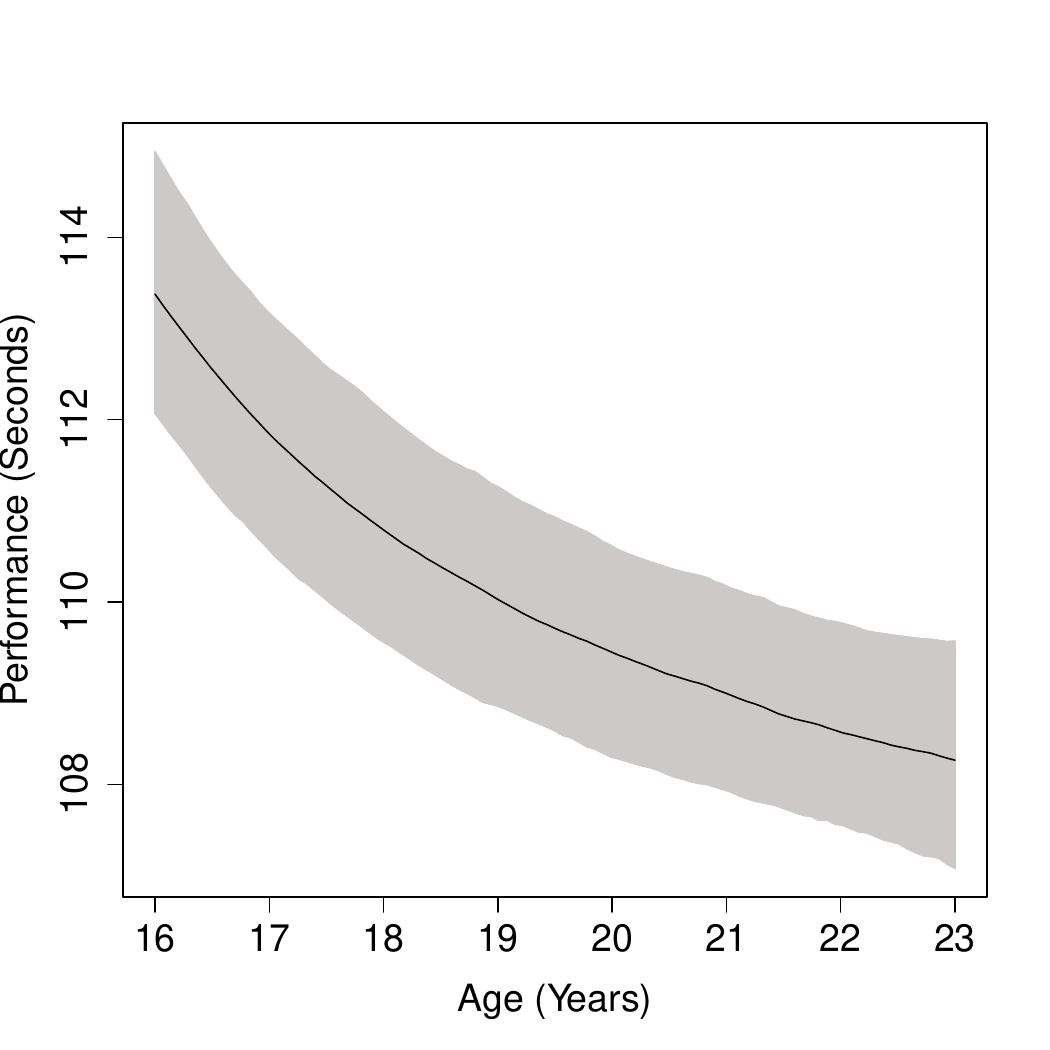} &
\includegraphics[scale=0.24]{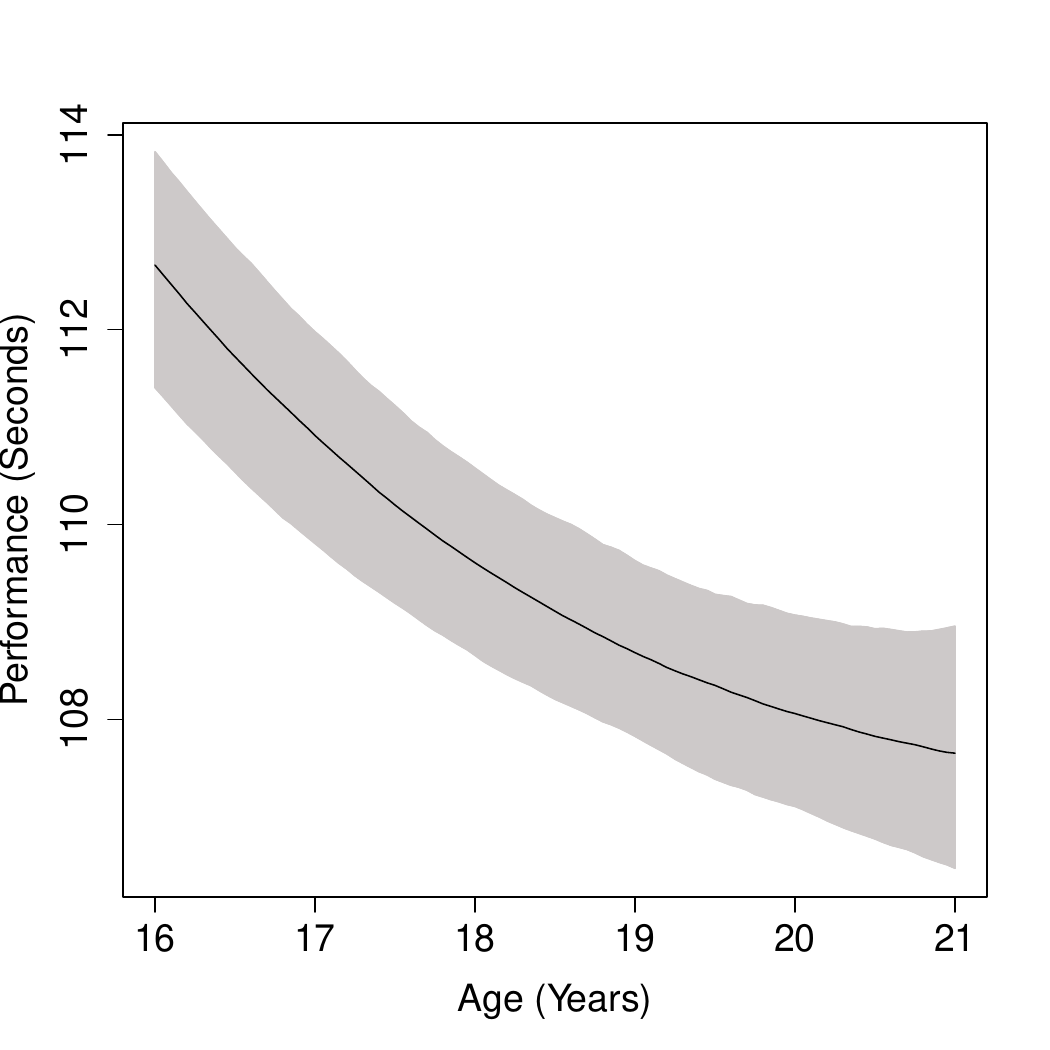}\\
\includegraphics[scale=0.24]{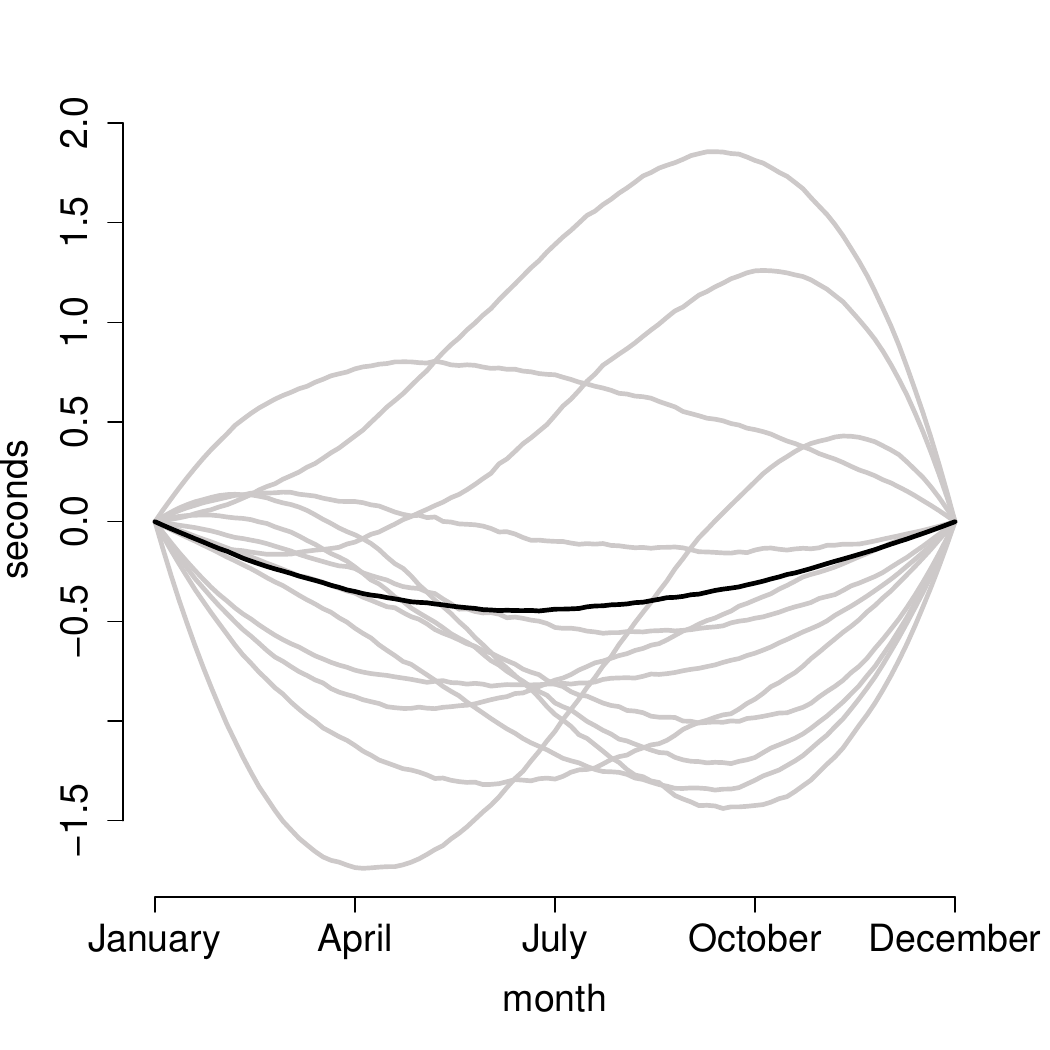}&
\includegraphics[scale=0.24]
{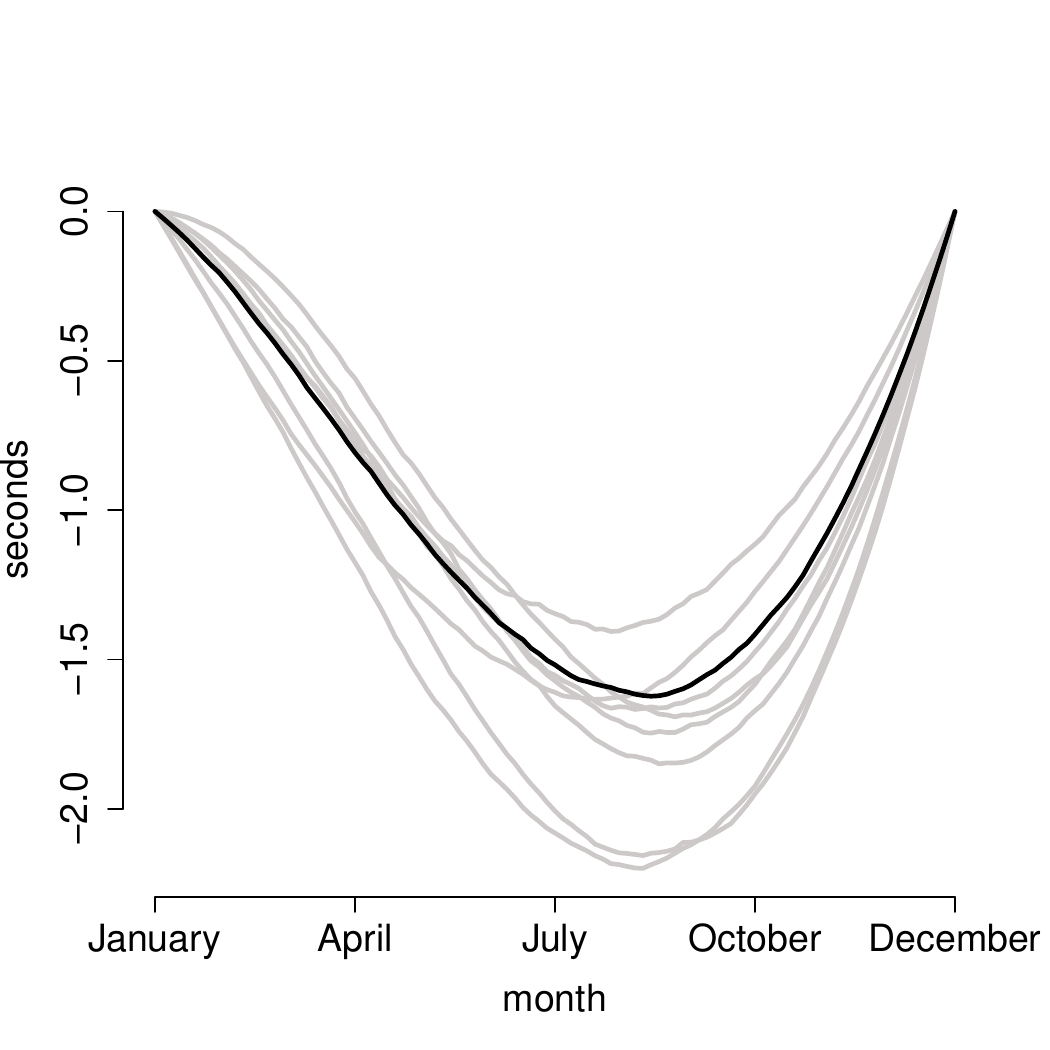}&
\includegraphics[scale=0.24]{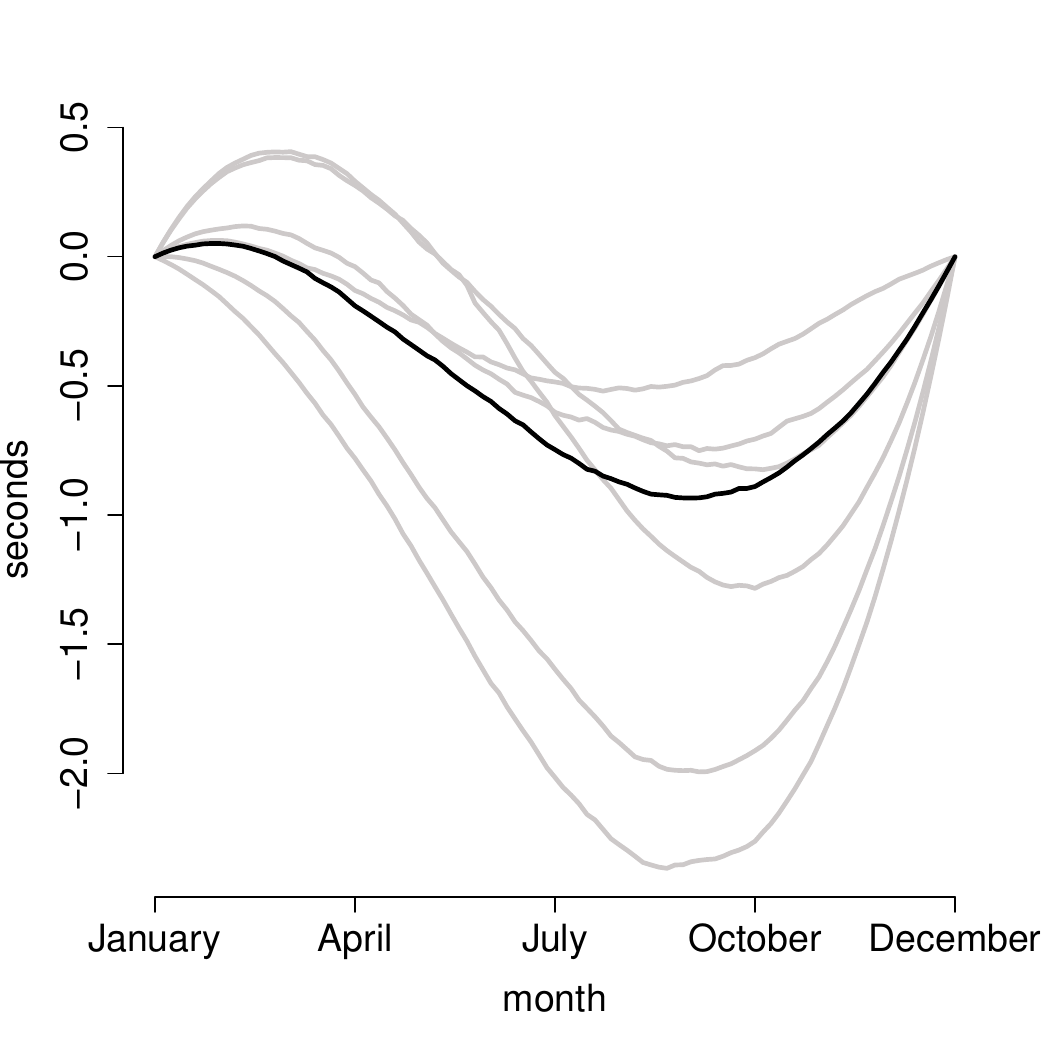}
\end{tabular}
\end{center}
\caption{\label{male2} 
Male 200 metres freestyle swimmers:
Individual performance trajectories (top row) and estimated individual trend performance trajectories (middle row). The trajectories are shown as posterior median (black line) and 95\% credible interval (grey shading)
with the observed performances (dots).  
 Bottom row: posterior median athlete-level (black line) and  within-season performance trajectories for each career season (grey lines).}
\end{figure}

\begin{figure}[!htbp]
\begin{center}
\begin{tabular}{c c c}
Swimmer 4 & Swimmer 5 & Swimmer 6\\
\includegraphics[scale=0.24]{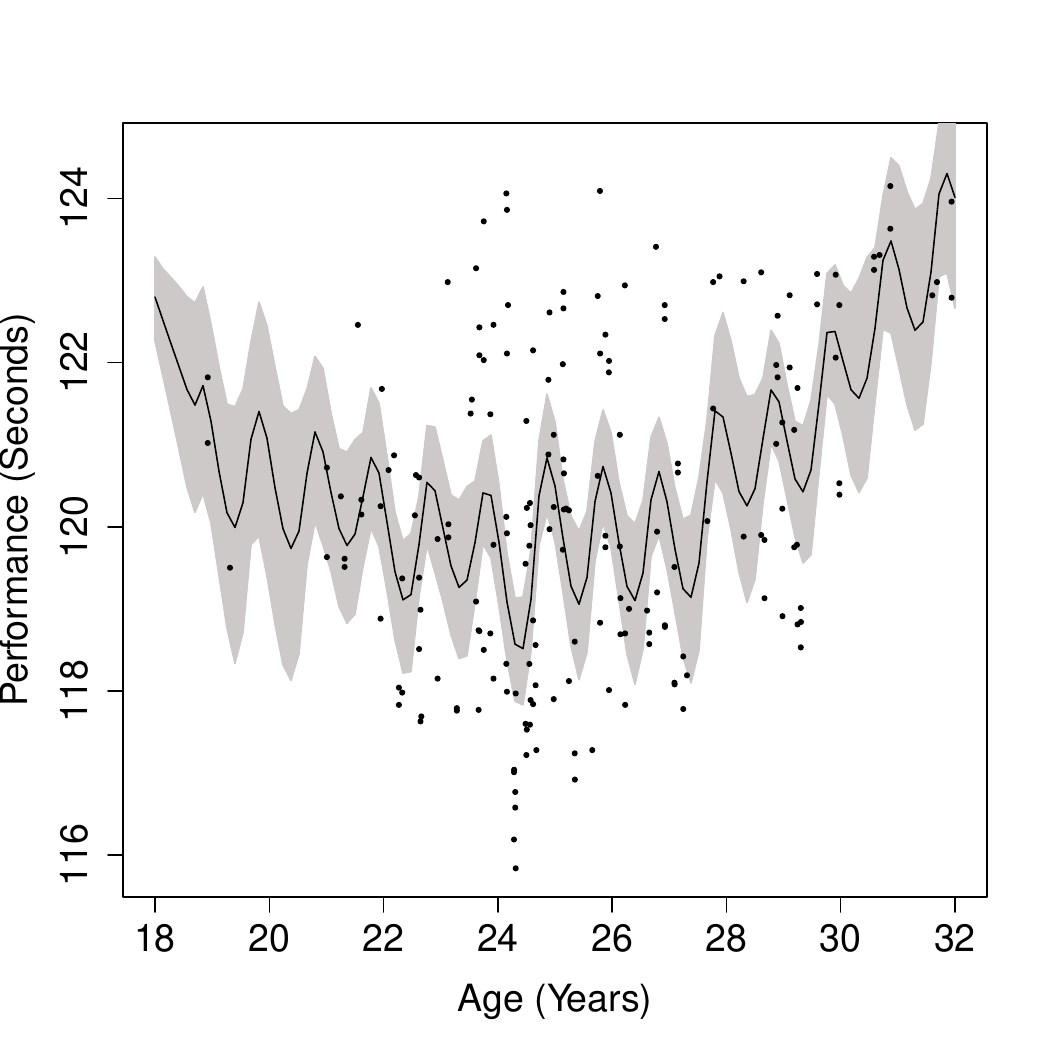}&
\includegraphics[scale=0.24]{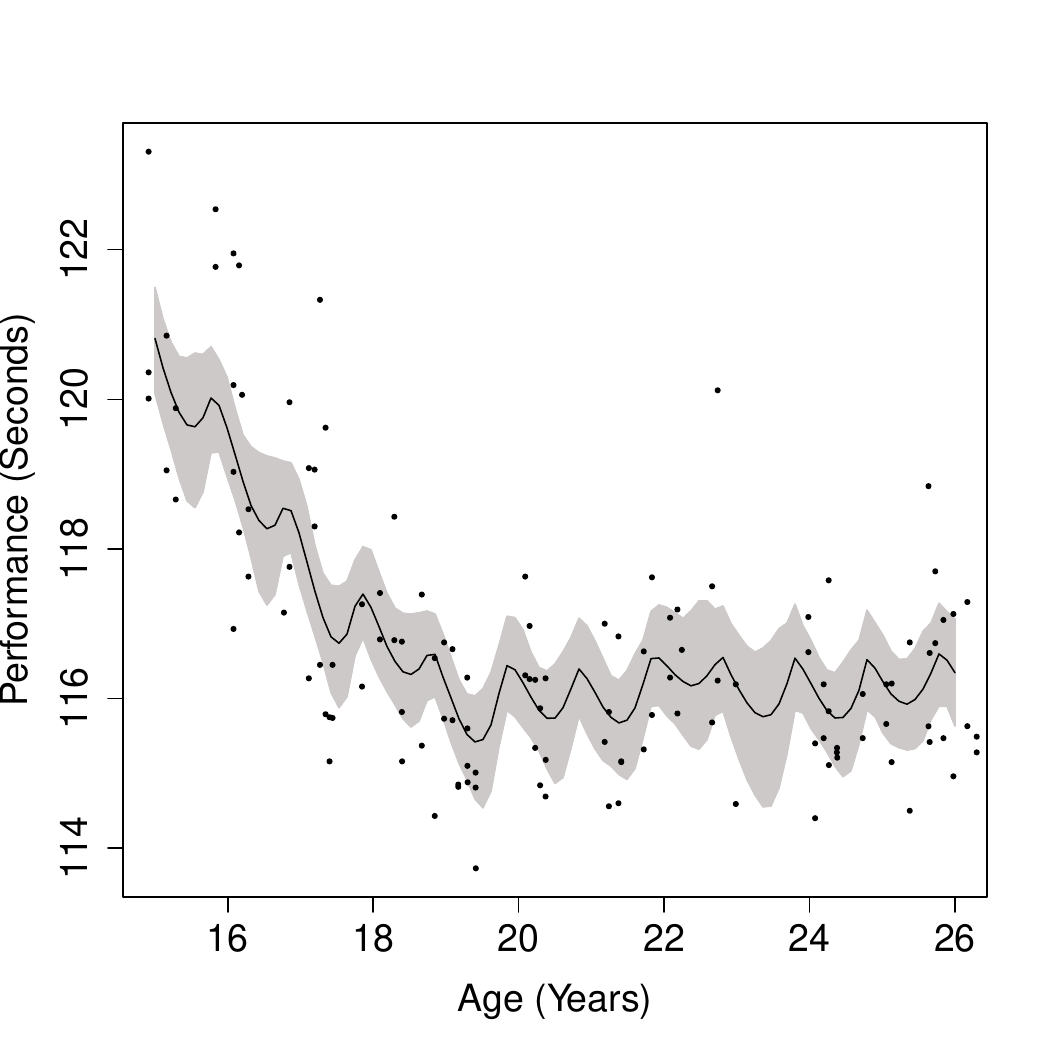}&
\includegraphics[scale=0.24]{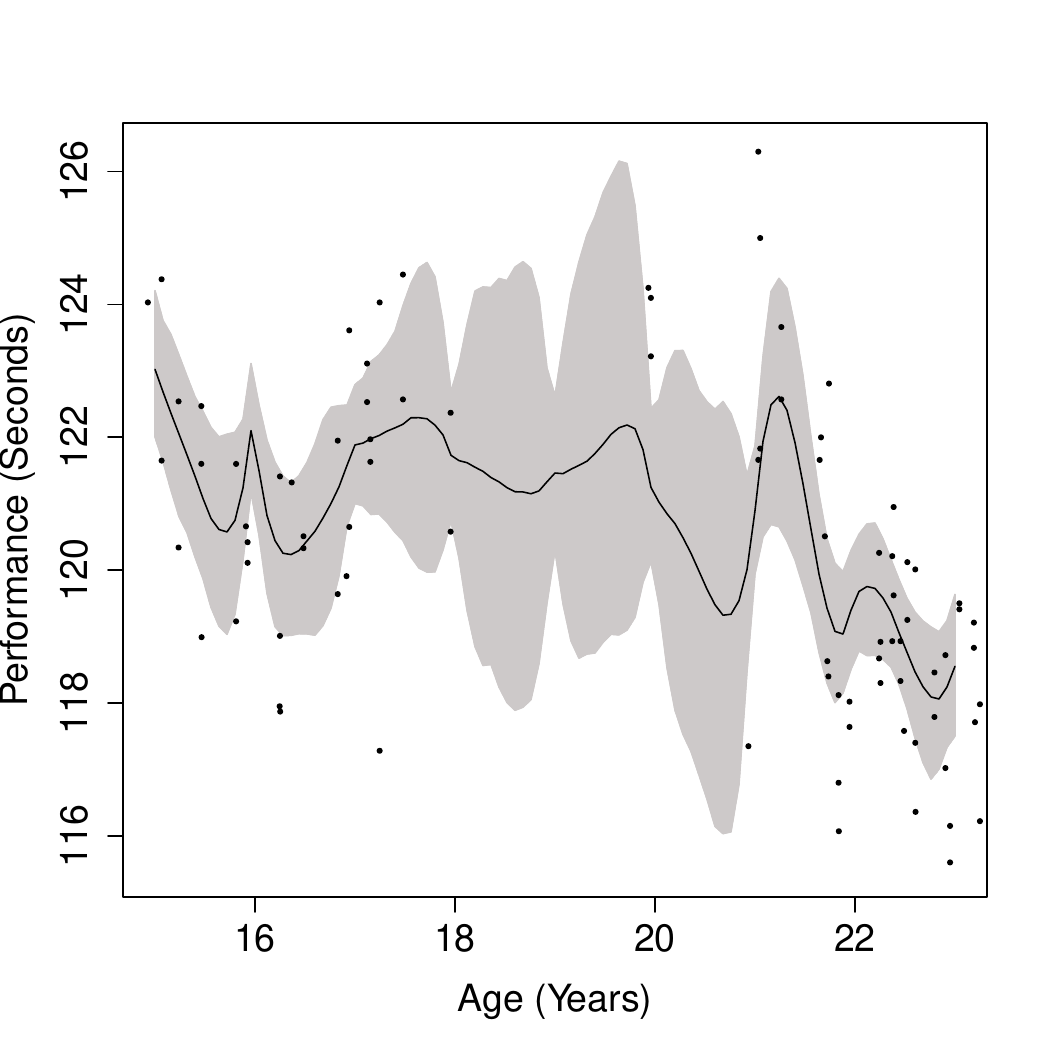}\\
\includegraphics[scale=0.24]{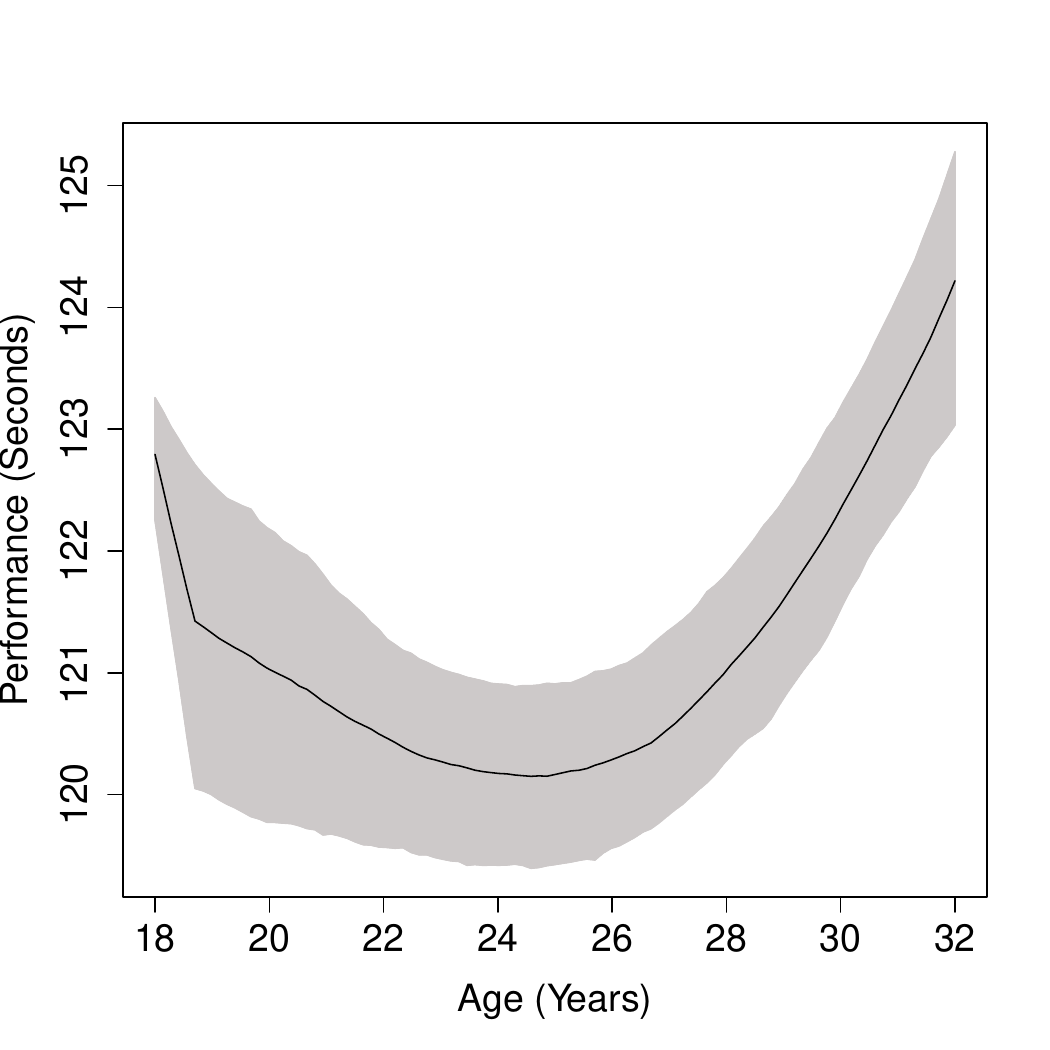} &
\includegraphics[scale=0.24]{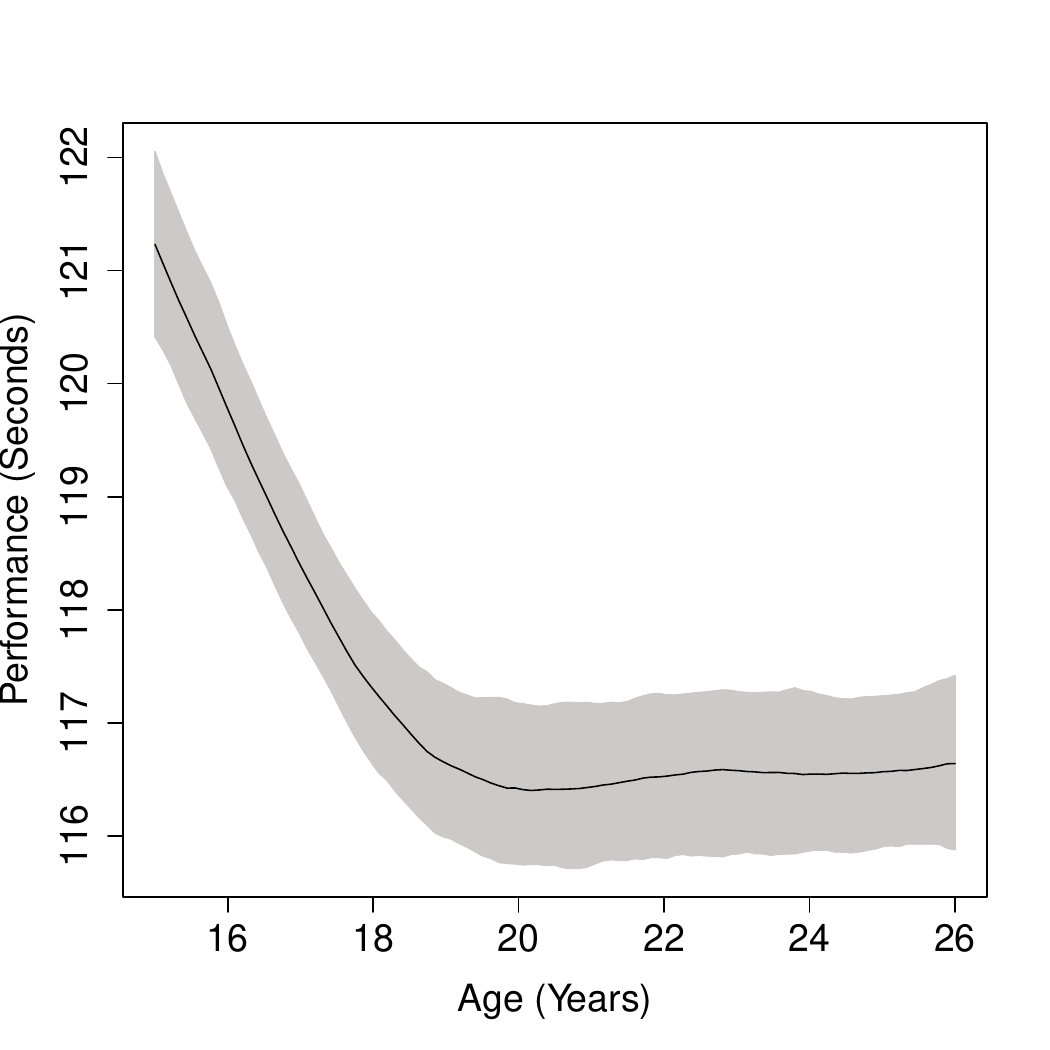} &
\includegraphics[scale=0.24]{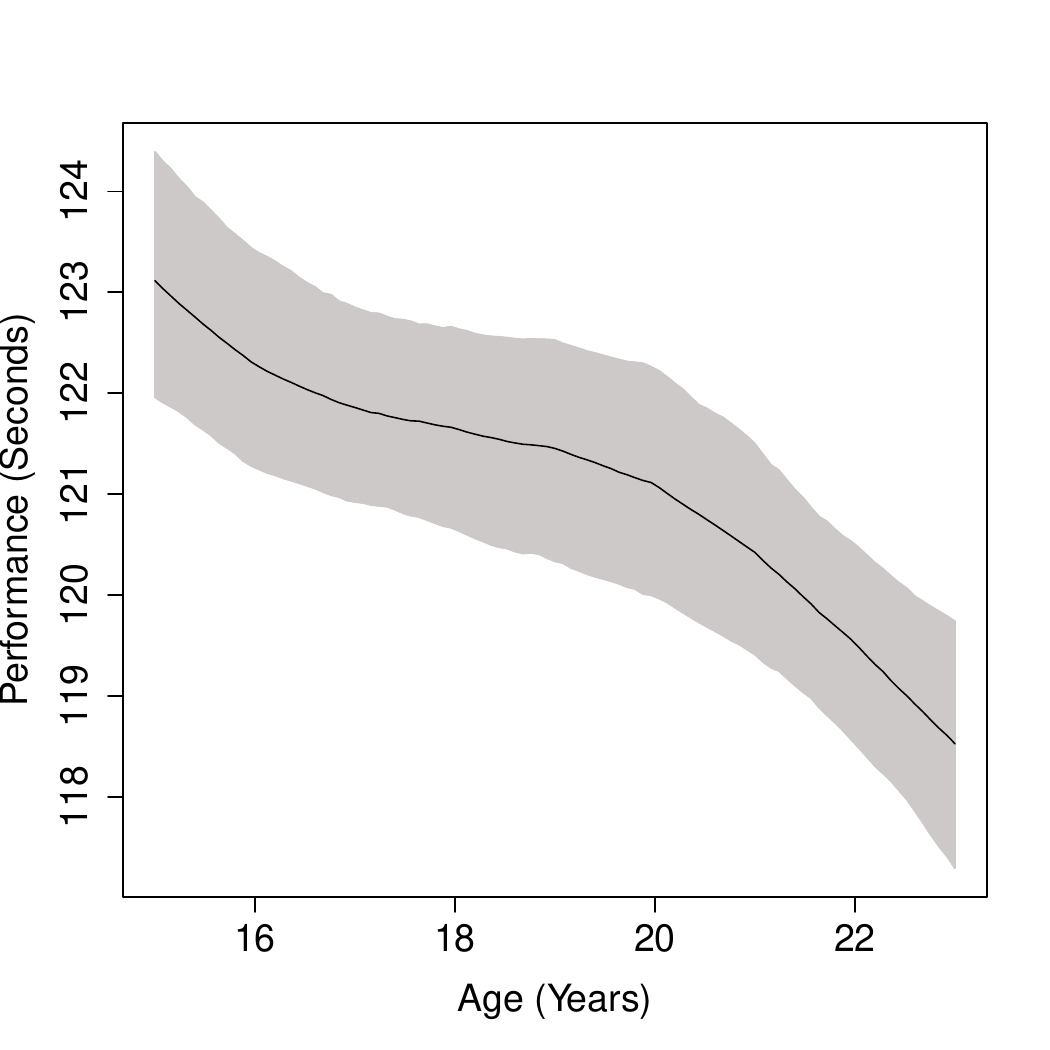}\\
\includegraphics[scale=0.24]{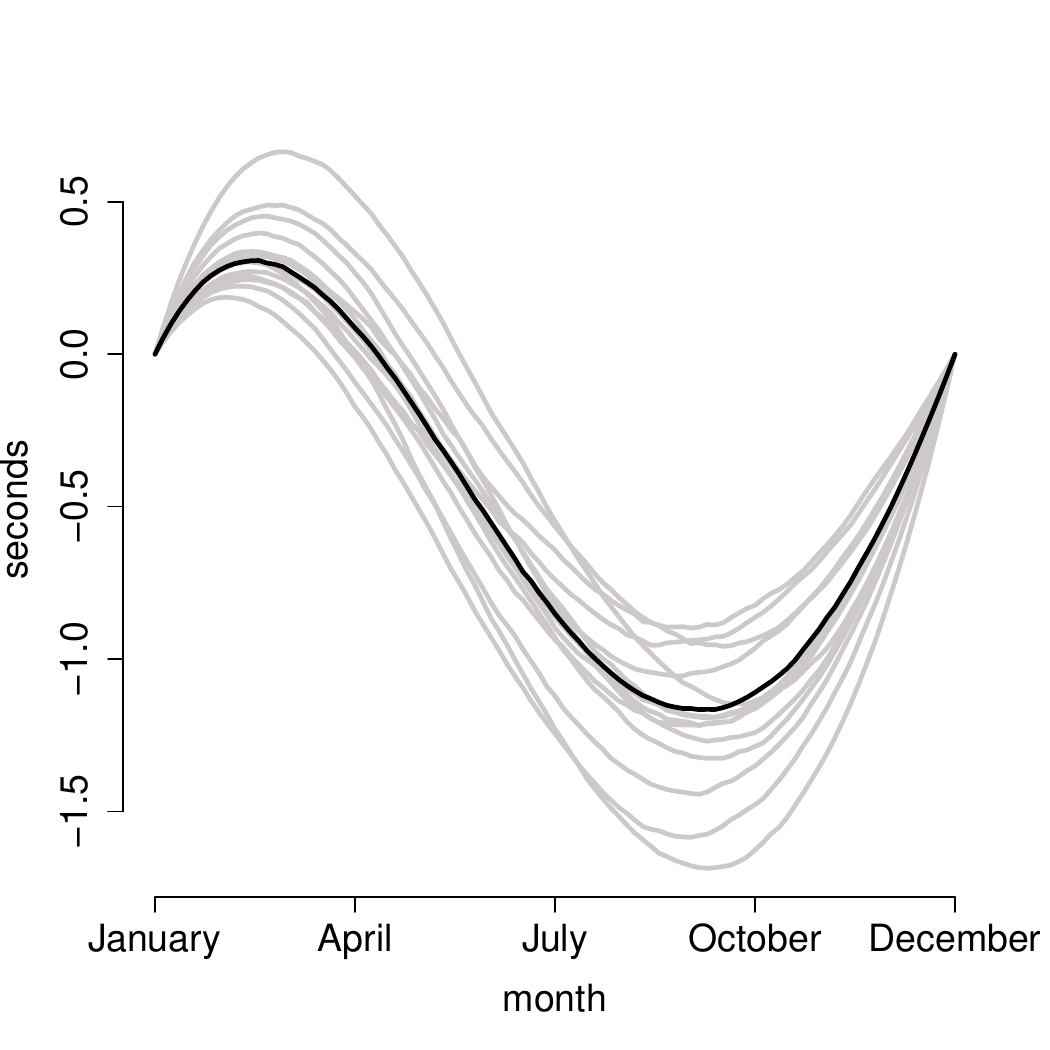}&
\includegraphics[scale=0.24]
{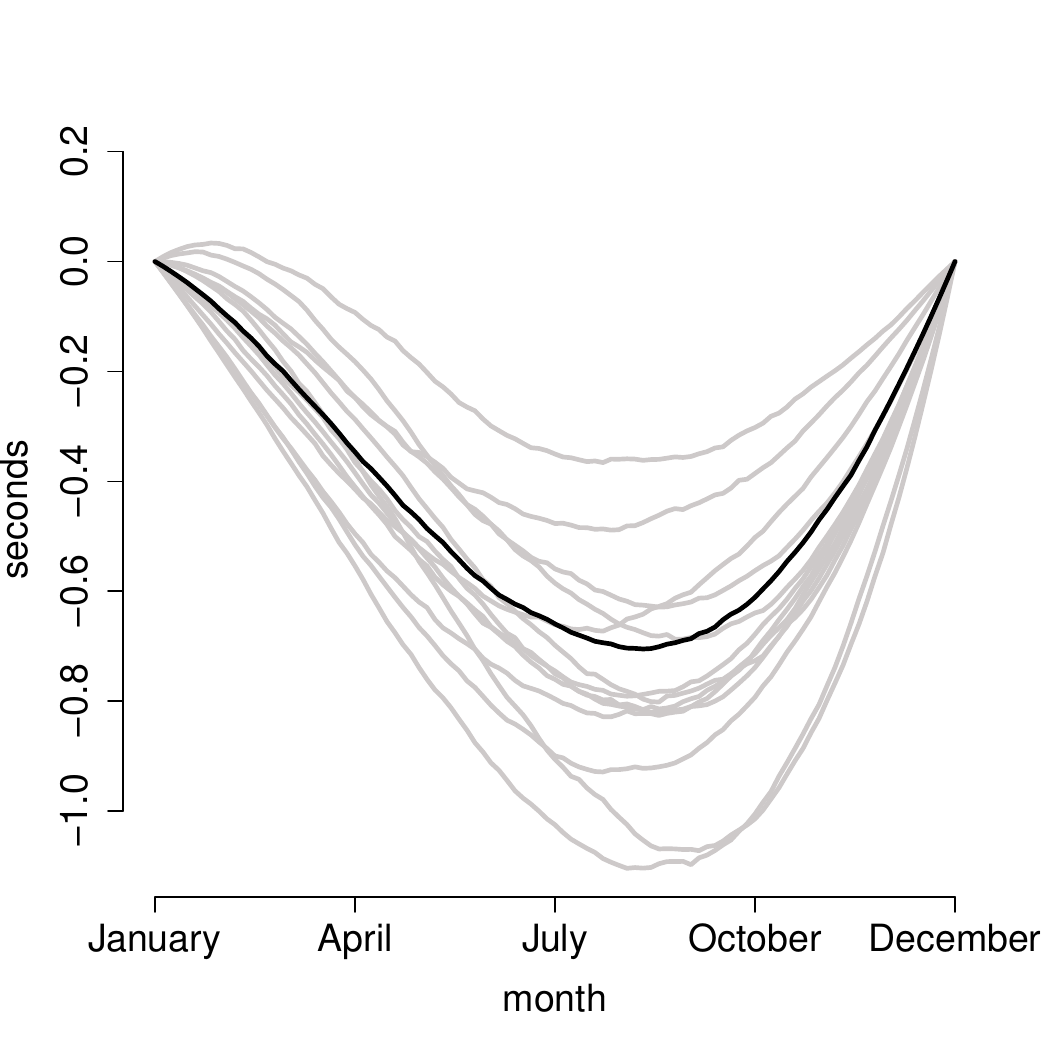}&
\includegraphics[scale=0.24]{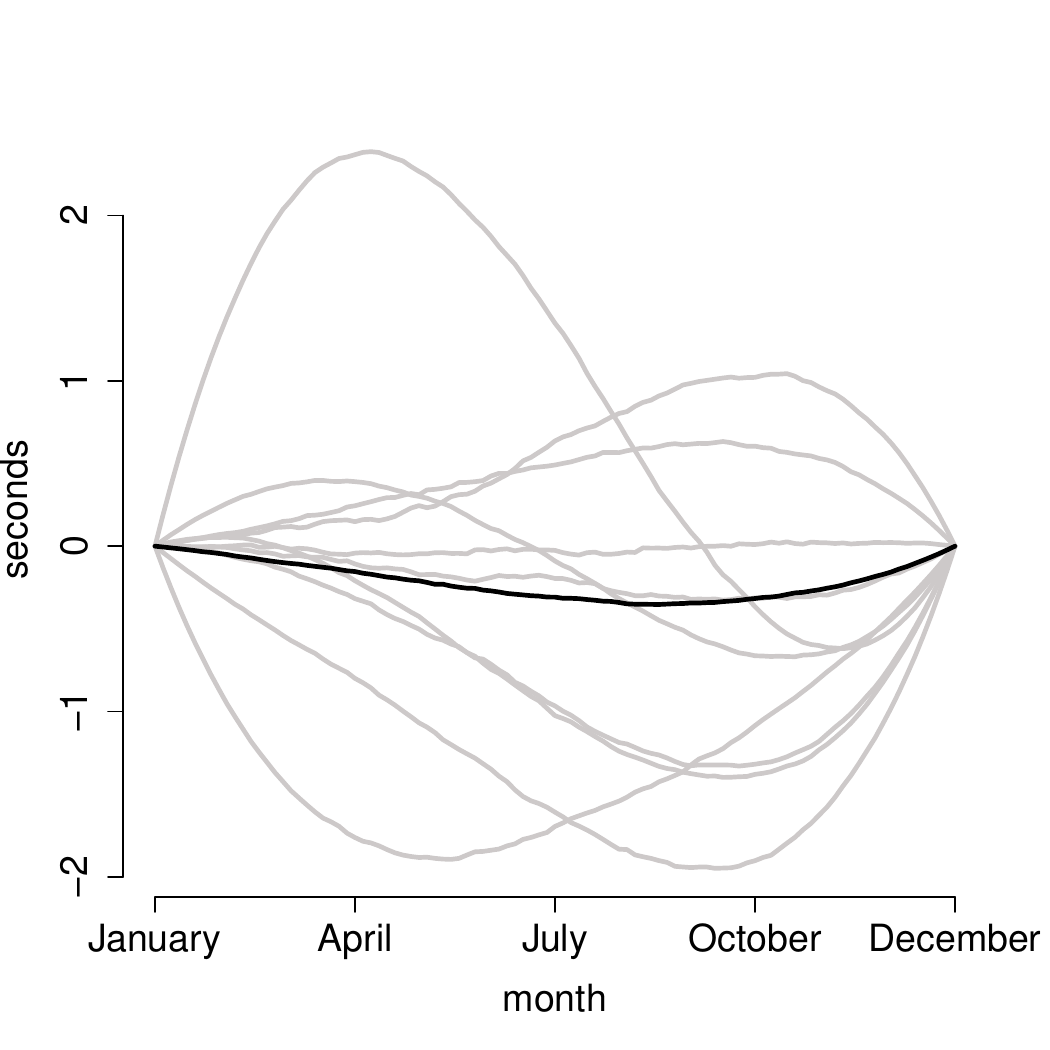}
\end{tabular}
\end{center}
\caption{\label{female2} 
Female 200 metres freestyle swimmers:
Individual performance trajectories (top row) and estimated individual trend performance trajectories (middle row). 
The trajectories are shown as posterior median (black line) and 95\% credible interval (grey shading)
with the observed performances (dots).  
 Bottom row: posterior median athlete-level (black line) and  within-season performance trajectories for each career season (grey lines).}
\end{figure}

\end{document}